\documentclass{mythesis}

\usepackage[T1]{fontenc}
\usepackage{times}
\usepackage{inconsolata}
\usepackage{pifont}

\usepackage{tikz}
\usetikzlibrary{arrows}
\usetikzlibrary{patterns}
\usetikzlibrary{calc}
\usetikzlibrary{shapes}
\usetikzlibrary{positioning,fit}

\usepackage{circuitikz}

\newcommand{\mrtwo}[1]{\multirow{2}{*}{#1}}
\newcommand{\mrthr}[1]{\multirow{3}{*}{#1}}

\newcommand{\vdd}{V$_{DD}$\xspace}
\newcommand{\halfvdd}{$\frac{1}{2}$V$_{DD}$\xspace}
\newcommand{\halfvddpd}{$\frac{1}{2}$V$_{DD}+\delta$\xspace}

\newcommand{\src}{\texttt{src}\xspace}
\newcommand{\dst}{\texttt{dst}\xspace}
\newcommand{\tmp}{\texttt{tmp}\xspace}

\newcommand{\cmdact}{\texttt{ACTIVATE}\xspace}
\newcommand{\cmdpre}{\texttt{PRECHARGE}\xspace}
\newcommand{\cmdwr}{\texttt{WRITE}\xspace}
\newcommand{\cmdrd}{\texttt{READ}\xspace}
\newcommand{\cmdtr}{\texttt{TRANSFER}\xspace}
\newcommand{\cmdacts}{\texttt{ACTIVATE}s\xspace}

\newcommand{\mcpy}{\texttt{memcopy}\xspace}
\newcommand{\minit}{\texttt{meminit}\xspace}

\newcommand{\forkbench}{\texttt{forkbench}\xspace}
\newcommand{\ubsize}{$\mathcal{S}$\xspace}

\newcommand{\gsdram}{{\sffamily{GS-DRAM}}\xspace}
\newcommand{\gsdramp}[3]{\mbox{{\sffamily GS-DRAM}$_{#1,#2,#3}$}\xspace}
\newcommand{\gsdrampf}[4]{\mbox{{\sffamily GS-DRAM}$_{#1,#2,#3,#4}$}\xspace}

\newcommand{\rowstore}{\mbox{\sffamily Row Store}\xspace}
\newcommand{\colstore}{\mbox{\sffamily Column Store}\xspace}

\newcommand{\ttvdd}{$\frac{2}{3}$V$_{DD}$\xspace}
\newcommand{\otvdd}{$\frac{1}{3}$V$_{DD}$\xspace}

\newcommand{\bbar}{$\overline{\textrm{bitline}}$\xspace}

\newcommand{\band}{\textrm{\texttt{and}}\xspace}
\newcommand{\bor}{\textrm{\texttt{or}}\xspace}
\newcommand{\bnot}{\textrm{\texttt{not}}\xspace}
\newcommand{\bxor}{\textrm{\texttt{xor}}\xspace}
\newcommand{\bnand}{\textrm{\texttt{nand}}\xspace}
\newcommand{\bnor}{\textrm{\texttt{nor}}\xspace}
\newcommand{\bxnor}{\textrm{\texttt{xnor}}\xspace}

\newcommand{\dwordline}{\emph{d-wordline}\xspace}
\newcommand{\nwordline}{\emph{n-wordline}\xspace}

\newcommand{\baddr}[1]{\texttt{B#1}\xspace}
\newcommand{\taddr}[1]{\texttt{T#1}\xspace}
\newcommand{\daddr}[1]{\texttt{D#1}\xspace}

\newcommand{\dccdz}{\texttt{DCC0}\xspace}
\newcommand{\dccnz}{$\overline{\textrm{\texttt{DCC0}}}$\xspace}

\newcommand{\dccdo}{\texttt{DCC1}\xspace}
\newcommand{\dccno}{$\overline{\textrm{\texttt{DCC1}}}$\xspace}

\newcommand{\dccd}{\texttt{DCC}\xspace}

\newcommand{\czero}{\texttt{C0}\xspace}
\newcommand{\cone}{\texttt{C1}\xspace}

\newcommand{\comment}[1]{\textcolor{black!80}{; #1}\xspace}
\newcommand{\tcomment}[1]{\textcolor{black!80}{\texttt{; #1}}\xspace}

\newcommand{\aap}{\texttt{AAP}\xspace}
\newcommand{\callaap}[2]{\aap \texttt{(#1, #2)}}

\newcommand{\ap}{\texttt{AP}\xspace}

\newcommand{\tras}{t$_{\textrm{\footnotesize{RAS}}}$\xspace}
\newcommand{\trp}{t$_{\textrm{\footnotesize{RP}}}$\xspace}
\newcommand{\twl}{t$_{\textrm{\footnotesize{WL}}}$\xspace}

\newcommand{\buddyao}{Buddy-AND/OR\xspace}
\newcommand{\buddynot}{Buddy-NOT\xspace}

\newcommand{\dbi}{DBI\xspace}

\begin{document}

\tikzset{no padding/.style={inner sep=0pt, outer sep=0pt}}
\tikzset{rounded/.style={rounded corners=2pt}}
\tikzset{rounded1/.style={rounded corners=1pt}}

\tikzset{page/.style={draw,minimum width=1.75cm,minimum height=2.25cm, rounded}}

\tikzset{value/.style={draw,minimum width=2mm,minimum height=2mm, rounded1, fill=white}}

\newcommand{\capacitor}[4]{
  \node (#1) at (#2) [anchor=east,no padding,xshift=1pt] {
    \begin{tikzpicture}[scale=#4]
      \draw [fill=black!30,black!30] (3mm,-3mm) rectangle ++(3mm,#3*6mm);
      \draw [thick] (3mm,0)
      ++(0,3mm) -- ++(0,-6mm)
      -- ++(3mm,0)
      -- ++(0,6mm) ++(0,-3mm)
      -- ++(2mm,0);
      \draw [->,thick] (3mm,0) -| (0,-3mm);
    \end{tikzpicture}
  };
}

\newcommand{\transtop}[4]{
  \node (#1) at (#2) [anchor=center,no padding] {};

  \coordinate (#1-trans-right) at ([xshift=-#4*3mm]#1);
  \coordinate (#1-trans-left) at ([xshift=-#4*6mm]#1-trans-right);
  \coordinate (#1-trans-gate) at ([xshift=-#4*3mm,yshift=#4*7mm]#1-trans-right);
  \coordinate (#1-trans-connect) at ([xshift=-#4*2mm]#1-trans-left);

  \draw [thick,#3] (#1) -- (#1-trans-right) -- ++(0,#4*3mm) -|
  (#1-trans-left) -- (#1-trans-connect);

  \draw [thick,#3] ([yshift=#4*4mm]#1-trans-right) --
  ([yshift=#4*4mm]#1-trans-left);
  \draw [thick,#3] (#1-trans-gate) -- ++(0,-#4*3mm);
}

\newcommand{\transbottom}[3]{
  \coordinate (#1-right) at (#2);
  \coordinate (#1-mid-right) at ([xshift=-#3*1mm]#1-right);
  \coordinate (#1-mid-left) at ([xshift=-#3*2mm]#1-mid-right);
  \coordinate (#1-contact) at ([xshift=-#3*1mm]#1-mid-left);

  \draw [thick] (#1-right) -- (#1-mid-right)
    -- ++(0,#3*-1mm) -| (#1-mid-left) -- (#1-contact);
  \draw [thick] ([yshift=-#3*1.25mm]#1-mid-left) --
    ([yshift=-#3*1.25mm]#1-mid-right);

  \coordinate (#1-gate) at
    ([yshift=#3*-2mm]$(#1-mid-right)!0.5!(#1-mid-left)$);
  \draw [thick] (#1-gate) -- ++(0,#3*0.75mm);
}

\newcommand{\invertortop}[3]{
  \node (#1) at (#2) {};

  \coordinate (#1-top) at ([yshift=#3*2mm]#1);
  \coordinate (#1-bottom) at ([yshift=-#3*2mm]#1);
  \coordinate (#1-left) at ([xshift=-#3*3mm]#1-bottom);
  \coordinate (#1-right) at ([xshift=#3*3mm]#1-bottom);

  \coordinate (#1-out) at ([yshift=#3*3mm]#1-top);
  \coordinate (#1-in) at ([yshift=-#3*2mm]#1-bottom);

  \coordinate (#1-enable) at ($(#1-top)!0.5!(#1-left)$);

  \draw [thick] (#1-top) -- (#1-left) -- (#1-right) -- (#1-top);
  \draw ([yshift=#3*1.5pt]#1-top) circle (#3*1.5pt);

  \draw [thick] (#1-in) -- (#1-bottom);
  \draw [thick] (#1-out) -- ([yshift=#3*3pt]#1-top);
}

\newcommand{\invertorbottom}[3]{
  \node (#1) at (#2) {};

  \coordinate (#1-top) at ([yshift=-#3*2mm]#1);
  \coordinate (#1-bottom) at ([yshift=#3*2mm]#1);
  \coordinate (#1-left) at ([xshift=-#3*3mm]#1-bottom);
  \coordinate (#1-right) at ([xshift=#3*3mm]#1-bottom);

  \coordinate (#1-out) at ([yshift=-#3*3mm]#1-top);
  \coordinate (#1-in) at ([yshift=#3*2mm]#1-bottom);

  \coordinate (#1-enable) at ($(#1-top)!0.5!(#1-left)$);

  \draw [thick] (#1-top) -- (#1-left) -- (#1-right) -- (#1-top);
  \draw ([yshift=-#3*1.5pt]#1-top) circle (#3*1.5pt);

  \draw [thick] (#1-in) -- (#1-bottom);
  \draw [thick] (#1-out) -- ([yshift=-#3*3pt]#1-top);
}

\newcommand{\senseampfull}[4]{
  \node (#1) at (#2) [no padding] {};

  \invertortop{#1-inv1}{[xshift=-#4*7mm]#1}{#4};
  \invertorbottom{#1-inv2}{[xshift=#4*7mm]#1}{#4};

  \coordinate (#1-top) at ([yshift=#4*6mm]#1-inv2-in);
  \coordinate (#1-bottom) at ([yshift=-#4*6mm]#1-inv1-in);
  \coordinate (#1-enable) at ([xshift=-#4*5mm]#1-inv1-enable);
  \coordinate (#1-center) at (#1.center|-#1-inv2-enable);

  \draw [thick] (#1-inv2-in) -- (#1-top) ++(0,-#4*3mm) -| (#1-inv1-out);
  \draw [thick] (#1-inv1-in) -- (#1-bottom) ++(0,#4*3mm) -| (#1-inv2-out);

  \draw [thick,#3] (#1-enable) -- (#1-inv1-enable) ++(-#4*2mm,0)
    -- ++(0,#4*5mm) -| (#1-center) -- (#1-inv2-enable);

  \draw [fill] (#1-inv1-enable) ++(-#4*2mm,0) circle(#4*1pt);
}

\newcommand{\dramcell}[5]{
  \node (#1) at (#2) [anchor=center,no padding] {};

  \coordinate (#1-trans-right) at ([xshift=-#5*3mm]#1);
  \coordinate (#1-trans-left) at ([xshift=-#5*6mm]#1-trans-right);
  \coordinate (#1-trans-gate) at ([xshift=-#5*3mm,yshift=#5*7mm]#1-trans-right);
  \coordinate (#1-trans-connect) at ([xshift=-#5*2mm]#1-trans-left);

  \draw [thick,#4] (#1) -- (#1-trans-right) -- ++(0,#5*3mm) -|
  (#1-trans-left) -- (#1-trans-connect);

  \draw [thick,#4] ([yshift=#5*4mm]#1-trans-right) --
  ([yshift=#5*4mm]#1-trans-left);
  \draw [thick,#4] (#1-trans-gate) -- ++(0,-#5*3mm);
  
  \coordinate (#1-cap-left) at (#1-trans-connect);

  \draw (#1-cap-left) -- (#1-trans-connect);

  \capacitor{#1-cap}{#1-cap-left}{#3}{#5};
}

\newcommand{\dramcellwide}[5]{
  \node (#1) at (#2) [anchor=center,no padding] {};

  \coordinate (#1-trans-right) at ([xshift=-#5*3mm]#1);
  \coordinate (#1-trans-left) at ([xshift=-#5*6mm]#1-trans-right);
  \coordinate (#1-trans-gate) at ([xshift=-#5*3mm,yshift=#5*7mm]#1-trans-right);
  \coordinate (#1-trans-connect) at ([xshift=-#5*2mm]#1-trans-left);

  \draw [thick,#4] (#1) -- (#1-trans-right) -- ++(0,#5*3mm) -|
  (#1-trans-left) -- (#1-trans-connect);

  \draw [thick,#4] ([yshift=#5*4mm]#1-trans-right) --
  ([yshift=#5*4mm]#1-trans-left);
  \draw [thick,#4] (#1-trans-gate) -- ++(0,-#5*3mm);
  
  \coordinate (#1-cap-left) at ([xshift=-3mm]#1-trans-connect);

  \draw (#1-cap-left) -- (#1-trans-connect);

  \capacitor{#1-cap}{#1-cap-left}{#3}{#5};
}

\newcommand{\cellsa}[7]{
  \senseampfull{#1-sa}{#2}{#3}{#6};

  \coordinate (#1-bit-top) at ([yshift=1.25cm]#1-sa-top);
  \draw (#1-sa-top) -- (#1-bit-top);

  \coordinate (#1-cell-contact) at ([yshift=0.5cm]#1-sa-top);
  \dramcell{#1-cell}{#1-cell-contact}{#4}{#5}{#7};

  \coordinate (#1-word-left) at ([xshift=-1.2cm]#1-cell-trans-gate);
  \draw [thick,#5] (#1-word-left) -- (#1-cell-trans-gate) -- ++(#7*4mm,0);
  
  \draw [fill] (#1-cell-contact) circle (#6*1pt);
  \draw [fill] (#1-cell-trans-gate) circle (#6*1pt);
}

\newcommand{\labelcellsa}[6]{
  \node at (#1-word-left) [anchor=east] {#2};
  \node at (#1-sa-enable) [anchor=east] {#3};
  \node at (#1-bit-top) [anchor=west] {#4};
  \node at (#1-sa-bottom) [anchor=north] {#5};
  \draw [fill=black,text=white] (#1.south east) circle (7pt) node {#6};
}

\newcommand{\twocellsa}[9]{
  \senseampfull{#1-sa}{#2}{#3}{#8};

  \coordinate (#1-bit-top) at ([yshift=2cm]#1-sa-top);
  \draw (#1-sa-top) -- (#1-bit-top);

  \coordinate (#1-cell1-contact) at ([yshift=0.4cm]#1-sa-top);
  \dramcell{#1-cell1}{#1-cell1-contact}{#4}{#5}{#9};

  \coordinate (#1-cell2-contact) at ([yshift=1.8cm]#1-sa-top);
  \dramcell{#1-cell2}{#1-cell2-contact}{#6}{#7}{#9};

  \coordinate (#1-word1-left) at ([xshift=-1.2cm]#1-cell1-trans-gate);
  \draw [thick,#5] (#1-word1-left) -- (#1-cell1-trans-gate) -- ++(#9*4mm,0);
  
  \coordinate (#1-word2-left) at ([xshift=-1.2cm]#1-cell2-trans-gate);
  \draw [thick,#7] (#1-word2-left) -- (#1-cell2-trans-gate) -- ++(#9*4mm,0);
  
  \draw [fill] (#1-cell1-contact) circle (#8*1pt);
  \draw [fill] (#1-cell1-trans-gate) circle (#8*1pt);
  \draw [fill] (#1-cell2-contact) circle (#8*1pt);
  \draw [fill] (#1-cell2-trans-gate) circle (#8*1pt);
}

\newcommand{\labeltwocellsa}[7]{
  \node at (#1-word1-left) [anchor=east] {#2};
  \node at (#1-word2-left) [anchor=east] {#3};
  \node at (#1-sa-enable) [anchor=east] {#4};
  \node at (#1-bit-top) [anchor=west] {#5};
  \node at (#1-sa-bottom) [anchor=north] {#6};
  \draw [fill=black,text=white] (#1.south east) circle (7pt) node {#7};
}

\newcommand{\threecellsa}[7]{
  \senseampfull{#1-sa}{#2}{#3}{0.7};

  \coordinate (#1-bit-top) at ([yshift=2cm]#1-sa-top);
  \draw (#1-sa-top) -- (#1-bit-top);

  \coordinate (#1-cell1-contact) at ([yshift=0.3cm]#1-sa-top);
  \dramcell{#1-cell1}{#1-cell1-contact}{#4}{#7}{0.6};
  \coordinate (#1-cell2-contact) at ([yshift=1cm]#1-sa-top);
  \dramcell{#1-cell2}{#1-cell2-contact}{#5}{#7}{0.6};
  \coordinate (#1-cell3-contact) at ([yshift=1.7cm]#1-sa-top);
  \dramcell{#1-cell3}{#1-cell3-contact}{#6}{#7}{0.6};

  \foreach \i in {1,2,3} {
  \coordinate (#1-word\i-left) at ([xshift=-1cm]#1-cell\i-trans-gate);
  \draw [thick,#7] (#1-word\i-left) -- (#1-cell\i-trans-gate) --
  ++(0.6*4mm,0);
  \draw [fill] (#1-cell\i-contact) circle (0.7*1pt);
  \draw [fill] (#1-cell\i-trans-gate) circle (0.7*1pt);
  }
  
}

\newcommand{\labelthreecellsa}[6]{
  \foreach \i in {1,2,3} {
    \node at (#1-word\i-left) [anchor=east,scale=0.8] {#2};
  }
  \node at (#1-sa-enable) [anchor=east,scale=0.8] {#3};
  \node at (#1-bit-top) [anchor=west,scale=0.8] {#4};
  \node at (#1-sa-bottom) [anchor=north,scale=0.8] {#5};
  \draw [fill=black,text=white] (#1.south east) circle (5pt) node {\small{#6}};
}

\newcommand{\dccnot}[7]{
  \senseampfull{#1-sa}{#2}{#3}{#7}

  \coordinate (#1-bit-top) at ([yshift=#7*23mm]#1-sa-top);
  \draw (#1-sa-top) -- (#1-bit-top);

  \coordinate (#1-cell-contact) at ([yshift=#7*11mm]#1-sa-top);
  \dramcellwide{#1-cell}{#1-cell-contact}{#4}{#5}{0.7*#7};
  
  \coordinate (#1-nt-loc) at ([yshift=-#7*11mm,xshift=#7*2mm]#1-cell-cap-left);
  \transtop{#1-nt}{#1-nt-loc}{#6}{0.7*#7};
  \draw (#1-nt-loc) -- (#1-nt-loc|-#1-cell-cap-left);

  \draw (#1-nt-trans-connect) -- ++(-3*#7 mm,0) |- (#1-sa-south);
  \draw [fill] (#1-sa-south) circle(#7 pt);

  \coordinate (#1-word-left) at ([xshift=-#7*20mm]#1-cell-trans-gate);
  \draw [thick,#5] (#1-word-left) -- (#1-cell-trans-gate) -- ++(#7*3mm,0);

  \coordinate (#1-nword-left) at (#1-word-left|-#1-nt-trans-gate);
  \draw [thick,#6] (#1-nword-left) -- (#1-nt-trans-gate) -- ++(#7*3mm,0);

  \coordinate (#1-enable) at (#1-word-left|-#1-sa-enable);
  \draw [thick,#3] (#1-enable) -- (#1-sa-enable);
}

\newcommand{\bitwisenot}[9]{
  \dccnot{#1}{#2}{#3}{#4}{#5}{#6}{#9}

  \coordinate (#1-rcell-contact) at ([yshift=#9*20mm]#1-sa-top);
  \dramcellwide{#1-rcell}{#1-rcell-contact}{#7}{#8}{#9*0.7};

  \coordinate (#1-rword-left) at (#1-word-left|-#1-rcell-trans-gate);
  \draw [thick,#8] (#1-rword-left) -- (#1-rcell-trans-gate) -- ++(#9*3mm,0);
}

\newcommand{\labelbitwisenot}[8]{
  \node at (#1-rword-left) [anchor=east,scale=0.8] {#2};
  \node at (#1-word-left) [anchor=east,scale=0.8] {#3};
  \node at (#1-nword-left) [anchor=east,scale=0.8] {#4};
  \node at (#1-enable) [anchor=east,scale=0.8] {#5};
  \node at (#1-bit-top) [anchor=west,scale=0.8] {#6};
  \node at (#1-sa-bottom) [anchor=north,scale=0.8] {#7};
  \draw [fill=black,text=white] (#1.south east) circle (5pt) node {\small{#8}};
}

\thispagestyle{empty}

\begin{center}
  \begin{Large}
    {\bf Simple DRAM and Virtual Memory Abstractions\\
      to Enable Highly Efficient Memory Subsystems}\\
  \end{Large}
  \vspace{10mm}
  \begin{Large}
    Vivek Seshadri\vspace{5mm}\\
  \end{Large}
    CMU-CS-16-106\\
    April 2016\\
  \vspace{8mm}
  School of Computer Science\\
  Computer Science Department\\
  Carnegie Mellon University\\
  Pittsburgh, PA 15213\\
  \vspace{8mm}
  \textbf{Thesis Committee}\\
  Todd C. Mowry (Co-chair)\\
  Onur Mutlu (Co-chair)\\
  David Andersen\\
  Phillip B. Gibbons\\
  Rajeev Balasubramonian, University of Utah\\
  \vspace{8mm}
  \emph{Submitted in partial fulfillment of the requirements\\
    for the degree of Doctor of Philosophy.}\\
  \vspace{5mm}
  Copyright \copyright\ 2016 Vivek Seshadri\\
\end{center}

\begin{spacing}{0.85}
\begin{footnotesize}
\noindent This research was sponsored by the National Science
Foundation under grant numbers CNS-0720790, EEC-0951919,
CCF-1116898, CCF-1212962, CNS-1320531, CNS-1423172, CCF-0953246,
CCF-1147397, and CNS-1409723, the Defense Advanced Research
Projects Agency under grant number FA8721-5-C-0003, Princeton
University under grant number 00001751, Semiconductor Research
Corporation under grant number 2012HJ2320, Intel ISTC-CC, Intel
URO Memory Hierarchy Program, Gigascale Systems Research Center,
AMD, Google, IBM, Intel, Nvidia, Oracle, Samsung, and
VMware.\vspace{2mm}
\end{footnotesize}
\end{spacing}
\begin{spacing}{0.85}
\begin{footnotesize}
\noindent The views and conclusions contained in this document are those of
the author and should not be interpreted as representing the
official policies, either expressed or implied, of any sponsoring
institution, the U.S. government or any other entity.
\end{footnotesize}
\end{spacing}

\newpage
\thispagestyle{empty}
\vspace*{\fill}
\noindent \textbf{Keywords:} Efficiency, Memory Subsystem, Virtual
Memory, DRAM, Data Movement, Processing in Memory, Fine-grained
Memory Management

\frontmatter
\chapter*{Abstract}
\addcontentsline{toc}{chapter}{Abstract}

In most modern systems, the memory subsystem is managed and accessed at
multiple different granularities at various resources. The software stack
typically accesses data at a \emph{word} granularity (typically 4 or 8
bytes). The on-chip caches store data at a \emph{cache line} granularity
(typically 64 bytes). The commodity off-chip memory interface is optimized
to fetch data from main memory at a cache line granularity. The main memory
capacity itself is managed at a \emph{page} granularity using virtual
memory (typically 4KB pages with support for larger super pages). The
off-chip commodity DRAM architecture internally operates at a \emph{row}
granularity (typically 8KB). In this thesis, we observe that this
\emph{curse of multiple granularities} results in significant inefficiency
in the memory subsystem.

We identify three specific problems. First, page-granularity virtual memory
unnecessarily triggers large memory operations. For instance, with the
widely-used copy-on-write technique, even a single byte update to a virtual
page results in a full 4KB copy operation. Second, with existing off-chip
memory interfaces, to perform any operation, the processor must first read
the source data into the on-chip caches and write the result back to main
memory. For bulk data operations, this model results in a large amount of
data transfer back and forth on the main memory channel. Existing systems
are particularly inefficient for bulk operations that do not require any
computation (e.g., data copy or initialization). Third, for operations that
do not exhibit good spatial locality, e.g., non-unit strided access
patterns, existing cache-line-optimized memory subsystems unnecessarily
fetch values that are not required by the application over the memory
channel and store them in the on-chip cache. All these problems result in
high latency, and high (and often unnecessary) memory bandwidth and energy
consumption.

To address these problems, we present a series of techniques in this
thesis. First, to address the inefficiency of existing page-granularity
virtual memory systems, we propose a new framework called \emph{page
  overlays}. At a high level, our framework augments the existing virtual
memory framework with the ability to track a new version of a subset of
cache lines within each virtual page. We show that this simple extension is
very powerful by demonstrating its benefits on a number of different
applications.

Second, we show that the analog operation of DRAM can perform more complex
operations than just store data. When combined with the row granularity
operation of commodity DRAM, we can perform these complex operations
efficiently in bulk. Specifically, we propose \emph{RowClone}, a mechanism
to perform bulk data copy and initialization operations completely inside
DRAM, and \emph{Buddy RAM}, a mechanism to perform bulk bitwise logical
operations using DRAM. Both these techniques achieve an order-of-magnitude
improvement in performance and energy-efficiency of the respective
operations.

Third, to improve the performance of non-unit strided access patterns, we
propose \emph{Gather-Scatter DRAM} (\gsdram), a technique that exploits the
module organization of commodity DRAM to effectively gather or scatter
values with any power-of-2 strided access pattern. For these access
patterns, \gsdram achieves near-ideal bandwidth and cache utilization,
without increasing the latency of fetching data from memory.

Finally, to improve the performance of the protocol to maintain
the coherence of dirty cache blocks, we propose the
\emph{Dirty-Block Index} (DBI), a new way of tracking dirty blocks
in the on-chip caches. In addition to improving the efficiency of
bulk data coherence, DBI has several applications, including
high-performance memory scheduling, efficient cache lookup
bypassing, and enabling heterogeneous ECC for on-chip caches.

\chapter*{Acknowledgments}
\addcontentsline{toc}{chapter}{Acknowledgments}
\vspace{-5mm}
\newcolumntype{L}[1]{>{\let\newline\\\arraybackslash\hspace{0pt}}m{#1}}
\newcolumntype{C}[1]{>{\centering\let\newline\\\arraybackslash\hspace{0pt}}m{#1}}
\newcolumntype{R}[1]{>{\raggedleft\let\newline\\\arraybackslash\hspace{0pt}\em}m{#1}}

\begin{figure}[b!]
  \hrule\vspace{1mm}
  \begin{footnotesize}
    The last chapter of this dissertation acknowledges the
    collaborations with my fellow graduate students. These
    collaborations have been critical in keeping my morale high
    throughout the course of my Ph.D.
  \end{footnotesize}
\end{figure}

Many people have directly or indirectly played a significant role
in the success of my Ph.D. Words alone may not be sufficient to
express my gratitude to these people. Yet, I would like to thank
\vspace{-2mm}
\begin{center}
  \setlength{\extrarowheight}{6pt}
  \begin{tabular}{R{1.75in}L{4in}}
    Todd \& Onur, & for trusting me and giving me freedom, despite some
    tough initial years, for constantly encouraging me to work
    hard, and for keeping me focused\\
    Mike \& Phil, & for being more than just collaborators\\
    Dave \& Rajeev, & for making the ending as less painful as
    possible\\
    Deb, & for ensuring that I did not have to worry about anything
    but research\\
    Donghyuk \& Yoongu, & for teaching me everything I know about
    DRAM\\
    Lavanya, & for being more than just a lab mate\\
    Michelle, Tunji, \& Evangelos, & for sharing the secrets of
    the trade\\
    Gena, Chris, \& Justin, & for giving me a different
    perspective in research and life\\
    other members of SAFARI, & for all the valuable and enjoyable
    discussions\\
    people @ PDL \& CALCM, & for all the feedback and comments\\
  \end{tabular}
\end{center}
    
\begin{center}
  \setlength{\extrarowheight}{6pt}
  \begin{tabular}{R{1.75in}L{4in}}
    Frank, Rick, Vyas, \& other members of AB league, & for
    making squash an integral part of my life\\
    Stan \& Patty, & for their warming smile every time I visit UC\\
    Gopu, &  for being my best roommate for the better part of my
    Ph.D.\\
    Shane, & for being a part of an important decade in my life\\
    G, RawT, KL, Coolie, \& BAP, & for making my life @ Pittsburgh enjoyable\\
    Vyas, & for teaching me (mostly necessary) life skills\\
    PJ, & for giving me hope\\
    Veda, Tomy, Murali, Hetu, \& Captain, & for all the gyan\\
    Ashwati, Usha, \& Shapadi, & for all the chai, and random life
    discussions\\
    James \& DSha, & for all the discussions, rants, and squash\\
    Ranjan, SKB, Sudha, \& Joy, & for the complete TeleTaaz
    experience\\
    Sharads, Sumon, Kaushik, \& KJ et al.,  & for food and company\\
    IGSA folks, & for the much-needed distraction\\
    Jyo, Pave, Sail, Vimal, Karthi, and Jayavel, & for always being the friends
    in need\\
    MahimaSri, NiMeg, Wassim, Yogi, \& Pro Club, & for making Redmond my
    second home in the US\\
    PK \& Kaavy, & for my go-to whatsapp group\\
    DHari, & for putting up with me\\
    Vikram, & for making my path in life smooth\\
    Appa \& Amma, & for freedom and support.
  \end{tabular}
\end{center}

\tableofcontents
\listoffigures
\listoftables

\mainmatter
\chapter{Introduction}
\label{chap:introduction}

In recent years, energy-efficiency has become a major design factor in
systems. This trend is fueled by the ever-growing use of battery-powered
hand-held devices on the one end, and large-scale data centers on the other
end. To ensure high energy-efficiency, all the resources in a system (e.g.,
processor, caches, memory) must be used efficiently.

To simplify system design, each resource typically exposes an
interface/abstraction to other resources in the system. Such
abstractions allow system designers to adopt newer technologies to
implement a resource \emph{without} modifying other
resources. However, a \emph{poor abstraction} to a resource that
does not expose all its capabilities can significantly limit the
overall efficiency of the system.

\section{Focus of This Dissertation: The Memory Subsystem}

\begin{figure}
  \centering
  \includegraphics{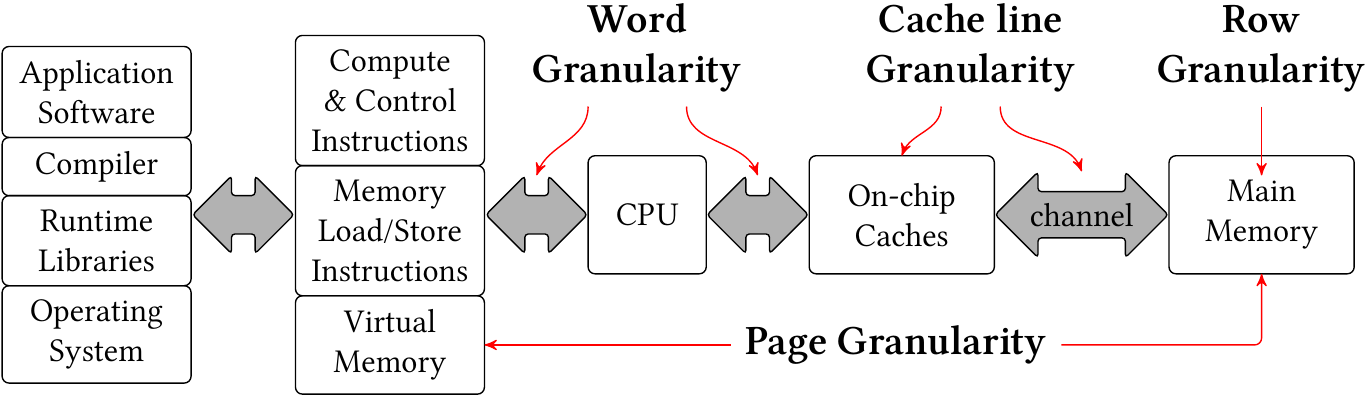}
  \caption[Interaction with memory resources]{Layers of the system
    stack and their interaction with memory resources.}
  \label{fig:system-stack}
\end{figure}

This dissertation focuses on the efficiency of the memory
subsystem. Main memory management in a modern system has two
components: 1)~memory mapping (affects capacity management,
protection, etc.), and 2)~memory access (reads, writes, etc.). We
observe that in existing systems, there is a mismatch in the
granularity at which memory is mapped and accessed at different
resources, resulting in significant
inefficiency. \Cref{fig:system-stack} shows the different layers
of the system stack and their interaction with different memory
resources.

\subsection{Different Granularities of Data Storage and Access}

First, most modern operating systems (OS) ubiquitously use virtual
memory~\cite{virtual-memory} to manage main memory capacity. To
map virtual memory to physical memory, virtual memory systems use
a set of mapping tables called \emph{page tables}. In order to
keep the overhead of the page tables low, most virtual memory
systems typically manage memory at a large granularity (\emph{4KB
  pages} or larger super pages). Second, to access memory, the
instruction set architecture (ISA) exposes a set of load and store
instructions to the software stack. To allow efficient
representation of various data types, such instructions typically
allow software to access memory at a small granularity (e.g.,
\emph{4B or 8B words}). Third, any memory request generated by
load/store instructions go through a hierarchy of \emph{on-chip
  caches} all the way to the \emph{off-chip main memory}. In order
to lower the cost of the cache tag stores and the memory
interface, the on-chip caches and the off-chip memory interface
are typically optimized to store and communicate data at a
granularity wider than a single word (e.g., \emph{64B cache
  lines}). Finally, to reduce cost-per-bit, commodity DRAM
architectures internally operate at a row granularity (typically
\emph{8KB}). It is clear that data are stored and accessed at
different granularities in different memory resources. We identify
three problems that result from this mismatch in granularity.

\subsection{The Curse of Multiple Granularities} 

First, we observe that the page-granularity virtual memory
management can result in unnecessary work. For instance, when
using the copy-on-write technique, even a write to a single byte
can trigger a full page copy operation. Second, existing off-chip
memory interfaces only expose a read-write abstraction to main
memory. As a result, to perform any operation, the processor must
read all the source data from main memory and write back the
results to main memory. For operations that involve a large amount
of data, i.e., bulk data operations, this approach results in a
large number of data transfers on the memory channel. Third, many
access patterns trigger accesses with poor spatial locality (e.g.,
non-unit strided accesses). With existing caches and off-chip
interfaces optimized for cache line granularity, such access
patterns fetch a large amount of data not required by the
application over the memory channel and store them in the on-chip
cache.

All these problems result in high latency, high (and often
unnecessary) memory bandwidth, and inefficient cache
utilization. As a result, they affect the performance of not only
the application performing the operations, but also the
performance of other co-running applications. Moreover, as data
movement on the memory channel consumes high
energy~\cite{bill-dally}, these operations also lower the overall
energy-efficiency of the system. \cref{chap:motivation} motivates
these problems in more detail using case studies.

\section{Related Work}

Several prior works have proposed mechanisms to improve memory
efficiency. In this section, we discuss some closely related prior
approaches. We group prior works based on their high level
approach and describe their shortcomings.

\subsection{New Virtual Memory Frameworks}

Page-granularity virtual memory hinders efficient implementation
of many techniques that require tracking memory at a fine
granularity (e.g., fine-grained memory deduplication, fine-grained
metadata management). Prior works have proposed new frameworks to
implement such techniques (e.g., HiCAMP~\cite{hicamp}, Mondrian
Memory Protection~\cite{mmp}, architectural support for shadow
memory~\cite{shadow-memory,ems,shadow-mem-check,umbra,iwatcher}). Unfortunately,
these mechanisms either significantly change the existing virtual
memory structure, thereby resulting in high cost, or introduce
significant changes solely for a specific functionality, thereby
reducing overall value.

\subsection{Adding Processing Logic Near Memory (DRAM)}

One of the primary sources of memory inefficiency in existing
systems is the data movement. Data has to travel off-chip buses
and multiple levels of caches before reaching the CPU. To avoid
this data movement, many works (e.g., Logic-in-Memory
Computer~\cite{lim-computer}, NON-VON Database
Machine~\cite{non-von-machine}, EXECUBE~\cite{execube},
Terasys~\cite{pim-terasys}, Intelligent RAM~\cite{iram}, Active
Pages~\cite{active-pages},
FlexRAM~\cite{flexram,programming-flexram}, Computational
RAM~\cite{cram}, DIVA~\cite{diva} ) have proposed mechanisms and
models to add processing logic close to memory. The idea is to
integrate memory and CPU on the same chip by designing the CPU
using the memory process technology. While the reduced data
movement allows these approaches to enable low-latency,
high-bandwidth, and low-energy data communication, they suffer
from two key shortcomings.

First, this approach of integrating processor on the same chip as
memory greatly increases the overall cost of the system. Second,
DRAM vendors use a high-density process to minimize
cost-per-bit. Unfortunately, high-density DRAM process is not
suitable for building high-speed logic~\cite{iram}. As a result,
this approach is not suitable for building a general purpose
processor near memory, at least with modern logic and high-density
DRAM technologies.

\subsection{3D-Stacked DRAM Architectures}

Some recent architectures~\cite{3d-stacking,hmc,hbm} use
3D-stacking technology to stack multiple DRAM chips on top of the
processor chip or a separate logic layer. These architectures
offer much higher bandwidth to the logic layer compared to
traditional off-chip interfaces. This enables an opportunity to
offload some computation to the logic layer, thereby improving
performance. In fact, many recent works have proposed mechanisms
to improve and exploit such architectures
(e.g.,~\cite{pim-enabled-insts,pim-graph,top-pim,nda,msa3d,spmm-mul-lim,data-access-opt-pim,tom,hrl,gp-simd,ndp-architecture,pim-analytics,nda-arch,jafar,data-reorg-3d-stack,smla}). Unfortunately,
despite enabling higher bandwidth compared to off-chip memory,
such 3D-stacked architectures are still require data to be
transferred outside the DRAM chip, and hence can be
bandwidth-limited. In addition, thermal constraints constrain the
number of chips that can be stacked, thereby limiting the memory
capacity. As a result, multiple 3D-stacked DRAMs are required to
scale to large workloads.

\subsection{Adding Logic to the Memory Controller}

Many prior works have proposed mechanisms to export certain memory
operations to the memory controller with the goal of improving the
efficiency of the operation (e.g., Copy Engine~\cite{copy-engine}
to perform bulk copy or initialization, Impulse~\cite{impulse} to
perform gather/scatter operations, Enhanced Memory
Controller~\cite{emc} to accelerate dependent cache
misses). Recent memory technologies which stack DRAM chips on top
of a logic layer containing the memory controller~\cite{hmc,hbm}
will likely make this approach attractive. Although these
mechanisms definitely reduce the pressure on the CPU and on-chip
caches, they still have to go through the cache-line-granularity
main memory interface, which is inefficient to perform these
operations.

\subsection{Supporting Fine-grained Memory Accesses in DRAM}

A number of works exploit the module-level organization of DRAM to
enable efficient fine-grained memory accesses (e.g.,
Mini-rank~\cite{mini-rank}, Multi-Core DIMM~\cite{mc-dimm},
Threaded memory modules~\cite{threaded-module}, Scatter/Gather
DIMMs~\cite{sg-dimm}). These works add logic to the DRAM module
that enables the memory controller to access data from individual
chips rather than the entire module. Unfortunately, such
interfaces 1)~are much costlier compared to existing memory
interfaces, 2)~potentially lower the DRAM bandwidth utilization,
and 3)~do not alleviate the inefficiency for bulk data operations.

\subsection{Improving DRAM Latency and Parallelism}

A number of prior works have proposed new DRAM microarchitectures
to lower the latency of DRAM access or enable more parallelism
within DRAM. Approaches employed by these works include
1)~introducing heterogeneity in access latency inside DRAM for a
low cost (e.g., Tiered-Latency DRAM~\cite{tl-dram}, Asymmetric
Banks~\cite{charm}, Dynamic Asymmetric
Subarrays~\cite{da-subarray}, Low-cost Interlinked
Subarrays~\cite{lisa}), 2)~improving parallelism within DRAM
(e.g., Subarray-level Parallelism~\cite{salp}, parallelizing
refreshes~\cite{dsarp}, Dual-Port DRAM~\cite{ddma}), 3)~exploiting
charge characteristics of cells to reduce DRAM latency (e.g.,
Multi-clone-row DRAM~\cite{mcr-dram}, Charge
Cache~\cite{chargecache}), 4)~reducing the granularity of internal
DRAM operation through microarchitectural changes (e.g.,
Half-DRAM~\cite{half-dram}, Sub-row
activation~\cite{rethinking-dram}), 5)~adding SRAM cache to DRAM
chips~\cite{cache-dram}, 6)~exploiting variation in DRAM (e.g.,
Adaptive Latency DRAM~\cite{al-dram}, FLY-DRAM~\cite{fly-dram}),
and 7)~better refresh scheduling and refresh reduction
(e.g.,~\cite{raidr,smart-refresh,elastic-refresh,refresh-nt,eskimo,opt-dram-refresh,dynamic-memory-design,avatar,efficacy-error-techniques}. While
many of these approaches will improve the performance of various
memory operations, they are still far from mitigating the
unnecessary bandwidth consumed by certain memory operations (e.g.,
bulk data copy, non-unit strided access).

\subsection{Reducing Memory Bandwidth Requirements}

Many prior works have proposed techniques to reduce memory
bandwidth consumption of applications. Approaches used by these
works include 1)~data compression
(e.g.,~\cite{bdi,lcp,camp,toggle-compression,rmc,mxt}), 2)~value
prediction (e.g.,~\cite{value-prediction,value-locality}), 3)~load
approximation (e.g.,~\cite{rollback-vp,load-approx}), 4)~adaptive
granularity memory systems (e.g.,~\cite{agms,dgms}), and 5)~better
caching to reduce the number of memory requests
(e.g.,~\cite{eaf,rrip,dip}). Some of these techniques require
significant changes to the hardware (e.g., compression, adaptive
granularity memory systems). Having said that, all these
approaches are orthogonal to the techniques proposed in this
dissertation.

\subsection{Mitigating Contention for Memory Bandwidth}

One of the problems that result from bulk data operations is the
contention for memory bandwidth, which can negatively affect the
performance of applications co-running in the system. A plethora
of prior works have proposed mechanisms to mitigate this
performance degradation using better memory request scheduling
(e.g.,~\cite{stfm,mpa,parbs,tcm,atlas,bliss,bliss-tpds,critical-scheduler,dram-aware-wb,prefetch-dram,dash,sms,somc,clams,medic,firm}). While
these works improve overall system performance and fairness, they
do not fundamentally reduce the bandwidth consumption of the
applications performing the bulk operations.



\section{Thesis Statement and Overview}

Our goal in this thesis is to improve the overall efficiency of
the memory subsystem without significantly modifying existing
abstractions and without degrading the performance/efficiency of
applications that do not use our proposed techniques. Towards this
end, our thesis is that,
\begin{quote}
  \emph{we can exploit the diversity in the granularity at which
    different hardware resources manage memory to mitigate the
    inefficiency that arises from that very diversity. To this
    end, we propose to augment existing processor and main memory
    architectures with some simple, low-cost features that bridge
    the gap resulting from the granularity mismatch.}
\end{quote}

Our proposed techniques are based on two observations. First,
modern processors are capable of tracking data at a cache-line
granularity. Therefore, even though memory capacity is managed at
a larger page granularity, using some simple features, it should
be possible to enable more efficient implementations of
fine-grained memory operations. Second, although off-chip memory
interfaces are optimized to access cache lines, we observe that
the commodity memory architecture has the ability to internally
operate at both a bulk row granularity and at a fine word
granularity.

We exploit these observations to propose a new virtual memory
framework that enables efficient fine-grained memory management,
and a series of techniques to exploit the commodity DRAM
architecture to efficiently perform bulk data operations and
accelerate memory operations with low spatial locality.

\section{Contributions}

This dissertation makes the following contributions.

\begin{enumerate}
\item We propose a new virtual memory framework called \emph{page
  overlays} that allows memory to be managed at a sub-page (cache
  line) granularity. The page overlays framework significantly
  improves the efficiency of several memory management techniques,
  e.g., copy-on-write, and super pages. \Cref{chap:page-overlays}
  describes our framework, its implementation, and applications in
  detail.

\item We observe that DRAM internally operates at a large, row
  granularity. Through simple changes to the DRAM architecture, we
  propose \emph{RowClone}, a mechanism that enables fast and
  efficient bulk copy operations completely within DRAM. We
  exploit RowClone to accelerate copy-on-write and bulk zeroing,
  two important primitives in modern systems. \Cref{chap:rowclone}
  describes RowClone and its applications in detail.

\item We observe that the analog operation of DRAM has the
  potential to efficiently perform bitwise logical operations. We
  propose \emph{Buddy RAM}, a mechanism that exploits this
  potential to enable efficient bulk bitwise operations completely
  with DRAM. We demonstrate the performance benefits of this
  mechanism using 1)~a database bitmap index library, and 2)~an
  efficient implementation of a set data
  structure. \Cref{chap:buddy} describes Buddy RAM in detail.

\item We observe that commodity DRAM architectures heavily
  interleave data of a single cache line across many DRAM devices
  and multiple arrays within each device. We propose
  \emph{Gather-Scatter DRAM} (GS-DRAM), which exploits this fact
  to enable the memory controller to gather or scatter data of
  common access patterns with near ideal efficiency. We propose
  mechanisms that use GS-DRAM to accelerate non-unit strided
  access patterns in many important applications, e.g.,
  databases. \Cref{chap:gsdram} describes GS-DRAM and its
  applications in detail.

\item Our mechanisms to perform operations completely in DRAM
  require appropriate dirty cache lines from the on-chip cache to
  be flushed. We propose the \emph{Dirty-Block Index} (DBI) that
  significantly improves the efficiency of this flushing
  operation. \Cref{chap:dbi} describes DBI and several
  other of its potential applications in detail.
\end{enumerate}

\chapter{The Curse of Multiple Granularities}
\label{chap:motivation}

As mentioned in \Cref{chap:introduction}, different memory
resources are managed and accessed at a different granularity ---
main memory capacity is managed at a page (typically 4KB)
granularity, on-chip caches and off-chip memory interfaces store
and access data at a cache line (typically 64B) granularity, DRAM
internally performs operations at a row granularity (typically
8KB), and the applications (and CPU) access data at a small word
(typically 4B or 8B) granularity. This mismatch results in a
significant inefficiency in the execution of two important classes
of operations: 1)~bulk data operations, and 2)~operations with low
spatial locality. In this chapter, we discuss the sources of this
inefficiency for each of these operations using one example
operation in each class.

\section{Bulk Data Operations}
\label{sec:bulk-data-problem}

A bulk data operation is one that involves a large amount of
data. In existing systems, to perform any operation, the
corresponding data must first be brought to the CPU L1
cache. Unfortunately, this model results in high inefficiency for
a bulk data operation, especially if the operation does not
involve any computation on the part of the processor (e.g., data
movement). To understand the sources of inefficiency in existing
systems, let us consider the widely-used copy-on-write~\cite{fork}
technique.

\subsection{The Copy-on-Write Technique}

\begin{figure}
  \centering
  \includegraphics{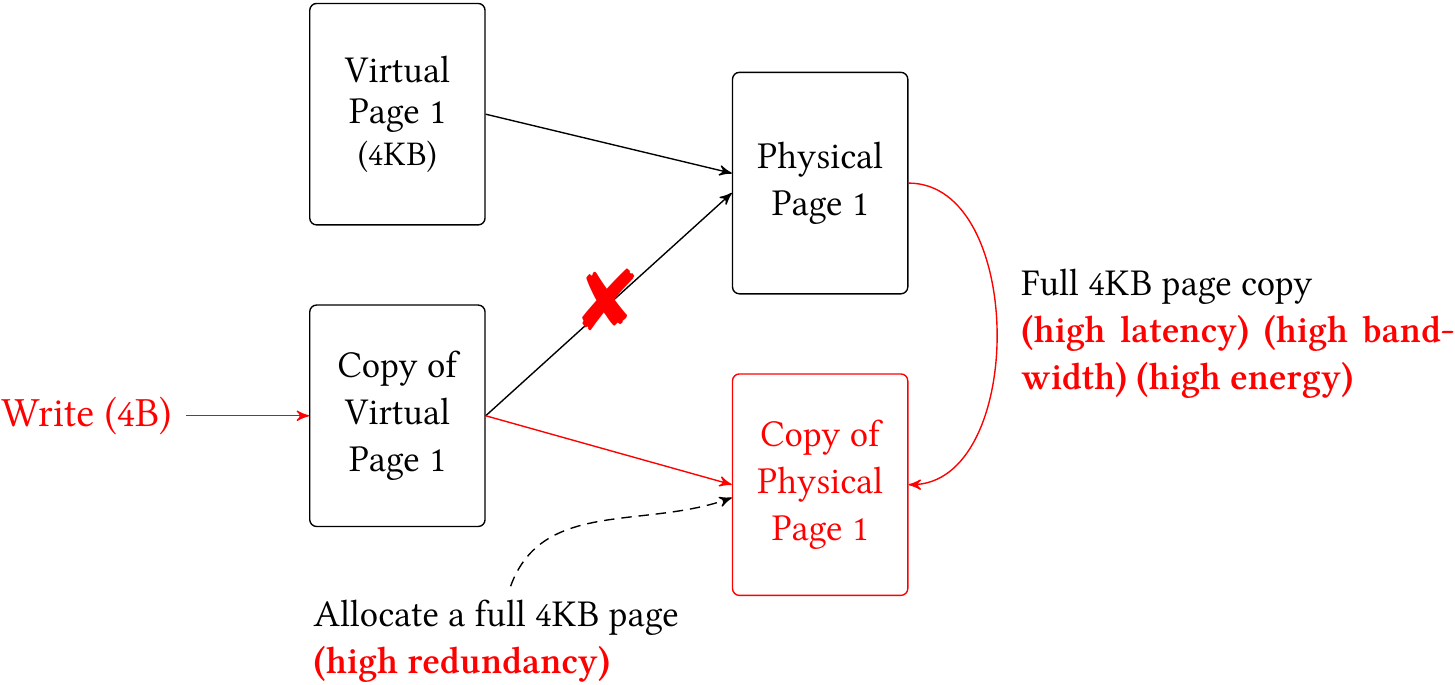}
  \caption[Copy-on-write: Mechanism and shortcomings]{The
    Copy-on-Write technique and shortcomings of existing systems}
  \label{fig:cow}
\end{figure}

\Cref{fig:cow} shows how the copy-on-write technique works.  When
the system wants to copy the data from the virtual page V1 to the
virtual page V2, it simply maps the page V2 to the same physical
page (P1) to which V1 is mapped to. Based on the semantics of
virtual memory, any read access to either virtual page is directed
to the same page, ensuring correct execution. In fact, if neither
of the two virtual pages are modified after the \emph{remap}, the
system would have avoided an unnecessary copy operation. However,
if one of the virtual pages, say V2, does receive a \emph{write},
the system must perform three steps. First, the operating system
must \emph{identify a new physical page} (P2) from the free page
list. Second, it must \emph{copy} the data from the original
physical page (P1) to the newly identified page (P2). Third, it
must \emph{remap} the virtual page that received the write (V2) to
the new physical page (P2). After these steps are completed, the
system can execute the write operation.

\subsection{Sources of Inefficiency in Executing Copy-on-Write}
\label{sec:cow-problems}

Existing interfaces to manage and access the memory subsystem
result in several sources of inefficiency in completing a
copy-on-write operation. First, existing virtual memory systems
manage main memory at a large page granularity. Therefore, even if
only a single byte or word is modified in the virtual page V2, the
system must allocate and copy a full physical page. This results
in \emph{high memory redundancy}. Second, the CPU accesses data at
a word or at best a vector register granularity. Existing systems
must therefore perform these copy operations one word or a vector
register at a time. This results in \emph{high latency} and
\emph{inefficient use of the CPU}. Third, all the cache lines
involved in the copy operation must be transferred from main
memory to the processor caches. These cache line transfers result
in \emph{high memory bandwidth} consumption and can potentially
cause \emph{cache pollution}. Finally, all the data movement
between the CPU, caches, and main memory consumes significant
amounts of energy.

Ideally, instead of copying an entire page of data, the system
should eliminate all the redundancy by remapping only the data
that is actually modified. In a case where the entire page needs
to be copied, the system should all the unnecessary data movement
by performing the copy operation completely in main memory.

\section{Fine-grained Operations with Low Spatial Locality}

As we mentioned in \Cref{chap:introduction}, the on-chip caches
and the off-chip memory interface are both optimized to store and
communicate wide cache lines (e.g., 64B). However, the data types
typically used by applications are much smaller than a cache
line. While access patterns with good spatial locality benefit
from the cache-line-granularity management, existing systems incur
high inefficiency when performing operations with low spatial
locality. Specifically, non-unit strided access patterns are
common in several applications, e.g., databases, scientific
computing, and graphics. To illustrate the shortcomings of
existing memory interfaces, we use an example of an in-memory
database table.

\subsection{Accessing a Column from a Row-Oriented Database Table}
\label{sec:non-unit-stride-problem}

In-memory databases are becoming popular among many
applications. A table in such a database consist of many records
(or rows). Each record in turn consists of many fields (or
columns). Typically, a table is stored either in the row-oriented
format or the column-oriented format. In the row-oriented format
or \emph{row store}, all fields of a record are stored
together. On the other hand, in the column-oriented format or
\emph{column store}, the values of each field from all records are
stored together. Depending on the nature of the query being
performed on the table, one format may be better suited than the
other. For example, the row store is better suited for inserting
new records or executing \emph{transactions} on existing
records. On the other hand, the column store is better suited for
executing \emph{analytical queries} that aim to extract aggregate
information from one or few fields of many records.

Unfortunately, neither organization is better suited for
\emph{both} transactions and analytical queries. With the recently
growing need for real-time analytics, workloads that run both
transactions and analytics on the same system, referred to as
Hybrid Transaction/Analytics Processing or HTAP~\cite{htap}, are
becoming important. Accessing a column of data from a row store
results in a strided access pattern.

\begin{figure}
  \centering
  \includegraphics{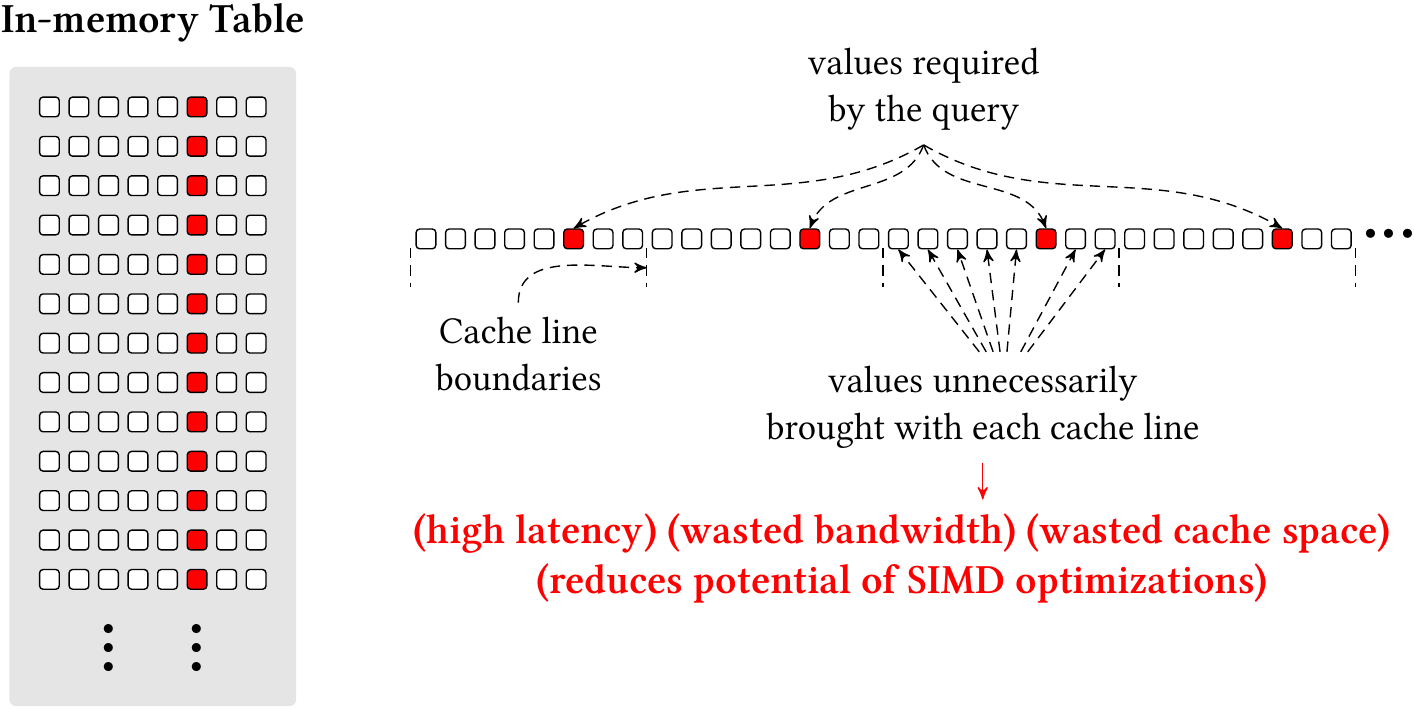}
  \caption[Strided access pattern in an in-memory table]{Accessing
    a column from a row store}
  \label{fig:col-row-store}
\end{figure}

\subsection{Shortcomings of Strided Access Patterns}

\Cref{fig:col-row-store} shows the shortcomings of accessing a
column of data from a database table that is stored as a row
store. For ease of explanation, we assume that each record fits
exactly into a cache line. As shown in the figure, each cache line
contains only one useful value. However, since the caches and
memory interface in existing systems are heavily optimized to
store and access cache lines, existing systems have to fetch more
data than necessary to complete the strided access operation. In
the example shown in the figure, the system has to bring eight
times more data than necessary to access a single column from a
row store. This amplification in the amount of data fetched
results in several problems. First, it significantly
\emph{increases the latency} to complete the operation, thereby
degrading the performance of the application. Second, it results
in inefficient use of memory bandwidth and on-chip cache
space. Finally, since different values of the strided access
pattern are stored in different cache lines, it is difficult to
enable SIMD (single instruction multiple data) optimizations for
the computation performing the strided access pattern.

Ideally, the memory system should be able to identify the strided
access pattern (either automatically or with the help of the
application), and fetch cache lines from memory that contain only
values from the access pattern. This will eliminate all the
inefficiency that results from the data overfetch and also
seamlessly enable SIMD optimizations.

\section{Goal of This Dissertation}

In this dissertation, our goal is to develop efficient solutions
to address the problems that result from the mismatch in the
granularity of memory management at different resources. To this
end, our approach is to exploit the untapped potential in various
hardware structures by introducing \emph{new virtual memory and
  DRAM abstractions} that mitigate the negative impact of multiple
granularities.

Specifically, first, we observe that modern processors can
efficiently track data at a cache line granularity using the
on-chip caches. We exploit this to propose a new virtual memory
abstraction called \emph{Page Overlays} to improve the efficiency
of many fine-grained memory operations. Second, we observe that
DRAM technology can be used to perform a variety of operations
rather than just store data. We exploit this potential to design
two mechanisms: \emph{RowClone} to perform bulk copy and
initialization operations inside DRAM, and \emph{Buddy RAM} to
perform bulk bitwise logical operations using DRAM. Third, we
observe that commodity DRAM modules interleave data across
multiple DRAM chips. We exploit this architecture to design
\emph{Gather-Scatter DRAM}, which efficiently gathers or scatters
data with access patterns that normally exhibit poor spatial
locality. Finally, we propose the \emph{Dirty-Block Index}, which
accelerates the coherence protocol that ensures the coherence of
data between the caches and the main memory.

The rest of the dissertation is organized as
follows. Chapter~\ref{chap:page-overlays} describes our new page
overlay framework. Chapter~\ref{chap:dram-background} provides a
detailed background on modern DRAM design and
architecture. Chapters~\ref{chap:rowclone}, \ref{chap:buddy}, and
\ref{chap:gsdram} describe RowClone, Buddy RAM, and Gather-Scatter
DRAM, respectively. Chapter~\ref{chap:dbi} describes the
Dirty-Block Index. Finally, we conclude the dissertation and
present some relevant future work in Chapter~\ref{chap:conclusion}.

\chapter{Page Overlays}
\label{chap:page-overlays}

\begin{figure}[b!]
  \hrule\vspace{2mm}
  \begin{footnotesize}
    Originally published as ``Page Overlays: An Enhanced Virtual Memory
    Framework to Enable Fine-grained Memory Management'' in the
    International Symposium on Computer Architecture,
    2015~\cite{page-overlays}
  \end{footnotesize}
\end{figure}

\newcommand{\obvect}{\texttt{OBitVector}\xspace}

\newcommand{\bench}[1]{\mbox{\emph{#1}}}
\newcommand{\OA}{Overlay Address}
\newcommand{\OMS}{Overlay Memory Store}
\newcommand{\OMSshort}{OMS}
\newcommand{\OMSA}{\OMS{} Address}
\newcommand{\OMSaddr}{\OMSshort{}addr}
\newcommand{\OMT}{Overlay Mapping Table}
\newcommand{\OMTshort}{OMT}
\newcommand{\OMTcache}{\OMTshort{} Cache}
\newcommand{\naive}{na\"{i}ve}

As described in \Cref{sec:cow-problems}, the large page
granularity organization of virtual memory results in significant
inefficiency for many operations (e.g., copy-on-write). The source
of this inefficiency is the fact that the large page granularity
(e.g., 4KB) amplifies the amount of work that needs to be done for
simple fine-granularity operations (e.g., few
bytes). \Cref{sec:cow-problems} explains this problem with the
example of the copy-on-write technique, wherein modification of a
small amount of data can trigger a full page copy operation. In
addition to copy-on-write, which is widely used in many
applications, we observe that the large page granularity
management hinders efficient implementation of several techniques
like fine-grained deduplication~\cite{de,hicamp}, fine-grained
data protection~\cite{mmp,legba}, cache-line-level
compression~\cite{lcp,rmc,pcm-compression}, and fine-grained
metadata management~\cite{ems,shadow-memory}. 

While managing memory at a finer granularity than pages enables
several techniques that can significantly boost system performance
and efficiency, simply reducing the page size results in an
unacceptable increase in virtual-to-physical mapping table
overhead and TLB pressure. Prior works to address this problem
either rely on software techniques~\cite{de} (high performance
overhead), propose hardware support specific to a particular
application~\cite{indra,shadow-memory,ems} (low value for cost),
or significantly modify the structure of existing virtual
memory~\cite{mmp,hicamp} (high cost for adoption). In this
chapter, we describe our new virtual memory framework, called
\emph{page overlays}, that enables efficient implementation of
several fine-grained memory management techniques.

\section{Page Overlays: Semantics and Benefits}
\label{sec:overlays-semantics}

We first present a detailed overview of the semantics of our
proposed virtual memory framework, page overlays. Then we describe
the benefits of overlays using the example of
copy-on-write. \Cref{sec:overlays-applications} describes
several applications of overlays.

\subsection{Overview of Semantics of Our Framework}
\label{sec:semantics-overview}

\begin{figure}[h]
  \centering
  \newcommand{\drawpage}[2]{
  \foreach \y in {1,...,#1} {
    \draw [rounded corners=2pt]($(#2) + (0,-0.3) + (0, 0.3*\y)$) rectangle ++(1.5, 0.3);
  }
}

\begin{tikzpicture}[>=stealth']

  \node (vp) at (0,0) [inner sep=0pt, outer sep=0pt] {
    \begin{tikzpicture}
      \coordinate (origin) at (0,0);
      \drawpage{4}{origin};
    \draw [rounded corners=2pt,fill=black!15]($(origin) + (0,-0.3) + (0, 0.3*4)$) rectangle ++(1.5, 0.3);
    \draw [rounded corners=2pt,fill=black!15]($(origin) + (0,-0.3) + (0, 0.3*2)$) rectangle ++(1.5, 0.3);
      \node at ($(origin) + (0.75, 0*0.3 + 0.15)$) {\footnotesize{\textbf{C4}}};
      \node at ($(origin) + (0.75, 1*0.3 + 0.15)$) {\footnotesize{\textbf{C3}}};
      \node at ($(origin) + (0.75, 2*0.3 + 0.15)$) {\footnotesize{\textbf{C2}}};
      \node at ($(origin) + (0.75, 3*0.3 + 0.15)$) {\footnotesize{\textbf{C1}}};
    \end{tikzpicture}
  };
  
  \node (pp) at ($(vp.east) + (2,0.75)$) [inner sep=0pt, outer sep=0pt] {
    \begin{tikzpicture}
      \coordinate (origin) at (0,0);
      \drawpage{4}{origin};
      \node at ($(origin) + (0.75, 0*0.3 + 0.15)$) {\footnotesize{\textbf{C4}}};
      \node at ($(origin) + (0.75, 1*0.3 + 0.15)$) {\footnotesize{\textbf{C3}}};
      \node at ($(origin) + (0.75, 2*0.3 + 0.15)$) {\footnotesize{\textbf{C2}}};
      \node at ($(origin) + (0.75, 3*0.3 + 0.15)$) {\footnotesize{\textbf{C1}}};
    \end{tikzpicture}
  };
  
  \node (op) at ($(vp.east) + (2,-0.5)$) [inner sep=0pt, outer sep=0pt] {
    \begin{tikzpicture}
      \coordinate (origin) at (0,0);
    \draw [rounded corners=2pt,fill=black!15]($(origin) + (0,-0.3) + (0, 0.3*2)$) rectangle ++(1.5, 0.3);
    \draw [rounded corners=2pt,fill=black!15]($(origin) + (0,-0.3) + (0, 0.3*1)$) rectangle ++(1.5, 0.3);
      \node at ($(origin) + (0.75, 0*0.3 + 0.15)$) {\footnotesize{\textbf{C3}}};
      \node at ($(origin) + (0.75, 1*0.3 + 0.15)$) {\footnotesize{\textbf{C1}}};
    \end{tikzpicture}
  };

  \draw [->] ($(vp.east) + (0,0*0.3+0.15)$) -- ($(pp.west) + (0,0*0.3+0.15)$);
  \draw [->] ($(vp.east) - (0,1*0.3+0.15)$) -- ($(pp.west) - (0,1*0.3+0.15)$);

  \draw [->,densely dashed] ($(vp.east) - (0,1*0.3-0.15)$) -- ($(op.west) - (0,1*0.15)$);
  \draw [->,densely dashed] ($(vp.east) + (0,1*0.3+0.15)$) -- ($(op.west) + (0,1*0.15)$);

  \node at (vp.west) [anchor=east,xshift=-1mm] {
    \begin{minipage}{2cm}
      \flushright
      \small
    Virtual\\Page
    \end{minipage}
  };
  \node at (pp.east) [anchor=west,xshift=1mm] {
    \begin{minipage}{2cm}
      \flushleft
      \small
      Physical\\Page
    \end{minipage}
  };
  \node at (op.east) [anchor=west,xshift=1mm] {
    \begin{minipage}{2cm}
      \flushleft
      \small
      Overlay
    \end{minipage}
  };

  

\end{tikzpicture}
  \caption[Page overlays: Semantics]{Semantics of our proposed framework}
  \label{fig:overlay-detailed}
\end{figure}
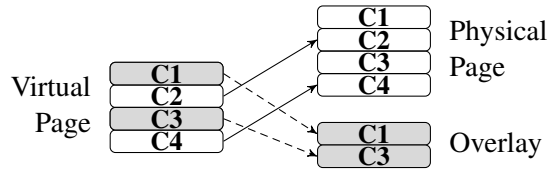

\Cref{fig:overlay-detailed} shows our proposed framework. The
figure shows a virtual page mapped to a physical page as in
existing frameworks. For ease of explanation, we assume each page
has only four cache lines. As shown in the figure, the virtual
page is also mapped to another structure referred to as
\emph{overlay}.  There are two aspects to a page overlay. First,
unlike the physical page, which has the same size as the virtual
page, the overlay of a virtual page contains only a \emph{subset
  of cache lines} from the page, and hence is smaller in size than
the virtual page. In the example in the figure, only two cache
lines (C1 and C3) are present in the overlay. Second, when a
virtual page has both a physical page and an overlay mapping, we
define the access semantics such that any cache line that is
present in the overlay is accessed from there. Only cache lines
that are \emph{not} present in the overlay are accessed from the
physical page.  In our example, accesses to C1 and C3 are mapped
to the overlay, and the remaining cache lines are mapped to the
physical page.

\subsection{Overlay-on-write: A More Efficient Copy-on-write}
\label{sec:overlay-applications-oow}

We described the copy-on-write technique and its shortcomings in
detail in \Cref{sec:bulk-data-problem}. Briefly, the copy-on-write
technique maps multiple virtual pages that contain the same data
to a single physical page in a read-only mode. When one of the
pages receive a write, the system creates a full copy of the
physical page and remaps the virtual page that received the write
to the new physical page in a read-write mode.

\begin{figure}
  \centering
  \begin{tikzpicture}[>=stealth']

  \input{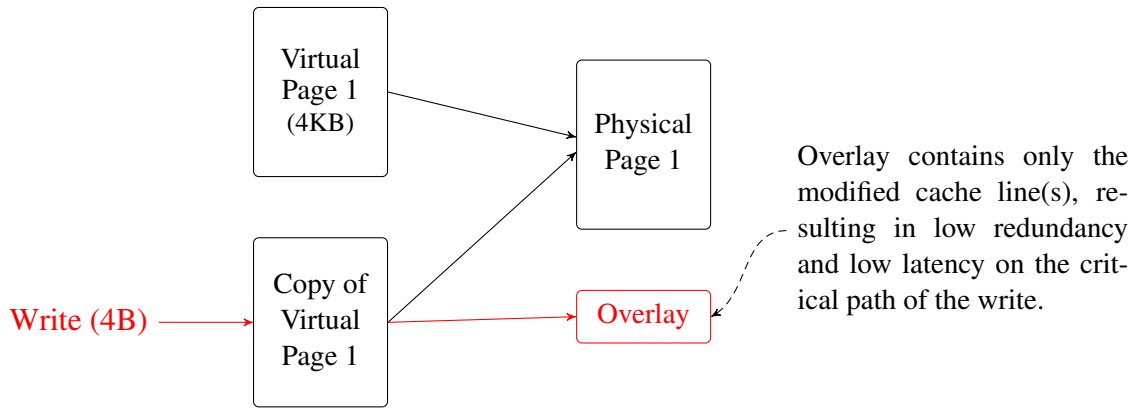}

  \node (v1) [page] {
    \begin{minipage}{1.5cm}\centering\small Virtual Page 1 \footnotesize{(4KB)}\end{minipage}};
  
  \node (v2) [page, anchor=north, yshift=-8mm] at (v1.south) {
    \begin{minipage}{1.5cm}\centering\small Copy of Virtual Page 1\end{minipage}};

  \node (p1) [page, anchor=west, yshift=-7mm, xshift=2.5cm] at (v1.east) {
    \begin{minipage}{1.5cm}\centering\small Physical Page 1\end{minipage}};

  \node (p2) [page, anchor=north, yshift=-8mm, red, minimum height=0.7cm] at (p1.south) {
    \begin{minipage}{1.5cm}\centering\small Overlay\end{minipage}};

  \draw [->] (v1.east) -- ([yshift=1mm]p1.west);
  \draw [->] (v2.east) -- ([yshift=-1mm]p1.west);

  \node (write) at (v2.west) [anchor=east,xshift=-1.25cm,red] {Write (4B)};
  \draw [->,red] (write.east) -- (v2.west);

  \draw [->,red] (v2.east) -- (p2.west);

  \node (copy) at ($(p1.east)!0.5!(p2.east)$) [xshift=1cm, anchor=west] {
    \begin{minipage}{4.4cm}
      \small
      Overlay contains only the modified cache line(s), resulting
      in low redundancy and low latency on the critical path of
      the write.
    \end{minipage}
  };

  \draw [->,densely dashed] (copy.west) to [out=180,in=30] (p2.east);
  
\end{tikzpicture}
  \caption[Overlay-on-write technique]{Overlay-on-Write: A more
    version of efficient copy-on-write}
  \label{fig:overlays-oow}
\end{figure}

Our page overlay framework enables a more efficient version of the
copy-on-write technique, which does not require a full page copy
and hence avoids all associated shortcomings. We refer to this
mechanism as \emph{overlay-on-write}.  \Cref{fig:overlays-oow}
shows how overlay-on-write works. When multiple virtual pages
share the same physical page, the OS explicitly indicates to the
hardware, through the page tables, that the cache lines of the
pages should be copied-on-write.  When one of the pages receives a
write, our framework first creates an overlay that contains
\emph{only the modified cache line}. It then maps the overlay to
the virtual page that received the write.

Overlay-on-write has many benefits over copy-on-write. First, it
avoids the need to copy the entire physical page before the write
operation, thereby significantly reducing the latency on the
critical path of execution (as well as the associated increase in
memory bandwidth and energy). Second, it allows the system to
eliminate significant redundancy in the data stored in main memory
because only the overlay lines need to be stored, compared to a
full page with copy-on-write.  Finally, as we describe in
\Cref{sec:overlaying-write-op}, our design exploits the
fact that only a single cache line is remapped from the source
physical page to the overlay to significantly reduce the latency
of the remapping operation.

Copy-on-write has a wide variety of applications (e.g., process
forking~\cite{fork}, virtual machine cloning~\cite{snowflock},
operating system
speculation~\cite{os-speculation-1,os-speculation-2,os-speculation-3},
deduplication~\cite{esx-server}, software
debugging~\cite{flashback},
checkpointing~\cite{checkpointing,hpc-survey}).  Overlay-on-write,
being a faster and more efficient alternative to copy-on-write,
can significantly benefit all these applications.

\subsection{Benefits of the Overlay Semantics}
\label{sec:overlay-benefits}

Our framework offers two distinct benefits over the existing
virtual memory frameworks.  First, our framework \textbf{reduces
  the amount of work that the system has to do}, thereby improving
system performance. For instance, in the overlay-on-write and
sparse data structure
(\Cref{sec:applications-sparse-data-structures}) techniques, our
framework reduces the amount of data that needs to be
copied/accessed.  Second, our framework \textbf{enables
  significant reduction in memory capacity requirements}. Each
overlay contains only a subset of cache lines from the virtual
page, so the system can reduce overall memory consumption by
compactly storing the overlays in main memory---i.e., for each
overlay, store only the cache lines that are actually present in
the overlay. We quantitatively evaluate these benefits in
\Cref{sec:overlays-applications} using two techniques and show
that our framework is effective.

\section{Overview of Design}
\label{sec:page-overlays-overview}

While our framework imposes simple access semantics, there are
several key challenges to efficiently implement the proposed
semantics.  In this section, we first discuss these challenges
with an overview of how we address them. We then
provide a full overview of our proposed mechanism that addresses
these challenges, thereby enabling a simple, efficient, and
low-overhead design of our framework.

\subsection{Challenges in Implementing Page Overlays}
\label{sec:challenges}

\textbf{Challenge 1:} \emph{Checking if a cache line is part of
  the overlay}. When the processor needs to access a virtual
address, it must first check if the accessed cache line is part of
the overlay. Since most modern processors use a physically-tagged
L1 cache, this check is on the critical path of the L1 access. To
address this challenge, we associate each virtual page with a bit
vector that represents which cache lines from the virtual page are
part of the overlay. We call this bit vector the \emph{overlay bit
  vector} (\obvect). We cache the \obvect in the processor TLB,
thereby enabling the processor to quickly check if the accessed
cache line is part of the overlay.

\textbf{Challenge 2:} \emph{Identifying the physical address of an
  overlay cache line}. If the accessed cache line is part of the
overlay (i.e., it is an \emph{overlay cache line}), the processor
must quickly determine the physical address of the overlay cache
line, as this address is required to access the L1 cache. The
simple approach to address this challenge is to store in the TLB
the base address of the region where the overlay is stored in main
memory (we refer to this region as the \emph{overlay
  store}). While this may enable the processor to identify each
overlay cache line with a unique physical address, this approach
has three shortcomings when overlays are stored compactly in main
memory.

First, the overlay store (in main memory) does \emph{not} contain
all the cache lines from the virtual page. Therefore, the
processor must explicitly compute the address of the accessed
overlay cache line. This will delay the L1 access. Second, most
modern processors use a virtually-indexed physically-tagged L1
cache to partially overlap the L1 cache access with the TLB
access. This technique requires the virtual index and the physical
index of the cache line to be the same. However, since the overlay
is smaller than the virtual page, the overlay physical index of a
cache line will likely not be the same as the cache line's virtual
index. As a result, the cache access will have to be delayed until
the TLB access is complete. Finally, inserting a new cache line
into an overlay is a relatively complex operation. Depending on
how the overlay is represented in main memory, inserting a new
cache line into an overlay can potentially change the addresses of
other cache lines in the overlay. Handling this scenario requires
a likely complex mechanism to ensure that the tags of these other
cache lines are appropriately modified.

In our design, we address this challenge by using two different
addresses for each overlay---one to address the processor caches,
called the {\em \OA{}}, and another to address main memory, called
the {\em \OMSA{}}. As we will describe shortly, this
\emph{dual-address design} enables the system to manage the
overlay in main memory independently of how overlay cache lines
are addressed in the processor caches, thereby overcoming the
above three shortcomings.

\textbf{Challenge 3:} \emph{Ensuring the consistency of the
  TLBs}. In our design, since the TLBs cache the \obvect, when a
cache line is moved from the physical page to the overlay or vice
versa, any TLB that has cached the mapping for the corresponding
virtual page should update its mapping to reflect the cache line
remapping. The \naive{} approach to addressing this challenge is
to use a TLB shootdown~\cite{tlb-consistency-1,tlb-consistency-2}, which
is expensive~\cite{didi,unitd}. Fortunately, in the above
scenario, the TLB mapping is updated only for a \emph{single cache
  line} (rather than an entire virtual page). We propose a simple
mechanism that exploits this fact and uses the cache coherence
protocol to keep the TLBs coherent
(Section~\ref{sec:overlaying-write-op}).

\subsection{Overview of Our Design}
\label{sec:final-implementation}

A key aspect of our dual-address design, mentioned above, is that
the address to access the cache (the {\em \OA{}}) is taken from an
\emph{address space} where the size of each overlay is the
\emph{same} as that of a regular physical page. This enables our
design to seamlessly address Challenge 2 (overlay cache line
address computation), without incurring the drawbacks of the \naive{}
approach to address the challenge (described in
Section~\ref{sec:challenges}).  The question is, \emph{from what
  address space is the \OA{} taken?}

Towards answering this question, we observe that only a small
fraction of the physical address space is backed by main memory
(DRAM) and a large portion of the physical address space is
\emph{unused}, even after a portion is consumed for memory-mapped
I/O~\cite{mmio} and other system constructs.  We propose to use
this unused physical address space for the overlay cache address
and refer to this space as the \emph{\OA{} Space}.\footnote{A
  prior work, the Impulse Memory Controller~\cite{impulse}, uses
  the unused physical address space to communicate gather/scatter
  access patterns to the memory controller. The goal of
  Impulse~\cite{impulse} is different from ours, and it is
  difficult to use the design proposed by Impulse to enable
  fine-granularity memory management.}

\begin{figure}
  \centering
  \input{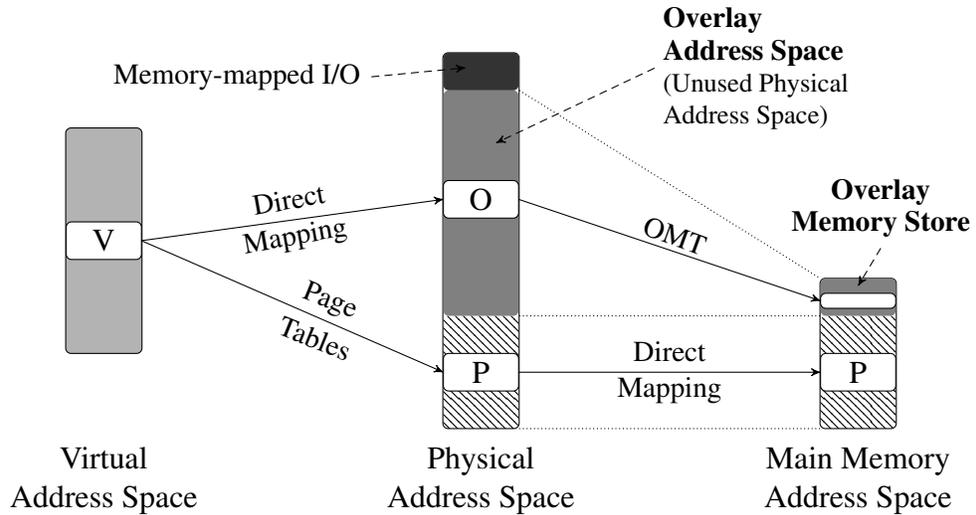}
  \caption[Page overlays: Design overview]{Overview of our design. ``Direct mapping''
    indicates that the corresponding mapping is implicit in the
    source address. OMT = Overlay Mapping Table
    (Section~\ref{sec:o2mmap}).}
  \label{fig:implementation-overview}
\end{figure}

Figure~\ref{fig:implementation-overview} shows the overview of our
design. There are three address spaces: the virtual address space,
the physical address space, and the main memory address space. The
main memory address space is split between regular physical pages
and the \emph{\OMS{}} (OMS), a region where the overlays are
stored compactly. In our design, to associate a virtual page with
an overlay, the virtual page is first mapped to a full size page
in the overlay address space using a direct mapping without any
translation or indirection (Section~\ref{sec:v2omap}). The overlay
page is in turn mapped to a location in the OMS using a mapping
table stored in the memory controller
(Section~\ref{sec:o2mmap}). We will describe the figure in more
detail in Section~\ref{sec:overlays-design-implementation}.

\subsection{Benefits of Our Design}
\label{sec:implementation-benefits}

There are three main benefits of our high-level design.  First,
our approach makes no changes to the way the existing VM framework
maps virtual pages to physical pages.  This is very important as
the system can treat overlays as an inexpensive feature that can
be turned on only when the application benefits from it. Second,
as mentioned before, by using two distinct addresses for each
overlay, our implementation decouples the way the caches are
addressed from the way overlays are stored in main memory. This
enables the system to treat overlay cache accesses very similarly
to regular cache accesses, and consequently requires very few
changes to the existing hardware structures (e.g., it works
seamlessly with virtually-indexed physically-tagged caches).
Third, as we will describe in the next section, in our design, the
Overlay Memory Store (in main memory) is accessed only when an
access completely misses in the cache hierarchy. This 1)~greatly
reduces the number of operations related to managing the OMS,
2)~reduces the amount of information that needs to be cached in
the processor TLBs, and 3)~more importantly, enables the memory
controller to completely manage the OMS with \emph{minimal}
interaction with the OS.

\section{Detailed Design and Implementation}
\label{sec:overlays-design-implementation}

To recap our high-level design (Figure~\ref{fig:implementation-overview}), 
each virtual page in the
system is mapped to two entities: 1)~a regular physical page,
which in turn directly maps to a page in main memory, and 2)~an
overlay page in the \OA{} space (which is not directly
backed by main memory). Each page in this space is
in turn mapped to a region in the \OMS{}, where the overlay
is stored compactly. Because our implementation does not modify the
way virtual pages are mapped to regular physical pages, we now
focus our attention on how virtual pages are mapped to overlays. 

\subsection{Virtual-to-Overlay Mapping}
\label{sec:v2omap}

The virtual-to-overlay mapping maps a virtual page to a page in
the \OA{} space. One simple approach to maintain this
mapping information is to store it in the page table and allow the
OS to manage the mappings (similar to regular physical
pages). However, this increases the overhead of the mapping table
and complicates the OS. We make a simple observation and impose a
constraint that makes the virtual-to-overlay mapping a direct 1-1
mapping.

Our \textbf{observation} is that since the \OA{} space is part of
the \emph{unused} physical address space, it can be
\emph{significantly larger} than the amount of main memory.  To
enable a 1-1 mapping between virtual pages and overlay pages, we
impose a simple \textbf{constraint} wherein no two virtual pages
can be mapped to the same overlay page.

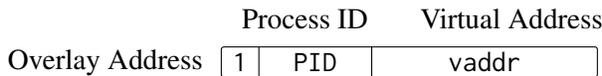
\begin{figure}[h]
  \centering
  \begin{tikzpicture}
  \tikzset{no padding/.style={inner sep=0pt, outer sep=0pt}};
  \tikzset{rounded/.style={rounded corners=1pt}};

  \node (pas) [no padding,draw,rounded,
  minimum height=0.4cm, minimum width=5cm] {};

  \node at (pas.west) [anchor=east,xshift=-1mm] {\small Overlay Address};

  \draw (pas.south west) ++ (0.5,0) -- +(0,0.4)
                         ++ (1.5,0) -- +(0,0.4);

  \node at ($(pas.south west) + (0.25,0.2)$) {\small\texttt{1}};
  \node at ($(pas.south west) + (0.5+0.75,0.2)$) {\small\texttt{PID}};
  \node at ($(pas.south west) + (2+1.5,0.2)$) {\small\texttt{vaddr}};

  \node at (pas.north) [anchor=south west,yshift=1mm] {\small{Virtual Address}};
  \node at (pas.north) [anchor=south east,yshift=1mm,xshift=-4mm] {\small{Process ID}};
\end{tikzpicture}
  \caption[Virtual address space to overlay address space
    mapping]{Virtual-to-Overlay Mapping. The MSB indicates if the
    physical address is part of the \OA{} space.}
  \label{fig:virtual-to-overlay-mapping}
\end{figure}

Figure~\ref{fig:virtual-to-overlay-mapping} shows how our design
maps a virtual address to the corresponding overlay address. Our
scheme widens the physical address space such that the overlay
address corresponding to the virtual address \texttt{vaddr} of a
process with ID \texttt{PID} is obtained by simply concatenating
an overlay bit (set to 1), \texttt{PID}, and \texttt{vaddr}.
Since two virtual pages cannot share an overlay, when data of a
virtual page is copied to another virtual page, the overlay cache
lines of the source page must be copied into the appropriate
locations in the destination page. While this approach requires a
slightly wider physical address space than in existing systems,
this is a more practical mechanism compared to storing this
mapping explicitly in a separate table, which can lead to much
higher storage and management overheads than our approach. With a
64-bit physical address space and a 48-bit virtual address space
per process, this approach can support $2^{15}$ different
processes.

Note that a similar approach \emph{cannot} be used to map virtual
pages to physical pages due to the \emph{synonym}
problem~\cite{virtual-caches}, which results from multiple virtual
pages being mapped to the same physical page. However, this
problem does not occur with the virtual-to-overlay mapping because
of the constraint we impose: no two virtual pages can map to the
same overlay page. Even with this constraint, our framework
enables many applications that can improve performance and reduce
memory capacity requirements
(Section~\ref{sec:overlays-applications}).

\subsection{\OA{} Mapping}
\label{sec:o2mmap}

Overlay cache lines tagged in the \OA{} space must be mapped into
an \OMS{} location upon eviction. In our design, since there is a
1-1 mapping between a virtual page and an overlay page, we could
potentially store this mapping in the page table along with the
physical page mapping. However, since many pages may not have an
overlay, we store this mapping information in a separate mapping
table similar to the page table.  This \emph{\OMT{}} (\OMTshort{})
is maintained and controlled fully by the memory controller with
minimal interaction with the OS. Section~\ref{sec:overlay-store}
describes \OMS{} management in detail.

\subsection{Microarchitecture and Memory Access Operations}
\label{sec:mem-ops}

\begin{figure}
  \centering
  \includegraphics[scale=0.9]{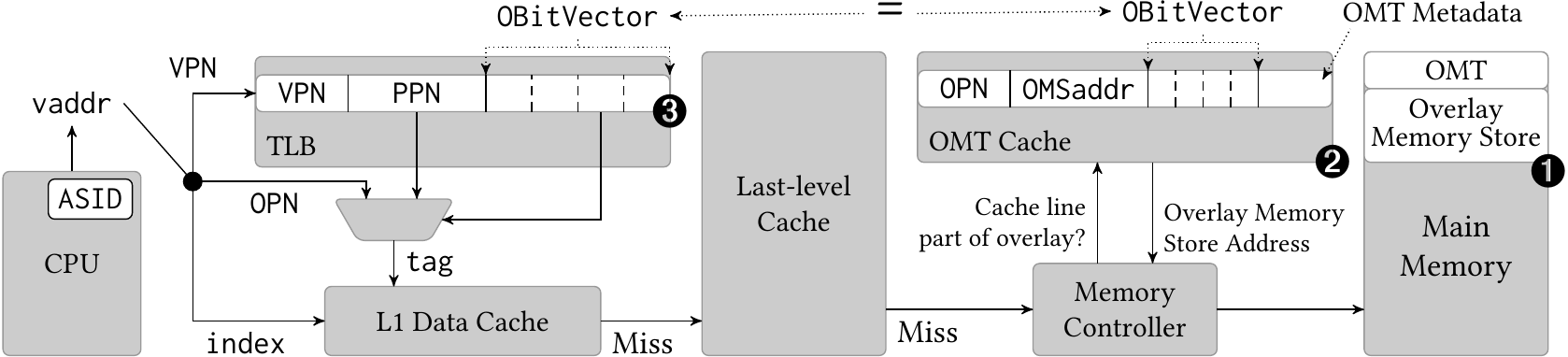}
  \caption[Page overlays: Microarchitectural
    implementation]{Microarchitectural details of our
    implementation. The main changes (\ding{202}, \ding{203} and
    \ding{204}) are described in Section~\ref{sec:mem-ops}.}
  \label{fig:end-to-end-operation}
\end{figure}

Figure~\ref{fig:end-to-end-operation} depicts the details of our
design. There are three main changes over the microarchitecture of
current systems. First (\ding{202} in the figure), main memory is
split into two regions that store 1)~regular physical pages and
2)~the \OMS{} (\OMSshort{}).  The \OMSshort{} stores both a
compact representation of the overlays and the \emph{\OMT{}}
(\OMTshort{}), which maps each page from the \OA{} Space to a
location in the \OMS{}. At a high level, each \OMTshort{} entry
contains 1) the \obvect, indicating if each cache line within the
corresponding page is present in the overlay, and
2)~\texttt{OMSaddr}, the location of the overlay in the
\OMSshort{}. Second \ding{203}, we augment the memory controller
with a cache called the \emph{\OMTcache{}}, which caches recently
accessed entries from the \OMTshort{}. Third \ding{204}, because
the TLB must determine if an access to a virtual address should be
directed to the corresponding overlay, we extend each TLB entry to
store the \obvect.  While this potentially increases the cost of
each TLB miss (as it requires the \obvect to be fetched from the
OMT), our evaluations (Section~\ref{sec:overlays-applications})
show that the performance benefit of using overlays more than
offsets this additional TLB fill latency.

To describe the operation of different memory accesses, we use
overlay-on-write (Section~\ref{sec:overlay-applications-oow}) as an
example. Let us assume that two virtual pages (V1 and V2) are
mapped to the same physical page in the copy-on-write mode, with a
few cache lines of V2 already mapped to the overlay. There are
three possible operations on V2: 1)~a read, 2)~a write to a cache
line already in the overlay (\emph{simple write}), and 3)~a write
to a cache line not present in the overlay (\emph{overlaying
  write}). We now describe each of these operations in detail.

\subsubsection{Memory Read Operation.}
\label{sec:memory-read-op}

When the page V2 receives a read request, the processor first
accesses the TLB with the corresponding page number (\texttt{VPN})
to retrieve the physical mapping (\texttt{PPN}) and the
\obvect. It generates the overlay page number (\texttt{OPN}) by
concatenating the address space ID (\texttt{ASID}) of the process
and the \texttt{VPN} (as described in
Section~\ref{sec:v2omap}). Depending on whether the accessed cache
line is present in the overlay (as indicated by the corresponding
bit in the \obvect), the processor uses either the \texttt{PPN} or
the \texttt{OPN} to generate the L1 cache \texttt{tag}. If the
access misses in the entire cache hierarchy (L1 through last-level
cache), the request is sent to the memory controller. The
controller checks if the requested address is part of the overlay
address space by checking the overlay bit in the physical
address. If so, it looks up the overlay store address
(\texttt{\OMSaddr{}}) of the corresponding overlay page from the
\OMTcache{}, and computes the exact location of the requested
cache line within main memory (as described later in
Section~\ref{sec:overlay-store}). It then accesses the cache line
from the main memory and returns the data to the cache hierarchy.

\subsubsection{Simple Write Operation.}
\label{sec:memory-write-op}

When the processor receives a write to a cache line already
present in the overlay, it simply has to update the cache line in
the overlay. This operation is similar to a read operation, except
the cache line is updated after it is read into the L1 cache.

\subsubsection{Overlaying Write Operation.}
\label{sec:overlaying-write-op}

An \emph{overlaying write} operation is a write to a cache line
that is \emph{not} already present in the overlay. Since the
virtual page is mapped to the regular physical page in the
copy-on-write mode, the corresponding cache line must be remapped
to the overlay (based on our semantics described in
Section~\ref{sec:overlay-applications-oow}). We complete the
overlaying write in three steps: 1)~copy the data of the cache
line in the regular physical page (\texttt{PPN}) to the
corresponding cache line in the Overlay Address Space page
(\texttt{OPN}), 2)~update all the TLBs and the \OMTshort{} to
indicate that the cache line is mapped to the overlay, and
3)~process the write operation.

The first step can be completed in hardware by reading the cache
line from the regular physical page and simply updating the cache
tag to correspond to the overlay page number (or by making an
explicit copy of the cache line). Na\"{i}vely implementing the second
step will involve a TLB shootdown for the corresponding virtual
page. However, we exploit three simple facts to use the cache
coherence network to keep the TLBs and the \OMTshort{} coherent:
i)~the mapping is modified \emph{only} for a single cache line,
and not an entire page, ii)~the overlay page address can be used
to uniquely identify the virtual page since no overlay is shared
between virtual pages, and iii)~the overlay address is part of the
physical address space and hence, part of the cache coherence
network. Based on these facts, we propose a new cache coherence
message called \emph{overlaying read exclusive}. When a core
receives this request, it checks if its TLB has cached the mapping
for the virtual page. If so, the core simply sets the bit for the
corresponding cache line in the \obvect. The \emph{overlaying read
  exclusive} request is also sent to the memory controller so that
it can update the \obvect of the corresponding overlay page in the
\OMTshort{} (via the \OMTcache{}). Once the remapping operation is
complete, the write operation (the third step) is processed
similar to the simple write operation.

Note that after an \emph{overlaying write}, the corresponding
cache line (which we will refer to as the \emph{overlay cache
  line}) is marked dirty. However, unlike copy-on-write, which
must allocate memory before the write operation, our mechanism
allocates memory space \emph{lazily} upon the eviction of the
dirty overlay cache line -- significantly improving performance.

\subsubsection{Converting an Overlay to a Regular Physical Page.}
\label{sec:converting}

Depending on the technique for which overlays are used,
maintaining an overlay for a virtual page may be unnecessary after
a point. For example, when using overlay-on-write, if most of the
cache lines within a virtual page are modified, maintaining them
in an overlay does not provide any advantage.  The system may take
one of three actions to promote an overlay to a physical page: The
\emph{copy-and-commit} action is one where the OS copies the data
from the regular physical page to a new physical page and updates
the data of the new physical page with the corresponding data from
the overlay.  The \emph{commit} action updates the data of the
regular physical page with the corresponding data from the
overlay.  The \emph{discard} action simply discards the overlay.

While the \emph{copy-and-commit} action is used with
overlay-on-write, the \emph{commit} and \emph{discard} actions are
used, for example, in the context of speculation, where our
mechanism stores speculative updates in the overlays
(Section~\ref{sec:applications-speculation}).  After any of these
actions, the system clears the \obvect of the corresponding
virtual page, and frees the overlay memory store space allocated
for the overlay (discussed next in
Section~\ref{sec:overlay-store}).

\subsection{Managing the \OMS{}}
\label{sec:overlay-store}

The \emph{\OMS{}} (\OMSshort{}) is the region in main memory where
all the overlays are stored. As described in
Section~\ref{sec:memory-read-op}, the \OMSshort{} is accessed
\emph{only} when an overlay access completely misses in the cache
hierarchy.  As a result, there are many simple ways to manage the
\OMSshort{}. One way is to have a small embedded core on the
memory controller that can run a software routine that manages the
\OMSshort{} (similar mechanisms are supported in existing systems,
e.g., Intel Active Management
Technology~\cite{intel-amt}). Another approach is to let the
memory controller manage the \OMSshort{} by using a full physical
page to store each overlay. While this approach will forgo the
memory capacity benefit of our framework, it will still obtain the
benefit of reducing overall work
(Section~\ref{sec:overlay-benefits}).

In this section, we describe a hardware mechanism that obtains
both the work reduction and the memory capacity reduction benefits
of using overlays. In our mechanism, the controller fully manages
the \OMSshort{} with minimal interaction with the OS. Managing the
\OMSshort{} has two aspects. First, because each overlay contains
only a subset of cache lines from the virtual page, we need a
\emph{compact representation for the overlay}, such that the
\OMSshort{} contains only cache lines that are actually present in
the overlay. Second, the memory controller must manage multiple
\emph{overlays of different sizes}. We need a mechanism to handle
such different sizes and the associated free space fragmentation
issues. Although operations that allocate new overlays or relocate
existing overlays are slightly complex, they are triggered only
when a dirty overlay cache line is written back to main
memory. Therefore, these operations are rare and are not on the
critical path of execution.

\subsubsection{Compact Overlay Representation.}
\label{sec:overlay-representation}
One approach to compactly maintain the overlays is to store the
cache lines in an overlay in the order in which they appear in the
virtual page. While this representation is simple, if a new cache
line is inserted into the overlay before other overlay cache
lines, then the memory controller must \emph{move} such cache
lines to create a slot for the inserted line. This is a
read-modify-write operation, which results in significant
performance overhead.

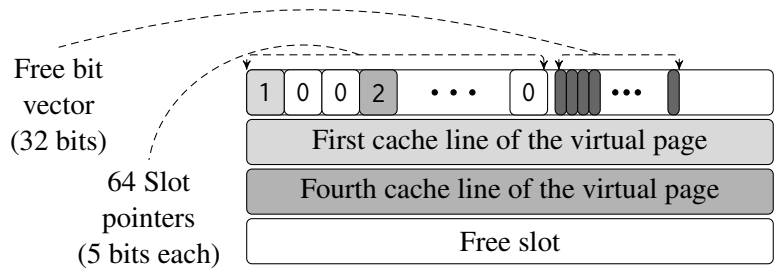
\begin{figure}[h]
  \centering
  \begin{tikzpicture}[>=stealth',rounded corners=2pt]
  \newcommand{\segheight}{0.6}
  \newcommand{\segwidth}{7}
  \newcommand{\ptrwidth}{0.5}
  \newcommand{\bitwidth}{0.15}
  \draw  (0,0) rectangle ++(\segwidth,\segheight);
  \draw  [fill=black!15] (0,-1.1*\segheight) rectangle ++(\segwidth,\segheight);
  \draw  [fill=black!30] (0,-2.2*\segheight) rectangle ++(\segwidth,\segheight);
  \draw  (0,-3.3*\segheight) rectangle ++(\segwidth,\segheight);

  \draw [fill=black!15] (0,0) rectangle ++(\ptrwidth,\segheight);
  \draw [fill=black!0] (1*\ptrwidth,0) rectangle ++(\ptrwidth,\segheight);
  \draw [fill=black!0] (2*\ptrwidth,0) rectangle ++(\ptrwidth,\segheight);
  \draw [fill=black!30] (3*\ptrwidth,0) rectangle ++(\ptrwidth,\segheight);
  \draw [fill=black!0] (7*\ptrwidth,0) rectangle ++(\ptrwidth,\segheight);

  \node at (0.5*\ptrwidth,0.5*\segheight) {\small{\texttt{1}}};
  \node at (1*\ptrwidth+0.5*\ptrwidth,0.5*\segheight) {\small{\texttt{0}}};
  \node at (2*\ptrwidth+0.5*\ptrwidth,0.5*\segheight) {\small{\texttt{0}}};
  \node at (3*\ptrwidth+0.5*\ptrwidth,0.5*\segheight) {\small{\texttt{2}}};
  \node at (7*\ptrwidth+0.5*\ptrwidth,0.5*\segheight) {\small{\texttt{0}}};

  \draw [fill=black] (5*\ptrwidth,0.5*\segheight) circle(1pt);
  \draw [fill=black] (5.5*\ptrwidth,0.5*\segheight) circle(1pt);
  \draw [fill=black] (6*\ptrwidth,0.5*\segheight) circle(1pt);

  \draw [fill=black!60] (8.2*\ptrwidth,0) rectangle ++(\bitwidth,\segheight);
  \draw [fill=black!60] (1*\bitwidth+8.2*\ptrwidth,0) rectangle ++(\bitwidth,\segheight);
  \draw [fill=black!60] (2*\bitwidth+8.2*\ptrwidth,0) rectangle ++(\bitwidth,\segheight);
  \draw [fill=black!60] (3*\bitwidth+8.2*\ptrwidth,0) rectangle ++(\bitwidth,\segheight);
  \draw [fill=black!60] (10*\bitwidth+8.2*\ptrwidth,0) rectangle ++(\bitwidth,\segheight);

  \draw [fill=black] (6*\bitwidth + 8*\ptrwidth,0.5*\segheight) circle (1pt);
  \draw [fill=black] (7*\bitwidth + 8*\ptrwidth,0.5*\segheight) circle (1pt);
  \draw [fill=black] (8*\bitwidth + 8*\ptrwidth,0.5*\segheight) circle (1pt);

  \draw [densely dashed,<->] (0,\segheight) |- (8*\ptrwidth-0.05,\segheight+0.2) -- ++(0,-0.2);
  \draw [densely dashed,<->] (8.2*\ptrwidth+0.05,\segheight) |- (11*\bitwidth+8.2*\ptrwidth,\segheight+0.2) -- ++(0,-0.2);

  \node at (0.5*\segwidth,-0.6*\segheight) {\small{First cache line of the virtual page}};
  \node at (0.5*\segwidth,-1.7*\segheight) {\small{Fourth cache line of the virtual page}};
  \node at (0.5*\segwidth,-2.8*\segheight) {\small{Free slot}};

  \node (slotlabel) at (-1.3,-1*\segheight) [anchor=north] {
    \begin{minipage}{2cm}
      \centering
      \small
      64 Slot pointers\\
      (5 bits each)
    \end{minipage}
  };

  \node (freelabel) at (-2.5,1.5*\segheight) [anchor=north] {
    \begin{minipage}{2cm}
      \centering
      \small
      Free bit vector\\
      (32 bits)
    \end{minipage}
  };

  \draw [densely dashed] (3*\ptrwidth,\segheight+0.2) to[out=160,in=80] (slotlabel.north);
  \draw [densely dashed] (8.2*\ptrwidth+4*\bitwidth,\segheight+0.2) to[out=165,in=15] (freelabel.north);
\end{tikzpicture}
  \caption[An example overlay segment in memory]{A 256B overlay
    segment (can store up to three overlay cache lines from the
    virtual page). The first line stores the metadata (array of
    pointers and the free bit vector).}
  \label{fig:segment}
\end{figure}

We propose an alternative mechanism, in which each overlay is
assigned a \emph{segment} in the \OMSshort{}. The overlay is
associated with an array of pointers---one pointer for each cache
line in the virtual page.  Each pointer either points to the slot
within the overlay segment that contains the cache line or is
invalid if the cache line is not present in the overlay. We store
this metadata in a single cache line at the head of the
segment. For segments less than 4KB size, we use 64 5-bit slot
pointers and a 32-bit vector indicating the free slots within a
segment---total of 352 bits. For a 4KB segment, we do not store
any metadata and simply store each overlay cache line at an offset
which is same as the offset of the cache line within the virtual
page. Figure~\ref{fig:segment} shows an overlay segment of size
256B, with only the first and the fourth cache lines of the
virtual page mapped to the overlay.

\subsubsection{Managing Multiple Overlay Sizes.}
\label{sec:multi-size-overlays}
Different virtual pages may contain overlays of different
sizes. The memory controller must store them efficiently in the
available space. To simplify this management, our mechanism splits
the available overlay space into segments of 5 fixed sizes: 256B,
512B, 1KB, 2KB, and 4KB. Each overlay is stored in the smallest
segment that is large enough to store the overlay cache lines.
When the memory controller requires a segment for a new overlay or
when it wants to migrate an existing overlay to a larger segment,
the controller identifies a free segment of the required size and
updates the \texttt{\OMSaddr{}} of the corresponding overlay page
with the base address of the new segment. Individual cache lines
are allocated their slots within the segment as and when they are
written back to main memory.

\subsubsection{Free Space Management.} To manage the free segments
within the \OMS{}, we use a simple linked-list based approach. For
each segment size, the memory controller maintains a memory
location or register that points to a free segment of that
size. Each free segment in turn stores a pointer to another free
segment of the same size or an invalid pointer denoting the end of
the list. If the controller runs out of free segments of a
particular size, it obtains a free segment of the next higher size
and splits it into two. If the controller runs out of free 4KB
segments, it requests the OS for an additional set of 4KB
pages. During system startup, the OS proactively allocates a chunk
of free pages to the memory controller. To reduce the number of
memory operations needed to manage free segments, we use a
\emph{grouped-linked-list} mechanism, similar to the one used by
some file systems~\cite{fs-free-space}.

\subsubsection{The \OMT{} (\OMTshort{}) and the \OMTcache{}.}
\label{sec:omt-cache}
The \OMTshort{} maps pages from the \OA{} Space to a specific
segment in the \OMS{}. For each page in the \OA{} Space (i.e., for
each \texttt{OPN}), the \OMTshort{} contains an entry with the
following pieces of information: 1)~the \obvect, indicating which
cache lines are present in the overlay, and 2)~the \OMSA{}
(\texttt{OMSaddr}), pointing to the segment that stores the
overlay. To reduce the storage cost of the \OMTshort{}, we store
it hierarchically, similar to the virtual-to-physical mapping
tables. The memory controller maintains the root address of the
hierarchical table in a register.

The \OMTcache{} stores the following details regarding
recently-accessed overlays: the \obvect, the \texttt{\OMSaddr{}},
and the overlay segment metadata (stored at the beginning of the
segment). To access a cache line from an overlay, the memory
controller consults the \OMTcache{} with the overlay page number
(\texttt{OPN}). In case of a hit, the controller acquires the
necessary information to locate the cache line in the overlay
memory store using the overlay segment metadata. In case of a
miss, the controller performs an \OMTshort{} walk (similar to a
page table walk) to look up the corresponding \OMTshort{} entry,
and inserts it in the \OMTcache{}. It also reads the overlay
segment metadata and caches it in the \OMTshort{} cache entry. The
controller may modify entries of the \OMTshort{}, as and when
overlays are updated. When such a modified entry is evicted from
the \OMTcache{}, the memory controller updates the corresponding
\OMTshort{} entry in memory.

\newcommand{\locality}{$\mathcal{L}$\xspace}
\newcommand{\fork}{\texttt{fork}\xspace}

\section{Applications and Evaluations}
\label{sec:overlays-applications}

We describe seven techniques enabled by our framework, and
quantitatively evaluate two of them.  For our evaluations, we use
memsim~\cite{memsim}, an event-driven multi-core simulator that
models out-of-order cores coupled with a DDR3-1066~\cite{ddr3}
DRAM simulator. All the simulated systems use a three-level cache
hierarchy with a uniform 64B cache line size. We do not enforce
inclusion in any level of the hierarchy. We use the
state-of-the-art DRRIP cache replacement policy~\cite{rrip} for
the last-level cache. All our evaluated systems use an aggressive
multi-stream prefetcher~\cite{fdp} similar to the one implemented
in IBM
Power~6~\cite{power6-prefetcher}. Table~\ref{table:parameters}
lists the main configuration parameters in detail.

\begin{table}
  \centering
  \newcommand{\tworow}[1]{\multirow{2}{*}{#1}}
\begin{tabular}{lp{0.7\textwidth}}
  \toprule
  \tworow{Processor} & 2.67 GHz, single issue, out-of-order, 64 entry instruction
  window, 64B cache lines\\
  TLB & 4K pages, 64-entry 4-way associative L1 (1 cycle),
  1024-entry L2 (10 cycles), TLB miss = 1000 cycles
  \\
  \tworow{L1 Cache} & 64KB, 4-way associative, tag/data latency = 1/2 cycles,
  parallel tag/data lookup, LRU policy
  \\
  \tworow{L2 Cache} & 512KB, 8-way associative,
  tag/data latency = 2/8 cycles,
  parallel tag/data lookup, LRU policy
  \\
  \tworow{Prefetcher} & Stream prefetcher~\cite{power6-prefetcher,fdp}, monitor L2 misses and prefetch into L3, 16 entries, degree = 4, distance = 24\\
  \tworow{L3 Cache} & 2MB, 16-way associative,
  tag/data latency = 10/24 cycles, serial tag/data lookup, DRRIP~\cite{rrip}
  policy\\
  \tworow{DRAM Controller} & Open row, FR-FCFS drain
  when full~\cite{dram-aware-wb}, 64-entry write buffer, 64-entry OMT cache, miss latency = 1000 cycles\\
  \tworow{DRAM and Bus} & DDR3-1066 MHz~\cite{ddr3}, 1 channel, 1 rank, 8
  banks, 8B-wide data bus, burst length = 8, 8KB row buffer\\
  \bottomrule
\end{tabular}

  \caption[Page Overlays: Simulation parameters]{Main parameters
    of our simulated system}
  \label{table:parameters}
\end{table}

\subsection{Overlay-on-write}
\label{sec:oow-evaluation}

As discussed in Section~\ref{sec:overlay-applications-oow},
overlay-on-write is a more efficient version of
copy-on-write~\cite{fork}: when multiple virtual pages share the
same physical page in the copy-on-write mode and one of them
receives a write, overlay-on-write simply moves the corresponding
cache line to the overlay and updates the cache line in the
overlay. 

We compare the performance of overlay-on-write with that of
copy-on-write using the \fork~\cite{fork} system call. \fork is a
widely-used system call with a number of different applications
including creating new processes, creating stateful threads in
multi-threaded applications, process
testing/debugging~\cite{flashback,self-test,hardware-bug}, and OS
speculation~\cite{os-speculation-1,os-speculation-2,os-speculation-3}. Despite
its wide applications, \fork is one of the most expensive system
calls~\cite{fork-exp}. When invoked, \fork creates a child process
with an identical virtual address space as the calling
process. \fork marks all the pages of both processes as
copy-on-write. As a result, when any such page receives a write,
the copy-on-write mechanism must copy the whole page and remap the
virtual page before it can proceed with the write.

Our evaluation models a scenario where a process is checkpointed
at regular intervals using the \fork system call. While we can
test the performance of fork with any application, we use a subset
of benchmarks from the SPEC CPU2006 benchmark
suite~\cite{spec2006}. Because the number of pages copied depends on
the \emph{write working set} of the application, we pick
benchmarks with three different types of write working sets:
1)~benchmarks with low write working set size, 2)~benchmarks for
which almost all cache lines within each modified page are
updated, and 3)~benchmarks for which only a few cache line within
each modified page are updated. We pick five benchmarks for each
type. For each benchmark, we fast forward the execution to its
representative portion (determined using
Simpoint~\cite{simpoints}), run the benchmark for 200 million
instructions (to warm up the caches), and execute a
\texttt{fork}. After the \texttt{fork}, we run the parent process
for another 300 million instructions, while the child process
idles.\footnote{While 300 million instructions might seem low,
  several prior works (e.g.,~\cite{self-test,hardware-bug}) argue
  for even shorter checkpoint intervals (10-100 million
  instructions).}

\begin{figure}[h]
  \centering
  \begin{minipage}{\linewidth}
    \centering
    \includegraphics[scale=1.2]{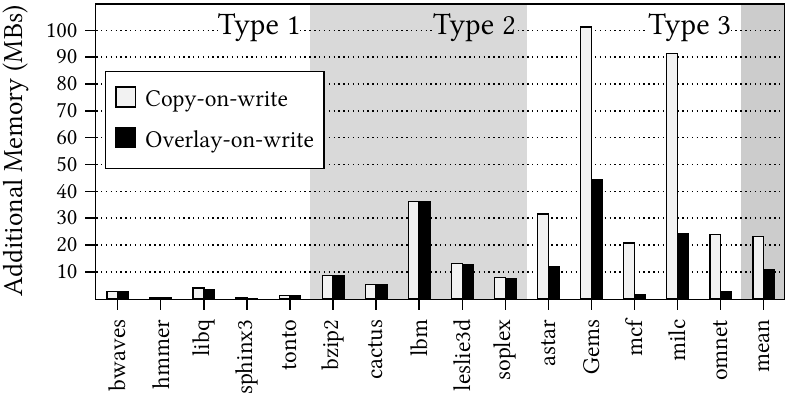}
    \caption[CoW vs. OoW: Additional memory consumption]{Additional memory consumed after a \fork}
    \label{plot:memory}
  \end{minipage}\vspace{2mm}
  \begin{minipage}{\linewidth}
    \centering
    \includegraphics[scale=1.2]{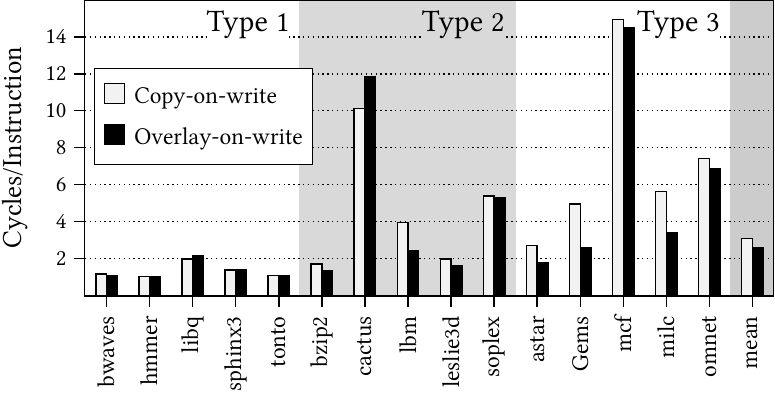}
    \caption[Cow vs. OoW: Performance]{Performance after a \fork
        (lower is better)}
    \label{plot:oow-perf}
  \end{minipage}
\end{figure}

Figure~\ref{plot:memory} plots the amount of additional memory
consumed by the parent process using copy-on-write and
overlay-on-write for the 300 million instructions after the \fork.
Figure~\ref{plot:oow-perf} plots the performance (cycles per
instruction) of the two mechanisms during the same period. We
group benchmarks based on their type. We draw three conclusions.

First, benchmarks with low write working set (Type 1) consume very
little additional memory after forking
(Figure~\ref{plot:memory}). As a result, there is not much
difference in the performance of copy-on-write and that of
overlay-on-write (Figure~\ref{plot:oow-perf}).

Second, for benchmarks of Type 2, both mechanisms consume almost
the same amount of additional memory. This is because for these
benchmarks, almost all cache lines within every modified page are
updated. However, with the exception of \bench{cactus},
overlay-on-write significantly improves performance for this type
of applications. Our analysis shows that the performance trends
can be explained by the distance in time when cache lines of each
page are updated by the application. When writes to different
cache lines within a page are close in time, copy-on-write
performs better than overlay-on-write. This is because
copy-on-write fetches \emph{all} the blocks of a page with high
memory-level parallelism. On the other hand, when writes to
different cache lines within a page are well separated in time,
copy-on-write may 1)~unnecessarily pollute the L1 cache with all
the cache lines of the copied page, and 2)~increase write
bandwidth by generating two writes for each updated cache line
(once when it is copied and again when the application updates the
cache line). Overlay-on-write has neither of these drawbacks, and
hence significantly improves performance over copy-on-write.

Third, for benchmarks of Type 3, overlay-on-write significantly
reduces the amount of additional memory consumed compared to
copy-on-write. This is because the write working set of these
applications are spread out in the virtual address space, and
copy-on-write unnecessarily copies cache lines that are actually
\emph{not} updated by the application. Consequently,
overlay-on-write significantly improves performance compared to
copy-on-write for this type of applications.

In summary, overlay-on-write reduces additional memory capacity
requirements by 53\% and improves performance by 15\% compared to
copy-on-write. Given the wide applicability of the \fork system
call, and the copy-on-write technique in general, we believe
overlay-on-write can significantly benefit a variety of such
applications.

\subsection{Representing Sparse Data Structures}
\label{sec:applications-sparse-data-structures}

A \emph{sparse} data structure is one with a significant fraction
of zero values, e.g., a sparse matrix. Since only non-zero values
typically contribute to computation, prior work developed many
software representations for sparse data structures
(e.g.,~\cite{yale-sm,bcsr-format}). One popular representation of
a sparse matrix is the Compressed Sparse Row (CSR)
format~\cite{bcsr-format}. To represent a sparse matrix, CSR
stores only the non-zero values in an array, and uses two arrays
of index pointers to identify the location of each non-zero value
within the matrix.

While CSR efficiently stores sparse matrices, the additional index
pointers maintained by CSR can result in inefficiency. First, the
index pointers lead to significant additional memory capacity
overhead (roughly 1.5 times the number of non-zero values in our
evaluation---each value is 8 bytes, and each index pointer is 4
bytes). Second, any computation on the sparse matrix requires
additional memory accesses to fetch the index pointers, which
degrades performance.

Our framework enables a very efficient hardware-based
representation for a sparse data structure: all virtual pages of
the data structure map to a zero physical page and each virtual
page is mapped to an overlay that contains only the \emph{non-zero
  cache lines} from that page. To avoid computation over zero
cache lines, we propose a new computation model that enables the
software to \emph{perform computation only on overlays}. When
overlays are used to represent sparse data structures, this model
enables the hardware to efficiently perform a computation only on
non-zero cache lines. Because the hardware is aware of the overlay
organization, it can efficiently prefetch the overlay cache lines
and hide the latency of memory accesses significantly.

Our representation stores non-zero data at a cache line
granularity. Hence, the performance and memory capacity benefits
of our representation over CSR depends on the spatial locality of
non-zero values within a cache line. To aid our analysis, we
define a metric called \emph{non-zero value locality} (\locality),
as the average number of non-zero values in each non-zero cache
line. On the one hand, when non-zero values have poor locality
(\locality $\approx$ 1), our representation will have to store a
significant number of zero values and perform redundant
computation over such values, degrading both memory capacity and
performance over CSR, which stores and performs computation on
only non-zero values. On the other hand, when non-zero values have
high locality (\locality $\approx$ 8---e.g., each cache line
stores 8 double-precision floating point values), our
representation is significantly more efficient than CSR as it
stores significantly less metadata about non-zero values than
CSR. As a result, it outperforms CSR both in terms of memory
capacity and performance.

\begin{figure}
  \centering
  \includegraphics[scale=1.2]{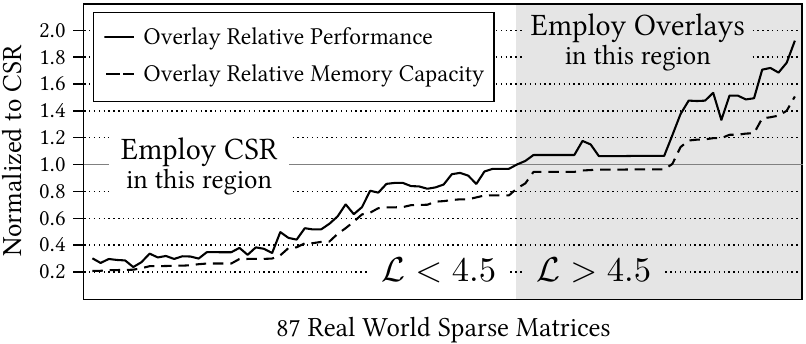}
  \caption[Overlays vs. CSR: SpMV performance and memory
    overhead]{SpMV multiplication: Performance of page overlays
    vs. CSR. \locality (non-zero value locality): Average \#
    non-zero values in each non-zero cache line.}\vspace{-1mm}
  \label{plot:perf-mem}
\end{figure}

We analyzed this trade-off using real-world sparse matrices of
double-precision floating point values obtained from the UF Sparse
Matrix Collection~\cite{ufspm}.  We considered all matrices with at
least 1.5 million non-zero values (87 in total).
Figure~\ref{plot:perf-mem} plots the memory capacity and
performance of one iteration of Sparse-Matrix Vector (SpMV)
multiplication of our mechanism normalized to CSR for each of
these matrices. The x-axis is sorted in the increasing order of
the \locality-value of the matrices.

The trends can be explained by looking at the extreme points. On
the left extreme, we have a matrix with \locality = 1.09 (\bench{poisson3Db}),
i.e., most non-zero cache lines have only one
non-zero value. As a result, our representation consumes 4.83
times more memory capacity and degrades performance by 70\%
compared to CSR. On the other extreme is a matrix with \locality = 8
(\bench{raefsky4}),
i.e., \emph{none} of the
non-zero cache lines have any zero value. As a result, our
representation is more efficient, reducing memory capacity by 34\%,
and improving performance by 92\% compared to CSR.

Our results indicate that even when a little more than half of the
values in each non-zero cache line are non-zero (\locality > 4.5),
overlays outperform CSR. For 34 of the 87 real-world matrices,
overlays reduce memory capacity by 8\% and improve
performance by 27\% on average compared to CSR.

In addition to the performance and memory capacity benefits, our
representation has several \textbf{other major advantages over
  CSR} (or any other software format). First, CSR is typically
helpful \emph{only} when the data structure is very sparse. In
contrast, our representation exploits a wider degree of sparsity
in the data structure. In fact, our simulations using
randomly-generated sparse matrices with varying levels of sparsity
(0\% to 100\%) show that our representation outperforms the
dense-matrix representation for all sparsity levels---the
performance gap increases linearly with the fraction of zero cache
lines in the matrix. Second, in our framework, dynamically
inserting non-zero values into a sparse matrix is as simple as
moving a cache line to the overlay. In contrast, CSR incur a high
cost to insert non-zero values. Finally, our computation model
enables the system to seamlessly use optimized dense matrix codes
on top of our representation. CSR, on the other hand, requires
programmers to rewrite algorithms to suit CSR.

\begin{figure}[h]
  \centering
  \includegraphics[scale=1.3]{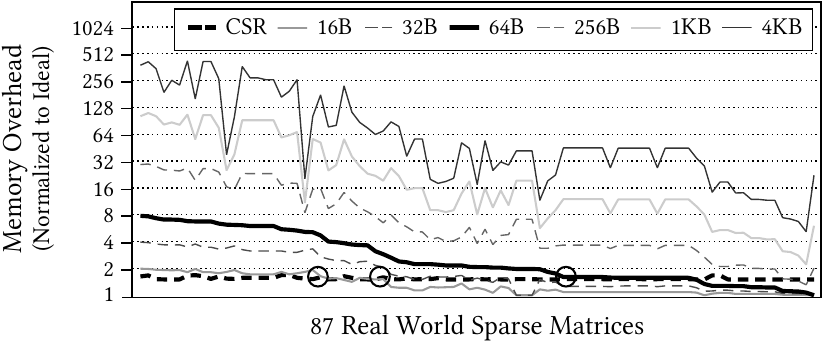}
  \caption[Overlays: Effect of cache line size]{Memory overhead
    of different cache line sizes over ``Ideal'' that stores only
    non-zero values. Circles indicate points where fine-grained
    management begins to outperform CSR.}
  \label{plot:block-sizes}
\end{figure}

\textbf{Sensitivity to Cache Line Size.} So far, we have described the
benefits of using overlays using 64B cache lines. However, one can
imagine employing our approach at a 4KB page granularity (i.e.,
storing only non-zero pages as opposed to non-zero cache
lines). To illustrate the benefits of fine-grained management, we
compare the memory overhead of storing the sparse matrices using
different cache line sizes (from 16B to
4KB). Figure~\ref{plot:block-sizes} shows the results. The memory
overhead for each cache line size is normalized to the ideal
mechanism which stores only the non-zero values. The matrices are
sorted in the same order as in Figure~\ref{plot:perf-mem}. We draw
two conclusions from the figure. First, while storing only
non-zero (4KB) pages may be a practical system to implement using
today's hardware, it increases the memory overhead by 53X on
average. It would also increase the amount of computation,
resulting in significant performance degradation. Hence, there is
significant benefit to the fine-grained memory management enabled
by overlays.  Second, the results show that a mechanism using a
finer granularity than 64B can outperform CSR on more matrices,
indicating a direction for future research on sub-block management
(e.g.,~\cite{amoeba}).

In summary, our overlay-based sparse matrix representation
outperforms the state-of-the-art software representation on many
real-world matrices, and consistently better than page-granularity
management. We believe our approach has much wider applicability
than existing representations.

\subsection{Other Applications of Our Framework}
\label{sec:other-applications}

We now describe five other applications that can be efficiently
implemented on top of our framework.  While prior works have
already proposed mechanisms for some of these applications, our
framework either enables a simpler mechanism or enables efficient
hardware support for mechanisms proposed by prior work. We
describe these mechanisms only at a high level, and defer more
detailed explanations to future work.

\subsubsection{Fine-grained Deduplication.}
\label{sec:applications-deduplication}

Gupta et al.~\cite{de} observe that in a system running multiple
virtual machines with the same guest operating system, there are a
number of pages that contain \emph{mostly same} data. Their
analysis shows that exploiting this redundancy can reduce memory
capacity requirements by 50\%. They propose the \emph{Difference
  Engine}, which stores such similar pages using small patches
over a common page. However, accessing such patched pages incurs
significant overhead because the OS must apply the patch before
retrieving the required data. Our framework enables a more
efficient implementation of the Difference Engine wherein cache
lines that are different from the base page can be stored in
overlays, thereby enabling seamless access to patched pages, while
also reducing the overall memory consumption.  Compared to
HICAMP~\cite{hicamp}, a cache line level deduplication mechanism
that locates cache lines based on their content, our framework
avoids significant changes to both the existing virtual memory
framework and programming model.

\subsubsection{Efficient Checkpointing.}
\label{sec:applications-checkpointing}

Checkpointing is an important primitive in high performance
computing applications where data structures are checkpointed at
regular intervals to avoid restarting long-running applications
from the beginning~\cite{hpc-survey,plfs,checkpointing}. However,
the frequency and latency of checkpoints are often limited by the
amount of memory data that needs to be written to the backing
store. With our framework, overlays could be used to capture
all the updates between two checkpoints.  Only these overlays
need to be written
to the backing store to take a new checkpoint,
reducing the
latency and bandwidth of checkpointing.  The overlays are then
committed (Section~\ref{sec:converting}), so that each checkpoint
captures precisely the delta since the last checkpoint.
In contrast to prior
works on efficient checkpointing such as INDRA~\cite{indra},
ReVive~\cite{revive}, and Sheaved Memory~\cite{sheaved-memory},
our framework is more flexible than INDRA and ReVive (which are
tied to recovery from remote attacks) and avoids the considerable
write amplification of Sheaved Memory (which can significantly
degrade overall system performance).

\subsubsection{Virtualizing Speculation.}
\label{sec:applications-speculation}

Several hardware-based speculative techniques (e.g., thread-level
speculation~\cite{tls,multiscalar}, transactional
memory~\cite{intel-htm,tm}) have been proposed to improve system
performance. Such techniques maintain speculative updates to
memory in the cache. As a result, when a speculatively-updated
cache line is evicted from the cache, these techniques must
necessarily declare the speculation as unsuccessful, resulting in
a potentially wasted opportunity. In our framework, these
techniques can store speculative updates to a virtual page in the
corresponding overlay. The overlay can be \emph{committed} or
\emph{discarded} based on whether the speculation succeeds or
fails. This approach is not limited by cache capacity and enables
potentially unbounded speculation~\cite{utm}.

\subsubsection{Fine-grained Metadata Management.}
\label{sec:applications-metadata}

Storing fine-grained (e.g., word granularity) metadata about data
has several applications (e.g., memcheck,
taintcheck~\cite{flexitaint}, fine-grained protection~\cite{mmp},
detecting lock violations~\cite{eraser}). Prior works
(e.g.,~\cite{ems,shadow-memory,mmp,flexitaint}) have proposed
frameworks to efficiently store and manipulate such
metadata. However, these mechanisms require hardware support
\emph{specific} to storing and maintaining metadata. In contrast,
with our framework, the system can potentially use overlays for
each virtual page to store metadata for the virtual page instead
of an alternate version of the data. In other words, \emph{the
  Overlay Address Space serves as shadow memory} for the virtual
address space. To access some piece of data, the application uses
the regular load and store instructions. The system would need new
\emph{metadata load} and \emph{metadata store} instructions to
enable the application to access the metadata from the overlays.

\subsubsection{Flexible Super-pages.}
\label{sec:applications-flexible-super-pages}

Many modern architectures support super-pages to reduce the number
of TLB misses. In fact, a recent prior work~\cite{direct-segment}
suggests that a single arbitrarily large super-page (direct
segment) can significantly reduce TLB misses for large
servers. Unfortunately, using super-pages reduces the flexibility
for the operating system to manage memory and implement techniques
like copy-on-write. For example, to our knowledge, there is no
system that shares a super-page across two processes in the
copy-on-write mode. This lack of flexibility introduces a
trade-off between the benefit of using super-pages to reduce TLB
misses and the benefit of using copy-on-write to reduce memory
capacity requirements.  Fortunately, with our framework, we can
apply overlays at higher-level page table entries to enable the OS
to manage super-pages at a finer granularity. In short, we
envision a mechanism that divides a super-page into smaller
segments (based on the number of bits available in the \obvect),
and allows the system to potentially remap a segment of the
super-page to the overlays. For example, when a super-page shared
between two processes receives a write, only the corresponding
segment is copied and the corresponding bit in the \obvect is
set. This approach can similarly be used to have multiple
protection domains within a super-page. Assuming only a few
segments within a super-page will require overlays, this approach
can still ensure low TLB misses while enabling more flexibility
for the OS.

\section{Summary}
\label{sec:overlays-summary}

In this chapter, we introduced a new, simple framework that
enables fine-grained memory management.  Our framework augments
virtual memory with a concept called \emph{overlays}. Each virtual
page can be mapped to both a physical page and an overlay. The
overlay contains only a subset of cache lines from the virtual
page, and cache lines that are present in the overlay are accessed
from there. We show that our proposed framework, with its simple
access semantics, enables several fine-grained memory management
techniques, without significantly altering the existing VM
framework. We quantitatively demonstrate the benefits of our
framework with two applications: 1)~\emph{overlay-on-write}, an
efficient alternative to copy-on-write, and 2)~an efficient
hardware representation of sparse data structures. Our evaluations
show that our framework significantly improves performance and
reduces memory capacity requirements for both applications (e.g.,
15\% performance improvement and 53\% memory capacity reduction,
on average, for \fork over traditional copy-on-write). Finally, we
discuss five other potential applications for the page overlays.

\chapter{Understanding DRAM}
\label{chap:dram-background}

In the second component of this dissertation, we propose a series
of techniques to improve the efficiency of certain key primitives
by exploiting the DRAM architecture. In this chapter, we will
describe the modern DRAM architecture and its implementation in
full detail. While we focus our attention primarily on commodity
DRAM design (i.e., the DDRx interface), most DRAM architectures
use very similar design approaches and vary only in higher-level
design choices. As a result, our mechanisms, which we describe in
the subsequent chapters, can be easily extended to any DRAM
architecture. We now describe the high-level organization of the
memory system.

\section{High-level Organization of the Memory System}

\begin{figure}[b]
  \centering
  \includegraphics{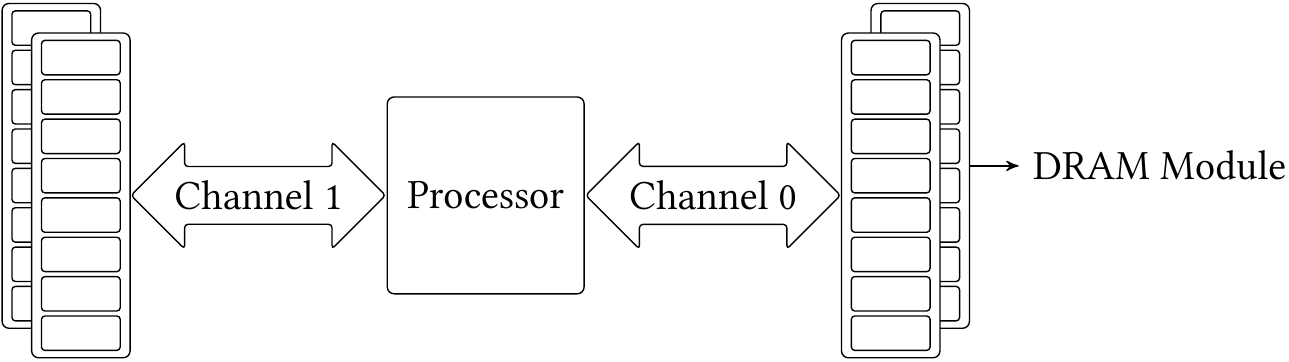}
  \caption{High-level organization of the memory subsystem}
  \label{fig:high-level-mem-org}
\end{figure}

Figure~\ref{fig:high-level-mem-org} shows the organization of the
memory subsystem in a modern system. At a high level, each
processor chip consists of one of more off-chip memory
\emph{channels}. Each memory channel consists of its own set of
\emph{command}, \emph{address}, and \emph{data} buses. Depending
on the design of the processor, there can be either an independent
memory controller for each memory channel or a single memory
controller for all memory channels. All modules connected to a
channel share the buses of the channel. Each module consists of
many DRAM devices (or chips). Most of this chapter
(Section~\ref{sec:dram-chip}) is dedicated to describing the
design of a modern DRAM chip. In Section~\ref{sec:dram-module}, we
present more details of the module organization of commodity DRAM.

\section{DRAM Chip}
\label{sec:dram-chip}

A modern DRAM chip consists of a hierarchy of structures: DRAM
\emph{cells}, \emph{tiles/MATs}, \emph{subarrays}, and
\emph{banks}. In this section, we will describe the design of a
modern DRAM chip in a bottom-up fashion, starting from a single
DRAM cell and its operation.

\subsection{DRAM Cell and Sense Amplifier}

\begin{figure}[h]
  \centering
  \includegraphics{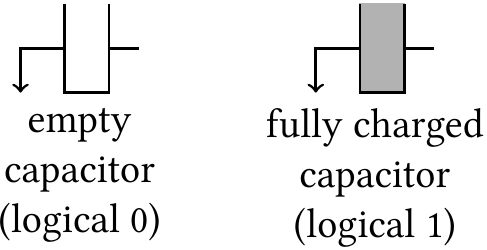}
  \caption[Capacitor]{Two states of a DRAM cell}
  \label{fig:cell-states}
\end{figure}

At the lowest level, DRAM technology uses capacitors to store
information. Specifically, it uses the two extreme states of a
capacitor, namely, the \emph{empty} and the \emph{fully charged}
states to store a single bit of information. For instance, an
empty capacitor can denote a logical value of 0, and a fully
charged capacitor can denote a logical value of 1.
Figure~\ref{fig:cell-states} shows the two extreme states of a
capacitor.

Unfortunately, the capacitors used for DRAM chips are small, and
will get smaller with each new generation. As a result, the amount
of charge that can be stored in the capacitor, and hence the
difference between the two states is also very small. In addition,
the capacitor can potentially lose its state after it is
accessed. Therefore, to extract the state of the capacitor, DRAM
manufactures use a component called \emph{sense amplifier}.

\begin{figure}[h]
  \centering
  \begin{minipage}{5cm}
    \centering
    \includegraphics{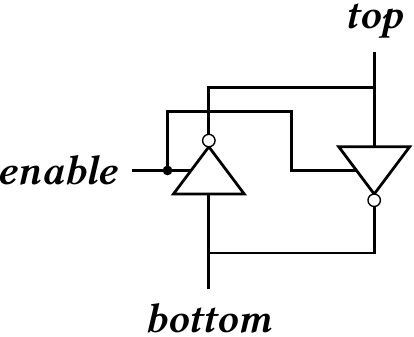}
    \caption{Sense amplifier}
    \label{fig:sense-amp}
  \end{minipage}\quad
  \begin{minipage}{9cm}
    \centering
    \includegraphics{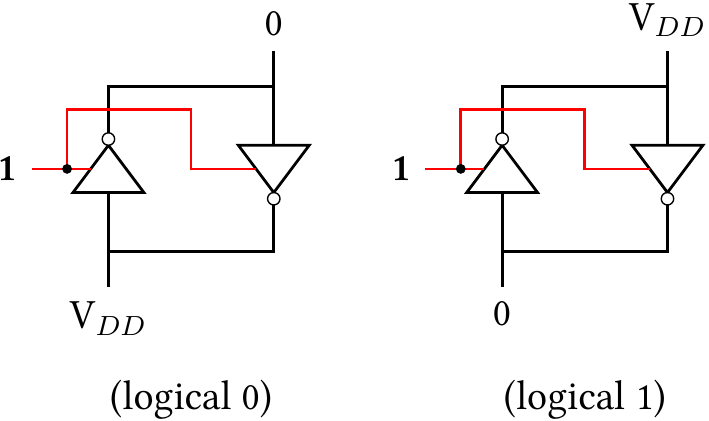}
    \caption{Stable states of a sense amplifier}
    \label{fig:sense-amp-states}
  \end{minipage}
\end{figure}

Figure~\ref{fig:sense-amp} shows a sense amplifier. A sense
amplifier contains two inverters which are connected together such
that the output of one inverter is connected to the input of the
other and vice versa. The sense amplifier also has an enable
signal that determines if the inverters are active. When enabled,
the sense amplifier has two stable states, as shown in
Figure~\ref{fig:sense-amp-states}. In both these stable states,
each inverter takes a logical value and feeds the other inverter
with the negated input.

\begin{figure}[h]
  \centering
  \includegraphics{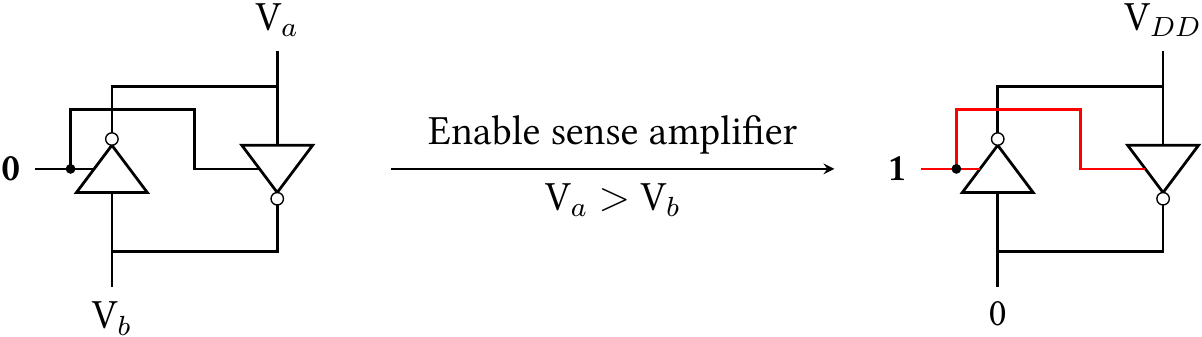}
  \caption{Operation of the sense amplifier}
  \label{fig:sense-amp-operation}
\end{figure}

Figure~\ref{fig:sense-amp-operation} shows the operation of the
sense amplifier from a disabled state. In the initial disabled
state, we assume that the voltage level of the top terminal
(V$_a$) is higher than that of the bottom terminal (V$_b$).
When the sense amplifier is enabled in this state, it
\emph{senses} the difference between the two terminals and
\emph{amplifies} the difference until it reaches one of the stable
state (hence the name ``sense amplifier'').  

\subsection{DRAM Cell Operation: The \texttt{ACTIVATE-PRECHARGE}
  cycle}
\label{sec:cell-operation}

\begin{figure}[h]
  \centering
  \includegraphics{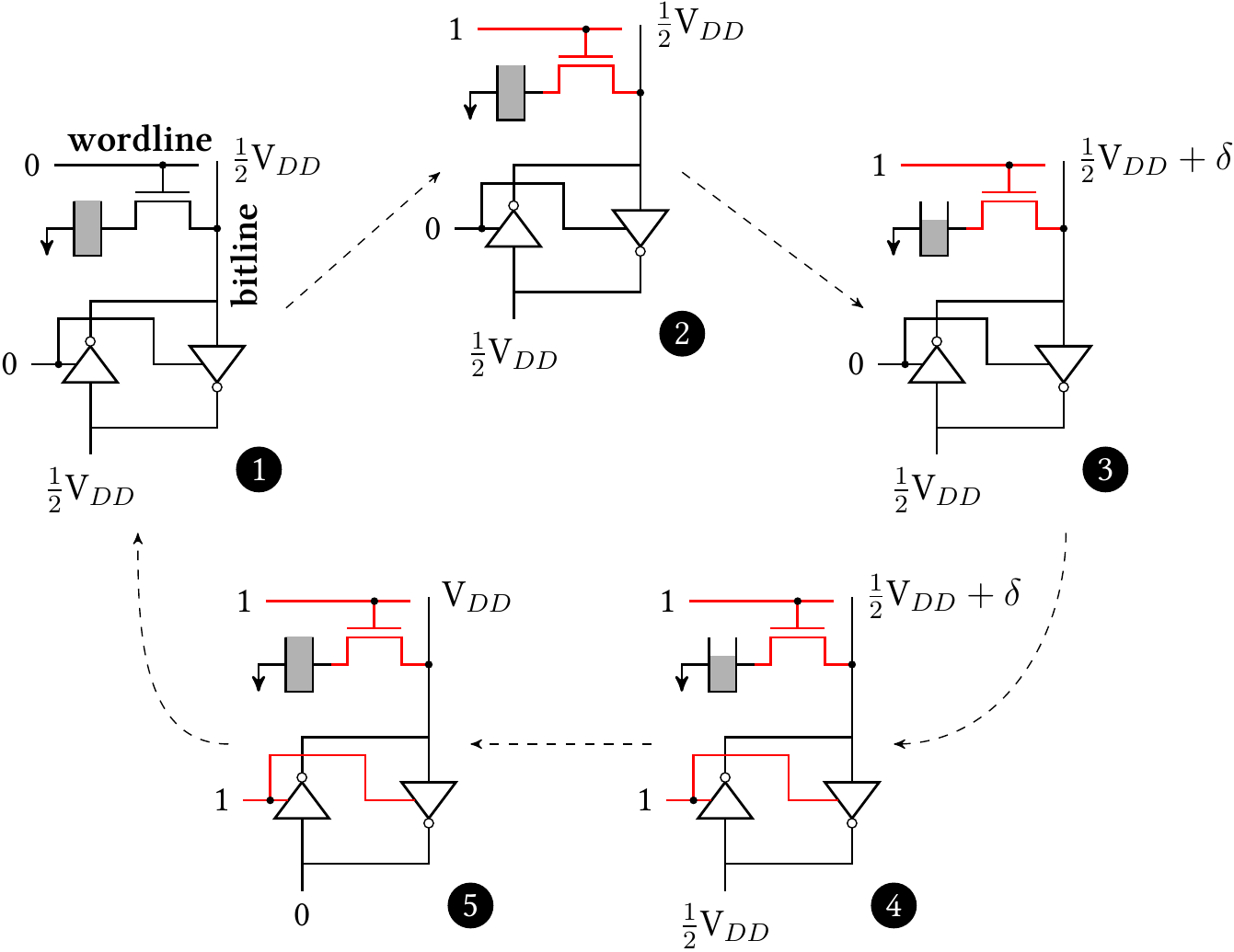}
  \caption{Operation of a DRAM cell and sense amplifier}
  \label{fig:cell-operation}
\end{figure}

DRAM technology uses a simple mechanism that converts the logical
state of a capacitor into a logical state of the sense
amplifier. Data can then be accessed from the sense amplifier
(since it is in a stable state). Figure~\ref{fig:cell-operation}
shows the connection between a DRAM cell and the sense amplifier
and the sequence of states involved in converting the cell state
to the sense amplifier state.

As shown in the figure (state \ding{202}), the capacitor is
connected to an access transistor that acts as a switch between
the capacitor and the sense amplifier. The transistor is
controller by a wire called \emph{wordline}. The wire that
connects the transistor to the top end of the sense amplifier is
called \emph{bitline}. In the initial state \ding{202}, the
wordline is lowered, the sense amplifier is disabled and both ends
of the sense amplifier are maintained at a voltage level of
\halfvdd. We assume that the capacitor is initially fully charged
(the operation is similar if the capacitor was empty). This state
is referred to as the \emph{precharged} state. An access to the
cell is triggered by a command called \cmdact. Upon receiving an
\cmdact, the corresponding wordline is first raised (state
\ding{203}). This connects the capacitor to the bitline. In the
ensuing phase called \emph{charge sharing} (state \ding{204}),
charge flows from the capacitor to the bitline, raising the
voltage level on the bitline (top end of the sense amplifier) to
\halfvddpd. After charge sharing, the sense amplifier is enabled
(state \ding{205}). The sense amplifier detects the difference in
voltage levels between its two ends and amplifies the deviation,
till it reaches the stable state where the top end is at \vdd
(state \ding{206}). Since the capacitor is still connected to the
bitline, the charge on the capacitor is also fully restored. We
will shortly describe how the data can be accessed form the sense
amplifier. However, once the access to the cell is complete, it is
taken back to the original precharged state using the command
called \cmdpre. Upon receiving a \cmdpre, the wordline is first
lowered, thereby disconnecting the cell from the sense
amplifier. Then, the two ends of the sense amplifier are driven to
\halfvdd using a precharge unit (not shown in the figure for
brevity).

\subsection{DRAM MAT/Tile: The Open Bitline Architecture}
\label{sec:dram-mat}

The goal of DRAM manufacturers is to maximize the density of the
DRAM chips while adhering to certain latency constraints
(described in Section~\ref{sec:dram-timing-constraints}). There
are two costly components in the setup described in the previous
section. The first component is the sense amplifier itself. Each
sense amplifier is around two orders of magnitude larger than a
single DRAM cell~\cite{rambus-power}. Second, the state of the
wordline is a function of the address that is currently being
accessed. The logic that is necessary to implement this function
(for each cell) is expensive.

\begin{figure}[h]
  \centering
  \includegraphics{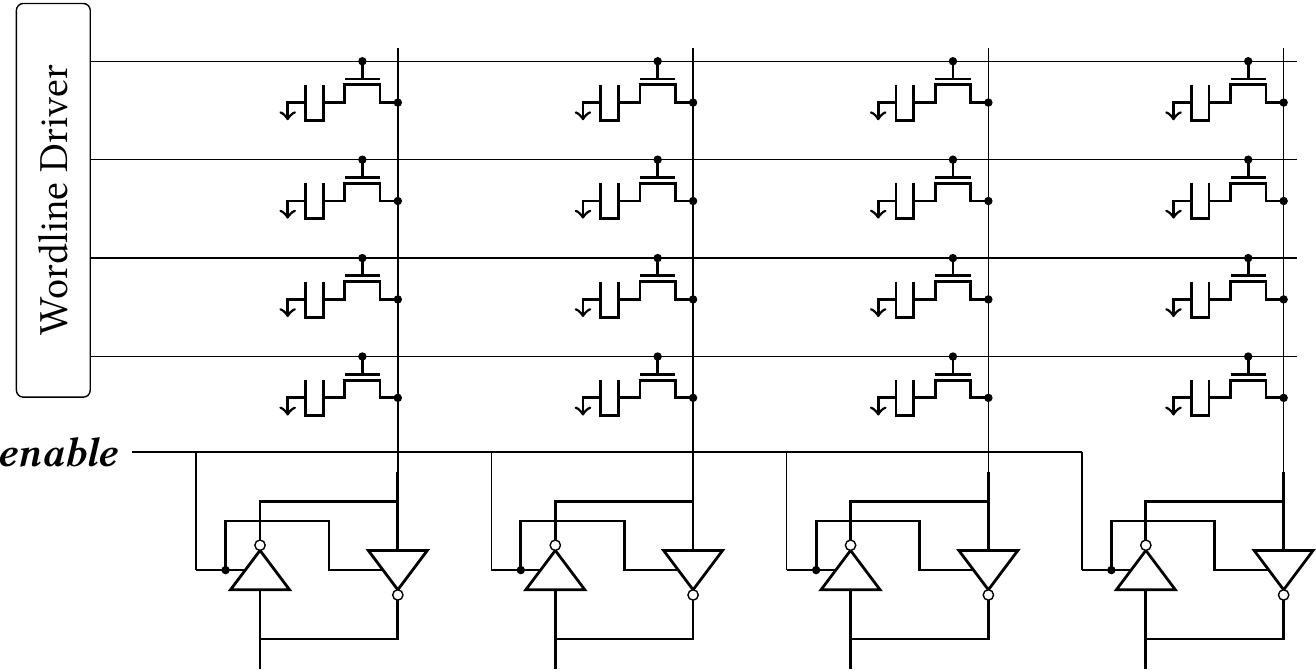}
  \caption{A 2-D array of DRAM cells}
  \label{fig:dram-mat}
\end{figure}

In order to reduce the overall cost of these two components, they
are shared by many DRAM cells. Specifically, each sense amplifier
is shared a column of DRAM cells. In other words, all the cells in
a single column are connected to the same bitline. Similarly, each
wordline is shared by a row of DRAM cells. Together, this
organization consists of a 2-D array of DRAM cells connected to a
row of sense amplifiers and a column of wordline
drivers. Figure~\ref{fig:dram-mat} shows this organization with a
$4 \times 4$ 2-D array.

\begin{figure}[h]
  \centering
  \includegraphics{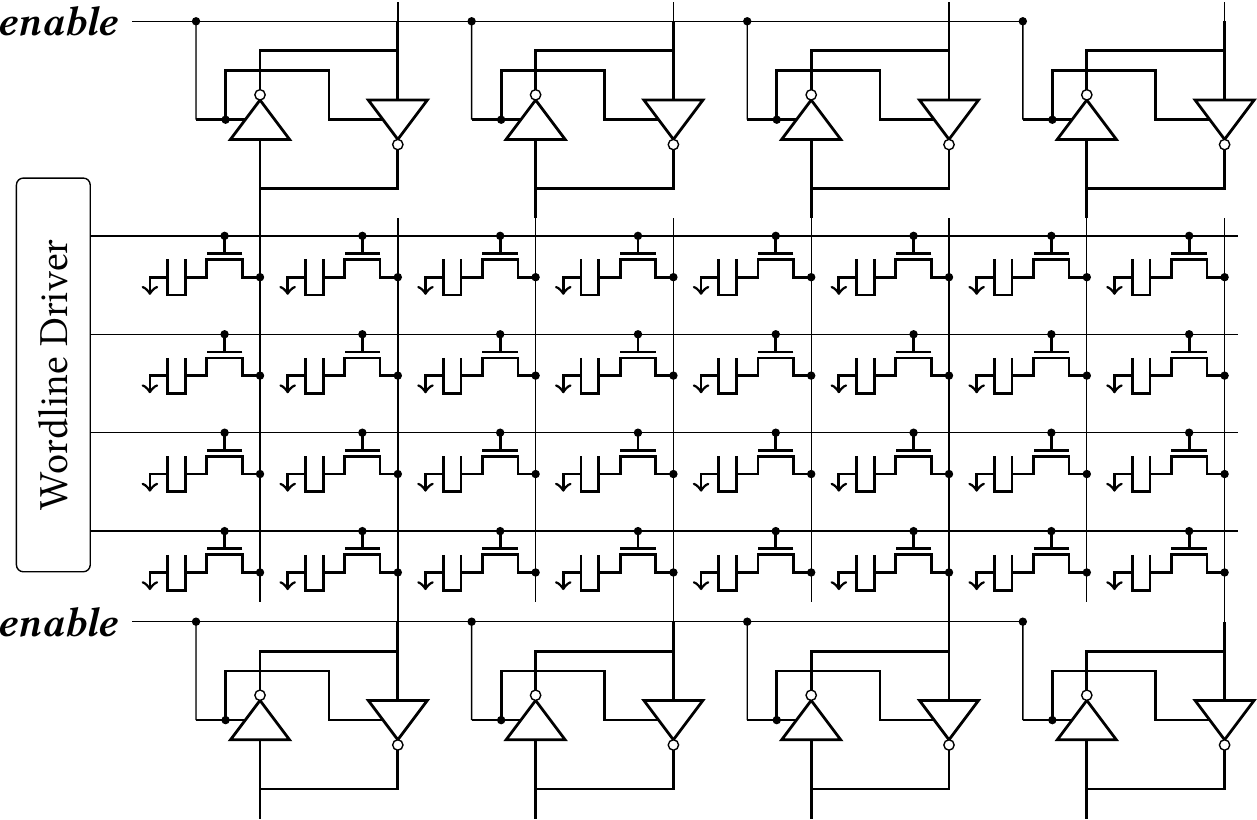}
  \caption{A DRAM MAT/Tile: The open bitline architecture}
  \label{fig:dram-mat-oba}
\end{figure}

To further reduce the overall cost of the sense amplifiers and the
wordline driver, modern DRAM chips use an architecture called the
\emph{open bitline architecture}. This architecture exploits two
observations. First, the sense amplifier is wider than the DRAM
cells. This difference in width results in a white space near each
column of cells. Second, the sense amplifier is
symmetric. Therefore, cells can also be connected to the bottom
part of the sense amplifier. Putting together these two
observations, we can pack twice as many cells in the same area
using the open bitline architecture, as shown in
Figure~\ref{fig:dram-mat-oba};

As shown in the figure, a 2-D array of DRAM cells is connected to
two rows of sense amplifiers: one on the top and one on the bottom
of the array. While all the cells in a given row share a common
wordline, half the cells in each row are connected to the top row
of sense amplifiers and the remaining half of the cells are
connected to the bottom row of sense amplifiers. This tightly
packed structure is called a DRAM
MAT/Tile~\cite{rethinking-dram,half-dram,salp}. In a modern DRAM
chip, each MAT typically is a $512 \times 512$ or $1024 \times
1024$ array. Multiple MATs are grouped together to form a larger
structure called a \emph{DRAM bank}, which we describe next.

\subsection{DRAM Bank}

\begin{figure}[h]
  \centering
  \includegraphics{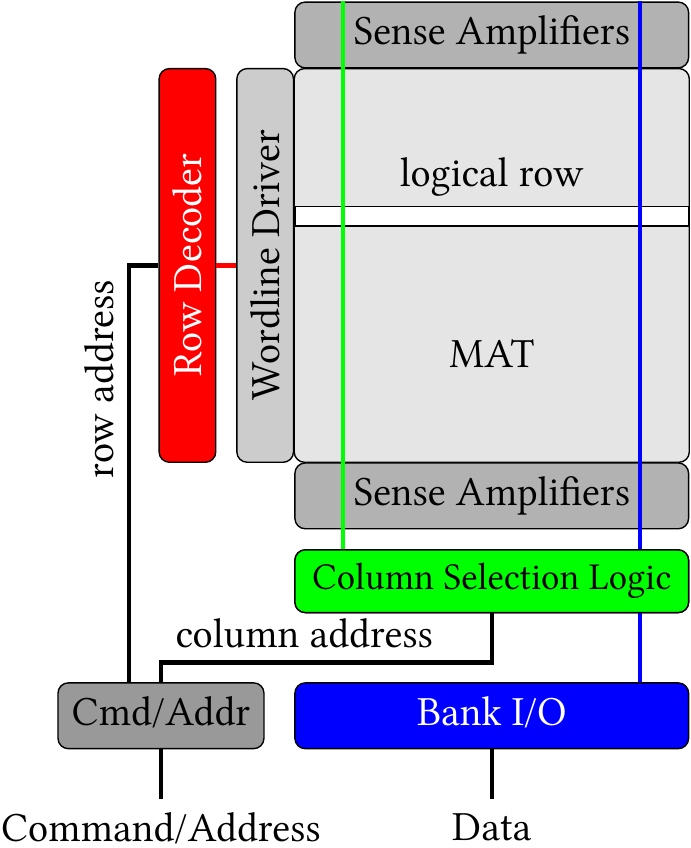}
  \caption{DRAM Bank: Logical view}
  \label{fig:dram-bank-logical}
\end{figure}

In most commodity DRAM interfaces~\cite{ddr3,ddr4}, a DRAM bank is
the smallest structure visible to the memory controller. All
commands related to data access are directed to a specific
bank. Logically, each DRAM bank is a large monolithic structure
with a 2-D array of DRAM cells connected to a single set of sense
amplifiers (also referred to as a row buffer). For example, in a
2Gb DRAM chip with 8 banks, each bank has $2^{15}$ rows and each
logical row has 8192 DRAM
cells. Figure~\ref{fig:dram-bank-logical} shows this logical view
of a bank.

In addition to the MAT, the array of sense amplifiers, and the
wordline driver, each bank also consists of some peripheral
structures to decode DRAM commands and addresses, and manage the
input/output to the DRAM bank. Specifically, each bank has a
\emph{row decoder} to decode the row address of row-level commands
(e.g., \cmdact). Each data access command (\cmdrd and \cmdwr)
accesses only a part of a DRAM row. Such individual parts are
referred to as \emph{columns}. With each data access command, the
address of the column to be accessed is provided. This address is
decoded by the \emph{column selection logic}. Depending on which
column is selected, the corresponding piece of data is
communicated between the sense amplifiers and the bank I/O
logic. The bank I/O logic intern acts as an interface between the
DRAM bank and the chip-level I/O logic.

\begin{figure}[h]
  \centering
  \includegraphics{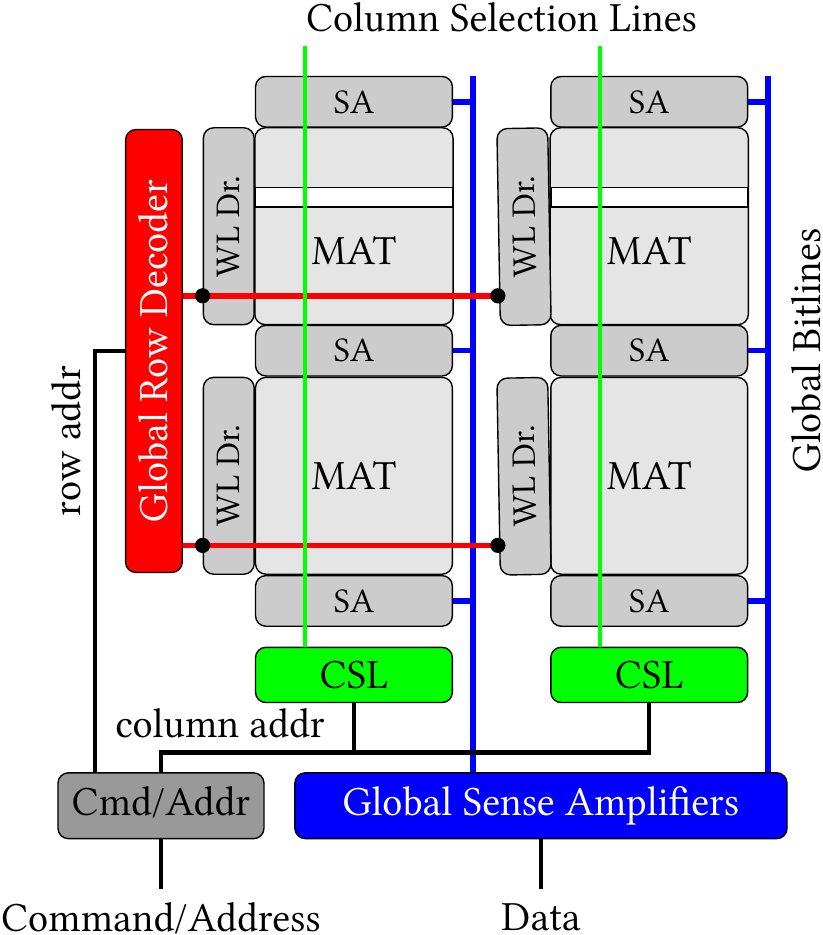}
  \caption{DRAM Bank: Physical implementation}
  \label{fig:dram-bank-physical}
\end{figure}

Although the bank can logically be viewed as a single MAT,
building a single MAT of a very large dimension is practically
not feasible as it will require very long bitlines and
wordlines. Therefore, each bank is physically implemented as a 2-D
array of DRAM MATs. Figure~\ref{fig:dram-bank-physical} shows a
physical implementation of the DRAM bank with 4 MATs arranged in
$2 \times 2$ array. As shown in the figure, the output of the
global row decoder is sent to each row of MATs.  The bank I/O
logic, also known as the \emph{global sense amplifiers}, are
connected to all the MATs through a set of \emph{global
  bitlines}. As shown in the figure, each vertical collection of
MATs consists of its own columns selection logic and global
bitlines. One implication of this division is that the data
accessed by any command is split equally across all the MATs in a
single row of MATs.

\begin{figure}[h]
  \centering
  \includegraphics{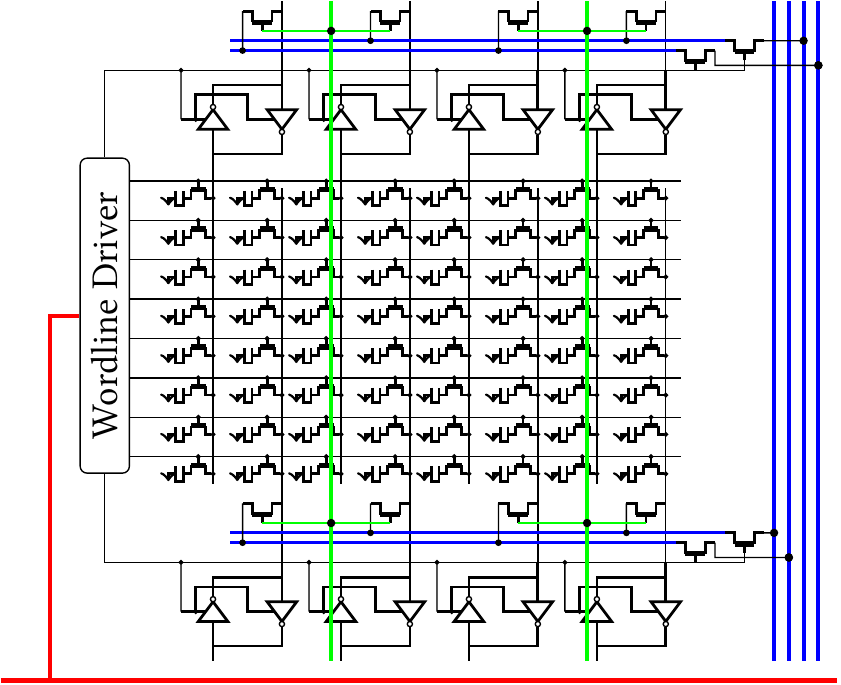}
  \caption{Detailed view of MAT}
  \label{fig:dram-mat-zoomed}
\end{figure}

Figure~\ref{fig:dram-mat-zoomed} shows the zoomed-in version of a
DRAM MAT with the surrounding peripheral logic. Specifically, the
figure shows how each column selection line selects specific sense
amplifiers from a MAT and connects them to the global bitlines. It
should be noted that the width of the global bitlines for each MAT
(typically 8/16) is much smaller than that of the width of the MAT
(typically 512/1024).

\begin{figure}[h]
  \centering
  \includegraphics{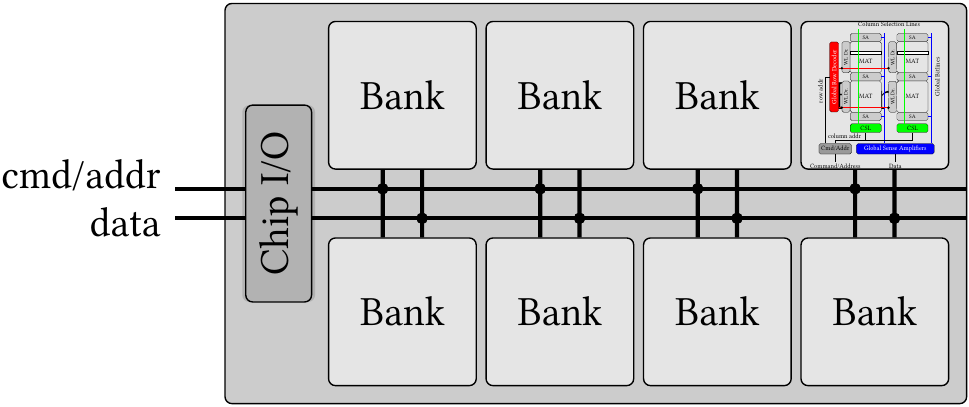}
  \caption{DRAM Chip}
  \label{fig:dram-chip}
\end{figure}

Each DRAM chip consist of multiple banks as shown in
Figure~\ref{fig:dram-chip}. All the banks share the chip's
internal command, address, and data buses. As mentioned before,
each bank operates mostly independently (except for operations
that involve the shared buses). The chip I/O manages the transfer
of data to and from the chip's internal bus to the channel. The
width of the chip output (typically 8 bits) is much smaller than
the output width of each bank (typically 64 bits). Any piece of
data accessed from a DRAM bank is first buffered at the chip I/O
and sent out on the memory bus 8 bits at a time. With the DDR
(double data rate) technology, 8 bits are sent out each half
cycle. Therefore, it takes 4 cycles to transfer 64 bits of data
from a DRAM chip I/O on to the memory channel.

\subsection{DRAM Commands: Accessing Data from a DRAM Chip}

To access a piece of data from a DRAM chip, the memory controller
must first identify the location of the data: the bank ID ($B$),
the row address ($R$) within the bank, and the column address
($C$) within the row. After identifying these pieces of
information, accessing the data involves three steps.

The first step is to issue a \cmdpre to the bank $B$. This step
prepares the bank for a data access by ensuring that all the sense
amplifiers are in the \emph{precharged} state
(Figure~\ref{fig:cell-operation}, state~\ding{202}). No wordline
within the bank is raised in this state.

The second step is to activate the row $R$ that contains the
data. This step is triggered by issuing a \cmdact to bank $B$ with
row address $R$. Upon receiving this command, the corresponding
bank feeds its global row decoder with the input $R$. The global
row decoder logic then raises the wordline of the DRAM row
corresponding to the address $R$ and enables the sense amplifiers
connected to that row. This triggers the DRAM cell operation
described in Section~\ref{sec:cell-operation}. At the end of the
activate operation the data from the entire row of DRAM cells is
copied to the corresponding array of sense amplifiers.

Finally, the third step is to access the data from the required
column. This is done by issuing a \cmdrd or \cmdwr command to the
bank with the column address $C$. Upon receiving a \cmdrd or
\cmdwr command, the corresponding address is fed to the column
selection logic. The column selection logic then raises the column
selection lines (Figure~\ref{fig:dram-mat-zoomed}) corresponding
the address $C$, thereby connecting those sense amplifiers to the
global sense amplifiers through the global bitlines. For a read
access, the global sense amplifiers sense the data from the MAT's
local sense amplifiers and transfer that data to chip's internal
bus. For a write access, the global sense amplifiers read the data
from the chip's internal bus and force the MAT's local sense
amplifiers to the appropriate state.

Not all data accesses require all three steps. Specifically, if
the row to be accessed is already activated in the corresponding
bank, then the first two steps can be skipped and the data can be
directly accessed by issuing a \cmdrd or \cmdwr to the bank. For
this reason, the array of sense amplifiers are also referred to as
a \emph{row buffer}, and such an access that skips the first two
steps is called a \emph{row buffer hit}. Similarly, if the bank is
already in the precharged state, then the first step can be
skipped. Such an access is referred to as a \emph{row buffer
  miss}. Finally, if a different row is activated within the bank,
then all three steps have to be performed. Such a situation is
referred to as a \emph{row buffer conflict}.

\subsection{DRAM Timing Constraints}
\label{sec:dram-timing-constraints}

Different operations within DRAM consume different amounts of
time. Therefore, after issuing a command, the memory controller
must wait for a sufficient amount of time before it can issue the
next command. Such wait times are managed by what are called the
\emph{timing constraints}. Timing constraints essentially dictate
the minimum amount of time between two commands issued to the same
bank/rank/channel. Table~\ref{table:timing-constraints} describes
some key timing constraints along with their values for the
DDR3-1600 interface.

\begin{table}[h]\small
  \centering
  \begin{tabular}{lrclp{2.2in}r}
  \toprule
  Name & \multicolumn{3}{c}{Constraint} & Description & Value (ns)\\
  \toprule
  \mrtwo{tRAS} & \mrtwo \cmdact & \mrtwo{$\rightarrow$} & \mrtwo \cmdpre & Time taken to complete a row
  activation operation in a bank & \mrtwo{35}\\
  \midrule
  \mrtwo{tRCD} & \mrtwo \cmdact & \mrtwo{$\rightarrow$} & \mrtwo {\cmdrd/\cmdwr} & Time between an activate
  command and column command to a bank & \mrtwo{15}\\
  \midrule
  \mrtwo{tRP} & \mrtwo \cmdpre & \mrtwo{$\rightarrow$} & \mrtwo \cmdact & Time taken to complete a precharge
  operation in a bank & \mrtwo{15}\\
  \midrule
  \mrthr{tWR} & \mrthr \cmdwr & \mrthr{$\rightarrow$} & \mrthr \cmdpre & Time taken to ensure that data is
  safely written to the DRAM cells after a write operation (\emph{write recovery}) & \mrthr{15}\\
  \bottomrule
\end{tabular}

  \caption[DDR3-1600 DRAM timing constraints]{Key DRAM timing constraints with their values for DDR3-1600}
  \label{table:timing-constraints}
\end{table}

\section{DRAM Module}
\label{sec:dram-module}

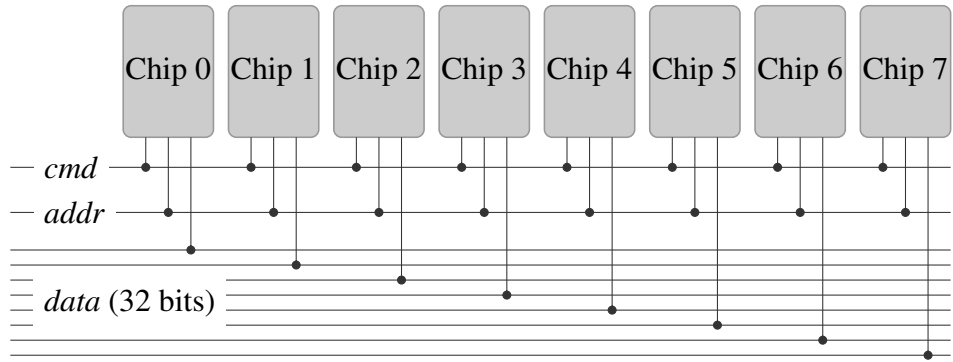
\begin{figure}
  \centering
  \begin{tikzpicture}[semithick]

  \tikzset{chip/.style={draw,rounded corners=3pt,black!40,fill=black!20,
     minimum width=1.2cm, minimum height=1.75cm,outer sep=0pt,anchor=west}};
  \tikzset{wire/.style={thin,black!80}};
  
  \node (chip0) [chip] at (0,0) {};
  \foreach \x [count=\i] in {0,...,6} {
    \node (chip\i) [chip, xshift=2mm] at (chip\x.east) {};
  }

  \foreach \i in {0,...,7} {
    \coordinate (cmd\i) at ($(chip\i.south) + (-0.3,0)$);
    \coordinate (addr\i) at (chip\i.south);
    \coordinate (data\i) at ($(chip\i.south) + (0.3,0)$);
  }

  \coordinate (cmdorigin) at ($(chip0.south west) + (-1.5,-0.4)$);
  \coordinate (addrorigin) at ($(chip0.south west) + (-1.5,-1)$);
  \draw [wire] (cmdorigin) -- (cmdorigin-|chip7.south east);
  \draw [wire] (addrorigin) -- (addrorigin-|chip7.south east);

  \foreach \i in {0,...,7} {
    \coordinate (dataorg\i) at ($(chip0.south west) + (-1.5,-1.5 - 0.2*\i)$);
    \draw [wire] (dataorg\i) -- (dataorg\i-|chip7.south east);
  }

  \foreach \i in {0,...,7} {
    \node at (chip\i) {Chip \i};
    \draw [wire,fill] (cmd\i) -- (cmd\i|-cmdorigin) circle(1.5pt);
    \draw [wire,fill] (addr\i) -- (addr\i|-addrorigin) circle(1.5pt);
    \draw [wire,fill] (data\i) -- (data\i|-dataorg\i) circle(1.5pt);
  }

  \node at (cmdorigin) [xshift=3mm,fill=white,anchor=west] {\emph{cmd}};
  \node at (addrorigin) [xshift=3mm,fill=white,anchor=west] {\emph{addr}};
  \node at ($(dataorg1)!0.5!(dataorg6)$) [xshift=3mm,fill=white,anchor=west] {\emph{data} (32 bits)};

\end{tikzpicture}
  \caption{Organization of a DRAM rank}
  \label{fig:dram-rank}
\end{figure}

As mentioned before, each \cmdrd or \cmdwr command for a single
DRAM chip typically involves only 64 bits. In order to achieve
high memory bandwidth, commodity DRAM modules group several DRAM
chips (typically 4 or 8) together to form a \emph{rank} of DRAM
chips. The idea is to connect all chips of a single rank to the
same command and address buses, while providing each chip with an
independent data bus. In effect, all the chips within a rank
receive the same commands with same addresses, making the rank a
logically wide DRAM chip. Figure~\ref{fig:dram-rank} shows the
logical organization of a DRAM rank.

Most commodity DRAM ranks consist of 8 chips. Therefore, each
\cmdrd or \cmdwr command accesses 64 bytes of data, the typical
cache line size in most processors.

\section{Summary}

In this section, we summarize the key takeaways of the DRAM design
and operation.
\begin{enumerate}\itemsep0pt
\item To access data from a DRAM cell, DRAM converts the state of
  the cell into one of the stable states of the sense
  amplifier. The \emph{precharged} state of the sense amplifier,
  wherein both the bitline and the \bbar are charged to a voltage
  level of \halfvdd, is key to this state transfer, as the DRAM
  cell is large enough to perturb the voltage level on the
  bitline.

\item The DRAM cell is not strong enough to switch the sense
  amplifier from one stable state to another. If a cell is
  connected to a stable sense amplifier, the charge on the cell
  gets overwritten to reflect the state of the sense amplifier.

\item In the DRAM cell operation, the final state of the sense
  amplifier after the amplification phase depends solely on the
  deviation on the bitline after \emph{charge sharing}. If the
  deviation is positive, the sense amplifier drives the bitline to
  \vdd. Otherwise, if the deviation is negative, the sense
  amplifier drives the bitline to 0.

\item In commodity DRAM, each \cmdact command simultaneously
  activates an entire row of DRAM cells. In a single chip, this
  typically corresponds to 8 Kbits of cells. Across a rank with 8
  chips, each \cmdact activates 8 KB of data.

\item In a commodity DRAM module, the data corresponding to each
  \cmdrd or \cmdwr is equally distributed across all the chips in
  a rank. All the chips share the same command and address bus,
  while each chip has an independent data bus.
\end{enumerate}
All our mechanisms are built on top of these observations. We will
recap these observations in the respective chapters.

\chapter{RowClone}
\label{chap:rowclone}

\begin{figure}[b!]
  \hrule\vspace{2mm}
  \begin{footnotesize}
    Originally published as ``RowClone: Fast and Energy-efficient
    In-DRAM Bulk Data Copy and Initialization'' in the
    International Symposium on Microarchitecture,
    2013~\cite{rowclone}
  \end{footnotesize}
\end{figure}

In Section~\ref{sec:cow-problems}, we described the source of
inefficiency in performing a page copy operation in existing
systems. Briefly, in existing systems, a page copy operation (or
any bulk copy operation) is at best performed one cache line at a
time. The operation requires a large number of cache lines to be
transferred back and forth on the main memory channel. As a
result, a bulk copy operation incurs high latency, high bandwidth
consumption, and high energy consumption. 

In this chapter, we present RowClone, a mechanism that can perform
bulk copy and initialization operations completely inside DRAM. We
show that this approach obviates the need to transfer large
quantities of data on the memory channel, thereby significantly
improving the efficiency of a bulk copy operation. As bulk data
initialization (specifically bulk zeroing) can be viewed as a
special case of a bulk copy operation, RowClone can be easily
extended to perform such bulk initialization operations with high
efficiency.

\section{The RowClone DRAM Substrate}
\label{sec:rowclone-overview}

RowClone consists of two independent mechanisms that exploit
several observations about DRAM organization and operation.  Our
first mechanism efficiently copies data between two rows of DRAM
cells that share the same set of sense amplifiers (i.e., two rows
within the same subarray). We call this mechanism the \emph{Fast
  Parallel Mode} (FPM). Our second mechanism efficiently copies
cache lines between two banks within a module in a pipelined
manner. We call this mechanism the \emph{Piplines Serial Mode}
(PSM). Although not as fast as FPM, PSM has fewer constraints and
hence is more generally applicable. We now describe these two
mechanism in detail.

\subsection{Fast-Parallel Mode}
\label{sec:rowclone-fpm}

The Fast Parallel Mode (FPM) is based on the following three
observations about DRAM.
\begin{enumerate}
\item In a commodity DRAM module, each \cmdact command transfers
  data from a large number of DRAM cells (multiple kilo-bytes) to
  the corresponding array of sense amplifiers
  (Section~\ref{sec:dram-module}).
\item Several rows of DRAM cells share the same set of sense
  amplifiers (Section~\ref{sec:dram-mat}).
\item A DRAM cell is not strong enough to flip the state of the
  sense amplifier from one stable state to another stable
  state. In other words, if a cell is connected to an already
  activated sense amplifier (or bitline), then the data of the
  cell gets overwritten with the data on the sense amplifier.
\end{enumerate}

\begin{figure}[h]
  \centering
  \includegraphics{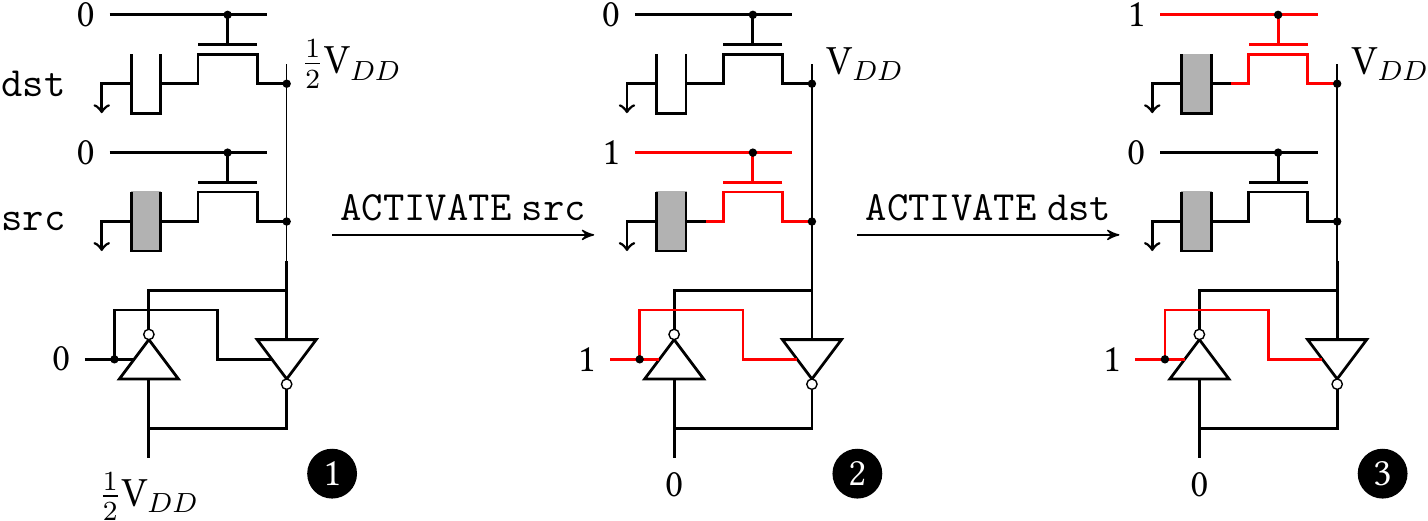}
  \caption{RowClone: Fast Parallel Mode}
  \label{fig:cell-fpm}
\end{figure}

While the first two observations are direct implications from the
design of commodity DRAM, the third observation exploits the fact
that DRAM cells are large enough to cause only a small
perturbation on the bitline voltage. Figure~\ref{fig:cell-fpm}
pictorially shows how this observation can be used to copy data
between two cells that share a sense amplifier.

The figure shows two cells (\src and \dst) connected to a single
sense amplifier. In the initial state, we assume that \src is
fully charged and \dst is fully empty, and the sense amplifier is
in the precharged state (\ding{202}). In this state, FPM issues an
\cmdact to \src. At the end of the activation operation, the sense
amplifier moves to a stable state where the bitline is at a
voltage level of \vdd and the charge in \src is fully restored
(\ding{203}). FPM follows this operation with an \cmdact to \dst,
without an intervening \cmdpre. This operation lowers the wordline
corresponding to \src and raises the wordline of \dst, connecting
\dst to the bitline.  Since the bitline is already fully
activated, even though \dst is initially empty, the perturbation
caused by the cell is not sufficient to flip the state of the
bitline. As a result, sense amplifier continues to drive the
bitline to \vdd, thereby pushing \dst to a fully charged state
(\ding{204}).

It can be shown that regardless of the initial state of \src and
\dst, the above operation copies the data from \src to
\dst. Given that each \cmdact operates on an entire row of DRAM
cells, the above operation can copy multiple kilo bytes of data
with just two back-to-back \cmdact operations.

Unfortunately, modern DRAM chips do not allow another \cmdact to
an already activated bank -- the expected result of such an action
is undefined. This is because a modern DRAM chip allows at most
one row (subarray) within each bank to be activated.  If a bank
that already has a row (subarray) activated receives an \cmdact to
a different subarray, the currently activated subarray must first
be precharged~\cite{salp}.\footnote{Some DRAM manufacturers design
  their chips to drop back-to-back {\cmdact}s to the same bank.}

To support FPM, we propose the following change to the DRAM chip
in the way it handles back-to-back {\cmdact}s. When an already
activated bank receives an \cmdact to a row, the chip processes
the command similar to any other \cmdact if and only if the
command is to a row that belongs to the currently activated
subarray. If the row does not belong to the currently activated
subarray, then the chip takes the action it normally does with
back-to-back {\cmdact}s---e.g., drop it.  Since the logic to
determine the subarray corresponding to a row address is already
present in today's chips, implementing FPM only requires a
comparison to check if the row address of an \cmdact belongs to
the currently activated subarray, the cost of which is almost
negligible.

\textbf{Summary.} To copy data from \src to \dst within the same
subarray, FPM first issues an \cmdact to \src. This copies the
data from \src to the subarray row buffer. FPM then issues an
\cmdact to \dst. This modifies the input to the subarray
row-decoder from \src to \dst and connects the cells of \dst row
to the row buffer. This, in effect, copies the data from the sense
amplifiers to the destination row.  As we show in
Section~\ref{sec:rowclone-analysis}, with these two steps, FPM
copies a 4KB page of data 11.6x faster and with 74.4x less energy
than an existing system.

\textbf{Limitations.} FPM has two constraints that limit its
general applicability. First, it requires the source and
destination rows to be within the same subarray (i.e., share the
same set of sense amplifiers). Second, it cannot partially copy
data from one row to another. Despite these limitations, we show
that FPM can be immediately applied to today's systems to
accelerate two commonly used primitives in modern systems --
Copy-on-Write and Bulk Zeroing (Section~\ref{sec:applications}).
In the following section, we describe the second mode of RowClone
-- the Pipelined Serial Mode (PSM). Although not as fast or
energy-efficient as FPM, PSM addresses these two limitations of
FPM.

\subsection{Pipelined Serial Mode}
\label{sec:rowclone-psm}

The Pipelined Serial Mode efficiently copies data from a source
row in one bank to a destination row in a \emph{different}
bank. PSM exploits the fact that a single internal bus that is
shared across all the banks is used for both read and write
operations. This enables the opportunity to copy an arbitrary
quantity of data one cache line at a time from one bank to another
in a pipelined manner.

To copy data from a source row in one bank to a destination row in
a different bank, PSM first activates the corresponding rows in
both banks. It then puts the source bank in the read mode, the
destination bank in the write mode, and transfers data one cache
line (corresponding to a column of data---64 bytes) at a time. For
this purpose, we propose a new DRAM command called \cmdtr. The
\cmdtr command takes four parameters: 1)~source bank index,
2)~source column index, 3)~destination bank index, and
4)~destination column index. It copies the cache line
corresponding to the source column index in the activated row of
the source bank to the cache line corresponding to the destination
column index in the activated row of the destination bank.

\begin{figure}[h]
  \centering
  \includegraphics[angle=90]{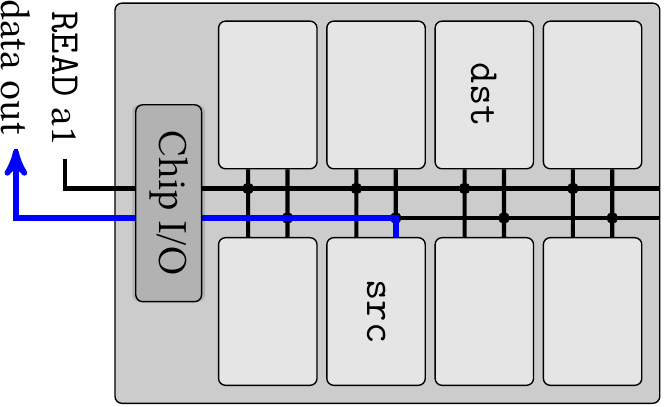}
  \includegraphics[angle=90]{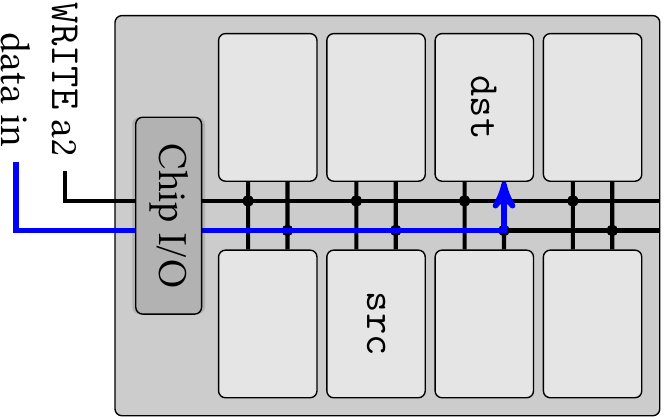}
  \includegraphics[angle=90]{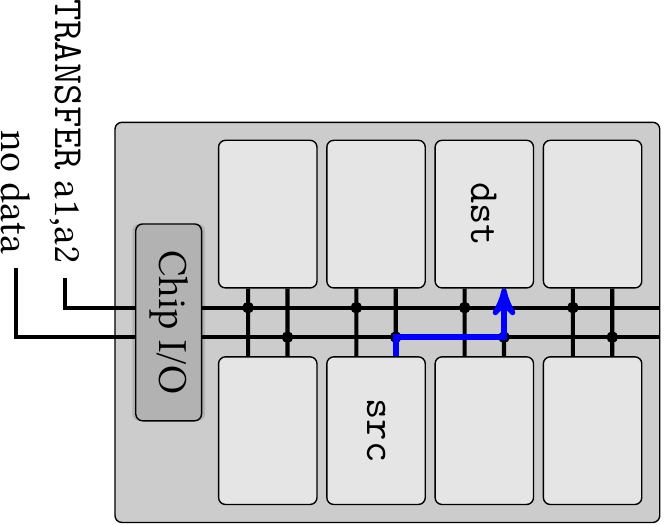}
  \caption{RowClone: Pipelined Serial Mode}
  \label{fig:psm}
\end{figure}

Unlike \cmdrd/\cmdwr which interact with the memory channel
connecting the processor and main memory, \cmdtr does not transfer
data outside the chip. Figure~\ref{fig:psm} pictorially compares
the operation of the \cmdtr command with that of \cmdrd and
\cmdwr. The dashed lines indicate the data flow corresponding to
the three commands. As shown in the figure, in contrast to the
\cmdrd or \cmdwr commands, \cmdtr does not transfer data from or
to the memory channel.

\subsection{Mechanism for Bulk Data Copy}
\label{sec:bulk-copy}

When the data from a source row (\src) needs to be copied to a
destination row (\dst), there are three possible cases depending
on the location of \src and \dst: 1)~\src and \dst are within the
same subarray, 2)~\src and \dst are in different banks, 3)~\src
and \dst are in different subarrays within the same bank. For case
1 and case 2, RowClone uses FPM and PSM, respectively, to complete
the operation (as described in Sections~\ref{sec:rowclone-fpm} and
\ref{sec:rowclone-psm}).

For the third case, when \src and \dst are in different subarrays
within the same bank, one can imagine a mechanism that uses the
global bitlines (shared across all subarrays within a bank --
described in \cite{salp}) to copy data across the two rows in
different subarrays. However, we do not employ such a mechanism
for two reasons.  First, it is not possible in today's DRAM chips
to activate multiple subarrays within the same bank
simultaneously.  Second, even if we enable simultaneous activation
of multiple subarrays, as in~\cite{salp}, transferring data from
one row buffer to another using the global bitlines requires the
bank I/O circuitry to switch between read and write modes for each
cache line transfer. This switching incurs significant latency
overhead.  To keep our design simple, for such an intra-bank copy
operation, our mechanism uses PSM to first copy the data from \src
to a temporary row (\tmp) in a different bank. It then uses PSM
again to copy the data back from \tmp to \dst. The capacity lost
due to reserving one row within each bank is negligible (0.0015\%
for a bank with 64k rows).

\subsection{Mechanism for Bulk Data Initialization}
\label{sec:bulk-initialization}

Bulk data initialization sets a large block of memory to a
specific value. To perform this operation efficiently, our
mechanism first initializes a single DRAM row with the
corresponding value. It then uses the appropriate copy mechanism
(from Section~\ref{sec:bulk-copy}) to copy the data to the other
rows to be initialized.

Bulk Zeroing (or BuZ), a special case of bulk initialization, is a
frequently occurring operation in today's
systems~\cite{bulk-copy-initialize,why-nothing-matters}. To
accelerate BuZ, one can reserve one row in each subarray that is
always initialized to zero. By doing so, our mechanism can use FPM
to efficiently BuZ any row in DRAM by copying data from the
reserved zero row of the corresponding subarray into the
destination row. The capacity loss of reserving one row out of 512
rows in each subarray is very modest (0.2\%).

While the reserved rows can potentially lead to gaps in the
physical address space, we can use an appropriate memory
interleaving technique that maps consecutive rows to different
subarrays. Such a technique ensures that the reserved zero rows
are contiguously located in the physical address space. Note that
interleaving techniques commonly used in today's systems (e.g.,
row or cache line interleaving) have this property.

\section{End-to-end System Design}
\label{sec:system-design}

So far, we described RowClone, a DRAM substrate that can
efficiently perform bulk data copy and initialization. In this
section, we describe the changes to the ISA, the processor
microarchitecture, and the operating system that will enable the
system to efficiently exploit the RowClone DRAM substrate.

\subsection{ISA Support}
\label{sec:isa-changes}

To enable the software to communicate occurrences of bulk copy and
initialization operations to the hardware, we introduce two new
instructions to the ISA: \mcpy and \minit.
Table~\ref{tab:isa-semantics} describes the semantics of these two new
instructions.  We deliberately keep the semantics of the instructions
simple in order to relieve the software from worrying about
microarchitectural aspects of RowClone such as row size, alignment,
etc.~(discussed in Section~\ref{sec:offset-alignment-size}). Note that
such instructions are already present in some of the instructions sets
in modern processors -- e.g., \texttt{rep movsd}, \texttt{rep stosb},
\texttt{ermsb} in x86~\cite{x86-ermsb} and \texttt{mvcl} in IBM
S/390~\cite{s390}.

\begin{table}[h]
  \centering
  \begin{tabular}{lll}
  \toprule
  Instruction & Operands & Semantics\\
  \midrule
  \mcpy & \emph{src, dst, size} & Copy \emph{size} bytes from
  \emph{src} to \emph{dst}\\
  \minit & \emph{dst, size, val} & Set \emph{size} bytes to
  \emph{val} at \emph{dst}\\
  \bottomrule
\end{tabular}

  \caption{Semantics of the \mcpy and \minit instructions}
  \label{tab:isa-semantics}
\end{table}

There are three points to note regarding the execution semantics
of these operations. First, the processor does not guarantee
atomicity for both \mcpy and \minit, but note that existing
systems also do not guarantee atomicity for such
operations. Therefore, the software must take care of atomicity
requirements using explicit synchronization. However, the
microarchitectural implementation will ensure that any data in the
on-chip caches is kept consistent during the execution of these
operations (Section~\ref{sec:rowclone-cache-coherence}). Second, the
processor will handle any page faults during the execution of
these operations. Third, the processor can take interrupts during
the execution of these operations.

\subsection{Processor Microarchitecture Support}
\label{sec:uarch-changes}

The microarchitectural implementation of the new instructions, \mcpy
and \minit, has two parts. The first part determines if a particular
instance of \mcpy or \minit can be fully/partially accelerated by
RowClone. The second part involves the changes required to the cache
coherence protocol to ensure coherence of data in the on-chip
caches. We discuss these parts in this section.

\subsubsection{Source/Destination Alignment and Size}
\label{sec:offset-alignment-size}

For the processor to accelerate a copy/initialization operation
using RowClone, the operation must satisfy certain alignment and
size constraints. Specifically, for an operation to be accelerated
by FPM, 1)~the source and destination regions should be within the
same subarray, 2)~the source and destination regions should be
row-aligned, and 3)~the operation should span an entire row. On
the other hand, for an operation to be accelerated by PSM, the
source and destination regions should be cache line-aligned and the
operation must span a full cache line.

Upon encountering a \mcpy/\minit instruction, the processor
divides the region to be copied/initialized into three portions:
1)~row-aligned row-sized portions that can be accelerated using
FPM, 2)~cache line-aligned cache line-sized portions that can be
accelerated using PSM, and 3)~the remaining portions that can be
performed by the processor. For the first two regions, the
processor sends appropriate requests to the memory controller
which completes the operations and sends an acknowledgment back to
the processor. Since \cmdtr copies only a single cache line, a
bulk copy using PSM can be interleaved with other commands to
memory. The processor completes the operation for the third region
similarly to how it is done in today's systems. Note that the CPU
can offload all these operations to the memory controller. In such
a design, the CPU need not be made aware of the DRAM organization
(e.g., row size and alignment, subarray mapping, etc.).

\subsubsection{Managing On-Chip Cache Coherence}
\label{sec:rowclone-cache-coherence}

RowClone allows the memory controller to directly read/modify data
in memory without going through the on-chip caches. Therefore, to
ensure cache coherence, the controller appropriately handles cache
lines from the source and destination regions that may be present
in the caches before issuing the copy/initialization operations to
memory.

First, the memory controller writes back any dirty cache line from
the source region as the main memory version of such a cache line
is likely stale. Copying the data in-memory before flushing such
cache lines will lead to stale data being copied to the
destination region. Second, the controller invalidates any cache
line (clean or dirty) from the destination region that is cached
in the on-chip caches. This is because after performing the copy
operation, the cached version of these blocks may contain stale
data. The controller already has the ability to perform such
flushes and invalidations to support Direct Memory Access
(DMA)~\cite{intel-dma}. After performing the necessary flushes and
invalidations, the memory controller performs the
copy/initialization operation. To ensure that cache lines of the
destination region are not cached again by the processor in the
meantime, the memory controller blocks all requests (including
prefetches) to the destination region until the copy or
initialization operation is complete.

While performing the flushes and invalidates as mentioned above
will ensure coherence, we propose a  modified solution to
handle dirty cache lines of the source region to reduce memory
bandwidth consumption. When the memory controller identifies a
dirty cache line belonging to the source region while performing a
copy, it creates an in-cache copy of the source cache line with
the tag corresponding to the destination cache line. This has two
benefits. First, it avoids the additional memory flush required
for the dirty source cache line. Second and more importantly, the
controller does not have to wait for all the dirty source cache
lines to be flushed before it can perform the copy. In
Section~\ref{res:rowclone-ii-apps}, we will consider another
optimization, called RowClone-Zero-Insert, which inserts clean
zero cache lines into the cache to further optimize Bulk Zeroing.
This optimization does not require further changes to our proposed
modifications to the cache coherence protocol.

Although RowClone requires the controller to manage cache
coherence, it does not affect memory consistency --- i.e.,
concurrent readers or writers to the source or destination regions
involved in a bulk copy or initialization operation. As mentioned
before, such an operation is not guaranteed to be atomic even in
current systems, and the software needs to perform the operation
within a critical section to ensure atomicity.

\subsection{Software Support}
\label{sec:os-changes}

The minimum support required from the system software is the use
of the proposed \mcpy and \minit instructions to indicate bulk
data operations to the processor. Although one can have a working
system with just this support, maximum latency and energy benefits
can be obtained if the hardware is able to accelerate most copy
operations using FPM rather than PSM. Increasing the likelihood of
the use of the FPM mode requires further support from the
operating system (OS) on two aspects: 1)~page mapping, and
2)~granularity of copy/initialization.

\subsubsection{Subarray-Aware Page Mapping}
\label{sec:subarray-awareness}

The use of FPM requires the source row and the destination row of a
copy operation to be within the same subarray. Therefore, to maximize
the use of FPM, the OS page mapping algorithm should be aware of
subarrays so that it can allocate a destination page of a copy
operation in the same subarray as the source page. More specifically,
the OS should have knowledge of which pages map to the same subarray
in DRAM. We propose that DRAM expose this information to software
using the small EEPROM that already exists in today's DRAM
modules. This EEPROM, called the Serial Presence Detect
(SPD)~\cite{spd}, stores information about the DRAM chips that is read
by the memory controller at system bootup. Exposing the subarray
mapping information will require only a few additional bytes to
communicate the bits of the physical address that map to the subarray
index.\footnote{To increase DRAM yield, DRAM manufacturers design
  chips with spare rows that can be mapped to faulty
  rows~\cite{spare-row-mapping}. Our mechanism can work with this technique
  by either requiring that each faulty row is remapped to a spare row
  within the same subarray, or exposing the location of all faulty
  rows to the memory controller so that it can use PSM to copy data
  across such rows.}

Once the OS has the mapping information between physical pages and
subarrays, it maintains multiple pools of free pages, one pool for
each subarray. When the OS allocates the destination page for a
copy operation (e.g., for a \emph{Copy-on-Write} operation), it
chooses the page from the same pool (subarray) as the source
page. Note that this approach does not require contiguous pages to
be placed within the same subarray. As mentioned before, commonly
used memory interleaving techniques spread out contiguous pages
across as many banks/subarrays as possible to improve
parallelism. Therefore, both the source and destination of a bulk
copy operation can be spread out across many subarrays.

\pagebreak
\subsubsection{Granularity of Copy/Initialization}
\label{sec:memory-interleaving}

The second aspect that affects the use of FPM is the granularity
at which data is copied or initialized. FPM has a minimum
granularity at which it can copy or initialize data. There are two
factors that affect this minimum granularity: 1)~the size of each
DRAM row, and 2)~the memory interleaving employed by the
controller.

First, FPM copies \emph{all} the data of the source row to the
destination row (across the entire DIMM). Therefore, the minimum
granularity of copy using FPM is at least the size of the
row. Second, to extract maximum bandwidth, some memory
interleaving techniques map consecutive cache lines to different
memory channels in the system. Therefore, to copy/initialize a
contiguous region of data with such interleaving strategies, FPM
must perform the copy operation in each channel. The minimum
amount of data copied by FPM in such a scenario is the product of
the row size and the number of channels.

To maximize the likelihood of using FPM, the system or application
software must ensure that the region of data copied (initialized)
using the \mcpy (\minit) instructions is at least as large as this
minimum granularity. For this purpose, we propose to expose this
minimum granularity to the software through a special register,
which we call the \emph{Minimum Copy Granularity Register}
(MCGR). On system bootup, the memory controller initializes the
MCGR based on the row size and the memory interleaving strategy,
which can later be used by the OS for effectively exploiting
RowClone. Note that some previously proposed techniques such as
sub-wordline activation~\cite{rethinking-dram} or
mini-rank~\cite{threaded-module,mini-rank} can be combined with
RowClone to reduce the minimum copy granularity, further
increasing the opportunity to use FPM.

\section{Applications}
\label{sec:applications}

RowClone can be used to accelerate any bulk copy and
initialization operation to improve both system performance and
energy efficiency. In this paper, we quantitatively evaluate the
efficacy of RowClone by using it to accelerate two primitives
widely used by modern system software: 1)~Copy-on-Write and
2)~Bulk Zeroing. We now describe these primitives followed
by several applications that frequently trigger them.

\subsection{Primitives Accelerated by RowClone}
\label{sec:apps-primitives}

\emph{Copy-on-Write} (CoW) is a technique used by most modern
operating systems (OS) to postpone an expensive copy operation until
it is actually needed. When data of one virtual page needs to be
copied to another, instead of creating a copy, the OS points both
virtual pages to the same physical page (source) and marks the page as
read-only. In the future, when one of the sharers attempts to write to
the page, the OS allocates a new physical page (destination) for the
writer and copies the contents of the source page to the newly
allocated page. Fortunately, prior to allocating the destination page,
the OS already knows the location of the source physical
page. Therefore, it can ensure that the destination is allocated in
the same subarray as the source, thereby enabling the processor to use
FPM to perform the copy.

\emph{Bulk Zeroing} (BuZ) is an operation where a large block of
memory is zeroed out. As mentioned in
Section~\ref{sec:bulk-initialization}, our mechanism maintains a
reserved row that is fully initialized to zero in each
subarray. For each row in the destination region to be zeroed out,
the processor uses FPM to copy the data from the reserved zero-row
of the corresponding subarray to the destination row.

\subsection{Applications that Use CoW/BuZ}
\label{sec:apps-cow-zeroing}
We now describe seven example applications or use-cases that
extensively use the CoW or BuZ operations. Note that these are just a
small number of example scenarios that incur a large number of copy
and initialization operations.

\textit{Process Forking.} \fork is a frequently-used system call in
modern operating systems (OS). When a process (parent) calls \fork, it
creates a new process (child) with the exact same memory image and
execution state as the parent. This semantics of \fork makes it useful
for different scenarios. Common uses of the \fork system call are to
1)~create new processes, and 2)~create stateful threads from a single
parent thread in multi-threaded programs. One main limitation of \fork
is that it results in a CoW operation whenever the child/parent
updates a shared page. Hence, despite its wide usage, as a result of
the large number of copy operations triggered by \fork, it remains one
of the most expensive system calls in terms of memory
performance~\cite{fork-exp}.

\textit{Initializing Large Data Structures.} Initializing large
data structures often triggers Bulk Zeroing. In fact, many managed
languages (e.g., C\#, Java, PHP) require zero initialization of
variables to ensure memory safety~\cite{why-nothing-matters}. In
such cases, to reduce the overhead of zeroing, memory is
zeroed-out in bulk.

\textit{Secure Deallocation.} Most operating systems (e.g.,
Linux~\cite{linux-security}, Windows~\cite{windows-security}, Mac
OS X~\cite{macos-security}) zero out pages newly allocated to a
process. This is done to prevent malicious processes from gaining
access to the data that previously belonged to other processes or
the kernel itself. Not doing so can potentially lead to security
vulnerabilities, as shown by prior
works~\cite{shredding,sunshine,coldboot,disclosure}.

\textit{Process Checkpointing.} Checkpointing is an operation
during which a consistent version of a process state is backed-up,
so that the process can be restored from that state in the
future. This checkpoint-restore primitive is useful in many cases
including high-performance computing servers~\cite{plfs}, software
debugging with reduced overhead~\cite{flashback}, hardware-level
fault and bug tolerance mechanisms~\cite{self-test,hardware-bug},
and speculative OS optimizations to improve
performance~\cite{os-speculation-2,os-speculation-1}. However, to
ensure that the checkpoint is consistent (i.e., the original process
does not update data while the checkpointing is in progress), the
pages of the process are marked with copy-on-write. As a result,
checkpointing often results in a large number of CoW operations. 

\textit{Virtual Machine Cloning/Deduplication.}  Virtual machine
(VM) cloning~\cite{snowflock} is a technique to significantly
reduce the startup cost of VMs in a cloud computing
server. Similarly, deduplication is a technique employed by modern
hypervisors~\cite{esx-server} to reduce the overall memory
capacity requirements of VMs. With this technique, different VMs
share physical pages that contain the same data. Similar to
forking, both these operations likely result in a large number of
CoW operations for pages shared across VMs.

\textit{Page Migration.} Bank conflicts, i.e., concurrent requests to
different rows within the same bank, typically result in reduced row
buffer hit rate and hence degrade both system performance and energy
efficiency. Prior work~\cite{micropages} proposed techniques to
mitigate bank conflicts using page migration. The PSM mode of RowClone
can be used in conjunction with such techniques to 1)~significantly
reduce the migration latency and 2)~make the migrations more
energy-efficient.

\textit{CPU-GPU Communication.} In many current and future processors,
the GPU is or is expected to be integrated on the same chip with the
CPU. Even in such systems where the CPU and GPU share the same
off-chip memory, the off-chip memory is partitioned between the two
devices. As a consequence, whenever a CPU program wants to offload
some computation to the GPU, it has to copy all the necessary data
from the CPU address space to the GPU address
space~\cite{cpu-gpu}. When the GPU computation is finished, all the
data needs to be copied back to the CPU address space. This copying
involves a significant overhead. By spreading out the GPU address
space over all subarrays and mapping the application data
appropriately, RowClone can significantly speed up these copy
operations. Note that communication between different processors and
accelerators in a heterogeneous System-on-a-chip (SoC) is done
similarly to the CPU-GPU communication and can also be accelerated by
RowClone.

We now quantitatively compare RowClone to existing systems and
show that RowClone significantly improves both system performance
and energy efficiency.

\section{Methodology}
\label{sec:methodology}

\noindent\textbf{Simulation.} Our evaluations use an in-house
cycle-level multi-core simulator along with a cycle-accurate
command-level DDR3 DRAM simulator. The multi-core simulator models
out-of-order cores, each with a private last-level
cache.\footnote{Since our mechanism primarily affects off-chip memory
  traffic, we expect our results and conclusions to be similar with
  shared caches as well.} We integrate RowClone into the simulator at
the command-level. We use DDR3 DRAM timing constraints~\cite{ddr3} to
calculate the latency of different operations. Since \cmdtr operates
similarly to \cmdrd/\cmdwr, we assume \cmdtr to have the same latency
as \cmdrd/\cmdwr. For our energy evaluations, we use DRAM energy/power
models from Rambus~\cite{rambus-power} and
Micron~\cite{micron-power}. Although, in DDR3 DRAM, a row corresponds
to 8KB across a rank, we assume a minimum in-DRAM copy granularity
(Section~\ref{sec:memory-interleaving}) of 4KB -- same as the page
size used by the operating system (Debian Linux) in our
evaluations. For this purpose, we model a DRAM module with 512-byte
rows per chip (4KB across a rank). Table~\ref{tab:parameters}
specifies the major parameters used for our simulations.

\begin{table}[h]
  \centering
  \begin{tabular}{lp{0.73\textwidth}}
  \toprule
  \textbf{Component} & \textbf{Parameters}\\
  \toprule
  \multirow{1}{*}{Processor} & 1--8 cores, OoO 128-entry window, 3-wide issue, 8 MSHRs/core\\
  \midrule
  \multirow{1}{*}{Last-level Cache} & 1MB per core, private, 64-byte cache line,  16-way associative\\
  \midrule
  \multirow{1}{*}{Memory Controller} & One per channel, 64-entry
  read queue, 64-entry write queue\\
  \midrule
  \multirow{1}{*}{Memory System} & DDR3-1066 (8-8-8)~\cite{ddr3},
  2 channels, 1 rank/channel, 8 banks/rank\\
  \bottomrule
\end{tabular}

  \caption[RowClone: Simulation parameters]{Configuration of the simulated system}
  \label{tab:parameters}
\end{table}

\noindent\textbf{Workloads.} We evaluate the benefits of RowClone
using 1)~a case study of the \fork system call, an important
operation used by modern operating systems, 2)~six copy and
initialization intensive benchmarks: \emph{bootup},
\emph{compile}, \emph{forkbench},
\emph{memcached}~\cite{memcached}, \emph{mysql}~\cite{mysql}, and
\emph{shell} (Section~\ref{res:rowclone-ii-apps} describes these
benchmarks), and 3)~a wide variety of multi-core workloads
comprising the copy/initialization intensive applications running
alongside memory-intensive applications from the SPEC CPU2006
benchmark suite~\cite{spec2006}. Note that benchmarks such as SPEC
CPU2006, which predominantly stress the CPU, typically use a small
number of page copy and initialization operations and therefore
would serve as poor individual evaluation benchmarks for RowClone.

We collected instruction traces for our workloads using
Bochs~\cite{bochs}, a full-system x86-64 emulator, running a
GNU/Linux system. We modify the kernel's implementation of page
copy/initialization to use the \mcpy and \minit instructions and
mark these instructions in our traces.\footnote{For our \fork
  benchmark (described in Section~\ref{sec:res-fork}), we used the
  Wind River Simics full system simulator~\cite{simics} to collect
  the traces.} We collect 1-billion instruction traces of the
representative portions of these workloads. We use the instruction
throughput (IPC) metric to measure single-core performance. We
evaluate multi-core runs using the weighted speedup metric, a
widely-used measure of system throughput for multi-programmed
workloads~\cite{weighted-speedup}, as well as five other
performance/fairness/bandwidth/energy metrics, as shown in
Table~\ref{tab:multi-core-ws}.

\section{Evaluations}
\label{sec:results}

In this section, we quantitatively evaluate the benefits of
RowClone. We first analyze the raw latency and energy improvement
enabled by the DRAM substrate to accelerate a single 4KB copy and
4KB zeroing operation (Section~\ref{sec:rowclone-analysis}). We
then discuss the results of our evaluation of RowClone using \fork
(Section~\ref{sec:res-fork}) and six copy/initialization intensive
applications
(Section~\ref{res:rowclone-ii-apps}). Section~\ref{sec:multi-core}
presents our analysis of RowClone on multi-core systems and
Section~\ref{sec:mc-dma} provides quantitative comparisons to
memory controller based DMA engines.

\subsection{Latency and Energy Analysis}
\label{sec:rowclone-analysis}

\begin{figure}
  \centering
  \includegraphics[scale=1.3]{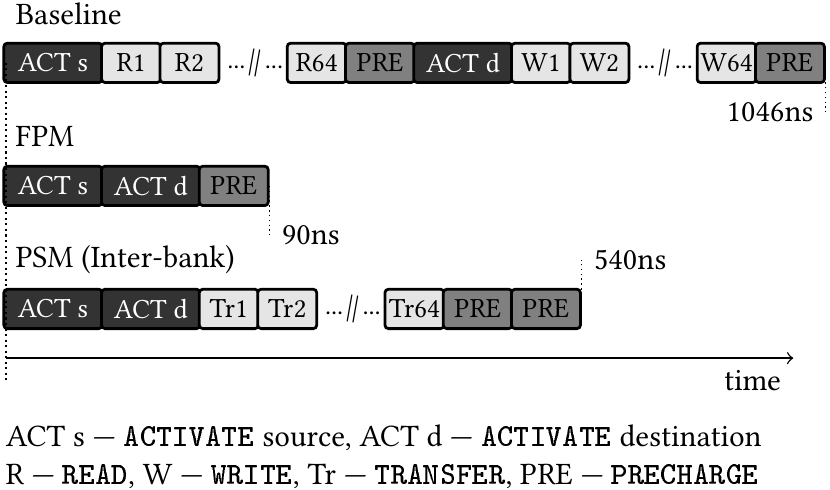}
  \caption[RowClone: Command sequence and latency
    comparison]{Command sequence and latency for Baseline, FPM,
    and Inter-bank PSM for a 4KB copy operation. Intra-bank PSM
    simply repeats the operations for Inter-bank PSM twice (source
    row to temporary row and temporary row to destination
    row). The figure is not drawn to scale.}
  \label{fig:timing}
\end{figure}

Figure~\ref{fig:timing} shows the sequence of commands issued by
the baseline, FPM and PSM (inter-bank) to perform a 4KB copy
operation. The figure also shows the overall latency incurred by
each of these mechanisms, assuming DDR3-1066 timing
constraints. Note that a 4KB copy involves copying 64 64B cache
lines. For ease of analysis, only for this section, we assume that
no cache line from the source or the destination region are cached
in the on-chip caches.  While the baseline serially reads each
cache line individually from the source page and writes it back
individually to the destination page, FPM parallelizes the copy
operation of all the cache lines by using the large internal
bandwidth available within a subarray. PSM, on the other hand,
uses the new \cmdtr command to overlap the latency of the read and
write operations involved in the page copy.

Table~\ref{tab:latency-energy} shows the reduction in latency and
energy consumption due to our mechanisms for different cases of 4KB
copy and zeroing operations.  To be fair to the baseline,
the results include only the energy consumed by the DRAM and the DRAM channel.
We draw two conclusions from our results.

\begin{table}[h!]
  \centering
  \begin{tabular}{crrrrr}
  \toprule
  & \multirow{4}{*}{\textbf{Mechanism}} &
  \multicolumn{2}{c}{\textbf{Absolute}} &
  \multicolumn{2}{c}{\textbf{Reduction}}\\
  & & & \small{Memory} & & \small{Memory}\vspace{-1mm}\\ 
  & & \small{Latency}  & \small{Energy} & \small{Latency} & \small{Energy}\\
  & & (ns) & ($\mu$J)\\
  \midrule
  \multirow{4}{*}{\rotatebox{90}{\textbf{Copy}}} & {Baseline} & 1046 & 3.6 & 1.00x & 1.0x\\
  & {FPM} & 90 & 0.04 & \textbf{11.62x} & \textbf{74.4x}\\
  & {Inter-Bank - PSM} & 540 & 1.1 & 1.93x & 3.2x\\
  & {Intra-Bank - PSM} & 1050 & 2.5 & 0.99x & 1.5x\\
  \toprule
  \multirow{2}{*}{\rotatebox{90}{\textbf{Zero}}} & {Baseline} & 546 & 2.0 & 1.00x  & 1.0x\\
  & {FPM} & 90 & 0.05 & \textbf{6.06x} & \textbf{41.5x}\\
  \bottomrule
\end{tabular}

  \caption[RowClone: Latency/energy reductions]{DRAM latency and memory energy reductions due to RowClone}
  \label{tab:latency-energy}
\end{table}

First, FPM significantly improves both the latency and the energy
consumed by bulk operations --- 11.6x and 6x reduction in latency
of 4KB copy and zeroing, and 74.4x and 41.5x reduction in memory
energy of 4KB copy and zeroing. Second, although PSM does not
provide as much benefit as FPM, it still reduces the latency and
energy of a 4KB inter-bank copy by 1.9x and 3.2x, while providing
a more generally applicable mechanism.

When an on-chip cache is employed, any line cached from the source or
destination page can be served at a lower latency than accessing main
memory. As a result, in such systems, the baseline will incur a lower
latency to perform a bulk copy or initialization compared to a system
without on-chip caches. However, as we show in the following sections
(\ref{sec:res-fork}--\ref{sec:multi-core}), \emph{even in the
  presence of on-chip caching}, the raw latency/energy improvement due
to RowClone translates to significant improvements in both overall
system performance and energy efficiency.

\subsection{The {\large{\fork}} System Call}
\label{sec:res-fork}

As mentioned in Section~\ref{sec:apps-cow-zeroing}, \fork is one
of the most expensive yet frequently-used system calls in modern
systems~\cite{fork-exp}. Since \fork triggers a large number of
CoW operations (as a result of updates to shared pages from the
parent or child process), RowClone can significantly improve the
performance of \fork.

The performance of \fork depends on two parameters: 1)~the size of
the address space used by the parent---which determines how much
data may potentially have to be copied, and 2)~the number of pages
updated after the \fork operation by either the parent or the
child---which determines how much data is actually copied. To
exercise these two parameters, we create a microbenchmark,
\forkbench, which first creates an array of size \ubsize and
initializes the array with random values. It then forks
itself. The child process updates $N$ random pages (by updating a
cache line within each page) and exits; the parent process waits
for the child process to complete before exiting itself.

As such, we expect the number of copy operations to depend on
$N$---the number of pages copied. Therefore, one may expect
RowClone's performance benefits to be proportional to $N$.
However, an application's performance typically depends on the
{\em overall memory access rate}~\cite{mise}, and RowClone can
only improve performance by reducing the {\em memory access rate
  due to copy operations}. As a result, we expect the performance
improvement due to RowClone to primarily depend on the
\emph{fraction} of memory traffic (total bytes transferred over
the memory channel) generated by copy operations. We refer to this
fraction as FMTC---Fraction of Memory Traffic due to Copies.

\begin{figure}[h]
  \centering
  \includegraphics{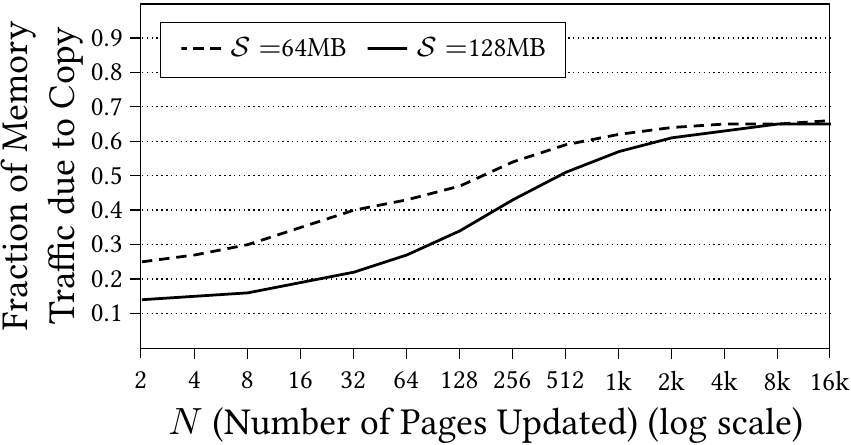}
  \caption[Memory traffic due to copy in \forkbench]{FMTC of
    \forkbench for varying \ubsize and $N$}
  \label{plot:fork-copy}
\end{figure}

Figure~\ref{plot:fork-copy} plots FMTC of \forkbench for different
values of \ubsize (64MB and 128MB) and $N$ (2 to 16k) in the baseline
system. As the figure shows, for both values of \ubsize, FMTC
increases with increasing $N$. This is expected as a higher $N$ (more
pages updated by the child) leads to more CoW
operations. However, because of the presence of other read/write
operations (e.g., during the initialization phase of the parent), for
a given value of $N$, FMTC is larger for \ubsize= 64MB compared to
\ubsize= 128MB.  Depending on the value of \ubsize and $N$, anywhere
between 14\% to 66\% of the memory traffic arises from copy
operations. This shows that accelerating copy operations using
RowClone has the potential to significantly improve the performance of
the \fork operation.

\begin{figure}[h]
  \centering
  \includegraphics{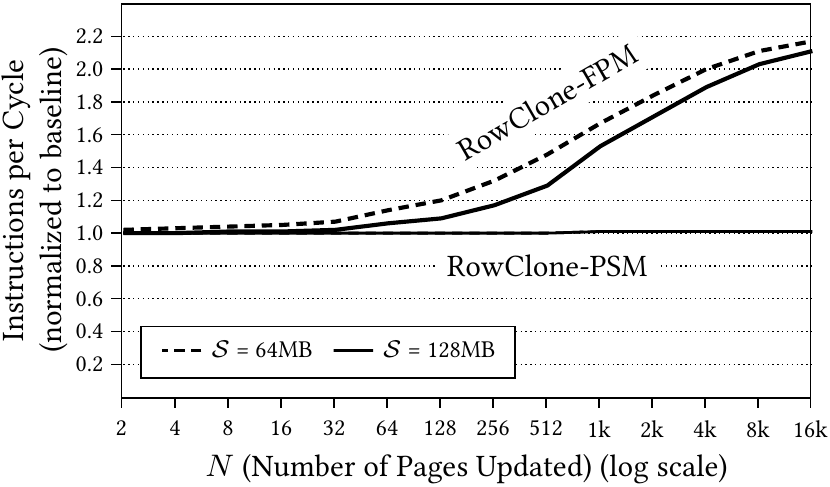}
  \caption[RowClone: \forkbench performance]{Performance
    improvement due to RowClone for \forkbench with different
    values of \ubsize and $N$}
  \label{plot:fork-perf}
\end{figure}

Figure~\ref{plot:fork-perf} plots the performance (IPC) of FPM and
PSM for \forkbench, normalized to that of the baseline system. We
draw two conclusions from the figure. First, FPM improves the
performance of \forkbench for both values of \ubsize and most
values of $N$. The peak performance improvement is 2.2x for $N$ =
16k (30\% on average across all data points). As expected, the
improvement of FPM increases as the number of pages updated
increases. The trend in performance improvement of FPM is similar
to that of FMTC (Figure~\ref{plot:fork-copy}), confirming our
hypothesis that FPM's performance improvement primarily depends on
FMTC. Second, PSM does not provide considerable performance
improvement over the baseline. This is because the large on-chip
cache in the baseline system buffers the writebacks generated by
the copy operations. These writebacks are flushed to memory at a
later point without further delaying the copy operation. As a
result, PSM, which just overlaps the read and write operations
involved in the copy, does not improve latency significantly in
the presence of a large on-chip cache. On the other hand, FPM, by
copying all cache lines from the source row to destination in
parallel, significantly reduces the latency compared to the
baseline (which still needs to read the source blocks from main
memory), resulting in high performance improvement.

\begin{figure}[h]
  \centering
  \includegraphics[scale=0.9]{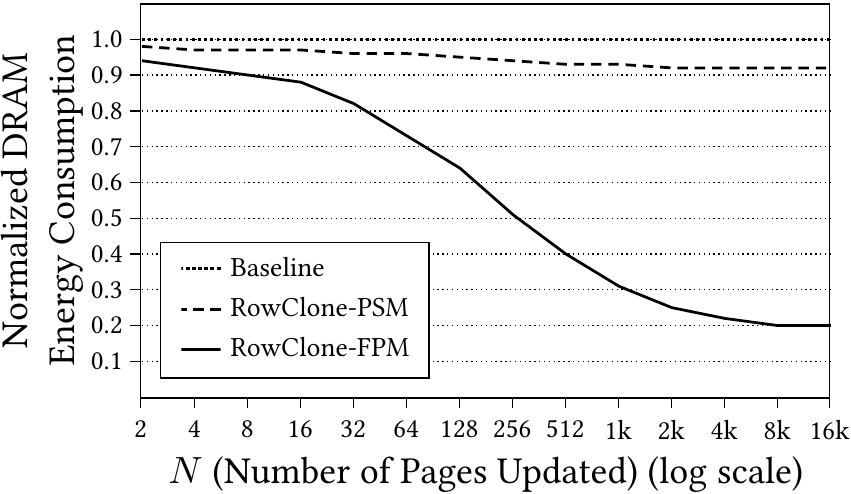}
  \caption[RowClone: Energy consumption for \forkbench]{Comparison
    of DRAM energy consumption of different mechanisms for
    \forkbench (\ubsize = 64MB)}
  \label{plot:fork-energy}
\end{figure}

Figure~\ref{plot:fork-energy} shows the reduction in DRAM energy
consumption (considering both the DRAM and the memory channel) 
of FPM and PSM modes of RowClone compared to that of
the baseline for \forkbench with \ubsize $=64$MB.
Similar to performance, the overall DRAM energy consumption also
depends on the total memory access rate. As a result, RowClone's
potential to reduce DRAM energy depends on the fraction of memory
traffic generated by copy operations. In fact, our results also
show that the DRAM energy reduction due to FPM and PSM correlate
well with FMTC (Figure~\ref{plot:fork-copy}).  By efficiently
performing the copy operations, FPM reduces DRAM energy
consumption by up to 80\% (average 50\%, across all data
points). Similar to FPM, the energy reduction of PSM also
increases with increasing $N$ with a maximum reduction of 9\% for
$N$=16k.

In a system that is agnostic to RowClone, we expect the
performance improvement and energy reduction of RowClone to be in
between that of FPM and PSM. By making the system software aware
of RowClone (Section~\ref{sec:os-changes}), we can approximate the
maximum performance and energy benefits by increasing the
likelihood of the use of FPM.

\subsection{Copy/Initialization Intensive Applications}
\label{res:rowclone-ii-apps}

In this section, we analyze the benefits of RowClone on six
copy/initialization intensive applications, including one instance of
the \forkbench described in the previous section. Table~\ref{tab:iias}
describes these applications.

\begin{table}[h]\small
  \centering
  \begin{tabular}{lp{0.87\textwidth}}
  \toprule
  \textbf{Name} & \textbf{Description}\\
  \toprule
  \multirow{1}{*}{\emph{bootup}} & A phase booting up the Debian
  operating system.\\
  \midrule
  \multirow{1}{*}{\emph{compile}} & The compilation phase from the GNU C compiler (while
  running \emph{cc1}).\\
  \midrule
  \multirow{1}{*}{\emph{forkbench}} & An instance of the \forkbench described in
  Section~\ref{sec:res-fork} with \mbox{\ubsize = 64MB} and $N$ = 1k.\\
  \midrule
  \multirow{2}{*}{\emph{mcached}} & Memcached~\cite{memcached}, a memory object
  caching system, a phase inserting many key-value pairs into
  the memcache.\\
  \midrule
  \multirow{1}{*}{\emph{mysql}} & MySQL~\cite{mysql}, an on-disk
  database system, a phase loading the sample \emph{employeedb}\\
  \midrule
  \multirow{2}{*}{\emph{shell}} & A Unix shell script running `find' on a directory
  tree with `ls' on each sub-directory (involves filesystem
  accesses and spawning new processes).\\
  \bottomrule
\end{tabular}

  \caption[RowClone: Copy/initialization-intensive benchmarks]{Copy/Initialization-intensive benchmarks}
  \label{tab:iias}
\end{table}

\begin{figure}[h]
  \centering
  \includegraphics{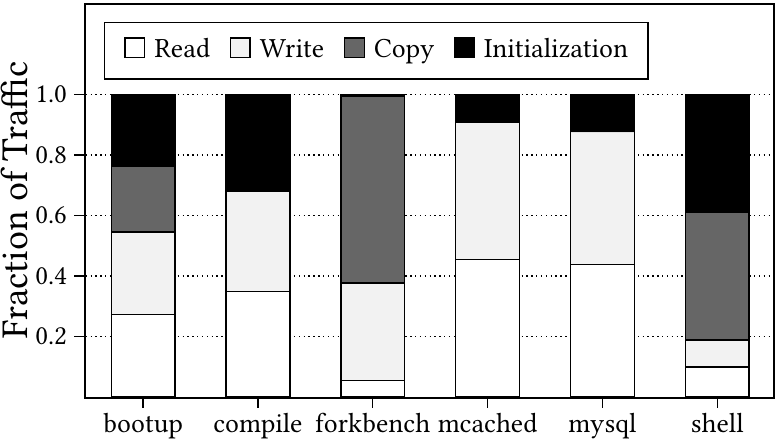}
  \caption[Copy/initialization intensive benchmark: Memory traffic
    breakdown]{Fraction of memory traffic due to read, write, copy
    and initialization}
  \label{plot:memfrac-apps}
\end{figure}

Figure~\ref{plot:memfrac-apps} plots the fraction of memory
traffic due to copy, initialization, and regular read/write
operations for the six applications. For these applications,
between 10\% and 80\% of the memory traffic is generated by copy
and initialization operations.

\begin{figure}[h]
  \centering
  \includegraphics{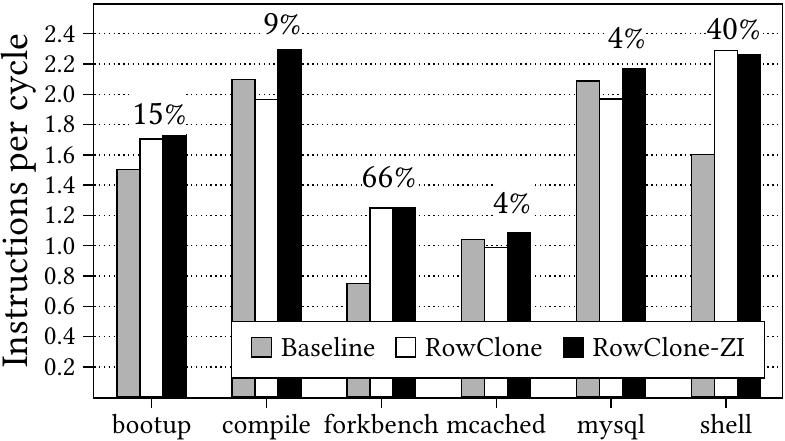}
  \caption[RowClone-ZI performance]{Performance improvement of
    RowClone and RowClone-ZI. \normalfont{Value on top indicates
      percentage improvement of RowClone-ZI over baseline.}}
  \label{plot:perf-apps}
\end{figure}

Figure~\ref{plot:perf-apps} compares the IPC of the baseline with
that of RowClone and a variant of RowClone, RowClone-ZI (described
shortly). The RowClone-based initialization mechanism slightly
degrades performance for the applications which have a negligible
number of copy operations (\emph{mcached}, \emph{compile}, and
\emph{mysql}).

Further analysis indicated that, for these applications, although the
operating system zeroes out any newly allocated page, the application
typically accesses almost all cache lines of a page immediately after
the page is zeroed out. There are two phases: 1)~the phase when the OS
zeroes out the page, and 2)~the phase when the application accesses
the cache lines of the page. While the baseline incurs cache misses
during phase 1, RowClone, as a result of performing the zeroing
operation completely in memory, incurs cache misses in phase
2. However, the baseline zeroing operation is heavily optimized for
memory-level parallelism (MLP)~\cite{effra,runahead}. In contrast, the cache
misses in phase 2 have low MLP. As a result, incurring the same misses
in Phase 2 (as with RowClone) causes higher overall stall time for the
application (because the latencies for the misses are serialized) than
incurring them in Phase 1 (as in the baseline), resulting in
RowClone's performance degradation compared to the baseline.

To address this problem, we introduce a variant of RowClone,
RowClone-Zero-Insert (RowClone-ZI). RowClone-ZI not only zeroes out a
page in DRAM but it also inserts a zero cache line into the
processor cache corresponding to each cache line in the page that is
zeroed out. By doing so, RowClone-ZI avoids the cache misses during
both phase 1 (zeroing operation) and phase 2 (when the application
accesses the cache lines of the zeroed page).  As a result, it improves
performance for all benchmarks, notably \forkbench (by 66\%) and
\emph{shell} (by 40\%), compared to the baseline.

Table~\ref{tab:energy-apps} shows the percentage reduction in DRAM
energy and memory bandwidth consumption with RowClone and RowClone-ZI
compared to the baseline. While RowClone significantly reduces both
energy and memory bandwidth consumption for \emph{bootup},
\emph{forkbench} and \emph{shell}, it has negligible impact on both
metrics for the remaining three benchmarks. The lack of energy and
bandwidth benefits in these three applications is due to serial
execution caused by the the cache misses incurred when the processor
accesses the zeroed out pages (i.e., {\em phase 2}, as described
above), which also leads to performance degradation in these workloads
(as also described above). RowClone-ZI, which eliminates the cache
misses in {\em phase 2}, significantly reduces energy consumption
(between 15\% to 69\%) and memory bandwidth consumption (between 16\%
and 81\%) for all benchmarks compared to the baseline. We conclude
that RowClone-ZI can effectively improve performance, memory energy,
and memory bandwidth efficiency in page copy and initialization
intensive single-core workloads.

\begin{table}[h]\small
  \centering
  \begin{tabular}{rrrrr}
  \toprule
  \multirow{2}{*}{\textbf{Application}} &
  \multicolumn{2}{c}{\textbf{Energy Reduction}} &
  \multicolumn{2}{c}{\textbf{Bandwidth Reduction}}\\
  \cmidrule{2-5}
  & \textbf{RowClone} & \textbf{+ZI}
  & \textbf{RowClone} & \textbf{+ZI}\\
  \midrule
  \emph{bootup} & 39\% & 40\% & 49\% & 52\%\vspace{1mm}\\
  \emph{compile} & -2\% & 32\% & 2\% & 47\%\vspace{1mm}\\
  \emph{forkbench} &  69\% & 69\% & 60\% & 60\%\vspace{1mm}\\
  \emph{mcached} & 0\% & 15\% & 0\% & 16\%\vspace{1mm}\\
  \emph{mysql} & -1\% & 17\%  & 0\% & 21\%\vspace{1mm}\\
  \emph{shell} & 68\% & 67\% & 81\% & 81\%\vspace{1mm}\\
  \bottomrule
\end{tabular}


  \caption[RowClone: DRAM energy/bandwidth reduction]{DRAM energy and bandwidth reduction due to RowClone and
  RowClone-ZI (indicated as +ZI)}
  \label{tab:energy-apps}
\end{table}

\subsection{Multi-core Evaluations}
\label{sec:multi-core}

As RowClone performs bulk data operations completely within DRAM, it
significantly reduces the memory bandwidth consumed by these
operations. As a result, RowClone can benefit other applications
running concurrently on the same system. We evaluate this benefit of
RowClone by running our copy/initialization-intensive applications
alongside memory-intensive applications from the SPEC CPU2006
benchmark suite~\cite{spec2006} (i.e., those applications with last-level
cache MPKI greater than 1). Table~\ref{tab:benchmarks} lists the set
of applications used for our multi-programmed workloads.

\begin{table}[h]\small
  \centering
  \begin{tabular}{p{0.85\textwidth}}
  \toprule
  \textbf{Copy/Initialization-intensive benchmarks}\vspace{0.5mm}\\
  \emph{bootup},
  \emph{compile}, \emph{forkbench}, \emph{mcached}, \emph{mysql}, \emph{shell}\vspace{1mm}\\
  \toprule
  \textbf{Memory-intensive benchmarks from SPEC CPU2006}\vspace{0.5mm}\\
  \emph{bzip2}, \emph{gcc}, \emph{mcf}, \emph{milc}, \emph{zeusmp},
  \emph{gromacs}, \emph{cactusADM}, \emph{leslie3d}, \emph{namd},
  \emph{gobmk}, \emph{dealII}, \emph{soplex}, \emph{hmmer},
  \emph{sjeng}, \emph{GemsFDTD}, \emph{libquantum}, \emph{h264ref},
  \emph{lbm}, \emph{omnetpp}, \emph{astar}, \emph{wrf},
  \emph{sphinx3}, \emph{xalancbmk}\vspace{1mm}\\
  \bottomrule
\end{tabular}

  \caption[RowClone: Benchmarks for multi-core evaluation]{List of benchmarks used for multi-core evaluation}
  \label{tab:benchmarks}
\end{table}

We generate multi-programmed workloads for 2-core, 4-core and
8-core systems. In each workload, half of the cores run
copy/initialization-intensive benchmarks and the remaining cores
run memory-intensive SPEC benchmarks. Benchmarks from each
category are chosen at random.

\begin{figure}[h]
  \centering
  \includegraphics[scale=0.9]{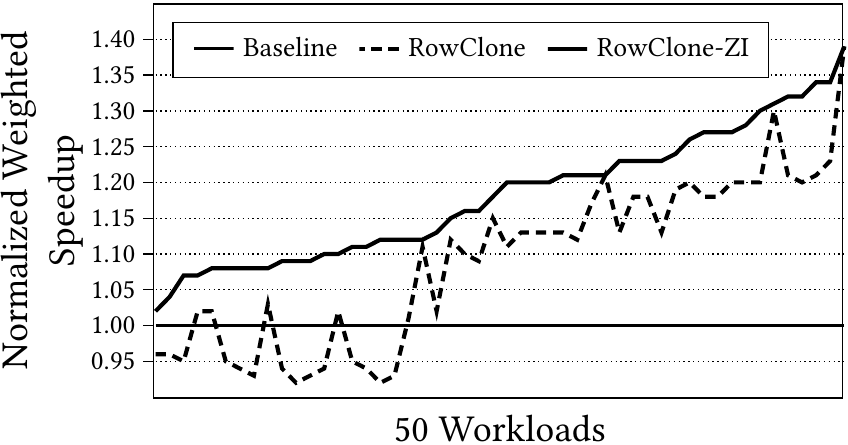}
  \caption[RowClone: 4-core performance]{System performance
    improvement of RowClone for 4-core workloads}
  \label{plot:s-curve}
\end{figure}

Figure~\ref{plot:s-curve} plots the performance improvement due to
RowClone and RowClone-ZI for the 50 4-core workloads we evaluated
(sorted based on the performance improvement due to RowClone-ZI).
Two conclusions are in order. First, although RowClone degrades
performance of certain 4-core workloads (with \emph{compile},
\emph{mcached} or \emph{mysql} benchmarks), it significantly
improves performance for all other workloads (by 10\% across all
workloads). Second, like in our single-core evaluations
(Section~\ref{res:rowclone-ii-apps}), RowClone-ZI eliminates the
performance degradation due to RowClone and consistently
outperforms both the baseline and RowClone for all workloads (20\%
on average).

Table~\ref{tab:multi-core-ws} shows the number of workloads and six
metrics that evaluate the performance, fairness, memory bandwidth and
energy efficiency improvement due to RowClone compared to the baseline
for systems with 2, 4, and 8 cores. For all three systems, RowClone
significantly outperforms the baseline on all metrics.

\begin{table}[h]\small
  \centering
  \begin{tabular}{rrrr}
  \toprule
  \textbf{Number of Cores} & \multicolumn{1}{c}{2} & \multicolumn{1}{c}{4} & \multicolumn{1}{c}{8}\\
  \midrule
  \textbf{Number of Workloads} & 138 & 50 & 40\vspace{1mm}\\
  \textbf{Weighted Speedup~\cite{weighted-speedup} Improvement} & 15\% & 20\% & 27\%\vspace{1mm}\\
  \textbf{Instruction Throughput Improvement} & 14\% & 15\% & 25\%\vspace{1mm}\\
  \textbf{Harmonic Speedup~\cite{hmean} Improvement} & 13\% & 16\% & 29\%\vspace{1mm}\\
  \textbf{Maximum Slowdown~\cite{atlas,tcm} Reduction} & 6\% & 12\% & 23\%\vspace{1mm}\\
  \textbf{Memory Bandwidth/Instruction~\cite{fdp} Reduction} & 29\% & 27\% & 28\%\vspace{1mm}\\
  \textbf{Memory Energy/Instruction Reduction} & 19\% & 17\% & 17\%\vspace{1mm}\\
  \bottomrule
\end{tabular}

  \caption[RowClone: Multi-core results]{Multi-core performance, fairness,
  bandwidth, and energy}
  \label{tab:multi-core-ws}
\end{table}

\begin{figure}[h!]
  \centering
  \includegraphics[scale=0.9]{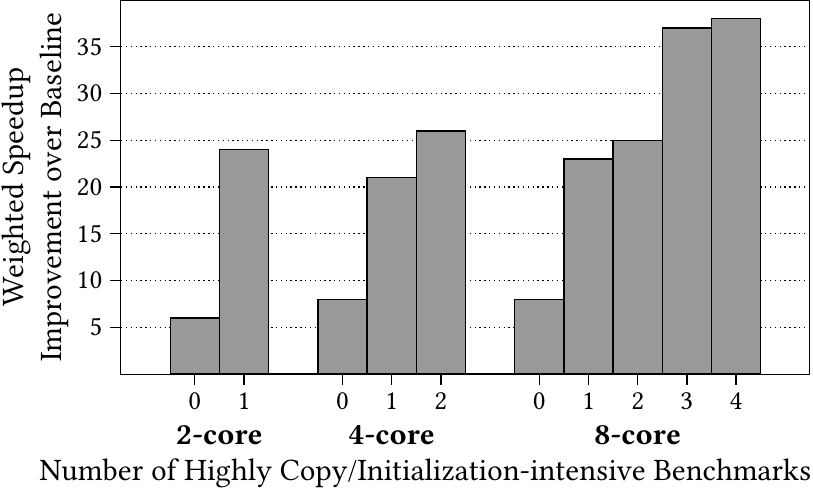}
  \caption[RowClone: Effect of increasing copy/initialization
    intensity]{Effect of increasing copy/initialization intensity}
  \label{plot:multi-trend}
\end{figure}

To provide more insight into the benefits of RowClone on multi-core
systems, we classify our copy/initialization-intensive benchmarks into
two categories: 1) Moderately copy/initialization-intensive
(\emph{compile}, \emph{mcached}, and \emph{mysql}) and highly
copy/initialization-intensive (\emph{bootup}, \emph{forkbench}, and
\emph{shell}). Figure~\ref{plot:multi-trend} shows the average
improvement in weighted speedup for the different multi-core
workloads, categorized based on the number of highly
copy/initialization-intensive benchmarks. As the trends indicate, the
performance improvement increases with increasing number of such
benchmarks for all three multi-core systems, indicating the
effectiveness of RowClone in accelerating bulk copy/initialization
operations.

We conclude that RowClone is an effective mechanism to improve
system performance, energy efficiency and bandwidth efficiency of
future, bandwidth-constrained multi-core systems.

\subsection{Memory-Controller-based DMA}
\label{sec:mc-dma}

One alternative way to perform a bulk data operation is to use the
memory controller to complete the operation using the regular DRAM
interface (similar to some prior
approaches~\cite{bulk-copy-initialize,copy-engine}). We refer to
this approach as the memory-controller-based DMA (MC-DMA).  MC-DMA
can potentially avoid the cache pollution caused by inserting
blocks (involved in the copy/initialization) unnecessarily into
the caches. However, it still requires data to be transferred over
the memory bus. Hence, it suffers from the large latency,
bandwidth, and energy consumption associated with the data
transfer. Because the applications used in our evaluations do not
suffer from cache pollution, we expect the MC-DMA to perform
comparably or worse than the baseline. In fact, our evaluations
show that MC-DMA degrades performance compared to our baseline by
2\% on average for the six copy/initialization intensive
applications (16\% compared to RowClone). In addition, the MC-DMA
does not conserve any DRAM energy, unlike RowClone.

\section{Summary}

In this chapter, we introduced RowClone, a technique for exporting
bulk data copy and initialization operations to DRAM.  Our fastest
mechanism copies an entire row of data between rows that share a
row buffer, with very few changes to the DRAM architecture, while
leading to significant reduction in the latency and energy of
performing bulk copy/initialization. We also propose a more
flexible mechanism that uses the internal bus of a chip to copy
data between different banks within a chip. Our evaluations using
copy and initialization intensive applications show that RowClone
can significantly reduce memory bandwidth consumption for both
single-core and multi-core systems (by 28\% on average for 8-core
systems), resulting in significant system performance improvement
and memory energy reduction (27\% and 17\%, on average, for 8-core
systems).

We conclude that our approach of performing bulk copy and
initialization completely in DRAM is effective in improving both
system performance and energy efficiency for future,
bandwidth-constrained, multi-core systems. We hope that greatly
reducing the bandwidth, energy and performance cost of bulk data
copy and initialization can lead to new and easier ways of writing
applications that would otherwise need to be designed to avoid
bulk data copy and initialization operations.

\chapter{Buddy RAM}
\label{chap:buddy}

\begin{figure}[b!]
  \hrule\vspace{2mm}
  \begin{footnotesize}
    A part of this chapter is originally published as ``Fast Bulk
    Bitwise AND and OR in DRAM'' in IEEE Computer Architecture
    Letters, 2015~\cite{buddy-cal}
  \end{footnotesize}
\end{figure}

In the line of research aiming to identify primitives that can be
efficiently performed inside DRAM, the second mechanism we explore
in this thesis is one that can perform bitwise logical operations
completely inside DRAM. Our mechanism \emph{uses} the internal
analog operation of DRAM to efficiently perform bitwise
operations. For this reason, we call our mechanism \emph{Buddy RAM}
or {Bitwise-ops Using DRAM} (BU-D-RAM).

Bitwise operations are an important component of modern day
programming. They have a wide variety of applications, and can
often replace arithmetic operations with more efficient
algorithms~\cite{btt-knuth,hacker-delight}. In fact, many modern
processors provide support for accelerating a variety of bitwise
operations (e.g., Intel Advance Vector
eXtensions~\cite{intel-avx}).

We focus our attention on bitwise operations on large amounts of
input data. We refer to such operations as \emph{bulk bitwise
  operations}. Many applications trigger such bulk bitwise
operations. For example, in databases, bitmap
indices~\cite{bmide,bmidc} can be more efficient than
commonly-used B-trees for performing range queries and
joins~\cite{bmide,fastbit,bicompression}. In fact, bitmap indices
are supported by many real-world implementations (e.g.,
Redis~\cite{redis}, Fastbit~\cite{fastbit}). Improving the
throughput of bitwise operations can boost the performance of such
bitmap indices and many other primitives (e.g., string matching,
bulk hashing).

As bitwise operations are computationally inexpensive, in existing
systems, the throughput of bulk bitwise operations is limited by
the available memory bandwidth. This is because, to perform a bulk
bitwise operation, existing systems must first read the source
data from main memory into the processor caches. After performing
the operation at the processor, they may have to write the result
back to main memory. As a result, this approach requires a large
amount of data to be transferred back and forth on the memory
channel, resulting in high latency, bandwidth, and energy
consumption.

Our mechanism, Buddy RAM, consist of two components: one to
perform bitwise AND/OR operations (\buddyao), and the second
component to perform bitwise NOT operations (\buddynot). Both
components heavily exploit the operation of the sense amplifier
and the DRAM cells (described in
Section~\ref{sec:cell-operation}). In the following sections, we
first provide an overview of both these mechanisms, followed by a
detailed implementation of Buddy that requires minimal changes to
the internal design and the external interface of commodity DRAM.

\section{\buddyao}
\label{sec:bitwise-and-or}

As described in Section~\ref{sec:cell-operation}, when a DRAM cell
is connected to a bitline precharged to \halfvdd, the cell induces
a deviation on the bitline, and the deviation is amplified by the
sense amplifier. \emph{\buddyao} exploits the following fact about
the cell operation.
\begin{quote}
  The final state of the bitline after amplification is determined
  solely by the deviation on the bitline after the charge sharing
  phase (after state \ding{204} in
  Figure~\ref{fig:cell-operation}). If the deviation is positive
  (i.e., towards \vdd), the bitline is amplified to
  \vdd. Otherwise, if the deviation is negative (i.e., towards
  $0$), the bitline is amplified to $0$.
\end{quote}

\subsection{Triple-Row Activation}
\label{sec:triple-row-activation}

\buddyao simultaneously connects three cells to a sense
amplifier. When three cells are connected to the bitline, the
deviation of the bitline after charge sharing is determined by the
\emph{majority value} of the three cells. Specifically, if at
least two cells are initially in the charged state, the effective
voltage level of the three cells is at least \ttvdd. This results
in a positive deviation on the bitline. On the other hand, if at
most one cell is initially in the charged state, the effective
voltage level of the three cells is at most \otvdd. This results
in a negative deviation on the bitline voltage. As a result, the
final state of the bitline is determined by the logical majority
value of the three cells.

\begin{figure}[t]
  \centering
  \includegraphics{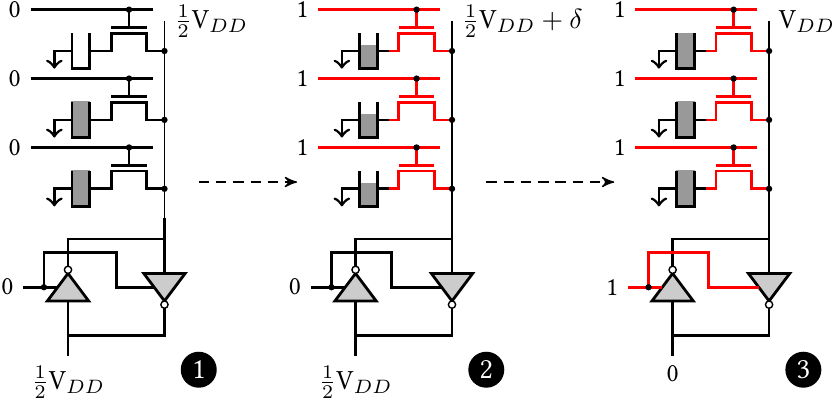}
  \caption[Triple-row activation in DRAM]{Triple-row activation}
  \label{fig:triple-row-activation}
\end{figure}

Figure~\ref{fig:triple-row-activation} shows an example of
activating three cells simultaneously. In the figure, we assume
that two of the three cells are initially in the charged state and
the third cell is in the empty state \ding{202}. When the
wordlines of all the three cells are raised simultaneously
\ding{203}, charge sharing results in a positive deviation on the
bitline.

More generally, if the cell's capacitance is $C_c$, the the
bitline's is $C_b$, and if $k$ of the three cells are initially in
the charged state, based on the charge sharing
principles~\cite{dram-cd}, the deviation $\delta$ on the bitline
voltage level is given by,
\begin{eqnarray}
  \delta &=& \frac{k.C_c.V_{DD} + C_b.\frac{1}{2}V_{DD}}{3C_c + C_b}
  - \frac{1}{2}V_{DD}\nonumber\\
  &=& \frac{(2k - 3)C_c}{6C_c + 2C_b}V_{DD}\label{eqn:delta}
\end{eqnarray}
From the above equation, it is clear that $\delta$ is positive for
$k = 2,3$, and $\delta$ is negative for $k = 0,1$. Therefore,
after amplification, the final voltage level on the bitline is
\vdd for $k = 2,3$ and $0$ for $k = 0,1$.

If $A$, $B$, and $C$ represent the logical values of the three
cells, then the final state of the bitline is $AB + BC + CA$ (i.e., at
least two of the values should be $1$ for the final state to be
$1$). Importantly, using simple boolean algebra, this expression
can be rewritten as $C(A + B) + \overline{C}(AB)$. In other words,
if the initial state of $C$ is $1$, then the final state of the
bitline is a bitwise OR of $A$ and $B$. Otherwise, if the initial
state of $C$ is $0$, then the final state of the bitline is a
bitwise AND of $A$ and $B$. Therefore, by controlling the value of
the cell C, we can execute a bitwise AND or bitwise OR operation
of the remaining two cells using the sense amplifier. Due to the
regular bulk operation of cells in DRAM, this approach naturally
extends to an entire row of DRAM cells and sense amplifiers,
enabling a multi-kilobyte-wide bitwise AND/OR
operation.\footnote{Note that the triple-row activation by itself
  can be useful to implement a \emph{bitwise majority}
  primitive. However, we do not explore this path in this thesis.}

\subsection{Challenges}
\label{sec:and-or-challenges}

There are two challenges in our approach. First,                                   
Equation~\ref{eqn:delta} assumes that the cells involved in the                    
triple-row activation are either fully charged or fully                            
empty. However, DRAM cells leak charge over time. Therefore, the                   
triple-row activation may not operate as expected. This problem                    
may be exacerbated by process variation in DRAM cells. Second, as                  
shown in Figure~\ref{fig:triple-row-activation}                                    
(state~\ding{204}), at the end of the triple-row activation, the                   
data in all the three cells are overwritten with the final state                   
of the bitline. In other words, our approach overwrites the source                 
data with the final value. In the following sections, we propose a                 
simple implementation that addresses these challenges.

\subsection{Overview of Implementation of \buddyao}
\label{sec:and-or-mechanism}

To ensure that the source data does not get modified, our
mechanism first \emph{copies} the data from the two source rows to
two reserved temporary rows ($T1$ and $T2$). Depending on the
operation to be performed (AND or OR), our mechanism initializes a
third reserved temporary row $T3$ to ($0$ or $1$). It then
simultaneously activates the three rows $T1$, $T2$, and $T3$. It
finally copies the result to the destination row. For example, to
perform a bitwise AND of two rows $A$ and $B$ and store the result
in row $R$, our mechanism performs the following steps.
\begin{enumerate}\itemsep0pt\parsep0pt\parskip0pt
\item \emph{Copy} data of row $A$ to row $T1$
\item \emph{Copy} data of row $B$ to row $T2$
\item \emph{Initialize} row $T3$ to $0$
\item \emph{Activate} rows $T1$, $T2$, and $T3$ simultaneously
\item \emph{Copy} data of row $T1$ to row $R$
\end{enumerate}

While the above mechanism is simple, the copy operations, if
performed naively, will nullify the benefits of our
mechanism. Fortunately, we can use RowClone (described in
Chapter~\ref{chap:rowclone}), to perform row-to-row copy
operations quickly and efficiently within DRAM. To recap, RowClone
consists of two techniques. The first technique, RowClone-FPM
(Fast Parallel Mode), which is the fastest and the most efficient,
copies data within a subarray by issuing two back-to-back
{\cmdact}s to the source row and the destination row, without an
intervening \cmdpre. The second technique, RowClone-PSM (Pipelined
Serial Mode), efficiently copies data between two banks by using
the shared internal bus to overlap the read to the source bank
with the write to the destination bank.

With RowClone, all three copy operations (Steps 1, 2, and 5) and
the initialization operation (Step 3) can be performed efficiently
within DRAM. To use RowClone for the initialization operation, we
reserve two additional rows, $C0$ and $C1$. $C0$ is
pre-initialized to $0$ and $C1$ is pre-initialized to 1. Depending
on the operation to be performed, our mechanism uses RowClone to
copy either $C0$ or $C1$ to $T3$. Furthermore, to maximize the use
of RowClone-FPM, we reserve five rows in each subarray to serve as
the temporary rows ($T1$, $T2$, and $T3$) and the control rows
($C0$ and $C1$).

In the best case, when all the three rows involved in the                          
operation ($A$, $B$, and $R$) are in the same subarray, our                        
mechanism can use RowClone-FPM for all copy and initialization                     
operations. However, if the three rows are in different                            
banks/subarrays, some of the three copy operations have to use                     
RowClone-PSM. In the worst case, when all three copy operations                    
have to use RowClone-PSM, our approach will consume higher latency                 
than the baseline. However, when only one or two RowClone-PSM                      
operations are required, our mechanism will be faster and more                     
energy-efficient than existing systems. As our goal in this paper                  
is to demonstrate the power of our approach, in the rest of the                    
paper, we will focus our attention on the case when all rows                       
involved in the bitwise operation are in the same subarray.

\subsection{Reliability of Our Mechanism}                                          
                                                                                   
While our mechanism trivially addresses the second challenge                       
(modification of the source data), it also addresses the first                     
challenge (DRAM cell leakage). This is because, in our approach,                   
the source (and the control) data are copied to the rows $T1$,                     
$T2$ and $T3$ \emph{just} before the triple-row activation. Each                   
copy operation takes much less than $1~{\mu}s$, which is five                      
\emph{orders} of magnitude less than the typical refresh interval                  
($64~ms$). Consequently, the cells involved in the triple-row                      
activation are very close to the fully refreshed state before the                  
operation, thereby ensuring reliable operation of the triple-row                   
activation. Having said that, an important aspect of our mechanism                 
is that a chip that fails the tests for triple-row activation                      
(e.g., due to process variation) \emph{can still be used as a                      
  regular DRAM chip}. As a result, our approach is likely to have                  
little impact on the overall yield of DRAM chips, which is a major                 
concern for manufacturers.

\section{\buddynot}
\label{sec:bitwise-not}

\buddynot exploits the fact that the sense amplifier itself
consists of two inverters and the following observation about the
sense amplifier operation.
\begin{quote}
  At the end of the sense amplification process, while the bitline
  voltage reflects the logical value of the cell, the voltage
  level of the \bbar corresponds to the negation of the logical
  value of the cell.
\end{quote}

\subsection{Dual-Contact Cell}
\label{sec:dcc}
\begin{figure}[h]
  \centering
  \includegraphics{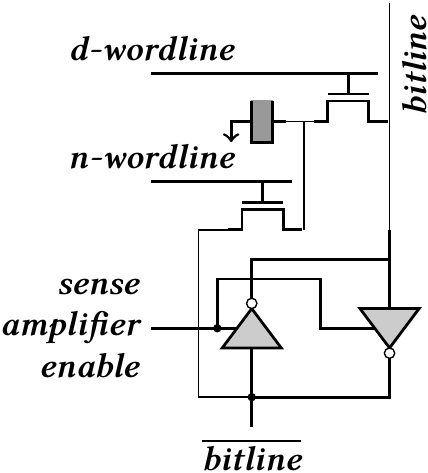}
  \caption[Dual-contact cell]{A dual-contact cell connected to
    both ends of a sense amplifier}
  \label{fig:dcc-not}
\end{figure}

Our high-level idea is to transfer the data on the \bbar to a cell
that can be connected to the bitline. For this purpose, we
introduce a special DRAM cell called \emph{dual-contact cell}. A
dual-contact cell (DCC) is a DRAM cell with two transistors and
one capacitor.  For each DCC, one transistor connects the DCC to
the bitline and the other transistor connects the DCC to the
\bbar. Each of the two transistors is controlled by a different
wordline. We refer to the wordline that controls the connection
between the DCC and the bitline as the \dwordline (or data
wordline). We refer to the wordline that controls the connection
between the DCC and the \bbar as the \nwordline (or negation
wordline). Figure~\ref{fig:dcc-not} shows one DCC connected to a
sense amplifier. In our mechanism, we use two DCCs for each sense
amplifier, one on each side of the sense amplifier.

\subsection{Exploiting the Dual-Contact Cell}
\label{sec:dcc-exploit}

Since the DCC is connected to both the bitline and the \bbar, we
can use a RowClone-like mechanism to transfer the negation of some
source data on to the DCC using the \nwordline. The negated data
can be transferred to the bitline by activating the \dwordline of
the DCC, and can then be copied to the destination cells using
RowClone.

\begin{figure}[h]
  \centering
  \includegraphics{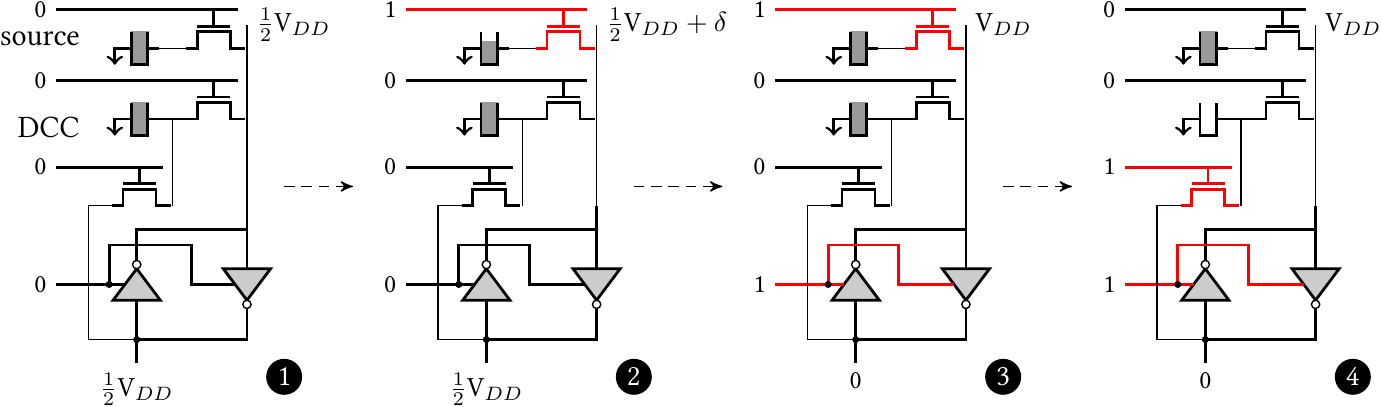}
  \caption{Bitwise NOT using a dual-contact cell}
  \label{fig:bitwise-not}
\end{figure}

Figure~\ref{fig:bitwise-not} shows the sequence of steps involved
in transferring the negation of a source cell on to the DCC. The
figure shows a \emph{source} cell and a DCC connected to the same
sense amplifier \ding{202}. Our mechanism first activates the
source cell \ding{203}. At the end of the activation process, the
bitline is driven to the data corresponding to the source cell,
\vdd in this case \ding{204}. More importantly, for the purpose of
our mechanism, the \bbar is driven to $0$. In this state, our
mechanism activates the \emph{n-wordline}, enabling the transistor
that connects the DCC to the \bbar~\ding{205}. Since the \bbar is
already at a stable voltage level of $0$, it overwrites the value
in the cell with $0$, essentially copying the negation of the
source data into the DCC. After this step, the negated data can be
efficiently copied into the destination cell using RowClone.
Section~\ref{sec:command-sequence} describes the sequence of
commands required to perform a bitwise NOT.

\section{Implementation and Hardware Cost}
\label{sec:buddy-implementation}

Besides the addition of a dual-contact cell on either side of each
sense amplifier, Buddy primarily relies on different variants and
sequences of row activations to perform bulk bitwise
operations. As a result, the main changes to the DRAM chip
introduced by Buddy are to the row decoding logic. While the
operation of \buddynot is similar to a regular cell operation,
\buddyao requires three rows to be activated
simultaneously. Activating three arbitrary rows within a subarray
1)~requires the memory controller to first communicate the three
addresses to DRAM, and 2)~requires DRAM to simultaneously decode
the three addresses. Both of these requirements incur huge cost,
i.e., wide address buses and three full row decoders to decode the
three addresses simultaneously.

In this work, we propose an implementation with much lower
cost. At a high level, we reserve a small fraction of row
addresses within each subarray for triple-row activation. Our
mechanism maps each reserved address to a pre-defined set of three
wordlines instead of one. With this approach, the memory
controller can perform a triple-row activation by issuing an
\cmdact with a \emph{single row address}. We now describe how we
exploit certain properties of Buddy to realize this
implementation.

\subsection{Row Address Grouping}
\label{sec:address-grouping}

\begin{figure}[h]
  \centering
  \includegraphics{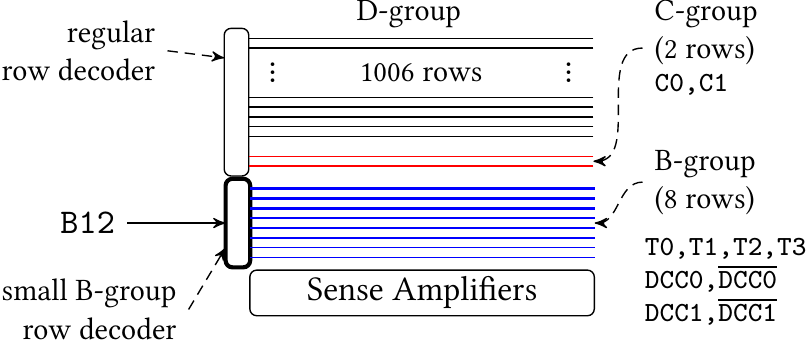}
  \caption[Logical subarray row address grouping]{Logical subarray
    address grouping. As an example, the figure shows how the
    B-group row decoder simultaneously activates rows \taddr{0},
    \taddr{1}, and \taddr{2} (highlighted in thick lines), with a
    single address \baddr{12}.}
  \label{fig:row-address-grouping}
\end{figure}

Before performing the triple-row activation, our mechanism copies
the source data and the control data to three temporary rows
(Section~\ref{sec:and-or-mechanism}). If we choose these temporary
rows at \emph{design time}, then these rows can be controlled
using a separate small row decoder. To exploit this idea, we
divide the space of row addresses within each subarray into three
distinct groups (B, C, and D), as shown in
Figure~\ref{fig:row-address-grouping}.

The \emph{B-group} (or the \emph{bitwise} group) corresponds to
rows that are used to perform the bitwise operations. This group
contains $16$ addresses that map to $8$ physical wordlines. Four
of the eight wordlines are the \emph{d-}and-\emph{n-wordlines}
that control the two rows of dual-contact cells. We will refer to
the {\dwordline}s of the two rows as \dccdz and \dccdo, and the
corresponding {\nwordline}s as \dccnz and \dccno. The remaining
four wordlines control four temporary rows of DRAM cells that will
be used by various bitwise operations. We refer to these rows as
\taddr{0}---\taddr{3}. While some B-group addresses activate
individual wordlines, others activate multiple wordlines
simultaneously.  Table~\ref{table:b-group-mapping} lists the
mapping between the 16 addresses and the wordlines. Addresses
\baddr{0}---\baddr{7} individually activate one of the $8$
physical wordlines. Addresses \baddr{12}---\baddr{15} activate
three wordlines simultaneously. These addresses will be used by
the memory controller to trigger bitwise AND or OR
operations. Finally, addresses \baddr{8}---\baddr{11} activate two
wordlines. As we will show in the next section, these addresses
will be used to copy the result of an operation simultaneously to
two rows (e.g., zero out two rows simultaneously).

\begin{table}[h]
  \centering
  \begin{tabular}{rl}
  \toprule
  \textbf{Addr.} & \textbf{Wordline(s)}\\
  \toprule
  \baddr{0} & \taddr{0}\\
  \baddr{1} & \taddr{1}\\
  \baddr{2} & \taddr{2}\\
  \baddr{3} & \taddr{3}\\
  \baddr{4} & \dccdz\\
  \baddr{5} & \dccnz\\
  \baddr{6} & \dccdo\\
  \baddr{7} & \dccno\\
  \bottomrule
\end{tabular}\quad
\begin{tabular}{rl}
  \toprule
  \textbf{Addr.} & \textbf{Wordline(s)}\\
  \toprule
  \baddr{8} & \dccnz, \taddr{0}\\
  \baddr{9} & \dccno, \taddr{1}\\
  \baddr{10} & \taddr{2}, \taddr{3}\\
  \baddr{11} & \taddr{0}, \taddr{3}\\
  \baddr{12} & \taddr{0}, \taddr{1}, \taddr{2}\\
  \baddr{13} & \taddr{1}, \taddr{2}, \taddr{3}\\
  \baddr{14} & \dccdz, \taddr{1}, \taddr{2}\\
  \baddr{15} & \dccdo, \taddr{0}, \taddr{3}\\
  \bottomrule
\end{tabular}

  \caption[Buddy: B-group address mapping]{Mapping of B-group addresses}
  \label{table:b-group-mapping}
\end{table}

The \emph{C-group} (or the \emph{control} group) contains the rows
that store the pre-initialized values for controlling the bitwise
AND/OR operations. Specifically, this group contains only two
addresses: \czero and \cone. The rows corresponding to \czero and
\cone are initialized to all-zeros and all-ones, respectively.

The \emph{D-group} (or the \emph{data} group) corresponds to the
rows that store regular user data. This group contains all the
addresses in the space of row addresses that are neither in the
\emph{B-group} nor in the \emph{C-group}. Specifically, if each
subarray contains $1024$ rows, then the \emph{D-group} contains
$1006$ addresses, labeled \daddr{0}---\daddr{1005}.

With these different address groups, the memory controller can
simply use the existing command interface and use the \cmdact
commands to communicate all variants of the command to the DRAM
chips. Depending on the address group, the DRAM chips can
internally process the activate command appropriately, e.g.,
perform a triple-row activation. For instance, by just issuing an
\cmdact to address \baddr{12}, the memory controller can
simultaneously activate rows \taddr{0}, \taddr{1}, and \taddr{2},
as illustrated in Figure~\ref{fig:row-address-grouping}.

\subsection{Split Row Decoder}
\label{sec:split-row-decoder}

Our idea of \emph{split row decoder} splits the row decoder into
two parts. The first part controls addresses from only the
\emph{B-group}, and the second part controls addresses from the
\emph{C-group} and the \emph{D-group} (as shown in
Figure~\ref{fig:row-address-grouping}). There are two benefits to
this approach. First, the complexity of activating multiple
wordlines is restricted to the small decoder that controls only
the \emph{B-group}. In fact, this decoder takes only a 4-bit input
(16 addresses) and generates a 8-bit output (8 wordlines). In
contrast, as described in the beginning of this section, a naive
mechanism to simultaneously activate three arbitrary rows incurs
high cost. Second, as we will describe in
Section~\ref{sec:command-sequence}, the memory controller must
perform several back-to-back {\cmdact}s to execute various bitwise
operations. In a majority of cases, the two rows involved in each
back-to-back {\cmdact}s are controlled by different decoders. This
enables an opportunity to overlap the two {\cmdact}s, thereby
significantly reducing their latency. We describe this
optimization in detail in Section~\ref{sec:accelerating-aap}.
Although the groups of addresses and the corresponding row
decoders are logically split, the physical implementation can use
a single large decoder with the wordlines from different groups
interleaved, if necessary.

\subsection{Executing Bitwise Ops: The AAP Primitive}
\label{sec:command-sequence}

To execute each bitwise operations, the memory controller must
send a sequence of commands. For example, to perform the bitwise
NOT operation, \daddr{k} \texttt{=} \bnot \daddr{i}, the memory
controller sends the following sequence of commands.

\begin{enumerate}\itemsep0pt\parskip0pt
\item \cmdact \daddr{i} \comment{Activate the source row}
\item \cmdact \baddr{5} \comment{Activate the \nwordline of \dccdz}
\item \cmdpre
\item \cmdact \baddr{4} \comment{Activate the \dwordline of \dccdz}
\item \cmdact \daddr{k} \comment{Activate the destination row}
\item \cmdpre
\end{enumerate}

Step 1 transfers the data from the source row to the array of
sense amplifiers. Step 2 activates the \nwordline of one of the
DCCs, which connects the dual-contact cells to the corresponding
{\bbar}s. As a result, this step stores the negation of the source
cells into the corresponding DCC row (as described in
Figure~\ref{fig:bitwise-not}). After the precharge operation in
Step 3, Step 4 activates the \dwordline of the DCC row,
transferring the negation of the source data on to the
bitlines. Finally, Step 5 activates the destination row. Since the
sense amplifiers are already activated, this step copies the data
on the bitlines, i.e., the negation of the source data, to the
destination row. Step 6 completes the negation operation by
precharging the bank.

If we observe the negation operation, it consists of two steps of
\cmdact-\cmdact-\cmdpre operations. We refer to this sequence of
operations as the \aap primitive. Each \aap takes two addresses as
input. \texttt{\aap(row1, row2)} corresponds to the following
sequence of commands:\\
\centerline{\texttt{\cmdact row1; \cmdact row2; \cmdpre;}} 

With the \aap primitive, the \bnot operation, \daddr{k}
\texttt{=} \bnot \daddr{i}, can be rewritten as,
\begin{enumerate}\itemsep0pt\parskip0pt
\item \callaap{\daddr{i}}{\baddr{5}} \tcomment{\dccd = \bnot \daddr{i}}
\item \callaap{\baddr{4}}{\daddr{k}} \tcomment{\daddr{k} = \dccd}
\end{enumerate}

\begin{figure}[h]
  \centering
  \includegraphics{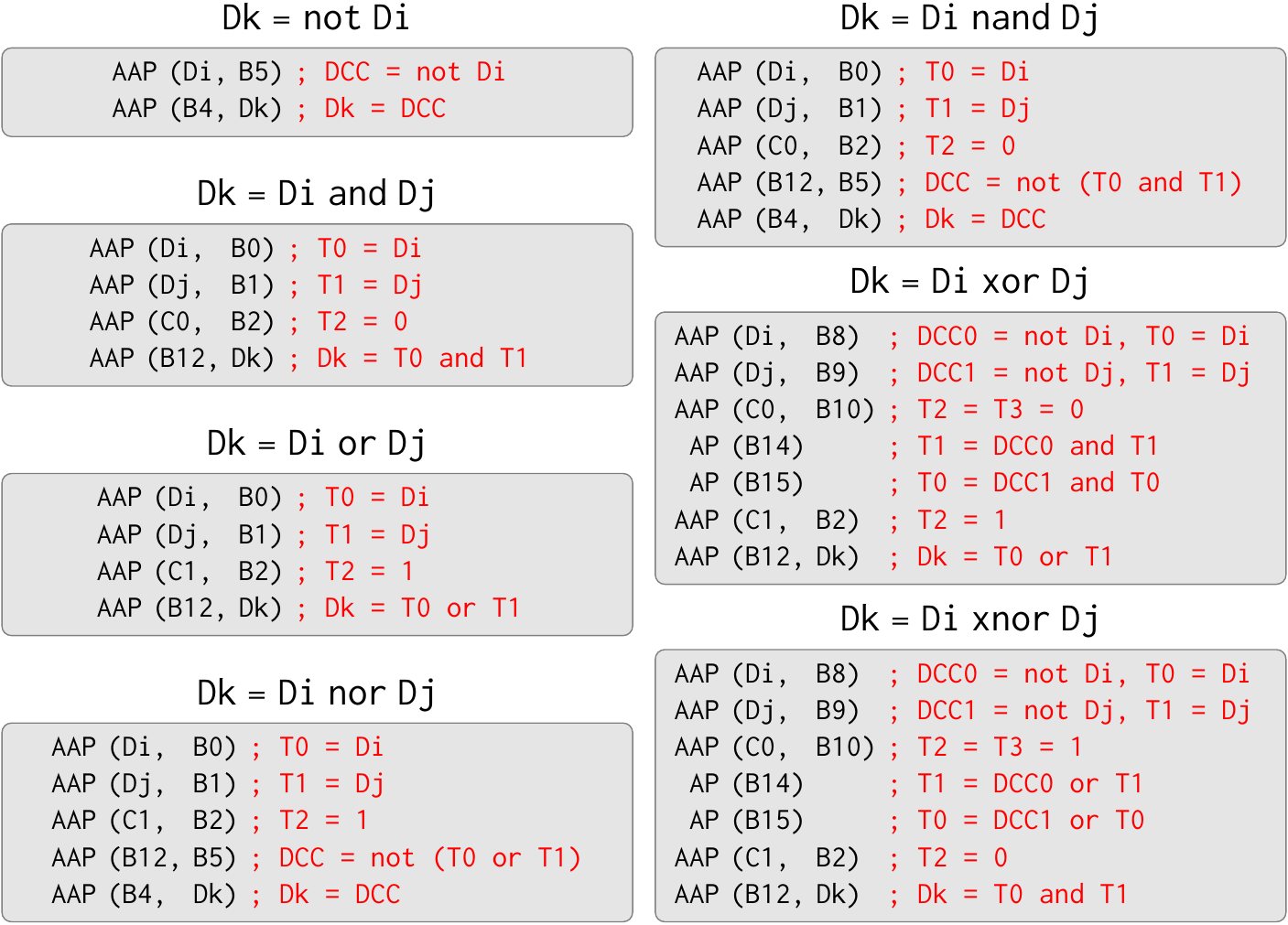}
  \caption{Command sequences for different bitwise operations}
  \label{fig:command-sequences}
\end{figure}

In fact, we observe that all the bitwise operations mainly involve
a sequence of \aap operations. Sometimes, they require a regular
\cmdact followed by a \cmdpre operation. We will use \ap to refer
to such operations.  Figure~\ref{fig:command-sequences} shows the
sequence of steps taken by the memory controller to execute seven
bitwise operations: \bnot, \band, \bor, \bnand, \bnor, \bxor, and
\bxnor. Each step is annotated with the logical result of
performing the step.

As an illustration, let us consider the \band operation, \daddr{k}
= \daddr{i} \band \daddr{j}. The first step
(\callaap{\daddr{i}}{\baddr{0}}) first activates the source row
\daddr{i}, followed by the temporary row \taddr{0} (which
corresponds to address \baddr{0}). As a result, this operation
copies the data of \daddr{i} to the temporary row \taddr{0}.
Similarly, the second step (\callaap{\daddr{j}}{\baddr{1}}) copies
the data of the source row \daddr{j} to the temporary row
\taddr{1}, and the third step \callaap{\czero}{\baddr{2}} copies
the data of the control row ``0'' to the temporary row
\taddr{2}. Finally, the last step
(\callaap{\baddr{12}}{\daddr{k}}) first issues an \cmdact to
address \baddr{12}. As described in
Table~\ref{table:b-group-mapping} and illustrated in
Figure~\ref{fig:row-address-grouping}, this command simultaneously
activates the rows \taddr{0}, \taddr{1}, and \taddr{2}, resulting
in a \band operation of the values of rows \taddr{0} and
\taddr{1}. As this command is immediately followed by an \cmdact
to \daddr{k}, the result of the \band operation is copied to the
destination row \daddr{k}. 

\subsection{Accelerating the AAP Primitive}
\label{sec:accelerating-aap}

It is clear from Figure~\ref{fig:command-sequences} that the
latency of executing any bitwise operation using Buddy depends on
the latency of executing the \aap primitive. The latency of the
\aap primitive in turn depends on the latency of the \cmdact and
the \cmdpre operations. In the following discussion, we assume
DDR3-1600 (8-8-8) timing parameters~\cite{ddr3-1600}. For these
parameters, the latency of an \cmdact operation is \tras = 35~ns,
and the latency of a \cmdpre operation is \trp = 10~ns.

\subsubsection{Naive Execution of AAP.}

The naive approach is to perform the three operations involved in
\aap serially one after the other. Using this simple approach, the
latency of the \aap operation is 2\tras + \trp =
\textbf{80~ns}. While Buddy outperforms existing systems even with
this naive approach, we exploit some properties of \aap to further
reduce its latency.

\subsubsection{Shortening the Second \cmdact.}

We observe that the second \cmdact operation is issued when the
bank is already activated. As a result, this \cmdact does not
require the full sense-amplification process, which is the
dominant source of the latency of an \cmdact. In fact, the second
\cmdact of an \aap only requires the corresponding wordline to be
raised, and the bitline data to overwrite the cell data. We
introduce a new timing parameter called \twl, to capture the
latency of these steps. With this optimization, the latency of
\aap is \tras + \twl + \trp.

\subsubsection{Overlapping the Two {\cmdact}s.}

For all the bitwise operations
(Figure~\ref{fig:command-sequences}), with the exception of one
\aap each in \bnand and \bnor, \emph{exactly one} of the two
\cmdacts in each \aap is to a \emph{B-group} address. Since the
wordlines in \emph{B-group} are controlled by a different row
decoder (Section~\ref{sec:split-row-decoder}), we can overlap the
two \cmdacts of the \aap primitive. More precisely, if the second
\cmdact is issued after the first activation has sufficiently
progressed, the sense amplifiers will force the data of the second
row to the result of the first activation. This operation is
similar to the inter-segment copy operation in Tiered-Latency
DRAM~\cite{tl-dram} (Section~4.4). Based on SPICE simulations, the
latency of the executing both the \cmdacts is 4~ns larger than
\tras. Therefore, with this optimization, the latency of \aap is
\tras + 4~ns + \trp = \textbf{49~ns}.

\subsection{DRAM Chip and Controller Cost}
\label{sec:hardware-cost}

Buddy has three main sources of cost to the DRAM chip. First,
Buddy requires changes to the row decoding logic. Specifically,
the row decoding logic must distinguish between the \emph{B-group}
addresses and the remaining addresses. Within the \emph{B-group},
it must implement the mapping between the addresses and the
wordlines described in Table~\ref{table:b-group-mapping}. As
described in Section~\ref{sec:address-grouping}, the
\emph{B-group} contains only 16 addresses that are mapped to 8
wordlines. As a result, we expect the complexity of the changes to
the row decoding logic to be low.

The second source of cost is the design and implementation of the
dual-contact cells (DCCs). In our design, each sense amplifier has
only one DCC on each side, and each DCC has two wordlines
associated with it. Consequently, there is enough space to
implement the second transistor that connects the DCC to the
corresponding \bbar. In terms of area, the cost of each DCC is
roughly equivalent to two regular DRAM cells. As a result, we can
view each row of DCCs as two rows of regular DRAM cells.

The third source of cost is the capacity lost as a result of
reserving the rows in the \emph{B-group} and \emph{C-group}. The
rows in these groups are reserved for the memory controller to
perform bitwise operations and cannot be used to store application
data (Section~\ref{sec:address-grouping}). Our proposed
implementation of Buddy reserves 18 addresses in each subarray for
the two groups. For a typical subarray size of 1024 rows, the loss
in memory capacity is $\approx$1\%.

On the controller side, Buddy requires the memory controller to                    
1)~store information about different address groups, 2)~track the                  
timing for different variants of the \cmdact (with or without the                  
optimizations), and 3)~track the status of different on-going                      
bitwise operations. While scheduling different requests, the                       
controller 1)~adheres to power constraints like tFAW which limit                   
the number of full row activations during a given time window, and                 
2)~can interleave the multiple AAP commands to perform a bitwise                   
operation with other requests from different applications. We                      
believe this modest increase in the DRAM chip/controller                           
complexity and capacity cost is negligible compared to the                         
improvement in throughput and performance enabled by Buddy.

\section{End-to-end System Support}
\label{sec:support}

We envision two distinct ways of integrating Buddy with the rest                   
of the system. The first way is a loose integration, where Buddy                   
is treated as an accelerator (similar to a GPU). The second way is                 
a much tighter integration, where Buddy is supported by the main                   
memory. In this section, we discuss these two ways along with                      
their pros and cons.                                                     
                                                                                   
\subsection{Buddy as an Accelerator}                                               
                                                                                   
Treating Buddy as an accelerator is probably the simplest way of
integrating Buddy into a system. In this approach, the
manufacturer of Buddy RAM designs the accelerator that can be
plugged into the system as a separate device (e.g., PCIe). While
this mechanism requires communication between the CPU and the
Buddy accelerator, there are benefits to this approach that lower
the cost of integration. First, a \emph{single} manufacturer can
design both the DRAM and the memory controller (which is not true
about commodity DRAM). Second, the details of the data mapping to
suit Buddy can be hidden behind the device driver, which can
expose a simple-to-use API to the applications.
                                                                                   
\subsection{Integrating Buddy with System Main Memory}                             
                                                                                   
A tighter integration of Buddy with the system main memory                         
requires support from different layers of the system stack, which                  
we discuss below.

\subsubsection{ISA Support}

For the processor to exploit Buddy, it must be able to identify
and export instances of bulk bitwise operations to the memory
controller. To enable this, we introduce new instructions that
will allow software to directly communicate instances of bulk
bitwise operations to the processor. Each new instruction takes
the following form:\\
\centerline{\texttt{bop dst, src1, [src2], size}}

where \texttt{bop} is the bitwise operation to be performed, dst
is the address of the destination, src1 and src2 correspond to the
addresses of the source, and size denotes the length of the
vectors on which the bitwise operations have to be performed. The
microarchitectural implementation of these instructions will
determine whether each instance of these instructions can be
accelerated using Buddy.

\subsubsection{Implementing the New Buddy Instructions.}

The microarchitectural implementation of the new instructions will                 
determine whether each instance can be accelerated using Buddy.                    
Buddy imposes two constraints on the data it operates on. First,                   
both the source data and the destination data should be within the                 
same subarray. Second, all Buddy operations are performed on an                    
entire row of data. As a result, the source and destination data                   
should be row-aligned and the operation should span at least an                    
entire row. The microarchitecture ensures that these constraints                   
are satisfied before performing the Buddy                                          
operations. Specifically, if the source and destination rows are                   
not in the same subarray, the processor can either 1)~use                          
RowClone-PSM~\cite{rowclone} to copy the data into the same                        
subarray, or 2)~execute the operation using the CPU. This choice                   
can be dynamically made depending on the number of RowClone-PSM                    
operations required and the memory bandwidth contention. If the                    
processor cannot ensure data alignment, or if the size of the                      
operation is smaller than the DRAM row size, it can execute the                    
operations using the CPU. However, with careful application design                 
and operating system support, the system can maximize the use of                   
Buddy to extract its performance and efficiency benefits.

\subsubsection{Maintaining On-chip Cache Coherence.}

Buddy directly reads/modifies data in main memory. As a result, we
need a mechanism to ensure the coherence of data present in the
on-chip caches. Specifically, before performing any Buddy
operation, the memory controller must first flush any dirty cache
lines from the source rows and invalidate any cache lines from
destination rows. While flushing the dirty cache lines of the
source rows is on the critical path of any Buddy operation, we can
speed up using the Dirty-Block Index (described in
Chapter~\ref{chap:dbi}). In contrast, the cache lines of the
destination rows can be invalidated in parallel with the Buddy
operation. The mechanism required to maintain cache coherence
introduces a small performance overhead. However, for applications
with large amounts of data, since the cache contains only a small
fraction of the data, the performance benefits of our mechanism
significantly outweigh the overhead of maintaining coherence,
resulting in a net gain in both performance and efficiency.

\subsubsection{Software Support}

The minimum support that Buddy requires from software is for the
application to use the new Buddy instructions to communicate the
occurrences of bulk bitwise operations to the processor. However,
with careful memory allocation support from the operating system,
the application can maximize the benefits it can extract from
Buddy. Specifically, the OS must allocate pages that are likely to
be involved in a bitwise operation such that 1)~they are
row-aligned, and 2)~belong to the same subarray. Note that the OS
can still interleave the pages of a single data structure to
multiple subarrays. Implementing this support, requires the OS to
be aware of the subarray mapping, i.e., determine if two physical
pages belong to the same subarray or not. The OS must extract this
information from the DRAM modules with the help of the memory
controller.

\section{Analysis of Throughput \& Energy}
\label{sec:lte-analysis}

In this section, we compare the raw throughput and energy of
performing bulk bitwise operations of Buddy to an Intel Skylake
Core i7 processor using Advanced Vector
eXtensions~\cite{intel-avx}. The system contains a per-core 32 KB
L1 cache and 256 KB L2 cache, and a shared 8 MB L3 cache. The
off-chip memory consists of 2 DDR3-2133 memory channels with 8GB
of memory on each channel. We run a simple microbenchmark that
performs each bitwise operation on one or two vectors and stores
the result in a result vector. We vary the size of the vector, and
for each size we measure the throughput of performing each
operation with 1, 2, and 4 cores.

\begin{figure}[h]
  \centering
  \includegraphics{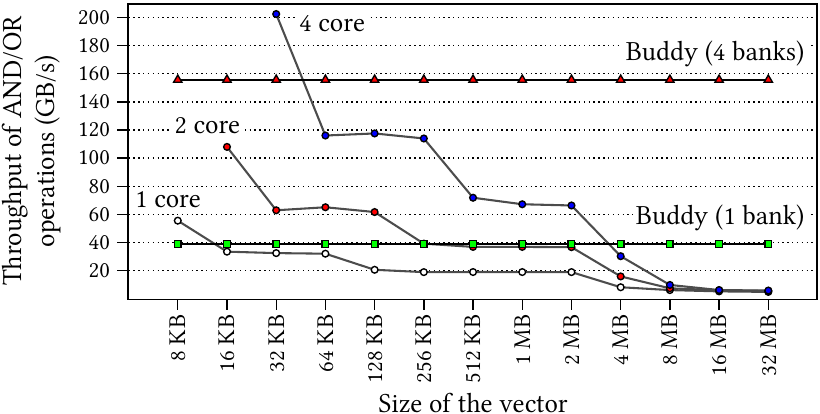}
  \caption{Comparison of throughput of AND/OR operations}
  \label{plot:buddy-throughput}
\end{figure}

Figure~\ref{plot:buddy-throughput} plots the results of this
experiment for bitwise AND/OR operations. The x-axis shows the
size of the vector, and the y-axis plots the corresponding
throughput (in terms of GB/s of computed result) for the Skylake
system with 1, 2, and 4 cores, and Buddy with 1 or 2 DRAM banks.

First, for each core count, the throughput of the Skylake drops
with increasing size of the vector. This is expected as the
working set will stop fitting in different levels of the on-chip
cache as we move to the right on the x-axis. Second, as long the
working set fits in some level of on-chip cache, running the
operation with more cores provides higher throughput. Third, when
the working set stops fitting in the on-chip caches (> 8MB per
vector), the throughput with different core counts is roughly the
same (5.8~GB/s of computed result). This is because, at this
point, the throughput of the operation is strictly limited by the
available memory bandwidth. Finally, for working sets larger than
the cache size, Buddy significantly outperforms the baseline
system. Even using only one bank, Buddy achieves a throughput of
38.92 GB/s, 6.7X better than the baseline. As modern DRAM chips
have abundant bank-level parallelism, Buddy can achieve much
higher throughput by using more banks (e.g., 26.8X better
throughput with 4 banks, compared to the baseline). In fact, while
the throughput of bitwise operations in existing systems is
limited by the memory bandwidth, the throughput enabled by Buddy
scales with the number of banks in the system.

Table~\ref{table:buddy-throughput-energy} shows the throughput and
energy results different bitwise operations for the 32MB input. We
estimate energy for DDR3-1333 using the Rambus
model~\cite{rambus-power}. Our energy numbers only include the
DRAM and channel energy, and does not include the energy spent at
the CPU and caches. For Buddy, some activate operations have rise
multiple wordlines and hence will consume higher energy. To
account for this, we increase the energy of the activate operation
by 22\% for each additional wordline raised. As a result, a
triple-row activation will consume 44\% more energy than a regular
activation.

\begin{table}[h]
  \centering
  \begin{tabular}{rrrrrrr}
  \toprule
  \multirow{2}{*}{Operation} & \multicolumn{3}{c}{Throughput (GB/s)} & \multicolumn{3}{c}{Energy (nJ)}\\
   & Base & Buddy & \multicolumn{1}{c}{($\uparrow$)} & Base & Buddy & \multicolumn{1}{c}{($\downarrow$)}\\
  \toprule
  \bnot & 7.7 & 77.8 & 10.1X & 749.6 & 12.6 & 59.5X\\
  \band/\bor & 5.8 & 38.9 & 6.7X & 1102.8 & 25.1 & 43.9X\\
  \bnand/\bnor & 5.8 & 31.1 & 5.3X & 1102.8 & 31.4 & 35.1X\\
  \bxor/\bxnor & 5.8 & 22.2 & 3.8X & 1102.8 & 44.0 & 25.1X\\
  \bottomrule  
\end{tabular}

  \caption[Buddy: Throughput/energy comparison]{Comparison of
    throughput and energy for various groups of bitwise
    operations. ($\uparrow$) and ($\downarrow$) respectively
    indicate the factor improvement and reduction in throughput
    and energy of Buddy (1 bank) over the baseline (Base).}
  \label{table:buddy-throughput-energy}
\end{table}

In summary, across all bitwise operations, Buddy reduces energy
consumption by at least 25.1X and up to 59.5X compared to the
baseline, and with just one bank, Buddy improves the throughput
by at least 3.8X and up to 10.1X compared to the baseline. In the
following section, we demonstrate the benefits of Buddy in some
real-world applications. 

\section{Effect on Real-world Applications}
\label{sec:applications}

To demonstrate the benefits of Buddy on real-world applications,
we implement Buddy in the Gem5~\cite{gem5} simulator. We implement
the new Buddy instructions using the pseudo instruction framework
in Gem5. We simulate an out-of-order, 4 GHz processor with 32 KB
L1 D-cache and I-cache, with a shared 2 MB L2 cache. All caches
use a 64B cache line size. We model a 1-channel, 1-rank, DDR4-2400
main memory.  In Section~\ref{sec:bitmap-indices}, we first show
that Buddy can significantly improve the performance of an
in-memory bitmap index. In Section~\ref{sec:bitset}, we show that
Buddy generally makes bitmaps more attractive for various set
operations compared to traditional red-black trees. In
Section~\ref{sec:other-apps}, we discuss other potential
applications that can benefit from Buddy.

\subsection{Bitmap Indices}
\label{sec:bitmap-indices}

Bitmap indices are an alternative to traditional B-tree indices
for databases. Compared to B-trees, bitmap indices can 1)~consume
less space, and 2)~improve performance of certain queries. There
are several real-world implementations of bitmap indices for
databases (e.g., Oracle~\cite{oracle}, Redis~\cite{redis},
Fastbit~\cite{fastbit}, rlite~\cite{rlite}). Several real
applications (e.g., Spool~\cite{spool}, Belly~\cite{belly},
bitmapist~\cite{bitmapist}, Audience Insights~\cite{ai}) use
bitmap indices for fast analytics.

Bitmap indices rely on fast bitwise operations on large bit
vectors to achieve high performance. Therefore, Buddy can
accelerate operations on bitmap indices, thereby improving overall
application performance.

To demonstrate this benefit, we use the following workload
representative of many applications. The application uses bitmap
indices to track users' characteristics (e.g., gender, premium)
and activities (e.g., did the user log in to the website on day
'X'?)  for $m$ users. The applications then uses bitwise
operations on these bitmaps to answer several different
queries. Our workload runs the following query: ``How many unique
users were active every week for the past $n$ weeks? and How many
premium users were active each of the past $n$ weeks?''  Executing
this query requires 6$n$ bitwise \bor, 2$n$-1 bitwise \band, and
$n$+1 bitcount operations.

\begin{figure}[h]
  \centering
  \includegraphics[scale=1.5]{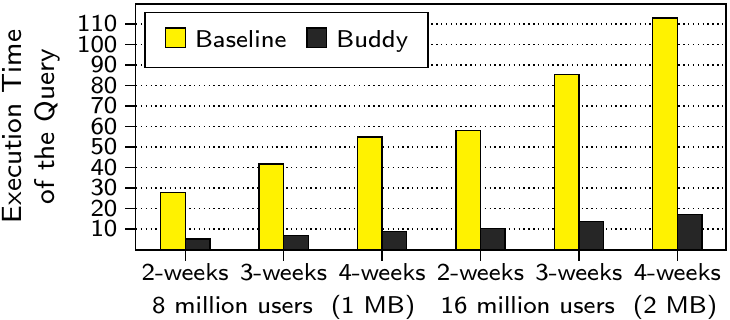}
  \caption{Performance of Buddy for bitmap indices}
  \label{fig:rlite}
\end{figure}

The size of each bitmap (and hence each bitwise operation) depends
on the number of users. For instance, a reasonably large
application that has 8 million users will require each bitmap to
be around 1 MB. Hence, these operations can easily be accelerated
using Buddy (the bitcount operations are performed by the CPU).
Figure~\ref{fig:rlite} shows the execution time of the baseline
and Buddy for the above experiment for various values of $m$
(number of users) and $n$ (number of days).

We draw two conclusions. First, as each query has $O(n)$ bitwise
operations and each bitwise operation takes $O(m)$ time, the
execution time of the query increases with increasing value
$mn$. Second, Buddy significantly reduces the query execution time
by 6X (on average) compared to the baseline.

While we demonstrate the benefits of Buddy using one query, as all
bitmap index queries involve several bitwise operations, Buddy
will provide similar performance benefits for any application
using bitmap indices.

\subsection{Bit Vectors vs. Red-Black Trees}
\label{sec:bitset}

A \emph{set} data structure is widely used in many
algorithms. Many libraries (e.g., C++ Standard Template
Library~\cite{stl}), use red-black trees~\cite{red-black-tree}
(RB-trees) to implement a set. While RB-trees are efficient when
the domain of elements is very large, when the domain is limited,
a set can be implemented using a bit vector. Bit vectors offer
constant time insert and lookup as opposed to RB-trees, which
consume $O(\log n)$ time for both operations. However, with bit
vectors, set operations like union, intersection, and difference
have to operate on the entire bit vector, regardless of whether
the elements are actually present in the set. As a result, for
these operations, depending on the number of elements actually
present in each set, bit vectors may outperform or perform worse
than a RB-trees. With support for fast bulk bitwise operations, we
show that Buddy significantly shifts the trade-off spectrum in
favor of bit vectors.

\begin{figure}[h]
  \centering
  \includegraphics[scale=1.5]{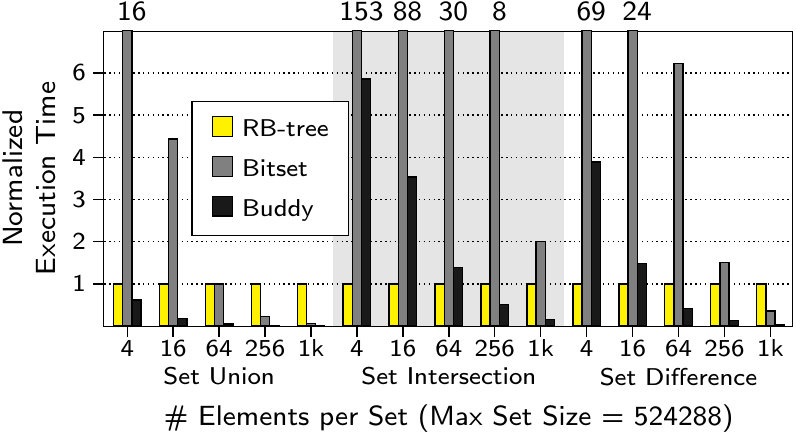}
  \caption{Comparison between RB-Tree, Bitset, and Buddy}
  \label{plot:set-results}
\end{figure}

To demonstrate this, we compare the performance of union,
intersection, and difference operations using three
implementations: RB-tree, bit vectors with SSE optimization
(Bitset), and bit vectors with Buddy. We run a microbenchmark that
performs each operation on 15 sets and stores the result in an
output set. Each set can contain elements between 1 and 524288
($2^{19}$). Therefore, the bit vector approaches requires 64~KB to
represent each set. For each operation, we vary the number of
elements present in the input sets. Figure~\ref{plot:set-results}
shows the results of this experiment. The figure plots the
execution time for each implementation normalized to RB-tree.

We draw three conclusions. First, by enabling much higher
throughput for bitwise operations, Buddy outperforms the baseline
bitset on all the experiments. Second, as expected, when the
number of elements in each set is very small (16 out of 524288),
RB-Tree performs better than the bit vector based
implementations. Third, even when each set contains only 1024 (out
of 524288) elements, Buddy significantly outperforms RB-Tree.

In summary, by performing bulk bitwise operations efficiently and
with much higher throughput compared to existing systems, Buddy
makes a bit-vector-based implementation of a set more attractive in
scenarios where red-black trees previously outperformed bit
vectors.

\subsection{Other Applications}
\label{sec:other-apps}

\subsubsection{Cryptography.} Many encryption algorithms in
cryptography heavily use bitwise operations (e.g.,
XOR)~\cite{xor1,xor2,enc1}. The Buddy support for fast and
efficient bitwise operations can i)~boost the performance of
existing encryption algorithms, and ii)~enable new encryption
algorithms with high throughput and efficiency.

\subsubsection{DNA Sequence Mapping.} DNA sequence mapping has
become an important problem, with applications in personalized
medicine. Most algorithms~\cite{dna-overview} rely on identifying
the locations where a small DNA sub-string occurs in the reference
gnome. As the reference gnome is large, a number of pre-processing
algorithms~\cite{dna-algo1,dna-algo2,dna-algo3,dna-algo4} have
been proposed to speedup this operation. Based on a prior
work~\cite{dna-our-algo}, we believe bit vectors with support for
fast bitwise operations using Buddy can enable an efficient
filtering mechanism.

\subsubsection{Approximate Statistics.} Certain large systems employ
probabilistic data structures to improve the efficiency of
maintaining statistics~\cite{summingbird}. Many such structures
(e.g., Bloom filters) rely on bitwise operations to achieve high
efficiency. By improving the throughput of bitwise operations,
Buddy can further improve the efficiency of such data structures,
and potentially enable the design of new data structures in this
space.

\section{Related Work}
\label{sec:buddy-related}

There are several prior works that aim to enable efficient
computation near memory. In this section, we qualitatively compare
Buddy to these prior works.

Some recent patents~\cite{mikamonu,mikamonu2} from Mikamonu
describe an architecture that employs a DRAM organization with
3T-1C cells and additional logic to perform NAND/NOR operations on
the data inside DRAM. While this architecture can perform bitwise
operations inside DRAM, it incurs significant additional cost to
the DRAM array due to the extra transistors, and hence reduces
overall memory density/capacity. In contrast, Buddy exploits
existing DRAM operation to perform bitwise operations efficiently
inside DRAM. As a result, it incurs much lower cost compared to
the Mikamonu architecture.

One main source of memory inefficiency in existing systems is data
movement. Data has to travel off-chip buses and multiple levels of
caches before reaching the CPU. To avoid this data movement, many
works (e.g., NON-VON Database Machine~\cite{non-von-machine},
DIVA~\cite{diva}, Terasys~\cite{pim-terasys}, Computational
RAM~\cite{cram}, FlexRAM~\cite{flexram,programming-flexram},
EXECUBE~\cite{execube}, Active Pages~\cite{active-pages},
Intelligent RAM~\cite{iram}, Logic-in-Memory
Computer~\cite{lim-computer}) have proposed mechanisms and models
to add processing logic close to memory. The idea is to integrate
memory and CPU on the same chip by designing the CPU using the
memory process technology. While the reduced data movement allows
these approaches to enable low-latency, high-bandwidth, and
low-energy data communication, they suffer from two key
shortcomings.  First, this approach of integrating processor on
the same chip as memory significantly deviates from existing
designs, and as a result, increases the overall cost of the
system. Second, DRAM vendors use a high-density process to
minimize cost-per-bit. Unfortunately, high-density DRAM process is
not suitable for building high-speed logic~\cite{iram}. As a
result, this approach is not suitable for building a general
purpose processor near memory.  In contrast, we restrict our focus
to bitwise operations, and propose a mechanism to perform them
efficiently inside DRAM with low cost.

Some recent DRAM architectures~\cite{3d-stacking,hmc,hbm} use
3D-stacking technology to stack multiple DRAM chips on top of the
processor chip or a separate logic layer. These architectures
offer much higher bandwidth to the logic layer compared to
traditional off-chip interfaces. This enables an opportunity to
offload some computation to the logic layer, thereby improving
performance. In fact, many recent works have proposed mechanisms
to improve and exploit such architectures
(e.g.,~\cite{pim-enabled-insts,pim-graph,top-pim,nda,msa3d,spmm-mul-lim,data-access-opt-pim,tom,hrl,gp-simd,ndp-architecture,pim-analytics,nda-arch,jafar,data-reorg-3d-stack,smla}). Unfortunately,
despite enabling higher bandwidth compared to off-chip memory,
such 3D-stacked architectures are still require data to be
transferred outside the DRAM chip, and hence can be
bandwidth-limited. However, since Buddy can be integrated easily
with such architectures, we believe the logic layer in such 3D
architectures should be used to implement \emph{more complex
  operations}, while Buddy can be used to efficiently implement
bitwise logical operations at low cost.

\section{Summary}
\label{sec:buddy-summary}

In this chapter, we introduced Buddy, a new DRAM substrate that
performs row-wide bitwise operations using DRAM
technology. Specifically, we proposed two component
mechanisms. First, we showed that simultaneous activation of three
DRAM rows that are connected to the same set of sense amplifiers
can be used to efficiently perform AND/OR operations. Second, we
showed that the inverters present in each sense amplifier can be
used to efficiently implement NOT operations. With these two
mechanisms, Buddy can perform any bulk bitwise logical operation
quickly and efficiently within DRAM. Our evaluations show that
Buddy enables an order-of-magnitude improvement in the throughput
of bitwise operations. This improvement directly translates to
significant performance improvement in the evaluated real-world
applications.  Buddy is generally applicable to any memory
architecture that uses DRAM technology, and we believe that the
support for fast and efficient bulk bitwise operations can enable
better design of applications that result in large improvements in
performance and efficiency.

\chapter{Gather-Scatter DRAM}
\label{chap:gsdram}

\begin{figure}[b!]
  \hrule\vspace{2mm}
  \begin{footnotesize}
    Originally published as ``Gather-Scatter DRAM: In-DRAM Address
    Translation to Improve the Spatial Locality of Non-unit
    Strided Accesses'' in the
    International Symposium on Microarchitecture,
    2015~\cite{gsdram}
  \end{footnotesize}
\end{figure}

In this chapter, we shift our focus to the problem of non-unit
strided access patterns. As described in
Section~\ref{sec:non-unit-stride-problem}, such access patterns
present themselves in many important applications such as
in-memory databases, scientific computation, etc. As illustrated
in that section, non-unit strided access patterns exhibit low
spatial locality. In existing memory systems that are optimized to
access and store wide cache lines, such access patterns result in
high inefficiency.

The problem presents itself at two levels. First, commodity DRAM
modules are designed to supply wide contiguous cache lines. As a
result, the cache lines fetched by the memory controller are only
partially useful---i.e., they contain many values that do not
belong to the strided access pattern. This results in both high
latency and wasted memory bandwidth. Second, modern caches are
optimized to store cache lines. Consequently, even the caches have
to store values that do not belong to the strided access. While
this results in inefficient use of the on-chip cache space, this
also negatively affects SIMD optimizations on strided data. The
application must first gather (using software or hardware) the
values of the strided access in a single vector register before it
can perform any SIMD operation. Unfortunately, this gather
involves multiple physical cache lines, and hence is a
long-latency operation.

Given the importance of strided access patterns, several prior
works (e.g., Impulse~\cite{impulse,impulse-journal},
Adaptive/Dynamic Granularity Memory Systems~\cite{agms,dgms}) have
proposed solutions to improve the performance of strided
accesses. Unfortunately, prior works~\cite{impulse,agms,dgms}
require the off-chip memory interface to support fine-grained
memory
accesses~\cite{mini-rank,mc-dimm,mc-dimm-cal,threaded-module,sg-dimm}
and, in some cases, a sectored
cache~\cite{sectored-cache,dscache}. These approaches
significantly increase the cost of the memory interface and the
cache tag store, and potentially lower the utilization of off-chip
memory bandwidth and on-chip cache space. We discuss these prior
works in more detail in Section~\ref{sec:gsdram-prior-work}.

Our goal is to design a mechanism that 1)~improves the performance
(cache hit rate and memory bandwidth consumption) of strided
accesses, and 2)~works with commodity DRAM modules and traditional
non-sectored caches with very few changes. To this end, we first
restrict our focus to power-of-2 strided access patterns and
propose the Gather-Scatter DRAM (\gsdram), a substrate that allows
the memory controller to gather or scatter data with any
power-of-2 stride efficiently with very few changes to the DRAM
module. In the following sections, we describe \gsdram in detail.

\section{The Gather-Scatter DRAM}
\label{sec:gsdram-mechanism}

\begin{figure}[h]
  \centering
  \includegraphics{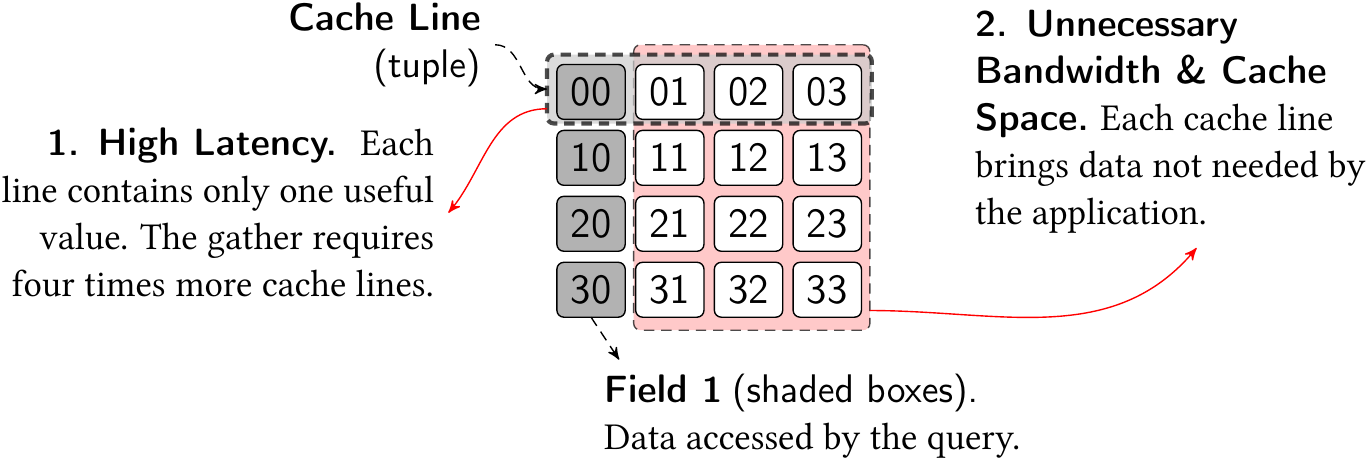}
  \caption[Shortcomings of strided access pattern]{Problems in
    accessing the first field (shaded boxes) from a table in a
    cache-line-optimized memory system. The box ``ij'' corresponds
    to the j$^{th}$ field of the i$^{th}$ tuple.}
  \label{fig:matrix}
\end{figure}

For the purpose of understanding the problems in existing systems
and understanding the challenges in designing in \gsdram, we use
the following database example. The database consists of a single
table with four fields. We assume that each tuple of the database
fits in a cache line. Figure~\ref{fig:matrix} illustrates the
two problems of accessing just the first field of the table. As
shown in the figure, this query results in high latency, and
wasted bandwidth and cache space.

Our \textbf{goal} is to design a DRAM substrate that will enable
the processor to access a field of the table (stored in
tuple-major order) across all tuples, without incurring the
penalties of existing interfaces. More specifically, if the memory
controller wants to read the first field of the first four tuples
of the table, it must be able to issue a \emph{single command}
that fetches the following gathered cache line:
\tikz[baseline={([yshift=-8pt]current bounding box.north)}]{
  \tikzset{col1/.style={draw,minimum height=2mm, minimum
      width=2mm,scale=0.8,inner sep=2pt, anchor=west,
      fill=black!30, rounded corners=2pt,outer
      sep=-0pt,xshift=1mm}}; \node (v1) [col1] {\sffamily 00}; \node (v2) at
  (v1.east) [col1] {\sffamily 10}; \node (v3) at (v2.east) [col1] {\sffamily 20};
  \node (v4) at (v3.east) [col1] {\sffamily 30};}. At the same time, the
controller must be able to read a tuple from memory (e.g.,
\tikz[baseline={([yshift=-8pt]current bounding box.north)}]{
  \tikzset{col1/.style={draw,minimum height=2mm, minimum
      width=2mm,inner sep=2pt, anchor=west, fill=black!30, rounded
      corners=2pt,outer sep=0pt,scale=0.8}};
  \tikzset{rest/.style={draw,minimum height=2mm, minimum
      width=5mm,inner sep=2pt, anchor=west, fill=white,rounded
      corners=2pt,outer sep=0pt,xshift=1mm, scale=0.8}}; \node
  (v1) [col1] {\sffamily 00}; \node (v2) at (v1.east) [rest] {\sffamily 01}; \node
  (v3) at (v2.east) [rest] {\sffamily 02}; \node (v4) at (v3.east) [rest]
          {\sffamily 03}; }) with a single command.

Our mechanism is based on the fact that modern DRAM modules
consist of many DRAM chips. As described in
Section~\ref{sec:dram-module}, to achieve high bandwidth, multiple
chips are grouped together to form a rank, and all chips within a
rank operate in unison. Our \textbf{idea} is to enable the
controller to access multiple values from a strided access from
\emph{different} chips within the rank with a single
command. However, there are two challenges in implementing this
idea. For the purpose of describing the challenges and our
mechanism, we assume that each rank consist of four DRAM
chips. However, \gsdram is general and can be extended to any rank
with power-of-2 number of DRAM chips.

\subsection{Challenges in Designing \gsdram}
\label{sec:gsdram-challenges}

\begin{figure}[h]
  \centering
  \includegraphics{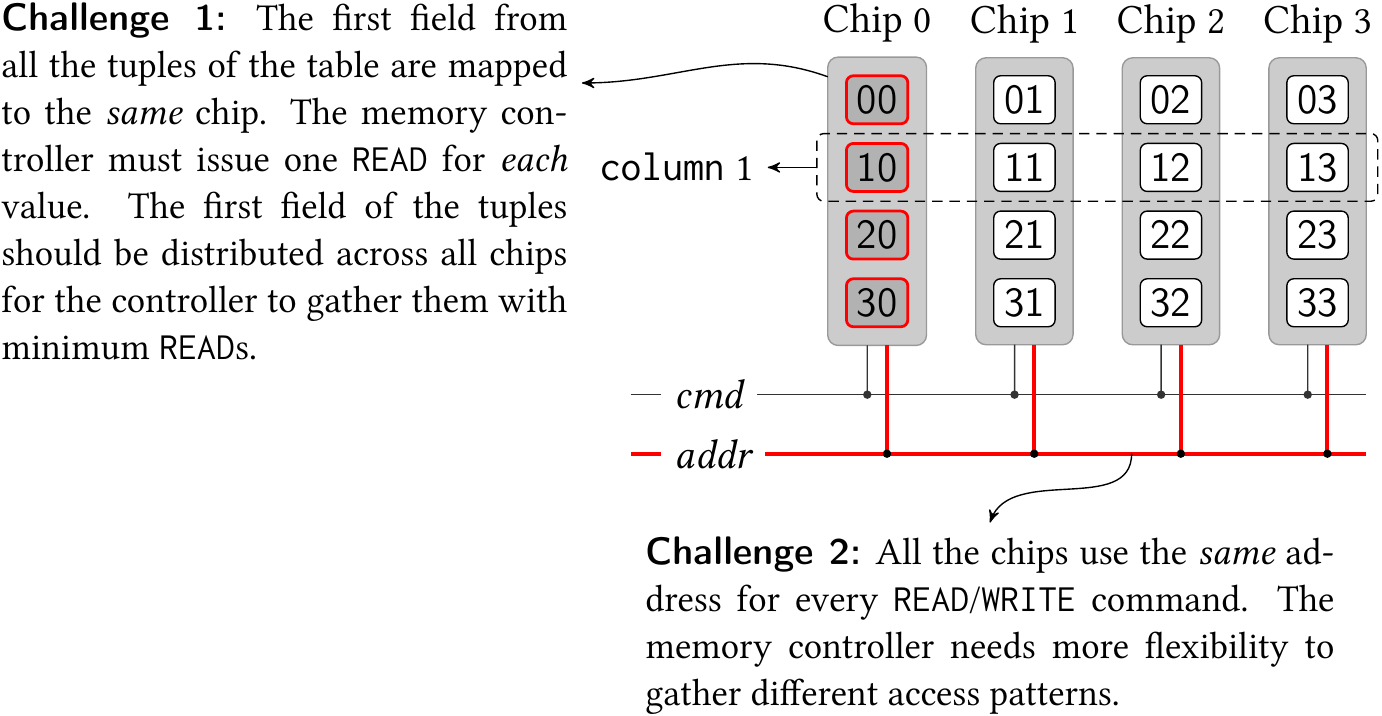}
  \caption[\gsdram: Challenges]{The two challenges in designing \gsdram.}
  \label{fig:gsdram-challenges}
\end{figure}

Figure~\ref{fig:gsdram-challenges} shows the two challenges. We assume
that the first four tuples of the table are stored from the
beginning of a DRAM row. Since each tuple maps to a single cache
line, the data of each tuple is split across all four chips.  The
mapping between different segments of the cache line and the chips
is controlled by the memory controller. Based on the mapping
scheme described in Section~\ref{sec:dram-module}, the $i^{th}$ 8
bytes of each cache line (i.e., the $i^{th}$ field of each tuple)
is mapped to the $i^{th}$ chip.

\noindent\textbf{\sffamily Challenge 1:} \emph{Reducing chip
  conflicts}. The simple mapping mechanism maps the first field of
\emph{all} the tuples to Chip 0. Since each chip can send out only
one field (8 bytes) per \cmdrd operation, gathering the first
field of the four tuples will necessarily require four
{\cmdrd}s. In a general scenario, different pieces of data that
are required by a strided access pattern will be mapped to
different chips. When two such pieces of data are mapped to the
same chip, it results in what we call a \emph{chip conflict}. Chip
conflicts increase the number of {\cmdrd}s required to gather all
the values of a strided access pattern. Therefore, we have to map
the data structure to the chips in a manner that minimizes the
number of chip conflicts for target access patterns.

\noindent\textbf{\sffamily Challenge 2:} \emph{Communicating the
  access pattern to the module}. As shown in
Figure~\ref{fig:gsdram-challenges}, when a column command is sent to a
rank, all the chips select the \emph{same} column from the
activated row and send out the data. If the memory controller
needs to access the first tuple of the table and the first field
of the four tuples each with a \emph{single} \cmdrd operation, we
need to break this constraint and allow the memory controller to
potentially read \emph{different} columns from different chips
using a single \cmdrd command. One naive way of achieving this
flexibility is to use multiple address buses, one for each
chip. Unfortunately, this approach is very costly as it
significantly increases the pin count of the memory
channel. Therefore, we need a simple and low cost mechanism to
allow the memory controller to efficiently communicate different
access patterns to the DRAM module.

In the following sections, we propose a simple mechanism to
address the above challenges with specific focus on power-of-2
strided access patterns. While non-power-of-2 strides (e.g., odd
strides) pose some additional challenges (e.g., alignment), a
similar approach can be used to support non-power-of-2 strides as
well.

\subsection{Column ID-based Data Shuffling}
\label{sec:shuffling}

To address challenge 1, i.e., to minimize chip conflicts, the
memory controller must employ a mapping scheme that distributes
data of each cache line to different DRAM chips with the following
three goals. First, the mapping scheme should be able to minimize
chip conflicts for a number of access patterns. Second, the memory
controller must be able to succinctly communicate an access
pattern along with a column command to the DRAM module. Third,
once the different parts of the cache line are read from different
chips, the memory controller must be able to quickly assemble the
cache line. Unfortunately, these goals are conflicting.

While a simple mapping scheme enables the controller to assemble a
cache line by concatenating the data received from different
chips, this scheme incurs a large number of chip conflicts for
many frequently occurring access patterns (e.g., any power-of-2
stride > 1). On the other hand, pseudo-random mapping
schemes~\cite{pr-interleaving} potentially incur a small number of
conflicts for almost any access pattern. Unfortunately, such
pseudo-random mapping schemes have two shortcomings. First, for
any cache line access, the memory controller must compute which
column of data to access from each chip and communicate this
information to the chips along with the column command. With a
pseudo random interleaving, this communication may require a
separate address bus for each chip, which would significantly
increase the cost of the memory channel. Second, after reading the
data, the memory controller must spend more time assembling the
cache line, increasing the overall latency of the \cmdrd
operation.

\begin{figure}[h]
  \centering
  \includegraphics[scale=0.95]{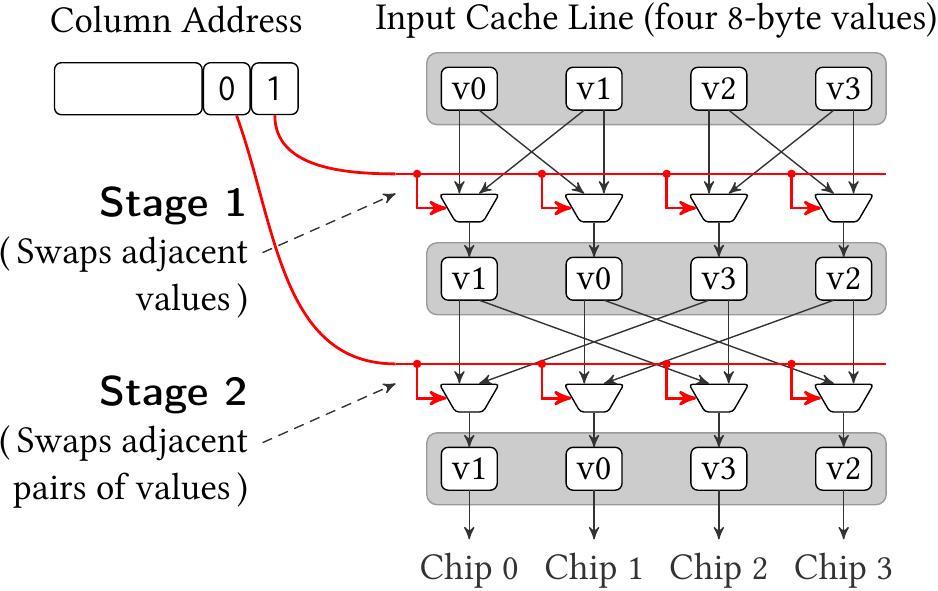}
  \caption[Column-ID based data shuffling]{2-stage shuffling
    mechanism that maps different 8-byte values within a cache
    line to a DRAM chip. For each mux, \texttt{0} selects the
    vertical input, and \texttt{1} selects the cross input.}
  \label{fig:shuffling}
\end{figure}

We propose a simple \emph{column ID-based data shuffling}
mechanism that achieves a sweet spot by restricting our focus to
power-of-2 strided access patterns. Our shuffling mechanism is
similar to a butterfly network~\cite{butterfly}, and is
implemented in the memory controller. To map the data of the cache
line with column address $C$ to different chips, the memory
controller inspects the $n$ \emph{least significant bits} (LSB) of
$C$. Based on these $n$ bits, the controller uses $n$ stages of
shuffling. Figure~\ref{fig:shuffling} shows an example of a
2-stage shuffling mechanism. In Stage 1
(Figure~\ref{fig:shuffling}), if the LSB is set, our mechanism
groups adjacent 8-byte values in the cache line into pairs and
swaps the values within each pair. In Stage 2
(Figure~\ref{fig:shuffling}), if the second LSB is set, our
mechanism groups the 8-byte values in the cache line into
quadruplets, and swaps the adjacent \emph{pairs} of values. The
mechanism proceeds similarly into the higher levels, doubling the
size of the group of values swapped in each higher stage.  The
shuffling mechanism can be enabled only for those data structures
that require our mechanism. Section~\ref{sec:gsdram-software}
discusses this in more detail.

With this simple multi-stage shuffling mechanism, the memory
controller can map data to DRAM chips such that \emph{any}
power-of-2 strided access pattern incurs minimal chip conflicts
for values within a single DRAM row.

\subsection{Pattern ID: Low-cost Column Translation}
\label{sec:pattern-id}

The second challenge is to enable the memory controller to
flexibly access different column addresses from different DRAM
chips using a single \cmdrd command. To this end, we propose a
mechanism wherein the controller associates a \emph{pattern ID}
with each access pattern. It provides this pattern ID with each
column command. Each DRAM chip then independently computes a new
column address based on 1)~the issued column address, 2)~the chip
ID, and 3)~the pattern ID. We refer to this mechanism as
\emph{column translation}.

\begin{figure}[h]
  \centering
  \includegraphics[scale=1]{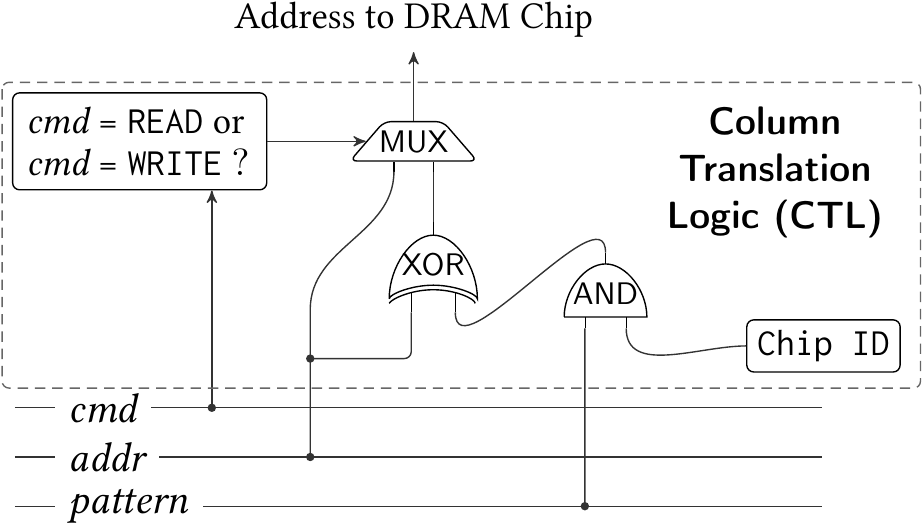}
  \caption[In-DRAM column translation logic]{Column Translation
    Logic (CTL). Each chip has its own CTL. The CTL can be
    implemented in the DRAM module (as shown in
    Figure~\ref{fig:overview}).  Each logic gate performs a
    bitwise operation of the input values.}
  \label{fig:ctl}
\end{figure}

Figure~\ref{fig:ctl} shows the column translation logic for a
single chip. As shown in the figure, our mechanism requires only
two bitwise operations per chip to compute the new column
address. More specifically, the output column address for each
chip is given by \mbox{(\texttt{Chip ID} \& \texttt{Pattern ID})
  $\oplus$ \texttt{Column ID}}, where \texttt{Column ID} is the
column address provided by the memory controller. In addition to
the logic to perform these simple bitwise operations, our
mechanism requires 1)~a register per chip to store the chip ID,
and 2)~a multiplexer to enable the address translation only for
column commands. While our column translation logic can be
combined with the column selection logic already present within
each chip, our mechanism can also be implemented within the DRAM
module with \emph{no} changes to the DRAM chips.

Combining this pattern-ID-based column translation mechanism with
the column-ID-based data shuffling mechanism, the memory
controller can gather or scatter any power-of-two strided access
pattern with no waste in memory bandwidth.

\subsection{\gsdram: Putting It All Together}
\label{sec:overview}

\begin{figure}[h]
  \centering
  \includegraphics[scale=1]{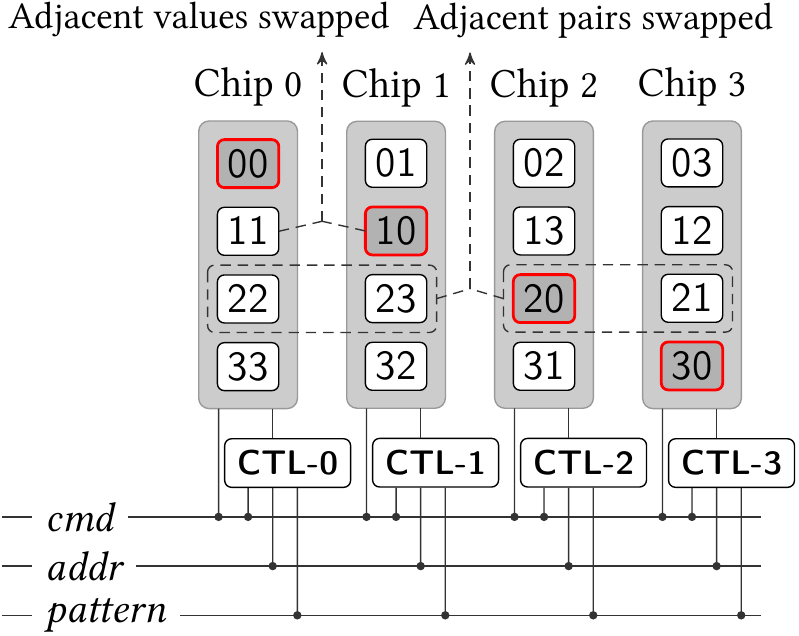}
  \caption[\gsdram Overview]{\gsdram Overview. CTL-i is the column translation logic
    with Chip ID $=$ i (as shown in Figure~\ref{fig:ctl}).}
  \label{fig:overview}
\end{figure}

Figure~\ref{fig:overview} shows the full overview of our \gsdram
substrate. The figure shows how the first four tuples of our
example table are mapped to the DRAM chips using our data
shuffling mechanism. The first tuple (column ID $=$ 0) undergoes
no shuffling as the two LSBs of the column ID are both 0 (see
Figure~\ref{fig:shuffling}). For the second tuple (column ID $=$
1), the adjacent values within each pairs of values are swapped
(Figure~\ref{fig:shuffling}, Stage 1). Similarly, for the third
tuple (column ID $=$ 2), adjacent pair of values are swapped
(Figure~\ref{fig:shuffling}, Stage 2). For the fourth tuple
(column ID $=$ 3), since the two LSBs of the column ID are both 1,
both stages of the shuffling scheme are enabled
(Figure~\ref{fig:shuffling}, Stages 1 and 2). As shown in shaded
boxes in Figure~\ref{fig:overview}, the first field of the four
tuples (i.e., \tikz[baseline={([yshift=-8pt]current bounding
    box.north)}]{ \tikzset{col1/.style={draw,minimum height=2mm,
      minimum width=2mm,inner sep=2pt, anchor=west, fill=black!30,
      rounded corners=2pt,outer sep=0pt,scale=0.8,xshift=1mm}};
  \tikzset{rest/.style={draw,minimum height=2mm, minimum
      width=5mm,inner sep=2pt, anchor=west, fill=white,rounded
      corners=2pt,outer sep=0pt,xshift=1mm, scale=0.8}}; \node
  (v1) [col1] {\sffamily 00}; \node (v2) at (v1.east) [col1] {\sffamily 10}; \node
  (v3) at (v2.east) [col1] {\sffamily 20}; \node (v4) at (v3.east) [col1]
          {\sffamily 30}; }) are mapped to \emph{different} chips, allowing
the memory controller to read them with a single \cmdrd
command. The same is true for the other fields of the table as
well (e.g., \tikz[baseline={([yshift=-8pt]current bounding
    box.north)}]{ \tikzset{col1/.style={draw,minimum height=2mm,
      minimum width=2mm,inner sep=2pt, anchor=west, fill=black!30,
      rounded corners=2pt,outer sep=0pt,scale=0.8}};
  \tikzset{rest/.style={draw,minimum height=2mm, minimum
      width=5mm,inner sep=2pt, anchor=west, fill=white,rounded
      corners=2pt,outer sep=0pt,xshift=1mm, scale=0.8}}; \node
  (v1) [rest] {\sffamily 01}; \node (v2) at (v1.east) [rest] {\sffamily 11}; \node
  (v3) at (v2.east) [rest] {\sffamily 21}; \node (v4) at (v3.east) [rest]
          {\sffamily 31}; })

The figure also shows the per-chip column translation logic. To
read a specific tuple from the table, the memory controller simply
issues a \cmdrd command with pattern ID $=$ 0 and an appropriate
column address. For example, when the memory controller issues the
\cmdrd for column ID 2 and pattern 0, the four chips return the
data corresponding to the columns (2 2 2 2), which is the data in
the third tuple of the table (i.e.,
\tikz[baseline={([yshift=-8pt]current bounding box.north)}]{
  \tikzset{col1/.style={draw,minimum height=2mm, minimum
      width=2mm,inner sep=2pt, anchor=west, fill=black!30, rounded
      corners=2pt,outer sep=0pt,scale=0.8}};
  \tikzset{rest/.style={draw,minimum height=2mm, minimum
      width=5mm,inner sep=2pt, anchor=west, fill=white,rounded
      corners=2pt,outer sep=0pt,xshift=1mm, scale=0.8}}; \node
  (v1) [rest] {\sffamily 22}; \node (v2) at (v1.east) [rest] {\sffamily 23}; \node
  (v3) at (v2.east) [rest] {\sffamily 20}; \node (v4) at (v3.east) [rest]
          {\sffamily 21}; }). In other words, pattern ID 0 allows the memory
controller to perform the default read operation. Hence, we refer
to pattern ID 0 as the \emph{default pattern}.

On the other hand, if the memory controller issues a \cmdrd for
column ID 0 and pattern 3, the four chips return the data
corresponding to columns (0 1 2 3), which precisely maps to the
first field of the table. Similarly, the other fields of the first
four tuples can be read from the database by varying the column ID
with pattern 3.

\subsection{\gsdram Parameters}
\label{sec:gsdramp}

\gsdram has three main parameters: 1)~the number of chips in each
module, 2)~the number of shuffling stages in the data shuffling
mechanism, and 3)~the number of bits of pattern ID. While the
number of chips determines the size of each cache line, the other
two parameters determine the set of access patterns that can be
efficiently gathered by \gsdram. We use the term \gsdramp{c}{s}{p}
to denote a \gsdram with $c$ chips, $s$ stages of shuffling, and
$p$ bits of pattern ID.

\begin{figure}[h]
  \centering
  \includegraphics[scale=0.9]{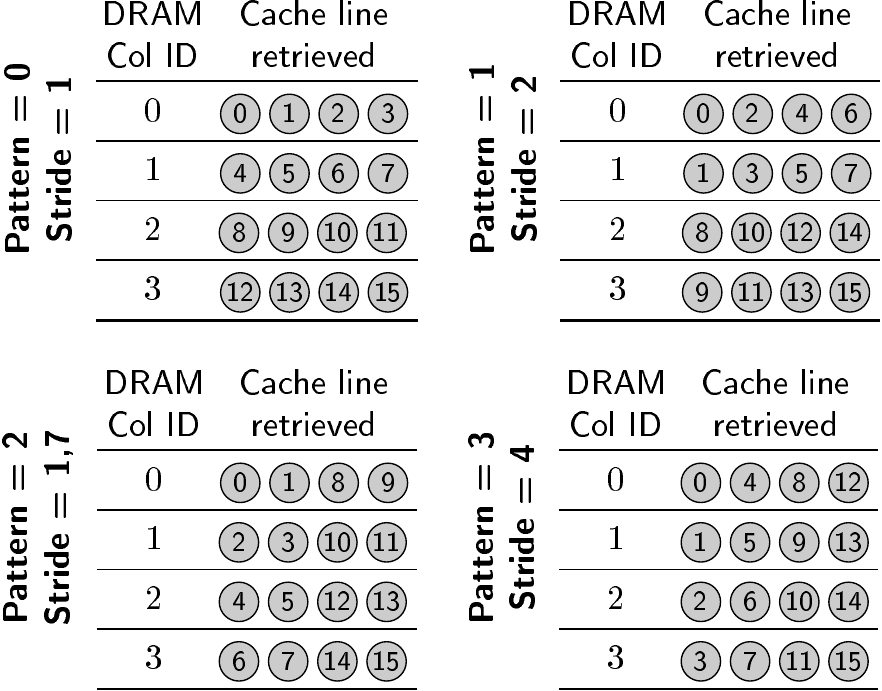}
  \caption[Patterns gathered by \gsdramp{4}{2}{2}]{Cache lines
    gathered by \gsdramp{4}{2}{2} for all patterns for column IDs
    0--3. Each circle contains the index of the 8-byte value
    inside the logical row buffer.}
  \label{fig:patterns}
\end{figure}

Figure~\ref{fig:patterns} shows all cache lines that can be
gathered by \gsdramp{4}{2}{2}, with the four possible patterns for
column IDs 0 through 3. For each pattern ID and column ID
combination, the figure shows the index of the four values within
the row buffer that are retrieved from the DRAM module. As shown,
pattern 0 retrieves contiguous values. Pattern 1 retrieves every
other value (stride = 2). Pattern 2 has a dual stride of
(1,7). Pattern 3 retrieves every 4th value (stride = 4). In
general, pattern $2^k - 1$ gathers data with a stride $2^k$.

While we showed a use case for pattern 3 (in our example), we
envision use-cases for other patterns as well. Pattern 1, for
instance, can be useful for data structures like key-value
stores. Assuming an 8-byte key and an 8-byte value, the cache line
(\texttt{Patt 0, Col 0}) corresponds to the first two key-value
pairs. However the cache line (\texttt{Patt 1, Col 0}) corresponds
to the first four keys, and (\texttt{Patt 1, Col 1}) corresponds
to the first four values. Similarly, pattern 2 can be use to fetch
odd-even pairs of fields from a data structure where each object
has 8 fields.

Our mechanism is general. For instance, with \gsdramp{8}{3}{3}
(i.e., 8 chips, 3 shuffling stages, and 3-bit pattern ID), the
memory controller can access data with seven different
patterns. Section~\ref{sec:gsdram-extensions} discusses other
simple extensions to our approach to enable more fine-grained
gather access patterns, and larger strides.

\subsection{Ease of Implementing \gsdram}
\label{sec:benefits}

In Section~\ref{sec:applications}, we will show that \gsdram has
compelling performance and energy benefits compared to existing
DRAM interfaces. These benefits are augmented by the fact that
\gsdram is simple to implement. First, our data shuffling
mechanism is simple and has low latency. Each stage involves only
data swapping and takes at most one processor cycle. Our
evaluations use \gsdramp{8}{3}{3}, thereby incurring 3 cycles of
additional latency to shuffle/unshuffle data for each DRAM
write/read. Second, for \gsdramp{*}{*}{p}, the column translation
logic requires only two p-bit bitwise operations, a p-bit register
to store the chip ID, and a p-bit multiplexer. In fact, this
mechanism can be implemented as part of the DRAM module
\emph{without} any changes to the DRAM chips themselves. Finally,
third, \gsdram requires the memory controller to communicate only
k bits of pattern ID to the DRAM module, adding only a few pins to
each channel. In fact, the column command in existing DDR DRAM
interfaces already has a few spare address pins that can
potentially be used by the memory controller to communicate the
pattern ID (e.g., DDR4 has two spare address pins for column
commands~\cite{ddr4}).

\section{End-to-end System Design}
\label{sec:end-to-end}

In this section, we discuss the support required from the rest of
the system stack to exploit the \gsdram substrate.  We propose a
mechanism that leverages support from different layers of the
system stack to exploit \gsdram: 1)~on-chip caches, 2)~the
instruction set architecture, and 3)~software. It is also possible
for the processor to dynamically identify different access
patterns present in an application and exploit \gsdram to
accelerate such patterns transparently to the application. As our
goal in this work is to demonstrate the benefits of \gsdram, we
leave the design of such an automatic mechanism for future
work. The following sections assume a \gsdramp{*}{*}{p}, i.e., a
p-bit pattern ID.

\subsection{On-Chip Cache Support}
\label{sec:on-chip-cache}

Our mechanism introduces two problems with respect to on-chip
cache management. First, when the memory controller gathers a
cache line from a non-zero pattern ID, the values in the cache
line are \emph{not} contiguously stored in physical memory. For
instance, in our example (Figure~\ref{fig:matrix}), although the
controller can fetch the first field of the first four tuples of
the table with a single \cmdrd, the first field of the table is
not stored contiguously in physical memory. Second, two cache
lines belonging to different patterns may have a partial
overlap. In our example (Figure~\ref{fig:matrix}), if the memory
controller reads the first tuple (pattern ID = 0, column ID = 0)
and the first field of the first four tuples (pattern ID = 3,
column ID = 0), the two resulting cache lines have a common value
(the first field of the first tuple, i.e.,
\tikz[baseline={([yshift=-8pt]current bounding box.north)}]{
  \tikzset{col1/.style={draw,minimum height=2mm, minimum
      width=2mm,scale=0.9,inner sep=2pt, anchor=west,
      fill=black!30, rounded corners=2pt,outer
      sep=-0pt,xshift=1mm}}; \node (v1) [col1] {\sffamily 00};}).

One simple way to avoid these problems is to store the individual
values of the gathered data in \emph{different} physical cache
lines by employing a sectored cache~\cite{sectored-cache} (for
example). However, with the off-chip interface to DRAM operating
at a wider-than-sector (i.e., a full cache line) granularity, such
a design will increase the complexity of the cache-DRAM
interface. For example, writebacks may require read-modify-writes
as the processor may not have the entire cache line. More
importantly, a mechanism that does not store the gathered values
in the same cache line cannot extract the full benefits of SIMD
optimizations because values that are required by a single SIMD
operation would now be stored in \emph{multiple} physical cache
lines. Therefore, we propose a simple mechanism that stores each
gathered cache line from DRAM in a single physical cache line in
the on-chip cache. Our mechanism has two aspects.\vspace{2mm}

\noindent\textbf{\sffamily 1. Identifying non-contiguous cache
  lines.} When a non-contiguous cache line is stored in the cache,
the cache controller needs a mechanism to identify the cache
line. We observe that, in our proposed system, each cache line can
be uniquely identified using the cache line address and the
pattern ID with which it was fetched from DRAM. Therefore, we
extend each cache line tag in the cache tag store with $p$
additional bits to store the pattern ID of the corresponding cache
line.\vspace{2mm}

\noindent\textbf{\sffamily 2. Maintaining cache coherence.} The
presence of overlapping cache lines has two implications on
coherence. First, before fetching a cache line from DRAM, the
controller must check if there are any dirty cache lines in the
cache which have a partial overlap with the cache line being
fetched. Second, when a value is modified by the processor, in
addition to invalidating the modified cache line from the other
caches, the processor must also invalidate all other cache lines
that contain the value that is being modified. With a number of
different available patterns, this operation can be a complex and
costly.

Fortunately, we observe that many applications that use strided
accesses require only two pattern IDs per data structure, the
default pattern and one other pattern ID.  Thus, as a trade-off to
simplify cache coherence, we restrict each data structure to use
only the zero pattern and one other pattern ID.  To implement this
constraint, we associate each virtual page with an additional
$p$-bit pattern ID. Any access to a cache line within the page can
use either the zero pattern or the page's pattern ID. If multiple
virtual pages are mapped to the same physical page, the OS must
ensure that the same alternate pattern ID is used for all
mappings.

Before fetching a cache line from DRAM with a pattern, the memory
controller must only look for dirty cache lines from the other
pattern. Since all these cache lines belong to the same DRAM row,
this operation is fast and can be accelerated using simple
structures like the Dirty-Block Index (described in
Chapter~\ref{chap:dbi} of this thesis). Similarly, when the
processor needs to modify a shared cache line, our mechanism
piggybacks the other pattern ID of the page along with the
\emph{read-exclusive} coherence request. Each cache controller
then locally invalidates the cache lines from the other pattern ID
that overlap with the cache line being modified. For
\gsdramp{c}{*}{*}, our mechanism requires $c$ additional
invalidations for each read-exclusive request.

\subsection{Instruction Set Architecture Support}
\label{sec:isa}

To enable software to communicate strided access patterns to the
processor, we introduce a new variant of the \texttt{load} and
\texttt{store} instructions, called \texttt{pattload} and
\texttt{pattstore}, that allow the code to specify the pattern
ID. These instructions takes the following
form:\\ \centerline{\texttt{pattload reg, addr, patt}}
\centerline{\texttt{pattstore reg, addr, patt}} where \texttt{reg}
is the source or destination register (depending on the
instruction type), \texttt{addr} is the address of the data, and
\texttt{patt} is the pattern ID.

To execute a \texttt{pattload} or \texttt{pattstore}, the
processor first splits the \texttt{addr} field into two parts: the
cache line address (\texttt{caddr}), and the offset within the
cache line (\texttt{offset}). Then the processor sends out a
request for the cache line with address-pattern combination
(\texttt{caddr}, \texttt{patt}). If the cache line is present in
the on-chip cache, it is sent to the processor. Otherwise, the
request reaches the memory controller. The memory controller
identifies the row address and the column address from
\texttt{caddr} and issues a \cmdrd command for a cache line with
pattern ID \texttt{patt}. If the memory controller interleaves
cache lines across multiple channels (or ranks), then it must
access the corresponding cache line within each channel (or rank)
and interleave the data from different channels appropriately
before obtaining the required cache line.  The cache line is then
stored in the on-chip cache and is also sent to the
processor. After receiving the cache line, the processor reads or
updates the data at the \texttt{offset} to or from the destination
or source register (\texttt{reg}).

Note that architectures like x86 allow instructions to directly
operate on memory by using different addressing modes to specify
memory operands~\cite{x86-addressing}. For such architectures, common
addressing modes may be augmented with a pattern ID field, or
instruction prefixes may be employed to specify the pattern.

\subsection{System and Application Software Support}
\label{sec:gsdram-software}

Our mechanism requires two pieces of information from the software
for each data structure: 1)~whether the data structure requires
the memory controller to use the shuffling mechanism
(Section~\ref{sec:shuffling}) (we refer to this as the
\emph{shuffle flag}), and 2)~the alternate pattern ID
(Section~\ref{sec:pattern-id}) with which the application will
access the data structure. To enable the application to specify
this information, we propose a new variant of the \texttt{malloc}
system call, called \texttt{pattmalloc}, which includes two
additional parameters: the shuffle flag, and the pattern ID. When
the OS allocates virtual pages for a \texttt{pattmalloc}, it also
updates the page tables with the shuffle flag and the alternate
pattern ID for those pages.

\begin{figure}[h]
  \centering
  \includegraphics[scale=1]{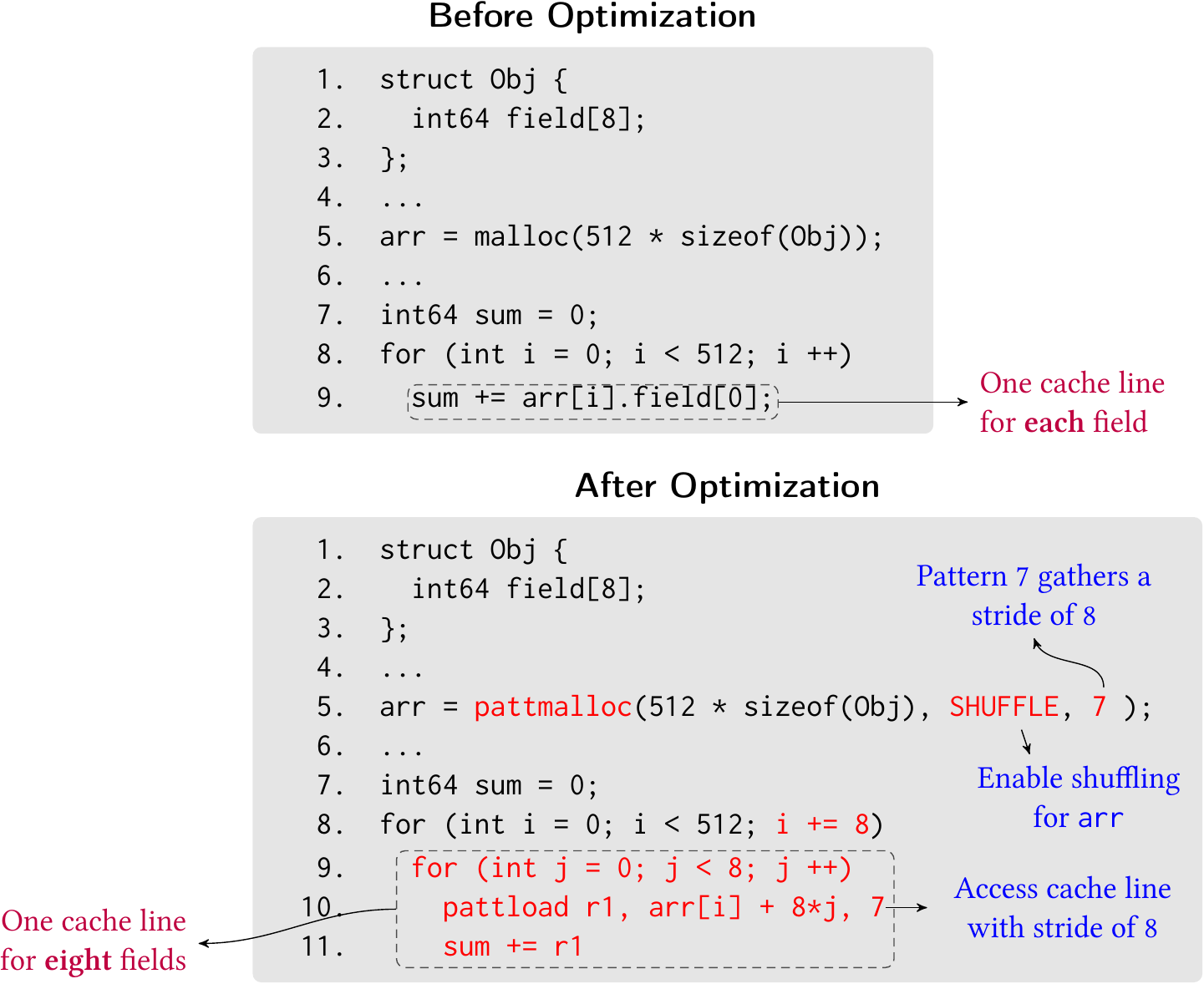}
  \caption[Optimizing programs for \gsdram]{Example code without
    and with our optimization.}
  \label{fig:example-code}
\end{figure}

Once the data structure is allocated with \texttt{pattmalloc}, the
application can use the \texttt{pattload} or \texttt{pattstore}
instruction to access the data structure efficiently with both the
zero pattern and the alternate access pattern. While we can
envision automating this process using a compiler optimization, we
do not explore that path in this
thesis. Figure~\ref{fig:example-code} shows an example piece of
code before and after our optimization. The original code (line 5)
allocates an array of 512 objects (each object with eight 8-byte
fields) and computes the sum of the first field of all the objects
(lines 8 and 9). The figure highlights the key benefit of our
approach.

In the program without our optimization
(Figure~\ref{fig:example-code}, left), each iteration of the loop
(line 9) fetches a different cache line. As a result, the entire
loop accesses 512 different cache lines. On the other hand, with
our optimization (Figure~\ref{fig:example-code}, right), the
program first allocates memory for the array using
\texttt{pattmalloc} (line 5), with the shuffle flag enabled and an
alternate pattern ID = 7 (i.e., stride of 8). The program then
breaks the loop into two parts. Each iteration of the outer loop
(line 8) fetches a single strided cache line that contains only
values from the first field. The loop skips the other fields
(\texttt{i += 8}). The inner loop (lines 9-11) iterates over
values within each strided cache line. In the first iteration of
the inner loop, the \texttt{pattload} instruction with pattern ID
7 fetches a cache line with a stride of 8. As a result, the
remaining seven iterations of the inner loop result in cache
hits. Consequently, with our optimization, the entire loop
accesses only 64 cache lines. As we will show in our evaluations,
this reduction in the number of accessed cache lines directly
translates to reduction in latency, bandwidth consumption, and
cache capacity consumption, thereby improving overall performance.

\subsection{Hardware Cost}
\label{sec:cost}

In this section, we quantify the changes required by \gsdram,
specifically \gsdramp{8}{3}{3} (Section~\ref{sec:gsdramp}), to
various hardware components. On the DRAM side, first, our
mechanism requires the addition of the column translation logic
(CTL) for each DRAM chip. Each CTL requires a 3-bit register for
the \texttt{Chip ID}, a 3-bit bitwise AND gate, a 3-bit bitwise
XOR gate and a 3-bit bitwise multiplexer. Even for a commodity
DRAM module with 8 chips, the overall cost is roughly 72 logic
gates and 24 bits of register storage, which is negligible
compared to the logic already present in a DRAM module. Second,
our mechanism requires a few additional pins on the DRAM interface
to communicate the pattern ID. However, existing DRAM interfaces
already have some spare address bits, which can be used to
communicate part of the pattern ID. Using this approach, a 3-bit
pattern ID requires only one additional pin for DDR4~\cite{ddr4}.

On the processor side, first, our mechanism requires the
controller to implement the shuffling logic. Second, our mechanism
augments each cache tag entry with the pattern ID. Each page table
entry and TLB entry stores the shuffle flag and the alternate
pattern ID for the corresponding page
(Section~\ref{sec:on-chip-cache}). For a 3-bit pattern ID, the
cost of this addition is less than 0.6\% of the cache
size. Finally, the processor must implement the \texttt{pattload}
and \texttt{pattstore} instructions, and the state machine for
invalidating additional cache lines on read-exclusive coherence
requests. The operation of \texttt{pattload}/\texttt{pattstore} is
not very different from that of a regular
\texttt{load}/\texttt{store} instruction. Therefore, we expect the
implementation of these new instructions to be simple. Similarly,
on a write, our mechanism has to check only eight cache lines (for
\gsdram with 8 chips) for possible overlap with the modified cache
line. Therefore, we expect the invalidation state machine to be
relatively simple. Note that a similar state machine has been used
to keep data coherent in a virtually-indexed physically-tagged
cache in the presence of synonyms~\cite{alpha-21264}.

\section{Applications and Evaluations}
\label{sec:applications}

To quantitatively evaluate the benefits of \gsdram, we implement
our framework in the Gem5 simulator~\cite{gem5}, on top of the x86
architecture. We implement the \texttt{pattload} instruction by
modifying the behavior of the \texttt{prefetch} instruction to
gather with a specific pattern into either the \texttt{rax}
register (8 bytes) or the \texttt{xmm0} register (16 bytes). None
of our evaluated applications required the \texttt{pattstore}
instruction. Table~\ref{table:gsdram-parameters} lists the main
parameters of the simulated system. All caches uniformly use
64-byte cache lines. While we envision several applications to
benefit from our framework, in this section, we primarily discuss
and evaluate two applications: 1)~an in-memory database workload,
and 2)~general matrix-matrix multiplication workload.

\begin{table}[h]\footnotesize
  \centering
  \begin{tabular}{ll}
  \toprule
  \sffamily Processor & 1-2 cores, x86, in-order, 4 GHz\\
  \midrule
  \sffamily L1-D Cache & Private, 32 KB, 8-way associative, LRU policy\\
  \midrule
  \sffamily L1-I Cache & Private, 32 KB, 8-way associative, LRU policy\\
  \midrule
  \sffamily L2 Cache & Shared, 2 MB, 8-way associative, LRU policy\\
  \midrule
  \multirow{2}{*}{\sffamily Memory} & DDR3-1600, 1 channel, 1 rank, 8 banks\\
  \cmidrule{2-2}
   & Open row, FR-FCFS~\cite{frfcfs,frfcfs-patent}, \gsdramp{8}{3}{3}\\
  \bottomrule
\end{tabular}

  \caption[GS-DRAM: Simulation parameters]{Main parameters of the simulated system.}
  \label{table:gsdram-parameters}
\end{table}

\subsection{In-Memory Databases}
\label{sec:in-memory-db}

In-memory databases (IMDB) (e.g.,~\cite{memsql,hstore,hyrise})
provide significantly higher performance than traditional
disk-oriented databases. Similar to any other database, an IMDB
may support two kinds of queries: \emph{transactions}, which
access many fields from a few tuples, and \emph{analytics}, which
access one or few fields from many tuples. As a result, the
storage model used for the database tables heavily impacts the
performance of transactions and analytical queries. While a
row-oriented organization (\emph{row store}) is better for
transactions, a column-oriented organization~\cite{c-store}
(\emph{column store}) is better for analytics. Increasing need for
both fast transactions and fast real-time analytics has given rise
to a new workload referred to as \emph{Hybrid
  Transaction/Analytical Processing} (HTAP)~\cite{htap}. In an
HTAP workload, both transactions and analytical queries are run on
the \emph{same version} of the database. Unfortunately, neither
the row store nor the column store provides the best performance
for both transactions and analytics.

With our \gsdram framework, each database table can be stored as a
row store in memory, but can be accessed at high performance
\emph{both} in the row-oriented access pattern \emph{and} the
field-oriented access pattern.\footnote{\gsdram requires the
  database to be structured (i.e., not have any variable length
  fields). This is fine for most high-performance IMDBs as they
  handle variable length fields using fixed size pointers for fast
  data retrieval~\cite{vlf1,vlf2}. \gsdram will perform at least as
  well as the baseline for unstructured databases.}  Therefore, we
expect \gsdram to provide the \emph{best of both row and column
  layouts} for both kinds of queries. We demonstrate this
potential benefit by comparing the performance of \gsdram with
both a row store layout (\rowstore) and a column store layout
(\colstore) on three workloads: 1)~a transaction-only workload,
2)~an analytics-only workload, and 3)~an HTAP workload. For our
experiments, we assume an IMDB with a single table with one
million tuples and no use of compression.  Each tuple contains
eight 8-byte fields, and fits exactly in a 64B cache line. (Our
mechanism naturally extends to any table with power-of-2 tuple
size.)

\begin{figure}[h]
  \centering
  \includegraphics{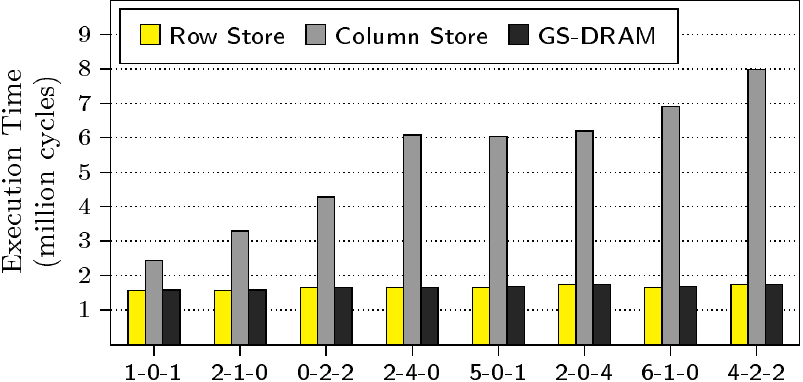}
  \caption[\gsdram: Transaction workload performance]{Transaction
    Workload Performance: Execution time for 10000
    transactions. The x-axis indicates the number of
    \emph{read-only}, \emph{write-only}, and \emph{read-write}
    fields for each workload.}
  \label{fig:transactions}
\end{figure}

\textbf{Transaction workload.} For this workload, each transaction
operates on a randomly-chosen tuple. All transactions access $i$,
$j$, and $k$ fields of the tuple in the read-only, write-only, and
read-write mode, respectively. Figure~\ref{fig:transactions}
compares the performance (execution time) of \gsdram, \rowstore,
and \colstore on the transaction workload for various values of
$i$, $j$, and $k$ (x-axis). The workloads are sorted based on the
total number of fields accessed by each transaction. For each
mechanism, the figure plots the execution time for running 10000
transactions.

We draw three conclusions. First, as each transaction accesses
only one tuple, it accesses only one cache line. Therefore, the
performance of \rowstore is almost the same regardless of the
number of fields read/written by each transaction. Second, the
performance of \colstore is worse than that of \rowstore, and
decreases with increasing number of fields. This is because
\colstore accesses a different cache line for each field of a
tuple accessed by a transaction, thereby causing a large number of
memory accesses. Finally, as expected, \gsdram performs as well as
\rowstore and 3X (on average) better than \colstore for the
transactions workload.

\begin{figure}[h]
  \centering
  \includegraphics{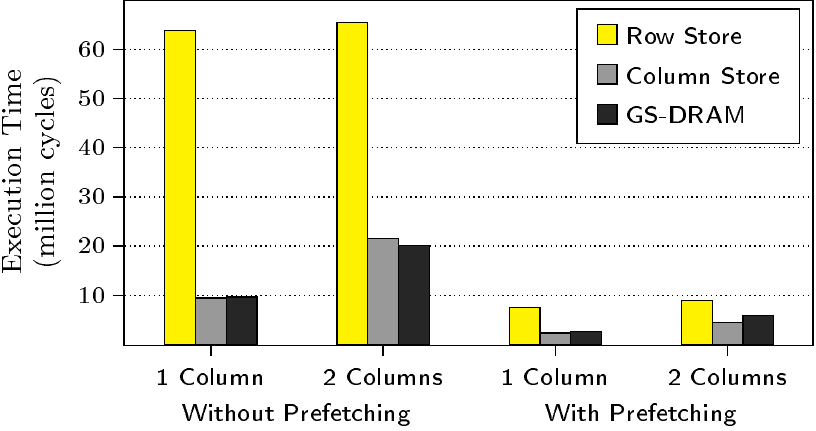}
  \caption[\gsdram: Analytics workload performance]{Analytics
    Workload Performance: Execution time for running an analytics
    query on 1 or 2 columns (without and with prefetching).}
  \label{fig:analytics}
\end{figure}

\textbf{Analytics workload.} For this workload, we measure the
time taken to run a query that computes the sum of $k$ columns
from the table. Figure~\ref{fig:analytics} compares the
performance of the three mechanisms on the analytics workload for
$k = 1$ and $k = 2$. The figure shows the performance of each
mechanism without and with prefetching. We use a PC-based stride
prefetcher~\cite{stride-prefetching} (with prefetching degree of
4~\cite{fdp}) that prefetches data into the L2 cache. We draw
several conclusions from the results.

First, prefetching significantly improves the performance of all
three mechanisms for both queries. This is expected as the
analytics query has a uniform stride for all mechanisms, which can
be easily detected by the prefetcher. Second, the performance of
\rowstore is roughly the same for both queries. This is because
each tuple of the table fits in a single cache line and hence, the
number of memory accesses for \rowstore is the same for both
queries (with and without prefetching). Third, the execution time
of \colstore increases with more fields. This is expected as
\colstore needs to fetch more cache lines when accessing more
fields from the table. Regardless, \colstore significantly
outperforms \rowstore for both queries, as it causes far fewer
cache line fetches compared to \rowstore. Finally, \gsdram, by
gathering the columns from the table as efficiently as \colstore,
performs similarly to \colstore and significantly better than
\rowstore both without and with prefetching (2X on average).

\begin{figure}[h]
  \centering
  \hspace{11mm}\includegraphics{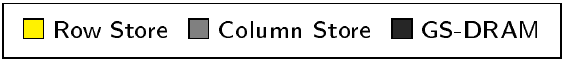}\vspace{1mm}\\
  \begin{subfigure}[b]{0.45\linewidth}
    \centering
    \includegraphics{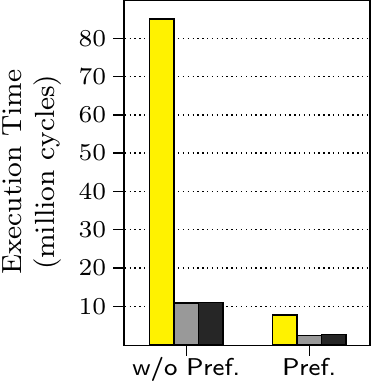}
    \caption{\footnotesize{Analytics Performance}}
    \label{plot:htap-anal}
  \end{subfigure}\quad
  \begin{subfigure}[b]{0.45\linewidth}
    \centering
    \includegraphics{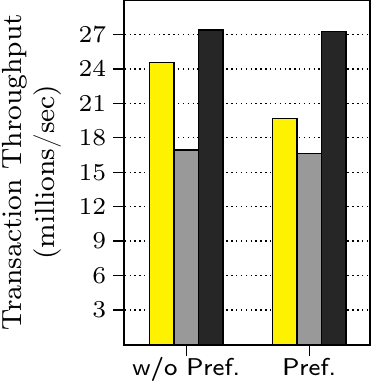}
    \caption{\footnotesize{Transaction Throughput}}
    \label{plot:htap-trans}
  \end{subfigure}
  \caption[\gsdram: HTAP performance]{HTAP (without and with
    prefetching) (transactions: 1 read-only, 1 write-only field;
    analytics: 1 column)}
  \label{plot:htap}
\end{figure}

\textbf{HTAP workload.} For this workload, we run one analytics
thread and one transactions thread concurrently on the same system
operating on the \emph{same} table. The analytics thread computes
the sum of a single column, whereas the transactions thread runs
transactions (on randomly chosen tuples with one read-only and one
write-only field). The transaction thread runs until the analytics
thread completes. We measure 1)~the time taken to complete the
analytics query, and 2)~the throughput of the transactions
thread. Figures~\ref{plot:htap-anal} and \ref{plot:htap-trans}
plot these results, without and with prefetching.

First, for analytics, prefetching significantly improves
performance for all three mechanisms. \gsdram performs as well as
\colstore. Second, for transactions, we find that \gsdram not only
outperforms \colstore, in terms of transaction throughput, but it
also performs better than \rowstore. We traced this effect back to
inter-thread contention for main memory bandwidth, a well-studied
problem
(e.g.,~\cite{bliss,critical-scheduler,tcm,stfm,parbs,atlas}).  The
FR-FCFS~\cite{frfcfs,frfcfs-patent} memory scheduler prioritizes
requests that hit in the row buffer. With \rowstore, the analytics
thread accesses all the cache lines in a DRAM row, thereby
starving requests of the transaction thread to the same bank
(similar to a memory performance hog program described
in~\cite{mpa}). In contrast, by fetching just the required field,
\gsdram accesses \emph{8 times fewer} cache lines per row. As a
result, it stalls the transaction thread for much smaller amount
of time, leading to higher transaction throughput than
\rowstore. The problem becomes worse for \rowstore with
prefetching, since the prefetcher makes the analytics thread run
even faster, thereby consuming a larger fraction of the memory
bandwidth.

\textbf{\sffamily Energy.} We use McPAT~\cite{mcpat} and
DRAMPower~\cite{drampower,drampower-paper} (integrated with
Gem5~\cite{gem5}) to estimate the processor and DRAM energy
consumption of the three mechanisms. Our evaluations show that,
for transactions, \gsdram consumes similar energy to \rowstore and
2.1X lower than \colstore. For analytics (with prefetching
enabled), \gsdram consumes similar energy to \colstore and 2.4X
lower energy (4X without prefetching) than \rowstore. (As
different mechanisms perform different amounts of work for the
HTAP workload, we do not compare energy for this workload.) The
energy benefits of \gsdram over prior approaches come from
1)~lower overall processor energy consumption due to reduced
execution time, and 2)~lower DRAM energy consumption due to
significantly fewer memory accesses.

\begin{figure}[h]
  \centering
  \hspace{11mm}\includegraphics{gsdram/plots/htap-legend}\vspace{1mm}\\
  \begin{subfigure}[b]{0.45\linewidth}
    \centering
    \includegraphics{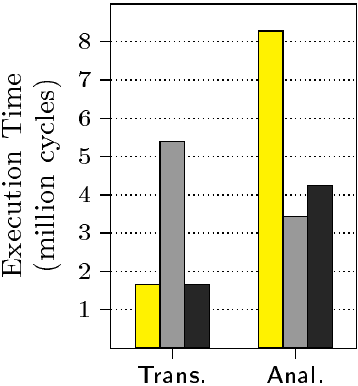}
    \caption{\footnotesize{Average Performance}}
    \label{plot:time-summary}
  \end{subfigure}\quad
  \begin{subfigure}[b]{0.45\linewidth}
    \centering
    \includegraphics{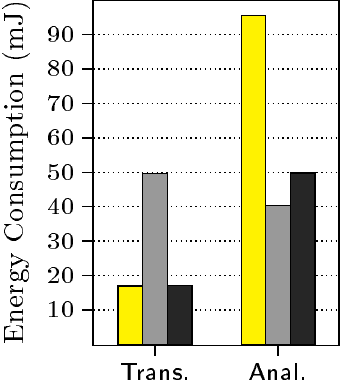}
    \caption{\footnotesize{Average Energy}}
    \label{plot:energy-summary}
  \end{subfigure}
  \caption[\gsdram: Performance and energy summary]{Summary of
    performance and energy consumption for the transactions and
    analytics workloads}
  \label{plot:summary}
\end{figure}

Figure~\ref{plot:summary} summarizes the performance and energy
benefits of \gsdram compared to \rowstore and \colstore for the
transactions and the analytics workloads. We conclude that \gsdram
provides the best of both the layouts.

\subsection{Scientific Computation: GEMM}
\label{sec:dgemm}

General Matrix-Matrix (GEMM) multiplication is an important kernel
in many scientific computations. When two $n \times n$ matrices
$A$ and $B$ are multiplied, the matrix $A$ is accessed in the
row-major order, whereas the matrix $B$ is accessed in the
column-major order. If both matrices are stored in row-major
order, a naive algorithm will result in poor spatial locality for
accesses to $B$. To mitigate this problem, matrix libraries use
two techniques. First, they split each matrix into smaller tiles,
converting the reuses of matrix values into L1 cache hits. Second,
they use SIMD instructions to speed up each vector dot product
involved in the operation.

Unfortunately, even after tiling, values of a column of matrix $B$
are stored in different cache lines. As a result, to exploit SIMD,
the software must gather the values of a column into a SIMD
register. In contrast, \gsdram can read each tile of the matrix in
the column-major order into the L1 cache such that each cache line
contains values gathered from one column. As a result, \gsdram
naturally enables SIMD operations, without requiring the software
to gather data into SIMD registers.

\begin{figure}
  \centering
  \includegraphics{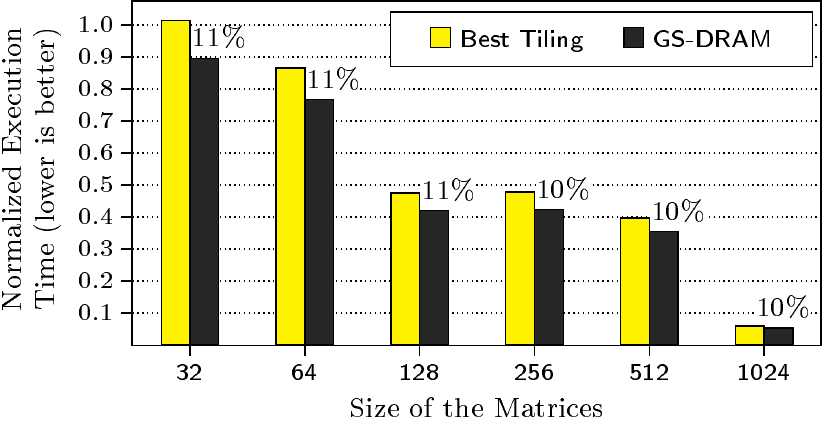}
  \caption[\gsdram: GEMM multiplication performance]{GEMM
    Multiplication: Performance of \gsdram and the best
    tiled-version (normalized to a non-tiled baseline). Values on
    top indicate percentage reduction in execution time of \gsdram
    compared to tiling.}
  \label{fig:dgemm}
\end{figure}

Figure~\ref{fig:dgemm} plots the performance of GEMM with \gsdram
and with the best tiled version normalized to a non-tiled version
for different sizes ($n$) of the input matrices. We draw two
conclusions. First, as the size of the matrices increases, tiling
provides significant performance improvement by eliminating many
memory references. Second, by seamlessly enabling SIMD operations,
\gsdram improves the performance of GEMM multiplication by 10\% on
average compared to the best tiled baseline.  Note that \gsdram
achieves 10\% improvement over a heavily-optimized tiled baseline
that spends most of its time in the L1 cache.

\subsection{Other Applications}
\label{sec:other-apps}

We envision \gsdram to benefit many other applications like
key-value stores, graph processing, and graphics. Key-value stores
have two main operations: \emph{insert} and \emph{lookup}. The
\emph{insert} operation benefits from both the key and value being
in the same cache line. On the other hand, the \emph{lookup}
operation benefits from accessing a cache line that contains only
keys. Similarly, in graph processing, operations that update
individual nodes in the graph have different access patterns than
those that traverse the graph. In graphics, multiple pieces of
information (e.g., RGB values of pixels) may be packed into small
objects. Different operations may access multiple values within an
object or a single value across a large number of objects. The
different access patterns exhibited by these applications have a
regular stride and can benefit significantly from \gsdram.

\section{Extensions to \gsdram}
\label{sec:gsdram-extensions}

In this section, we describe three simple extensions to \gsdram:
1)~programmable shuffling, 2)~wider pattern IDs, and 3)~intra-chip
column translation. These extensions (together or individually)
allow \gsdram to 1)~express more patterns (e.g., larger strides),
2)~gather or scatter data at a granularity smaller than 8 bytes,
and 3)~enable ECC support.

\subsection{Programmable Shuffling}
\label{sec:programmable-shuffling}

Although our shuffling mechanism uses the least significant bits
of the column ID to control the shuffling stages, there are two
simple ways of explicitly controlling which shuffling stages are
active. First, we can use a \emph{shuffle mask} to disable some
stages. For example, the shuffle mask \texttt{10} disables
swapping of adjacent values (Figure~\ref{fig:shuffling}, Stage
1). Second, instead of using the least significant bits to control
the shuffling stages, we can choose different combinations of bits
(e.g., XOR of multiple sets of
bits~\cite{power-of-2,xor-schemes}).  To enable programmable
shuffling, we add another parameter to \gsdram called the
\emph{shuffling function}, $f$. For \gsdrampf{c}{s}{p}{f}, the
function $f$ takes a column ID as input and generates an $n$-bit
value that is used as the control input to the $n$ shuffling
stages. The function $f$ can be application-specific, thereby
optimizing \gsdram for each application.

\subsection{Wider Pattern IDs}
\label{sec:wider-pattern}

Although a wide pattern ID comes at additional cost, using a wider
pattern ID allows the memory controller to express more access
patterns. However, the column translation logic (CTL) performs a
bitwise AND of the chip ID and the pattern ID to create a modifier
for the column address. As a result, even if we use a wide pattern
ID, a small chip ID disables the higher order bits of the pattern
ID. Specifically, for \gsdramp{c}{*}{p}, if $p > \log c$, the CTL
uses only the least significant $\log c$ bits of the pattern
ID. To enable wider pattern IDs, we propose to simply widen the
chip ID used by the CTL by repeating the physical chip ID multiple
times. For instance, with 8 chips and a 6-bit pattern ID, the chip
ID used by CTL for chip $3$ will be \texttt{011-011} (i.e.,
\texttt{011} repeated twice). With this simple extension, \gsdram
can enable more access patterns (e.g., larger strides).

\subsection{Intra-Chip Column Translation}
\label{sec:intra-chip-gather}

Although we have assumed that each DRAM bank has a single wide
row-buffer, in reality, each DRAM bank is a 2-D collection of
multiple small \emph{tiles} or
MATs~\cite{rethinking-dram,half-dram,salp}.  Similar to how each
chip within a rank contributes 64 bits to each cache line, each
tile contributes equally to the 64 bits of data supplied by each
chip.  We can use the column translation logic within each DRAM
chip to select different columns from different tiles for a single
\cmdrd or \cmdwr. This mechanism has two benefits. First, with the
support for intra-chip column translation, we can gather access
patterns at a granularity smaller than 8 bytes. Second, with DIMMs
that support ECC, \gsdram may incur additional bandwidth to read
all the required ECC values for non-zero patterns. However, if we
use a chip that supports intra-chip column selection for ECC,
accesses with non-zero patterns can gather the data from the eight
data chips and gather the ECC from the eight tiles within the ECC
chip, thereby seamlessly supporting ECC for all access patterns.

\section{Prior Work}
\label{sec:gsdram-prior-work}

Carter et al.~\cite{impulse} propose Impulse, a mechanism to
export gather operations to the memory controller. In their
system, applications specify a \emph{gather mapping} to the memory
controller (with the help of the OS). To perform a gather access,
the controller assembles a cache line with only the values
required by the access pattern and sends the cache line to the
processor, thereby reducing the bandwidth between the memory
controller and the processor. Impulse has two shortcomings. First,
with commodity DRAM modules, which are optimized for accessing
cache lines, Impulse cannot mitigate the wasted memory bandwidth
consumption between the memory controller and DRAM. Impulse
requires a memory interface that supports fine-grained accesses
(e.g.,~\cite{mini-rank,mc-dimm,mc-dimm-cal,threaded-module,sg-dimm}),
which significantly increases the system cost. Second, Impulse
punts the problem of maintaining cache coherence to software. In
contrast, \gsdram 1)~works with commodity modules with very few
changes, and 2)~provides coherence of gathered cache lines
transparent to software.

Yoon et al.~\cite{dgms,agms} propose the Dynamic Granularity
Memory System (DGMS), a memory interface that allows the memory
controller to dynamically change the granularity of memory
accesses in order to avoid unnecessary data transfers for accesses
with low spatial locality. Similar to Impulse, DGMS requires a
memory interface that supports fine-grained memory accesses
(e.g.,~\cite{mini-rank,mc-dimm,mc-dimm-cal,threaded-module,sg-dimm})
and a sectored cache~\cite{sectored-cache,dscache}. In contrast,
\gsdram works with commodity DRAM modules and conventionally-used
non-sectored caches with very few changes.

Prior works
(e.g.,~\cite{stride-prefetching,stride-prefetching-2,stride-stream-buffer,stride-pref-fp,fdp,ghb})
propose prefetching for strided accesses. While prefetching
reduces the latency of such accesses, it does not avoid the waste
in memory bandwidth and cache space.  He et
al.~\cite{gpu-gather-scatter} propose a model to analyze the
performance of gather-scatter accesses on a GPU. To improve cache
locality, their model splits gather-scatter loops into multiple
passes such that each pass performs only accesses from a small
group of values that fit in the cache.  This mechanism works only
when multiple values are \emph{actually} reused by the
application. In contrast, \gsdram fetches \emph{only} useful
values from DRAM, thereby achieving better memory bandwidth and
cache utilization.

\section{Summary}
\label{sec:gsdram-summary}

In this chapter, we introduced \emph{Gather-Scatter DRAM}, a
low-cost substrate that enables the memory controller to
efficiently gather or scatter data with different non-unit strided
access patterns. Our mechanism exploits the fact that multiple
DRAM chips contribute to each cache line access. \gsdram maps
values accessed by different strided patterns to different chips,
and uses a per-chip column translation logic to access data with
different patterns using significantly fewer memory accesses than
existing DRAM interfaces. Our framework requires no changes to
commodity DRAM chips, and very few changes to the DRAM module, the
memory interface, and the processor architecture. Our evaluations
show that \gsdram provides the best of both the row store and the
column store layouts for a number of in-memory database workloads,
and outperforms the best tiled layout on a well-optimized
matrix-matrix multiplication workload. Our framework can benefit
many other modern data-intensive applications like key-value
stores and graph processing. We conclude that the \gsdram
framework is a simple and effective way to improve the performance
of non-unit strided and gather/scatter memory accesses.

\chapter{The Dirty-Block Index}
\label{chap:dbi}

\begin{figure}[b!]
  \hrule\vspace{2mm}
  \begin{footnotesize}
    Originally published as ``The Dirty-Block Index'' in the
    International Symposium on Computer Architecture,
    2014~\cite{dbi}
  \end{footnotesize}
\end{figure}

In the previous three chapters, we described three mechanisms that
offload some key application-level primitives to DRAM. As
described in the respective chapters, these mechanisms directly
read/modify data in DRAM. As a result, they require the cache
coherence protocol to maintain the coherence of the data stored in
the on-chip caches. Specifically, for an in-DRAM operation, the
protocol must carry out two steps. First, any dirty cache line
that is directly read in DRAM should be flushed to DRAM
\emph{before} the operation. Second, any cache line that is
modified in DRAM should be invalidated from the caches.

While the second step can be performed in \emph{parallel} with the
in-DRAM operation, the first step, i.e., flushing the dirty cache
lines of the source data, is on the critical path of performing
the in-DRAM operation. In this chapter, we describe Dirty-Block
Index, a new way of tracking dirty blocks that can speed up
flushing dirty blocks of a DRAM row.

\section{DRAM-Aware Writeback}
\label{sec:dbi-dawb}

We conceived of the Dirty-Block Index based on a previously
proposed optimization called \emph{DRAM-Aware Writeback}. In this
section, we first provide a brief background on dirty blocks and the
interaction between the last-level cache and the memory
controller.  We then describe the optimization.

\subsection{Background on Dirty Block Management}
\label{sec:dbi-background}

Most modern high-performance systems use a writeback last-level
cache. When a cache block is modified by the CPU, it is marked
\emph{dirty} in the cache. For this purpose, each tag entry is
associated with a \emph{dirty} bit, which indicates if the block
corresponding to the tag entry is dirty. When a block is evicted,
if its dirty bit is set, then the cache sends the block to the
memory controller to be written back to the main memory.

The memory controller periodically writes back dirty blocks to
their locations in main memory. While such writes can be
interleaved with read requests at a fine granularity, their is a
penalty to switching the memory channel between the read and write
modes. As a result, most memory controllers buffer the writebacks
from the last-level cache in a write buffer. During this period,
which we refer to as the \emph{read phase}, the memory controller
only serves read requests. When the write buffer is close to
becoming full, the memory controller stops serving read requests
and starts flushing out the write buffer to main memory until the
write buffer is close to empty. We refer to this phase as the
\emph{writeback phase}. The memory controller then switches back
to serving read requests. This policy is referred to as
\emph{drain-when-full}~\cite{dram-aware-wb}.

\subsection{DRAM-Aware Writeback: Improving Write Locality}

In existing systems, the sequence with which dirty blocks are
evicted from the cache depends on primarily on the cache
replacement policy. As observed by two prior
works~\cite{dram-aware-wb,vwq}, this approach can fill the write buffer
with dirty blocks from many different rows in DRAM. As a result,
the writeback phase exhibits poor DRAM row buffer
locality. However, there could be other dirty blocks in the cache
which belong to the same DRAM row as those in the write
buffer.

\emph{DRAM-Aware Writeback} (DAWB) is a simple solution to counter
this problem. The idea is to writeback dirty blocks of the same
DRAM row together so as to improve the row buffer locality of the
writeback phase. This could reduce the time consumed by the
writeback phase, thereby allowing the memory controller to switch
to the read phase sooner. To implement this idea, whenever a dirty
block is evicted from the cache, DAWB checks if there are any
other dirty blocks in the cache that belong to the same DRAM
row. If such blocks exist, DAWB simply writes the contents of
those blocks to main memory and marks them as clean (i.e., clears
their dirty bits). Evaluations show that this simple optimization
can significantly improve the performance of many applications. 

\subsection{Inefficiency in Implementing DAWB}

Implementing DAWB with existing cache organizations requires the
cache to lookup each block of a DRAM row and determine if the
block is dirty. In a modern system with typical DRAM row buffer
size of 8KB and a cache block size of 64B, this operations
requires 128 cache lookups. With caches getting larger and more
cores sharing the cache, these lookups consume high latency and
also add to the contention for the tag store. To add to the
problem, many of these cache blocks may not be dirty to begin
with, making the lookups for those blocks unnecessary.

Ideally, as part of the DAWB optimization, the tag store should be
looked up only for those blocks from the DRAM row that are
actually dirty. In other words, the cache must be able to
efficiently identify the list of all cache blocks of a given DRAM
row (or region) that are dirty. This will not only enable a more
efficient implementation of the DAWB optimization, but also
address the cache coherence problem described in the beginning of
this chapter.

\section{The Dirty-Block Index}
\label{sec:dbi}

In our proposed system, we remove the dirty bits from the tag
store and organize them differently in a separate structure called
the Dirty-Block Index (DBI). At a high level, DBI organizes the
dirty bit information such that the dirty bits of all the blocks
of a DRAM row are stored together.

\subsection{\dbi Structure and Semantics}

\begin{figure}
  \centering
  \begin{subfigure}{0.55\textwidth}
    \centering
    \includegraphics[scale=0.8]{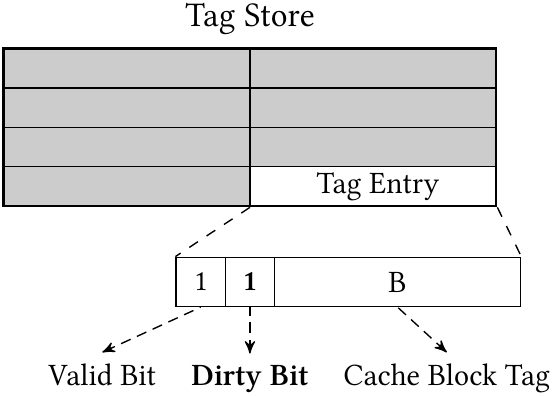}
    \caption{Conventional cache tag store}
    \label{fig:conventional}
  \end{subfigure}\vspace{2mm}
  \begin{subfigure}{0.55\textwidth}
    \centering
    \includegraphics[scale=0.8]{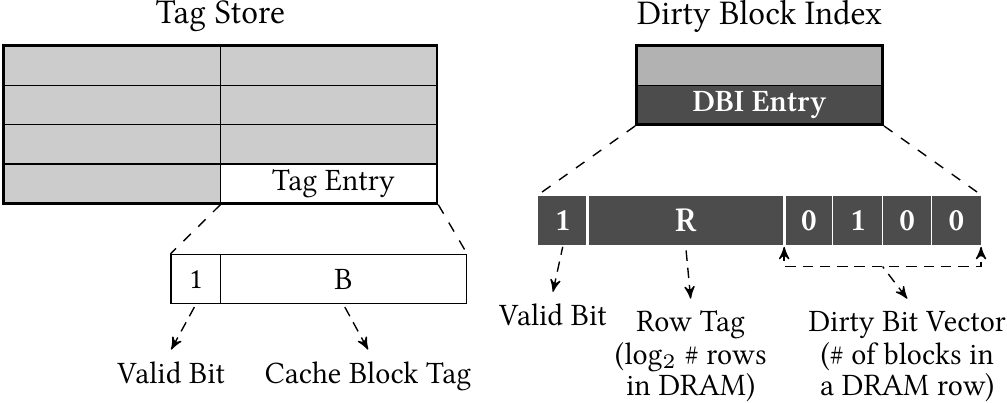}
    \caption{Cache tag store augmented with a DBI}
    \label{fig:dbi}
  \end{subfigure}
  \caption[DBI vs. conventional cache]{Comparison between conventional cache and a cache with DBI.}
  \label{fig:dbi-compare}
\end{figure}

Figure~\ref{fig:dbi-compare} compares the conventional tag store
with a tag store augmented with a \dbi. In the conventional
organization (shown in Figure~\ref{fig:conventional}), each tag
entry contains a dirty bit that indicates whether the
corresponding block is dirty or not. For example, to indicate that
a block B is dirty, the dirty bit of the corresponding tag entry
is set.

In contrast, in a cache augmented with a \dbi
(Figure~\ref{fig:dbi}), the dirty bits are removed from the main
tag store and organized differently in the \dbi. The organization
of \dbi is simple. It consists of multiple entries. Each entry
corresponds to some row in DRAM---identified using a \emph{row
  tag} present in each entry. Each \dbi entry contains a
\emph{dirty bit vector} that indicates if each block in the
corresponding DRAM row is dirty or not.

\textbf{\dbi Semantics.} A block in the cache is dirty \emph{if
  and only if} the \dbi contains a valid entry for the DRAM row
that contains the block and the bit corresponding to the block in
the bit vector of that \dbi entry is set. For example, assuming
that block B is the second block of DRAM row R, to indicate that
block B is dirty, the \dbi contains a valid entry for DRAM row R,
with the second bit of the corresponding bit vector set.

Note that the key difference between the DBI and the conventional
tag store is the \emph{logical organization} of the dirty bit
information. While some processors store the dirty bit information
in a separate physical structure, the logical organization of the
dirty bit information is same as the main tag store.

\subsection{\dbi Operation}
\label{sec:dbi-operation}

\begin{figure}[h]
  \centering
  \includegraphics{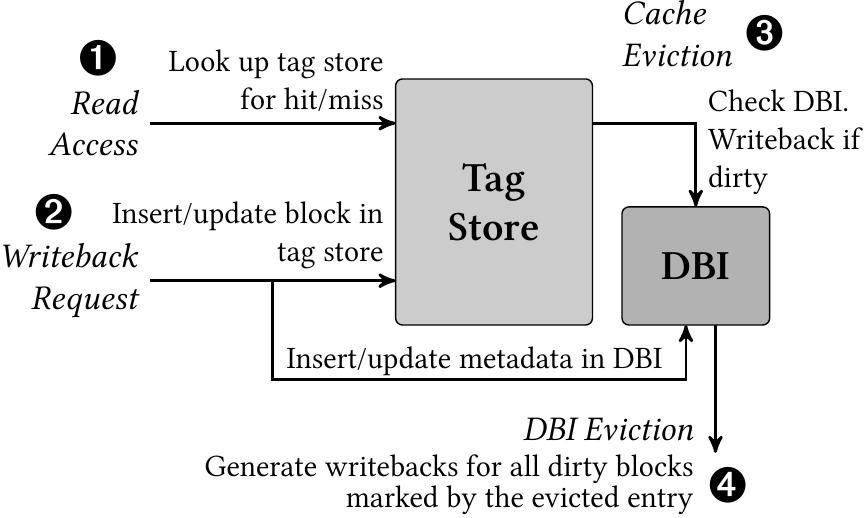}
  \caption{Operation of a cache with \dbi}
  \label{fig:dbi-operation}
\end{figure}

Figure~\ref{fig:dbi-operation} pictorially describes the operation
of a cache augmented with a \dbi. The focus of this work is on the
on-chip last-level cache (LLC). Therefore, for ease of
explanation, we assume that the cache does not receive any
sub-block writes and any dirty block in the cache is a result of a
writeback generated by the previous level of
cache.\footnote{Sub-block writes typically occur in the primary L1
  cache where writes are at a word-granularity, or at a cache
  which uses a larger block size than the previous level of
  cache. The \dbi operation described in this paper can be easily
  extended to caches with sub-block writes.}  There are four
possible operations, which we describe in detail below.

\subsubsection{Read Access to the Cache}
\label{sec:dbi-operation-read}

The addition of \dbi \emph{does not change} the path of a read
access in any way. On a read access, the cache simply looks up the
block in the tag store and returns the data on a cache
hit. Otherwise, it forwards the access to the memory controller.

\subsubsection{Writeback Request to the Cache}
\label{sec:dbi-operation-writeback}

In a system with multiple levels of on-chip cache, the LLC will
receive a writeback request when a dirty block is evicted from the
previous level of cache.  Upon receiving such a writeback request,
the cache performs two actions (as shown in
Figure~\ref{fig:dbi-operation}). First, it inserts the block into
the cache if it is not already present. This may result in a cache
block eviction (discussed in
Section~\ref{sec:dbi-operation-cache-eviction}). If the block is
already present in the cache, the cache just updates the data
store (not shown in the figure) with the new data. Second, the
cache updates the \dbi to indicate that the written-back block is
dirty. If the \dbi already has an entry for the DRAM row that
contains the block, the cache simply sets the bit corresponding to
the block in that \dbi entry. Otherwise, the cache inserts a new
entry into the \dbi for the DRAM row containing the block and with
the bit corresponding to the block set. Inserting a new entry into
the \dbi may require an existing \dbi entry to be
evicted. Section~\ref{sec:dbi-operation-dbi-eviction} discusses
how the cache handles such a \dbi eviction.

\subsubsection{Cache Eviction}
\label{sec:dbi-operation-cache-eviction}

When a block is evicted from the cache, it has to be written back
to main memory if it is dirty. Upon a cache block eviction, the
cache consults the \dbi to determine if the block is dirty. If so,
it first generates a writeback request for the block and sends it
to the memory controller. It then updates the \dbi to indicate
that the block is no longer dirty---done by simply resetting the
bit corresponding to the block in the bit vector of the \dbi
entry. If the evicted block is the last dirty block in the
corresponding \dbi entry, the cache invalidates the \dbi entry so
that the entry can be used to store the dirty block information of
some other DRAM row.

\subsubsection{DBI Eviction}
\label{sec:dbi-operation-dbi-eviction}

The last operation in a cache augmented with a \dbi is a
\emph{\dbi eviction}. Similar to the cache, since the \dbi has
limited space, it can only track the dirty block information for a
limited number of DRAM rows. As a result, inserting a new \dbi
entry (on a writeback request, discussed in
Section~\ref{sec:dbi-operation-writeback}) may require evicting an
existing \dbi entry. We call this event a \emph{\dbi
  eviction}. The \dbi entry to be evicted is decided by the \dbi
replacement policy (discussed in
Section~\ref{sec:dbi-replacement-policy}). When an entry is
evicted from the \dbi, \emph{all} the blocks indicated as dirty by
the entry should be written back to main memory. This is because,
once the entry is evicted, the \dbi can no longer maintain the
dirty status of those blocks. Therefore, not writing them back to
memory will likely lead to incorrect execution, as the version of
those blocks in memory is stale. Although a \dbi eviction may
require evicting many dirty blocks, with a small buffer to keep
track of the evicted \dbi entry (until all of its blocks are
written back to memory), the \dbi eviction can be interleaved with
other demand requests. Note that on a \dbi eviction, the
corresponding cache blocks need not be evicted---they only need to
be transitioned from the dirty state to clean state.

\subsection{Cache Coherence Protocols}
\label{sec:dbi-ccp}

Many cache coherence protocols implicitly store the dirty status
of cache blocks in the cache coherence states. For example, in the
MESI protocol~\cite{mesi}, the M (modified) state indicates that
the block is dirty. In the improved MOESI protocol~\cite{moesi},
both M (modified) and O (Owner) states indicate that the block is
dirty. To adapt such protocols to work with \dbi, we propose to
split the cache coherence states into multiple pairs---each pair
containing a state that indicates the block is dirty and the
non-dirty version of the same state. For example, we split the
MOESI protocol into three parts: (M, E), (O, S) and (I). We can
use a single bit to then distinguish between the two states in
each pair. This bit will be stored in the \dbi.

\section{\dbi Design Choices}
\label{sec:design-choices}

The \dbi design space can be defined using three key parameters:
1)~\emph{\dbi size}, 2)~\emph{\dbi granularity} and 3)~\emph{\dbi
  replacement policy}.\footnote{\dbi is also a set-associative
  structure and has a fixed associativity. However, we do not
  discuss the \dbi associativity in detail as its trade-offs are
  similar to any other set-associative structure.}  These
parameters determine the effectiveness of the three optimizations
discussed in the previous section. We now discuss these parameters
and their trade-offs in detail.

\subsection{\dbi Size}
\label{sec:dbi-size}

The \dbi size refers to the cumulative number of blocks tracked by
all the entries in the \dbi. For ease of analysis across systems
with different cache sizes, we represent the \dbi size as the
ratio of the cumulative number of blocks tracked by the \dbi and
the number of blocks tracked by the cache tag store. We denote
this ratio using $\alpha$. For example, for a 1MB cache with a
64B block size (16k blocks), a \dbi of size $\alpha =
\sfrac{1}{2}$ enables the \dbi to track 8k blocks.

The \dbi size presents a trade-off between the size of write
working set (set of frequently written blocks) that can be
captured by the \dbi, and the area, latency, and power cost of the
\dbi. A large \dbi has two benefits: 1)~it can track a larger
write working set, thereby reducing the writeback bandwidth
demand, and 2)~it gives more time for a \dbi entry to accumulate
writebacks to a DRAM row, thereby better exploiting the AWB
optimization. However, a large \dbi comes at a higher area,
latency and power cost. On the other hand, a smaller \dbi incurs
lower area, latency and power cost. This has two benefits:
1)~lower latency in the critical path for the CLB optimization and
2)~ECC storage for fewer dirty blocks. However, a small \dbi
limits the number of dirty blocks in the cache and thus, result in
premature \dbi evictions, reducing the potential to generate
aggressive writebacks. It can also potentially lead to thrashing
if the write working set is significantly larger than the number
of blocks tracked by the small \dbi.

\subsection{\dbi Granularity}
\label{sec:dbi-granularity}

The \dbi granularity refers to the number of blocks tracked by a
single \dbi entry. Although our discussion in
Section~\ref{sec:dbi} suggests that this is same as the
number of blocks in each DRAM row, we can design the \dbi to track
fewer blocks in each entry. For example, for a system with DRAM
row of size 8KB and cache block of size 64B, a natural choice for
the \dbi granularity is 8KB/64B = 128. Instead, we can design a
\dbi entry to track only 64 blocks, i.e. one half of a DRAM row.

The \dbi granularity presents another trade-off between the amount
of locality that can be extracted during the writeback phase
(using the AWB optimization) and the size of write working set
that can be captured using the \dbi. A large granularity leads to
better potential for exploiting the AWB optimization. However, if
writes have low spatial locality, a large granularity will result
in inefficient use of the \dbi space, potentially leading to write
working set thrashing.

\subsection{\dbi Replacement Policy}
\label{sec:dbi-replacement-policy}

The \dbi replacement policy determines which entry is evicted on a
\dbi eviction (Section~\ref{sec:dbi-operation-dbi-eviction}). A
\dbi eviction \emph{only writes back} the dirty blocks of the
corresponding DRAM row to main memory, and does not evict the
blocks themselves from the cache. Therefore, a \dbi eviction does
not affect the latency of future read requests for the
corresponding blocks. However, if the previous cache level
generates a writeback request for a block written back due to a
\dbi eviction, the block will have to be written back again,
leading to an additional write to main memory. Therefore, the goal
of the \dbi replacement policy is to ensure that blocks are not
prematurely written back to main memory.

The ideal policy is to evict the entry that has a writeback
request farthest into the future. However, similar to Belady's
optimal replacement policy~\cite{beladyopt}, this ideal policy is
impractical to implement in real systems. We evaluated five
practical replacement policies for \dbi: 1)~Least Recently Written
(LRW)---similar to the LRU policy for caches, 2)~LRW with Bimodal
Insertion Policy~\cite{dip}, 3)~Rewrite-interval prediction
policy---similar to the RRIP policy for caches~\cite{rrip},
4)~Max-Dirty---entry with the maximum number of dirty blocks,
5)~Min-Dirty---entry with the minimum number of dirty blocks. We
find that the LRW policy works comparably or better than the other
policies.

\section{Bulk Data Coherence with \dbi}

As mentioned in Section~\ref{sec:dbi}, we conceived of \dbi to
primarily improve the performance of the DRAM-Aware Writeback
optimization. However, we identified several other potential use
cases for \dbi, and quantitatively evaluated three such
optimizations (including DRAM-aware Writeback).  Since these
optimizations are not directly related to this thesis, we only
provide a brief summary of the evaluated optimizations and the
results in Section~\ref{sec:dbi-soe}. We direct the reader to our
paper published in ISCA 2014~\cite{dbi} for more details. In this
section, we describe how \dbi can be used to improve the
performance of certain cache coherence protocol operations,
specifically flushing bulk data.

\subsection{Checking If a Block is Dirty}

As described in Section~\ref{sec:dbi}, in the cache organization
used in existing systems, the cache needs to perform a full tag
store lookup to check if a block is dirty. Due to the large size
and associativity of the last-level cache, the latency of this
lookup is typically several processor cycles. In contrast, in a
cache augmented with \dbi, the cache only needs to perform a
single \dbi lookup to check if a block is dirty. Based on our
evaluations, even with just 64 \dbi entries (in comparison to
16384 entries in the main tag store), the cache with \dbi
outperforms the conventional cache organization. As a result,
checking if a block is dirty is significantly cheaper with the
\dbi organization.

This fast lookup for the dirty bit information can already make
many coherence protocol operations in multi-socket systems more
efficient. Specifically, when a processor wants to read a cache
line, it has to first check if any of the other processors contain
a dirty version of the cache line. With the \dbi, this operation
is more efficient than in existing systems. However, the major
benefit of \dbi is in accelerating bulk data flushing.

\subsection{Accelerating Bulk Data Flushing}

In many scenarios, the memory controller must check if a set of
blocks belonging to a region is dirty in the on-chip cache. For
instance, in Direct Memory Access, the memory controller may send
data to an I/O device by directly reading the data from
memory. Since these operations typically happen in bulk, the
memory controller may have to get all cache lines in an entire
page (for example) that are dirty in the on-chip cache. In fact,
all the DRAM-related mechanisms described in the previous three
chapters of this thesis require an efficient implementation of
this primitive.

With a cache augmented \dbi, this primitive can be implemented
with just a single \dbi lookup. When the controller wants to
extract all the dirty cache lines of a region, it looks up the
\dbi with the \dbi tag for the corresponding region, and extracts
the dirty bit vector, which indicates which blocks within the
region are dirty in the cache. If the size of the region is larger
than the segment tracked by each \dbi entry, the memory controller
must perform multiple lookups. However, with the \dbi, 1)~the
memory controller has to perform 64X or 128X (depending on the
\dbi granularity) fewer lookups compared to conventional cache
organization, 2)~each \dbi lookup is cheaper than a tag store
lookup, and 3)~the cache can lookup and flush \emph{only} blocks
that are actually dirty. With these benefits, \dbi can perform
flush data in bulk faster than existing organizations.

\section{Summary of Optimizations and Evaluations}
\label{sec:dbi-soe}

In this section, we provide a brief summary of the optimizations
enabled by \dbi and the corresponding evaluation. We refer the
reader to our ISCA paper~\cite{dbi} for more details. We
specifically evaluate the following three optimizations in detail:
1)~DRAM-aware writeback, 2)~cache lookup bypass, and
3)~heterogeneous ECC.

\subsection{Optimizations}

\subsubsection{Efficient DRAM-aware Writeback}

This is the optimization we described briefly in
Section~\ref{sec:dbi}. The key idea is to cluster
writebacks to dirty blocks of individual DRAM rows with the goal
of maximizing the write row buffer hit rate. Implementing this
optimization with DBI is straightforward. When a block is evicted
from the cache, the DBI tells the cache whether the block is dirty
or not. In addition, the bit vector in the corresponding DBI entry
also tells the cache the list of all other dirty blocks in the
corresponding DRAM row (assuming that each region in the DBI
corresponds to a DRAM row). The cache can then selectively lookup
only those dirty blocks and write them back to main memory. This
is significantly more efficient than the implementation with the
block-oriented organization. The best case for DBI is when no
other block from the corresponding DRAM row is dirty. In this
case, current implementations will have to look up every block in
the DRAM row unnecessarily, whereas DBI will perform zero
additional lookups.

\subsubsection{Bypassing Cache Lookups}

The idea behind this optimization is simple: \emph{If an access is
  likely to miss in the cache, then we can avoid the tag lookup
  for the access, reducing both latency and energy consumption of
  the access.}  In this optimization, the cache is augmented with
a miss predictor which can efficiently predict if an access will
hit or miss in the cache. If an access is predicted to miss, then
the request is directly sent to main memory. The main challenge
with this optimization is that an access to a dirty block should
not be bypassed. This restricts the range of miss predictors as
the predictor cannot falsely predict that an access will miss in
the cache~\cite{jsn}. Fortunately, with the DBI, when an access is
predicted to miss, the cache can first consult the DBI to ensure
that the block is not dirty before bypassing the tag lookup. As a
result, DBI enables very aggressive miss predictors (e.g., bypass
all accesses of an application~\cite{skipcache}).

\subsubsection{Reducing ECC Overhead}

The final optimization is again based on a simple idea:
\emph{clean blocks need only error detection; only dirty blocks
  need strong error correction.} Several prior works have propose
mechanisms to exploit this observation to reduce the ECC
overhead. However, they require complex mechanisms to handle the
case when an error is detected on a dirty block
(e.g.,~\cite{ecc-fifo}). In our proposed organization, since DBI
tracks dirty blocks, it is sufficient to store ECC only for the
blocks tracked by DBI. With the previously discussed
optimizations, we find that the DBI can get away with tracking far
fewer blocks than the main cache. As a result, DBI can seamlessly
reduce the ECC area overhead (8\% reduction in overall cache
area).

\subsection{Summary of Results}

\begin{figure}[b]
  \centering
  \includegraphics{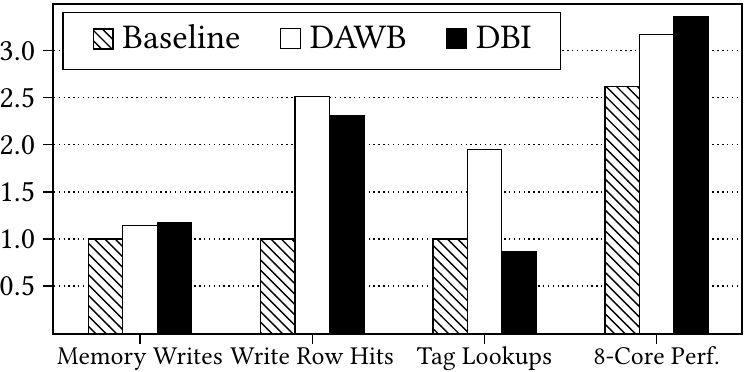}
  \caption[DBI: Summary of performance results]{Summary of
    Performance Results. The first three metrics are normalized to
    the Baseline.}
  \label{plot:dbi-perf}
\end{figure}

We refer the reader to our paper for full details on our
methodology.  Figure~\ref{plot:dbi-perf} briefly summarizes the
comparison between DBI (with the first two optimizations,
DRAM-aware writeback and cache lookup bypass) and the best
previous mechanism, DRAM-aware writeback~\cite{dram-aware-wb}
(DAWB).  As a result of proactively writing back blocks to main
memory, both mechanisms increase the number of memory
writes. However, for a small increase in the number of writes,
both mechanisms significantly improve the write row hit rate, and
hence also performance compared to the baseline. However, the key
difference between DAWB and DBI is that DAWB almost doubles the
number of tag lookups, whereas with both optimizations, DBI
actually reduces the number of tag lookups by 14\% compared to the
baseline. As a result, DBI improves performance by 6\% compared to
DAWB (31\% over baseline) across 120 workloads in an 8-core system
with 16MB shared cache.

\section{Summary}
\label{sec:dbi-summary}

In this chapter, we introduced the Dirty-Block Index (DBI), a
structure that aids the cache in efficiently responding to queries
regarding dirty blocks. Specifically, DBI can quickly list all
dirty blocks that belong to a contiguous region and, in general,
check if a block is dirty more efficiently than existing cache
organization. To achieve this, our mechanism removes the dirty
bits from the on-chip tag store and organizes them at a large
region (e.g., DRAM row) granularity in the DBI. DBI can be used to
accelerate the coherence protocol that ensures the coherence of
data between the on-chip caches and main memory in our in-DRAM
mechanisms. We described three other concrete use cases for the
DBI that can improve the performance and energy-efficiency of the
memory subsystem in general, and reduce the ECC overhead for
on-chip caches. This approach is an effective way of enabling
several other optimizations at different levels of caches by
organizing the DBI to cater to the write patterns of each cache
level. We believe this approach can be extended to more
efficiently organize other metadata in caches (e.g., cache
coherence states), enabling more optimizations to improve
performance and power-efficiency.

\chapter{Conclusions \& Future Work}
\label{chap:conclusion}

In modern systems, different resources of the memory subsystem
store and access data at different granularities. Specifically,
virtual memory manages main memory capacity at a (large) page
granularity; the off-chip memory interface access data from memory
at a cache line granularity; on-chip caches typically store and
access data at a cache line granularity; applications typically
access data at a (small) word granularity. We observe that this
mismatch in granularities results in significant inefficiency for
many memory operations. In Chapter~\ref{chap:motivation}, we
demonstrate this inefficiency using two example operations:
copy-on-write and non-unit strided access.

\section{Contributions of This Dissertation}

In this dissertation, we propose five distinct mechanisms to
address the inefficiency problem at various structures.  First, we
observe that page-granularity management of main memory capacity
can result in significant inefficiency in implementing many memory
management techniques. In Chapter~\ref{chap:page-overlays}, we
describe \emph{page overlays}, a new framework which augments the
existing virtual memory framework with a structure called
\emph{overlays}. In short, an overlay of a virtual page tracks a
newer version of a subset of segments from within the page. We
show that this simple framework is very powerful and enables many
applications. We quantitatively evaluate page overlays with two
mechanisms: \emph{overlay-on-write}, a more efficient version of
the widely-used copy-on-write technique, and an efficient
hardware-based representation for sparse data structures.

Second, we show that the internal organization and operation of a
DRAM chip can be used to transfer data quickly and efficiently
from one location to another. Exploiting this observation, in
Chapter~\ref{chap:rowclone}, we describe \emph{RowClone}, a
mechanism to perform bulk copy and initialization operations
completely inside DRAM. RowClone reduces the latency and energy of
performing a bulk copy operation by 11X and 74X, respectively,
compared to commodity DRAM interface. Our evaluations show that
this reduction significantly improves the performance of copy and
initialization intensive applications.

Third, we show that the analog operation of DRAM and the inverters
present in the sense amplifier can be used to perform bitwise
logical operations completely inside DRAM. In
Chapter~\ref{chap:buddy}, we describe \emph{Buddy RAM}, a
mechanism to perform bulk bitwise logical operations using
DRAM. Buddy improves the throughput of various bitwise logical
operations by between 3.8X (for bitwise XOR) and 10.1X (for
bitwise NOT) compared to a multi-core CPU, and reduces the energy
consumption of the respective operations by 25.1X and 59.5X. We
demonstrate the benefits of Buddy by using it to improve the
performance of set operations and in-memory bitmap indices.

Fourth, we show that the multi-chip module organization of
off-chip memory can be used to efficiently gather or scatter data
with strided access patterns. In Chapter~\ref{chap:gsdram}, we
describe \emph{Gather-Scatter DRAM} (\gsdram), a mechanism that
achieves near-ideal bandwidth utilization for any power-of-2
strided access pattern. We implement an in-memory database on top
of \gsdram and show that \gsdram gets the best of both a row store
layout and column store layout on a transactions workload,
analytics workload, and a hybrid transactions and analytics
workload.

Finally, in Chapter~\ref{chap:dbi}, we introduce the Dirty-Block
Index, a mechanism to improve the efficiency of the coherence
protocol that ensures the coherence of data across the on-chip
caches and main memory. DBI efficiently tracks spatially-collocated
dirty blocks, and has several applications in addition to
more efficient data coherence, e.g., efficient memory writeback,
efficient cache lookup bypass, and reducing cache ECC overhead.

\section{Future Research Directions}
\label{sec:future}

This dissertation opens up several avenues for research. In this
section, we describe six specific directions in which the ideas
and approaches proposed in this thesis can be extended to other
problems to improve the performance and efficiency of various
systems.

\subsection{Extending Overlays to Superpages}

In Chapter~\ref{chap:page-overlays}, we described and evaluated
the benefits of our page overlays mechanism for regular 4KB
pages. We believe that our mechanism can be easily extended to
superpages, which are predominantly used by many modern operating
systems. The primary benefit of using superpages is to increase
TLB reach and thereby reduce overall TLB misses. However, we
observe that superpages are ineffective for reducing memory
redundancy.

To understand this, let us consider the scenario when multiple
virtual superpages are mapped to the same physical page in the
copy-on-write mode (as a result of cloning a process or a virtual
machine). In this state, if any of the virtual superpages receives
a write, the operating system has two options: 1)~allocate a new
superpage and copy all the data from the old superpage to the
newly allocated superpage, and 2)~break the virtual superpage that
received the write and copy only the 4KB page that was
modified. While the former approach enables low TLB misses, it
results in significant redundancy in memory. On the other hand,
while the latter approach reduces memory redundancy, it sacrifices
the TLB miss reduction benefits of using a super page.

Extending overlays to superpages will allow the operating systems
to track small modifications to a superpage using an overlay. As a
result, our approach can potentially get both the TLB miss
reduction benefits of having a superpage mapping and the memory
redundancy benefits by using the overlays to track modifications.

\subsection{Using Overlays to Store and Operate on Metadata}

In the page overlays mechanism described in
Chapter~\ref{chap:page-overlays}, we used overlays to track a
newer version of data for each virtual page. Alternatively, each
overlay can be used to store metadata for the corresponding
virtual page instead of a newer version of the data. Since the
hardware is aware of overlays, it can provide an efficient
abstraction to the software for maintaining metadata for various
applications, e.g., memory error checking, security. We believe
using overlays to maintain metadata can also enable new and
efficient computation models, e.g., update-triggered computation.

\subsection{Efficiently Performing Reduction and Shift Operations in DRAM}

In Chapters~\ref{chap:rowclone} and \ref{chap:buddy}, we described
mechanisms to perform bulk copy, initialization, and bitwise
logical operations completely inside DRAM. These set of operations
will enable the memory controller to perform some primitive level
of bitwise computation completely inside DRAM. However, these
mechanisms lack support for two operations that are required by
many applications: 1)~data reduction, and 2)~bit shifting.

First, RowClone and Buddy operate at a row-buffer granularity. As
a result, to perform any kind of reduction within the row buffer
(e.g., bit counting, accumulation), the data must be read into the
processor. Providing support for such operations in DRAM will
further reduce the amount of bandwidth consumed by many queries.
While GS-DRAM (Chapter~\ref{chap:gsdram}) enables a simple form
of reduction (i.e., selection), further research is required to
enable support for more complex reduction operations.

Second, many applications, such as encryption algorithms in
cryptography, heavily rely on bit shifting operations. Enabling
support for bit shifting in DRAM can greatly improve the
performance of such algorithms. However, there are two main
challenges in enabling support for bit shifting in DRAM. First, we
need to design a low-cost DRAM substrate that can support moving
data between multiple bitlines. Given the rigid design of modern
DRAM chips, this can be difficult and tricky, especially when we
need support for multiple shifts. Second, the physical address
space is heavily interleaved on the DRAM
hierarchy~\cite{data-retention,parbor}). As a result, bits that
are adjacent in the physical address space may not map to adjacent
bitlines in DRAM, adding complexity to the data mapping mechanism.

\subsection{Designing Efficient ECC Mechanisms for Buddy/GS-DRAM}

Most server memory modules use error correction codes (ECC) to
protect their data. While RowClone works with such ECC schemes
without any changes, Buddy (Chapter~\ref{chap:buddy}) and GS-DRAM
(Chapter~\ref{chap:gsdram}) can break existing ECC
mechanisms. Designing low-cost ECC mechanisms that work with Buddy
and GS-DRAM will be critical to their adoption. While
Section~\ref{sec:gsdram-extensions} already describes a simple
way of extending GS-DRAM to support ECC, such a mechanism will
work only with a simple SECDED ECC mechanism. Further research is
required to design mechanisms that will 1)~enable GS-DRAM with
stronger ECC schemes and 2)~provide low-cost ECC support for
Buddy.

\subsection{Extending In-Memory Computation to Non-Volatile
  Memories}

Recently, many new non-volatile memory technologies (e.g., phase
change memory~\cite{pcm1,pcm2,pcm3,pcm4}, STT
MRAM~\cite{stt1,stt2,stt3,stt4}) have emerged as a scalable alternative
to DRAM. In fact, Intel has recently announced a real
product~\cite{3dcrosspoint} based on a non-volatile memory
technology. These new technologies are expected to have better
scalability properties than DRAM. In future systems, while we
expect DRAM to still play some role, bulk of the data may actually
be stored in non-volatile memory. This raises a natural question:
can we extend our in-memory data movement and computation
techniques to the new non-volatile memory technologies? Answering
this question requires a thorough understanding of these new
technologies (i.e., how they store data at the lowest level, what
is the architecture used to package them, etc.).

\subsection{Extending DBI to Other Caches and Metadata}

In Chapter~\ref{chap:dbi}, we described the Dirty-Block Index
(DBI), which reorganizes the way dirty blocks are tracked (i.e.,
the dirty bits are stored) to enable the cache to efficiently
respond to queries related to dirty blocks. In addition to the
dirty bit, the on-chip cache stores several other pieces of
metadata, e.g., valid bit, coherence state, and ECC, for each
cache block. In existing caches, all this information is stored in
the tag entry of the corresponding block. As a result, a query for
any metadata requires a full tag store lookup. Similar to the
dirty bits, these other pieces of metadata can potentially be
organized to suit the queries for each one, rather than organizing
them the way the tag stores do it today.

\section{Summary}

In this dissertation, we highlighted the inefficiency problem that
results from the different granularities at which different memory
resources (e.g., caches, DRAM) are managed and accessed. We
presented techniques that bridge this granularity mismatch for
several important memory operations: a new virtual memory
framework that enables memory capacity management at sub-page
granularity (Page Overlays), techniques to use DRAM to do more
than just store data (RowClone, Buddy RAM, and Gather-Scatter
DRAM), and a simple hardware structure for more efficient
management of dirty blocks (Dirty-Block Index). As we discussed in
Section~\ref{sec:future}, these works open up many avenues for new
research that can result in techniques to enable even higher
efficiency.

\chapter*{Other Works of the Author}
\label{chap:other-works}
\addcontentsline{toc}{chapter}{Other Works of the Author}

During the course of my Ph.D., I had the opportunity to
collaborate many of my fellow graduate students. These projects
were not only helpful in keeping my morale up, especially during
the initial years of my Ph.D., but also helped me in learning
about DRAM (an important aspect of this dissertation). In this
chapter, I would like to acknowledge these projects and also my
early works on caching, which kick started my Ph.D.

My interest in DRAM was triggered my work on subarray-level
parallelism (SALP)~\cite{salp} in collaboration with Yoongu
Kim. Since then, I have contributed to a number of projects on
low-latency DRAM architectures with Donghyuk Lee (Tiered-Latency
DRAM~\cite{tl-dram} and Adaptive-Latency DRAM~\cite{al-dram}), and
Hasan Hassan (Charge Cache~\cite{chargecache}). These works focus
on improving DRAM performance by either increasing parallelism or
lowering latency through low-cost modifications to the DRAM
interface and/or microarchitecture.

In collaboration with Gennady Pekhimenko, I have worked on
designing techniques to support data compression in a modern
memory hierarchy. Two contributions have resulted from this work:
1)~Base-Delta Immediate Compression~\cite{bdi}, an effective data
compression algorithm for on-chip caches, and 2)~Linearly
Compressed Pages~\cite{lcp}, a low-cost framework for storing
compressed data in main memory.

In collaboration with Lavanya Subramanian, I have worked on
techniques to quantify and mitigate slowdown in a multi-core
system running multi-programmed workloads. This line of work
started with MISE~\cite{mise}, a mechanism to estimate slowdown
induced by contention for memory bandwidth. Later, we extended
this with Application Slowdown Model~\cite{asm}, mechanism that
accounts for contention for on-chip cache capacity. Finally, we
propose the Blacklisting Memory Scheduler~\cite{bliss,bliss-tpds},
a simple memory scheduling algorithm to achieve high performance
and fairness with low complexity.

Finally, in the early years of my Ph.D., I have worked on
techniques to improve on-chip cache utilization using 1)~the
Evicted-Address Filter~\cite{eaf}, an improved cache insertion
policy to address pollution and thrashing, and 2)~ICP~\cite{icp},
a mechanism that better integrates caching policy for prefetched
blocks. We have released \texttt{memsim}~\cite{memsim}, the
simulator that we developed as part of the EAF work. The simulator
code can be found in github
(\texttt{github.com/CMU-SAFARI/memsim}). \texttt{memsim} has since
been used for many works~\cite{icp,bdi,dbi,lcp}.

\bibliographystyle{plain}
\bibliography{references}

\begin{thebibliography}{100}

\bibitem{belly}
Belly card engineering.
\newblock \url{https://tech.bellycard.com/}.

\bibitem{bitmapist}
{bitmapist: Powerful realtime analytics with Redis 2.6's bitmaps and Python}.
\newblock \url{https://github.com/Doist/bitmapist}.

\bibitem{bochs}
{Bochs} {IA-32} emulator project.
\newblock \url{http://bochs.sourceforge.net/}.

\bibitem{stl}
C++ containers libary, std::set.
\newblock \url{http://en.cppreference.com/w/cpp/container/set}.

\bibitem{fastbit}
{FastBit: An Efficient Compressed Bitmap Index Technology}.
\newblock \url{https://sdm.lbl.gov/fastbit/}.

\bibitem{hbm}
{High Bandwidth Memory DRAM}.
\newblock \url{http://www.jedec.org/standards-documents/docs/jesd235}.

\bibitem{x86-addressing}
{Intel 64 and IA-32 Architectures Software Developer's Manual}.
\newblock \url{http://download.intel.com/design/processor/manuals/253665.pdf},
  Vol. 1, Chap. 3.7.

\bibitem{memcached}
{Memcached: A high performance, distributed memory object caching system}.
\newblock \url{http://memcached.org}.

\bibitem{mysql}
{MySQL: An open source database}.
\newblock \url{http://www.mysql.com}.

\bibitem{redis}
Redis - bitmaps.
\newblock \url{http://redis.io/topics/data-types-intro#bitmaps}.

\bibitem{rlite}
{rlite: A Self-contained, Serverless, Zero-configuration, Transactional
  Redis-compatible Database Engine}.
\newblock \url{https://github.com/seppo0010/rlite}.

\bibitem{spool}
Spool.
\newblock \url{http://www.getspool.com/}.

\bibitem{simics}
{Wind River Simics full system simulation}.
\newblock \url{http://www.windriver.com/products/simics/}.

\bibitem{non-von-machine}
{The NON-VON Database Machine: An Overview}.
\newblock \url{http://hdl.handle.net/10022/AC:P:11530.}, 1981.

\bibitem{alpha-21264}
{Alpha 21264 Microprocessor Hardware Reference Manual}.
\newblock
  \url{http://h18000.www1.hp.com/cpq-alphaserver/technology/literature/21264hrm.pdf},
  1999.

\bibitem{memsim}
{Memsim}.
\newblock \url{http://safari.ece.cmu.edu/tools.html}, 2012.

\bibitem{htap}
{Hybrid Transaction/Analytical Processing Will Foster Opportunities for
  Dramatic Business Innovation}.
\newblock \texttt{https://www.gartner.com/doc/2657815/hybrid-}
  \texttt{transactionanalytical-processing-foster} \texttt{-opportunities},
  2014.

\bibitem{mc-dimm-cal}
J.~H. Ahn, J.~Leverich, R.~Schreiber, and N.~P. Jouppi.
\newblock {Multicore DIMM: An Energy Efficient Memory Module with Independently
  Controlled DRAMs}.
\newblock {\em IEEE CAL}, January 2009.

\bibitem{mc-dimm}
Jung~Ho Ahn, Norman~P. Jouppi, Christos Kozyrakis, Jacob Leverich, and
  Robert~S. Schreiber.
\newblock {Future Scaling of Processor-memory Interfaces}.
\newblock In {\em SC}, 2009.

\bibitem{pim-graph}
Junwhan Ahn, Sungpack Hong, Sungjoo Yoo, Onur Mutlu, and Kiyoung Choi.
\newblock {A Scalable Processing-in-memory Accelerator for Parallel Graph
  Processing}.
\newblock In {\em Proceedings of the 42nd Annual International Symposium on
  Computer Architecture}, ISCA '15, pages 105--117, New York, NY, USA, 2015.
  ACM.

\bibitem{pim-enabled-insts}
Junwhan Ahn, Sungjoo Yoo, Onur Mutlu, and Kiyoung Choi.
\newblock {PIM-enabled Instructions: A Low-overhead, Locality-aware
  Processing-in-memory Architecture}.
\newblock In {\em Proceedings of the 42nd Annual International Symposium on
  Computer Architecture}, ISCA '15, pages 336--348, New York, NY, USA, 2015.
  ACM.

\bibitem{mikamonu2}
A.~Akerib, O.~AGAM, E.~Ehrman, and M.~Meyassed.
\newblock {Using storage cells to perform computation}, December 2014.
\newblock US Patent 8,908,465.

\bibitem{mikamonu}
Avidan Akerib and Eli Ehrman.
\newblock {In-memory Computational Device, Patent No. 20150146491}, May 2015.

\bibitem{data-reorg-3d-stack}
Berkin Akin, Franz Franchetti, and James~C. Hoe.
\newblock {Data Reorganization in Memory Using 3D-stacked DRAM}.
\newblock In {\em Proceedings of the 42nd Annual International Symposium on
  Computer Architecture}, ISCA '15, pages 131--143, New York, NY, USA, 2015.
  ACM.

\bibitem{utm}
C.~Scott Ananian, Krste Asanovic, Bradley~C. Kuszmaul, Charles~E. Leiserson,
  and Sean Lie.
\newblock {Unbounded Transactional Memory}.
\newblock In {\em Proceedings of the 11th International Symposium on
  High-Performance Computer Architecture}, HPCA '05, pages 316--327,
  Washington, DC, USA, 2005. IEEE Computer Society.

\bibitem{sms}
Rachata Ausavarungnirun, Kevin Kai-Wei Chang, Lavanya Subramanian, Gabriel~H.
  Loh, and Onur Mutlu.
\newblock {Staged Memory Scheduling: Achieving High Performance and Scalability
  in Heterogeneous Systems}.
\newblock In {\em Proceedings of the 39th Annual International Symposium on
  Computer Architecture}, ISCA '12, pages 416--427, Washington, DC, USA, 2012.
  IEEE Computer Society.

\bibitem{medic}
Rachata Ausavarungnirun, Saugata Ghose, Onur Kayiran, Gavriel~H. Loh, Chita~R.
  Das, Mahmut~T. Kandemir, and Onur Mutlu.
\newblock {Exploiting Inter-Warp Heterogeneity to Improve GPGPU Performance}.
\newblock In {\em Proceedings of the 2015 International Conference on Parallel
  Architecture and Compilation (PACT)}, PACT '15, 2015.

\bibitem{jafar}
Oreoluwatomiwa~O. Babarinsa and Stratos Idreos.
\newblock {JAFAR: Near-Data Processing for Databases}.
\newblock In {\em Proceedings of the 2015 ACM SIGMOD International Conference
  on Management of Data}, SIGMOD '15, pages 2069--2070, New York, NY, USA,
  2015. ACM.

\bibitem{refresh-nt}
Seungjae Baek, Sangyeun Cho, and Rami Melhem.
\newblock {Refresh Now and Then}.
\newblock {\em IEEE Trans. Comput.}, 63(12):3114--3126, December 2014.

\bibitem{stride-prefetching}
Jean-Loup Baer and Tien-Fu Chen.
\newblock {Effective Hardware-Based Data Prefetching for High-Performance
  Processors}.
\newblock {\em IEEE TC}, 1995.

\bibitem{direct-segment}
Arkaprava Basu, Jayneel Gandhi, Jichuan Chang, Mark~D. Hill, and Michael~M.
  Swift.
\newblock {Efficient Virtual Memory for Big Memory Servers}.
\newblock In {\em Proceedings of the 40th Annual International Symposium on
  Computer Architecture}, ISCA '13, pages 237--248, New York, NY, USA, 2013.
  ACM.

\bibitem{beladyopt}
L.~A. Belady.
\newblock A study of replacement algorithms for a virtual-storage computer.
\newblock {\em IBM Systems Journal}, 5(2):78 --101, 1966.

\bibitem{plfs}
John Bent, Garth Gibson, Gary Grider, Ben McClelland, Paul Nowoczynski, James
  Nunez, Milo Polte, and Meghan Wingate.
\newblock {PLFS: A Checkpoint Filesystem for Parallel Applications}.
\newblock In {\em Proceedings of the Conference on High Performance Computing
  Networking, Storage and Analysis}, SC '09, pages 21:1--21:12, New York, NY,
  USA, 2009. ACM.

\bibitem{gem5}
Nathan Binkert, Bradford Beckmann, Gabriel Black, Steven~K. Reinhardt, Ali
  Saidi, Arkaprava Basu, Joel Hestness, Derek~R. Hower, Tushar Krishna, Somayeh
  Sardashti, Rathijit Sen, Korey Sewell, Muhammad Shoaib, Nilay Vaish, Mark~D.
  Hill, and David~A. Wood.
\newblock The gem5 simulator.
\newblock {\em SIGARCH Comput. Archit. News}, 39(2):1--7, August 2011.

\bibitem{tlb-consistency-2}
D.~L. Black, R.~F. Rashid, D.~B. Golub, and C.~R. Hill.
\newblock {Translation Lookaside Buffer Consistency: A Software Approach}.
\newblock In {\em Proceedings of the Third International Conference on
  Architectural Support for Programming Languages and Operating Systems},
  ASPLOS III, pages 113--122, New York, NY, USA, 1989. ACM.

\bibitem{linux-security}
D.~P. Bovet and M.~Cesati.
\newblock {\em Understanding the Linux Kernel}, page 388.
\newblock O'Reilly Media, 2005.

\bibitem{summingbird}
Oscar Boykin, Sam Ritchie, Ian O'Connell, and Jimmy Lin.
\newblock Summingbird: A framework for integrating batch and online mapreduce
  computations.
\newblock {\em Proc. VLDB Endow.}, 7(13):1441--1451, August 2014.

\bibitem{sg-dimm}
T.~M. Brewer.
\newblock {Instruction Set Innovations for the Convey HC-1 Computer}.
\newblock {\em IEEE Micro}, 30(2):70--79, March 2010.

\bibitem{impulse}
J.~Carter, W.~Hsieh, L.~Stoller, M.~Swanson, L.~Zhang, E.~Brunvand, A.~Davis,
  C.-C. Kuo, R.~Kuramkote, M.~Parker, L.~Schaelicke, and T.~Tateyama.
\newblock {Impulse: Building a Smarter Memory Controller}.
\newblock In {\em HPCA}, 1999.

\bibitem{virtual-caches}
Michel Cekleov and Michel Dubois.
\newblock Virtual-address caches part 1: Problems and solutions in
  uniprocessors.
\newblock {\em IEEE Micro}, 17(5):64--71, September 1997.

\bibitem{bmide}
Chee-Yong Chan and Yannis~E. Ioannidis.
\newblock Bitmap index design and evaluation.
\newblock In {\em Proceedings of the 1998 ACM SIGMOD International Conference
  on Management of Data}, SIGMOD '98, pages 355--366, New York, NY, USA, 1998.
  ACM.

\bibitem{drampower-paper}
K.~Chandrasekar, B.~Akesson, and K.~Goossens.
\newblock {Improved Power Modeling of DDR SDRAMs}.
\newblock In {\em DSD}, 2011.

\bibitem{drampower}
K.~Chandrasekar, C.~Weis, Y.~Li, S.~Goossens, M.~Jung, O.~Naji, B.~Akesson,
  N.~Wehn, and K.~Goossens.
\newblock {DRAMPower: Open-source DRAM Power \& Energy Estimation Tool}.
\newblock \url{http://www.drampower.info}.

\bibitem{os-speculation-2}
Fay Chang and Garth~A. Gibson.
\newblock {Automatic I/O Hint Generation Through Speculative Execution}.
\newblock In {\em Proceedings of the Third Symposium on Operating Systems
  Design and Implementation}, OSDI '99, pages 1--14, Berkeley, CA, USA, 1999.
  USENIX Association.

\bibitem{fly-dram}
Kevin~K. Chang, Abhijith Kashyap, Hasan Hassan, Saugata Ghose, Kevin Hsieh,
  Donghyuk Lee, Tianshi Li, Gennady Pekhimenko, Samira Khan, and Onur Mutlu.
\newblock {Understanding Latency Variation in Modern DRAM Chips: Experimental
  Characterization, Analysis, and Optimization}.
\newblock In {\em Sigmetrics}, 2016.

\bibitem{lisa}
Kevin~K Chang, Prashant~J Nair, Donghyuk Lee, Saugata Ghose, Moinuddin~K
  Qureshi, and Onur Mutlu.
\newblock {Low-Cost Inter-Linked Subarrays (LISA): Enabling Fast Inter-Subarray
  Data Movement in DRAM}.
\newblock In {\em HPCA}, 2016.

\bibitem{dsarp}
Kevin Kai-Wei Chang, Donghyuk Lee, Zeshan Chishti, Alaa~R Alameldeen, Chris
  Wilkerson, Yoongu Kim, and Onur Mutlu.
\newblock {Improving DRAM performance by parallelizing refreshes with
  accesses}.
\newblock In {\em 2014 IEEE 20th International Symposium on High Performance
  Computer Architecture (HPCA)}, pages 356--367. IEEE, 2014.

\bibitem{hicamp}
David Cheriton, Amin Firoozshahian, Alex Solomatnikov, John~P. Stevenson, and
  Omid Azizi.
\newblock {HICAMP: Architectural Support for Efficient Concurrency-safe Shared
  Structured Data Access}.
\newblock In {\em ASPLOS}, pages 287--300, New York, NY, USA, 2012. ACM.

\bibitem{mcr-dram}
Jungwhan Choi, Wongyu Shin, Jaemin Jang, Jinwoong Suh, Yongkee Kwon, Youngsuk
  Moon, and Lee-Sup Kim.
\newblock {Multiple Clone Row DRAM: A Low Latency and Area Optimized DRAM}.
\newblock In {\em Proceedings of the 42nd Annual International Symposium on
  Computer Architecture}, ISCA '15, pages 223--234, New York, NY, USA, 2015.
  ACM.

\bibitem{shredding}
Jim Chow, Ben Pfaff, Tal Garfinkel, and Mendel Rosenblum.
\newblock {Shredding Your Garbage: Reducing Data Lifetime Through Secure
  Deallocation}.
\newblock In {\em Proceedings of the 14th Conference on USENIX Security
  Symposium - Volume 14}, SSYM'05, pages 22--22, Berkeley, CA, USA, 2005.
  USENIX Association.

\bibitem{stt1}
K.~C. Chun, H.~Zhao, J.~D. Harms, T.~H. Kim, J.~P. Wang, and C.~H. Kim.
\newblock {A Scaling Roadmap and Performance Evaluation of In-Plane and
  Perpendicular MTJ Based STT-MRAMs for High-Density Cache Memory}.
\newblock {\em IEEE Journal of Solid-State Circuits}, 48(2):598--610, Feb 2013.

\bibitem{hardware-bug}
Kypros Constantinides, Onur Mutlu, and Todd Austin.
\newblock {Online Design Bug Detection: RTL Analysis, Flexible Mechanisms, and
  Evaluation}.
\newblock In {\em Proceedings of the 41st Annual IEEE/ACM International
  Symposium on Microarchitecture}, MICRO 41, pages 282--293, Washington, DC,
  USA, 2008. IEEE Computer Society.

\bibitem{self-test}
Kypros Constantinides, Onur Mutlu, Todd Austin, and Valeria Bertacco.
\newblock {Software-Based Online Detection of Hardware Defects Mechanisms,
  Architectural Support, and Evaluation}.
\newblock In {\em Proceedings of the 40th Annual IEEE/ACM International
  Symposium on Microarchitecture}, MICRO 40, pages 97--108, Washington, DC,
  USA, 2007. IEEE Computer Society.

\bibitem{intel-htm}
Intel Corporation.
\newblock {\em {Intel Architecture Instruction Set Extensions Programming
  Reference}}, chapter 8. Intel Transactional Synchronization Extensions.
\newblock Sep 2012.

\bibitem{spec2006}
Standard Performance~Evaluation Corporation.
\newblock {SPEC CPU2006 Benchmark Suite}.
\newblock \url{www.spec.org/cpu2006}, 2006.

\bibitem{butterfly}
W.~Dally and B.~Towles.
\newblock {\em {Principles and Practices of Interconnection Networks}}.
\newblock Morgan Kaufmann Publishers Inc., San Francisco, CA, USA, 2003.

\bibitem{bill-dally}
William Dally.
\newblock {GPU Computing to Exascale and Beyond}.
\newblock
  \url{http://www.nvidia.com/content/PDF/sc_2010/theater/Dally_SC10.pdf}.

\bibitem{ufspm}
Timothy~A. Davis and Yifan Hu.
\newblock {The University of Florida Sparse Matrix Collection}.
\newblock {\em ACM Trans. Math. Softw.}, 38(1):1:1--1:25, December 2011.

\bibitem{ai}
Demiz Denir, Islam AbdelRahman, Liang He, and Yingsheng Gao.
\newblock {Audience Insights query engine: In-memory integer store for social
  analytics}.
\newblock \texttt{https://code.facebook.com/posts/382299771946304/
  audience-insights-query-engine-in-memory-
  integer-store-for-social-analytics-/}.

\bibitem{virtual-memory}
Peter~J. Denning.
\newblock {Virtual Memory}.
\newblock {\em ACM Comput. Surv.}, 2(3):153--189, September 1970.

\bibitem{diva}
Jeff Draper, Jacqueline Chame, Mary Hall, Craig Steele, Tim Barrett, Jeff
  LaCoss, John Granacki, Jaewook Shin, Chun Chen, Chang~Woo Kang, Ihn Kim, and
  Gokhan Daglikoca.
\newblock {The Architecture of the DIVA Processing-in-memory Chip}.
\newblock In {\em Proceedings of the 16th International Conference on
  Supercomputing}, ICS '02, pages 14--25, New York, NY, USA, 2002. ACM.

\bibitem{sunshine}
Alan~M. Dunn, Michael~Z. Lee, Suman Jana, Sangman Kim, Mark Silberstein,
  Yuanzhong Xu, Vitaly Shmatikov, and Emmett Witchel.
\newblock Eternal sunshine of the spotless machine: Protecting privacy with
  ephemeral channels.
\newblock In {\em Proceedings of the 10th USENIX Conference on Operating
  Systems Design and Implementation}, OSDI'12, pages 61--75, Berkeley, CA, USA,
  2012. USENIX Association.

\bibitem{hpc-survey}
Ifeanyi~P. Egwutuoha, David Levy, Bran Selic, and Shiping Chen.
\newblock {A Survey of Fault Tolerance Mechanisms and Checkpoint/Restart
  Implementations for High Performance Computing Systems}.
\newblock {\em J. Supercomput.}, 65(3):1302--1326, September 2013.

\bibitem{yale-sm}
S.~C. Eisenstat, M.~C. Gursky, M.~H. Schultz, and A.~H. Sherman.
\newblock {Yale Sparse Matrix Package I: The Symmetric Codes}.
\newblock {\em IJNME}, 18(8), 1982.

\bibitem{rmc}
Magnus Ekman and Per Stenstrom.
\newblock {A Robust Main-Memory Compression Scheme}.
\newblock In {\em ISCA}, pages 74--85, Washington, DC, USA, 2005. IEEE Computer
  Society.

\bibitem{cram}
Duncan Elliott, Michael Stumm, W.~Martin Snelgrove, Christian Cojocaru, and
  Robert McKenzie.
\newblock {Computational RAM: Implementing Processors in Memory}.
\newblock {\em IEEE Des. Test}, 16(1):32--41, January 1999.

\bibitem{weighted-speedup}
Stijn Eyerman and Lieven Eeckhout.
\newblock {System-Level Performance Metrics for Multiprogram Workloads}.
\newblock {\em IEEE Micro}, 28(3):42--53, May 2008.

\bibitem{nda-arch}
A.~Farmahini-Farahani, J.~H. Ahn, K.~Morrow, and N.~S. Kim.
\newblock {NDA: Near-DRAM acceleration architecture leveraging commodity DRAM
  devices and standard memory modules}.
\newblock In {\em 2015 IEEE 21st International Symposium on High Performance
  Computer Architecture (HPCA)}, pages 283--295, Feb 2015.

\bibitem{nda}
A.~Farmahini-Farahani, Jung~Ho Ahn, K.~Morrow, and Nam~Sung Kim.
\newblock {NDA: Near-DRAM acceleration architecture leveraging commodity DRAM
  devices and standard memory modules}.
\newblock In {\em IEEE 21st International Symposium on High Performance
  Computer Architecture (HPCA), 2015}, pages 283--295, Feb 2015.

\bibitem{programming-flexram}
Basilio~B. Fraguela, Jose Renau, Paul Feautrier, David Padua, and Josep
  Torrellas.
\newblock {Programming the FlexRAM Parallel Intelligent Memory System}.
\newblock In {\em Proceedings of the Ninth ACM SIGPLAN Symposium on Principles
  and Practice of Parallel Programming}, PPoPP '03, pages 49--60, New York, NY,
  USA, 2003. ACM.

\bibitem{xor-schemes}
J.~M. Frailong, W.~Jalby, and J.~Lenfant.
\newblock {XOR-Schemes: A Flexible Data Organization in Parallel Memories}.
\newblock In {\em ICPP}, 1985.

\bibitem{stride-pref-fp}
J.~W.~C. Fu and J.~H. Patel.
\newblock {Data Prefetching in Multiprocessor Vector Cache Memories}.
\newblock In {\em ISCA}, 1991.

\bibitem{stride-prefetching-2}
J.~W.~C. Fu, J.~H. Patel, and B.~L. Janssens.
\newblock {Stride Directed Prefetching in Scalar Processors}.
\newblock In {\em MICRO}, 1992.

\bibitem{pim-analytics}
Mingyu Gao, Grant Ayers, and Christos Kozyrakis.
\newblock {Practical Near-Data Processing for In-Memory Analytics Frameworks}.
\newblock In {\em Proceedings of the 2015 International Conference on Parallel
  Architecture and Compilation (PACT)}, PACT '15, pages 113--124, Washington,
  DC, USA, 2015. IEEE Computer Society.

\bibitem{hrl}
Mingyu Gao and Christos Kozyrakis.
\newblock {HRL: Efficient and Flexible Reconfigurable Logic for Near-Data
  Processing}.
\newblock In {\em HPCA}, 2016.

\bibitem{critical-scheduler}
Saugata Ghose, Hyodong Lee, and Jos{\'e}~F. Mart\'{\i}nez.
\newblock Improving memory scheduling via processor-side load criticality
  information.
\newblock In {\em Proceedings of the 40th Annual International Symposium on
  Computer Architecture}, ISCA '13, pages 84--95, New York, NY, USA, 2013. ACM.

\bibitem{smart-refresh}
Mrinmoy Ghosh and Hsien-Hsin~S. Lee.
\newblock {Smart Refresh: An Enhanced Memory Controller Design for Reducing
  Energy in Conventional and 3D Die-Stacked DRAMs}.
\newblock In {\em Proceedings of the 40th Annual IEEE/ACM International
  Symposium on Microarchitecture}, MICRO 40, pages 134--145, Washington, DC,
  USA, 2007. IEEE Computer Society.

\bibitem{pim-terasys}
Maya Gokhale, Bill Holmes, and Ken Iobst.
\newblock {Processing in Memory: The Terasys Massively Parallel PIM Array}.
\newblock {\em Computer}, 28(4):23--31, April 1995.

\bibitem{fork}
Mel Gorman.
\newblock {\em {Understanding the Linux Virtual Memory Manager}}, chapter~4,
  page~57.
\newblock Prentice Hall, 2004.

\bibitem{hyrise}
Martin Grund, Jens Kr\"{u}ger, Hasso Plattner, Alexander Zeier, Philippe
  Cudre-Mauroux, and Samuel Madden.
\newblock Hyrise: A main memory hybrid storage engine.
\newblock {\em Proc. VLDB Endow.}, 4(2):105--116, November 2010.

\bibitem{red-black-tree}
Leo~J. Guibas and Robert Sedgewick.
\newblock {A Dichromatic Framework for Balanced Trees}.
\newblock In {\em Proceedings of the 19th Annual Symposium on Foundations of
  Computer Science}, SFCS '78, pages 8--21, Washington, DC, USA, 1978. IEEE
  Computer Society.

\bibitem{msa3d}
Q.~Guo et~al.
\newblock {3D-Stacked Memory-Side Acceleration: Accelerator and System Design}.
\newblock In {\em WoNDP}, 2013.

\bibitem{de}
Diwaker Gupta, Sangmin Lee, Michael Vrable, Stefan Savage, Alex~C. Snoeren,
  George Varghese, Geoffrey~M. Voelker, and Amin Vahdat.
\newblock {Difference Engine: Harnessing Memory Redundancy in Virtual
  Machines}.
\newblock In {\em OSDI}, pages 309--322, Berkeley, CA, USA, 2008. USENIX
  Association.

\bibitem{vlf2}
H-Store.
\newblock {Anti-Caching}.
\newblock
  \url{http://hstore.cs.brown.edu/documentation/deployment/anti-caching/},
  2015.

\bibitem{coldboot}
J.~Alex Halderman, Seth~D. Schoen, Nadia Heninger, William Clarkson, William
  Paul, Joseph~A. Calandrino, Ariel~J. Feldman, Jacob Appelbaum, and Edward~W.
  Felten.
\newblock {Lest We Remember: Cold-boot Attacks on Encryption Keys}.
\newblock {\em Commun. ACM}, 52(5):91--98, May 2009.

\bibitem{xor2}
Jong-Wook Han, Choon-Sik Park, Dae-Hyun Ryu, and Eun-Soo Kim.
\newblock {Optical image encryption based on XOR operations}.
\newblock {\em Optical Engineering}, 38(1):47--54, 1999.

\bibitem{disclosure}
K.~Harrison and Shouhuai Xu.
\newblock {Protecting Cryptographic Keys from Memory Disclosure Attacks}.
\newblock In {\em 37th Annual IEEE/IFIP International Conference on Dependable
  Systems and Networks}, pages 137--143, June 2007.

\bibitem{emc}
Milad Hashemi, Khubaib, Eiman Ebrahimi, Onur Mutlu, and Yale~N. Patt.
\newblock {Accelerating Dependent Cache Misses with an Enhanced Memory
  Controller}.
\newblock In {\em ISCA}, 2016.

\bibitem{chargecache}
Hasan Hassan, Gennady Pekhimenko, Nandita Vijaykumar, Vivek Seshadri, Donghyuk
  Lee, Oguz Ergin, and Onur Mutlu.
\newblock {ChargeCache: Reducing DRAM Latency by Exploiting Row Access
  Locality}.
\newblock In {\em HPCA}, 2016.

\bibitem{ndp-architecture}
Syed~Minhaj Hassan, Sudhakar Yalamanchili, and Saibal Mukhopadhyay.
\newblock {Near Data Processing: Impact and Optimization of 3D Memory System
  Architecture on the Uncore}.
\newblock In {\em Proceedings of the 2015 International Symposium on Memory
  Systems}, MEMSYS '15, pages 11--21, New York, NY, USA, 2015. ACM.

\bibitem{gpu-gather-scatter}
B.~He, N.~K. Govindaraju, Q.~Luo, and B.~Smith.
\newblock {Efficient Gather and Scatter Operations on Graphics Processors}.
\newblock In {\em SC}, 2007.

\bibitem{tm}
Maurice Herlihy and J.~Eliot~B. Moss.
\newblock {Transactional Memory: Architectural Support for Lock-free Data
  Structures}.
\newblock In {\em Proceedings of the 20th Annual International Symposium on
  Computer Architecture}, ISCA '93, pages 289--300, New York, NY, USA, 1993.
  ACM.

\bibitem{cache-dram}
Hideto Hidaka, Yoshio Matsuda, Mikio Asakura, and Kazuyasu Fujishima.
\newblock {The Cache DRAM Architecture: A DRAM with an On-Chip Cache Memory}.
\newblock {\em IEEE Micro}, 10(2):14--25, March 1990.

\bibitem{spare-row-mapping}
M.~Horiguchi and K.~Itoh.
\newblock {\em {Nanoscale Memory Repair}}.
\newblock Springer, 2011.

\bibitem{tom}
Kevin Hsieh, Eiman Ebrahimi, Gwangsun Kim, Niladrish Chatterjee, Mike O'Conner,
  Nandita Vijaykumar, Onur Mutlu, and Stephen~W. Keckler.
\newblock {Transparent Offloading and Mapping (TOM): Enabling
  Programmer-Transparent Near-Data Processing in GPU Systems}.
\newblock In {\em ISCA}, 2016.

\bibitem{s390}
{IBM Corporation}.
\newblock {Enterprise Systems Architecture/390 Principles of Operation}, 2001.

\bibitem{intel-amt}
Intel.
\newblock {Architecture Guide: Intel Active Management Technology}.
\newblock
  \url{https://software.intel.com/en-us/articles/architecture-guide-intel-active-management-technology/}.

\bibitem{intel-avx}
Intel.
\newblock {Intel Instruction Set Architecture Extensions}.
\newblock \url{https://software.intel.com/en-us/intel-isa-extensions}.

\bibitem{3dcrosspoint}
Intel.
\newblock {Non-Volatile Memory}.
\newblock
  \url{http://www.intel.com/content/www/us/en/architecture-and-technology/non-volatile-memory.html}.

\bibitem{bcsr-format}
Intel.
\newblock {Sparse Matrix Storage Formats, Intel Math Kernel Library}.
\newblock \url{https://software.intel.com/en-us/node/471374}.

\bibitem{x86-ermsb}
{Intel}.
\newblock {\em {Intel 64 and IA-32 Architectures Optimization Reference
  Manual}}.
\newblock April 2012.

\bibitem{intel-dma}
{Intel}.
\newblock {\em {Intel 64 and IA-32 Architectures Software Developer's Manual}},
  volume~3A, chapter~11, page~12.
\newblock April 2012.

\bibitem{somc}
Engin Ipek, Onur Mutlu, Jos{\'e}~F. Mart\'{\i}nez, and Rich Caruana.
\newblock {Self-Optimizing Memory Controllers: A Reinforcement Learning
  Approach}.
\newblock In {\em Proceedings of the 35th Annual International Symposium on
  Computer Architecture}, ISCA '08, pages 39--50, Washington, DC, USA, 2008.
  IEEE Computer Society.

\bibitem{eskimo}
Ciji Isen and Lizy John.
\newblock {ESKIMO: Energy Savings Using Semantic Knowledge of Inconsequential
  Memory Occupancy for DRAM Subsystem}.
\newblock In {\em Proceedings of the 42nd Annual IEEE/ACM International
  Symposium on Microarchitecture}, MICRO 42, pages 337--346, New York, NY, USA,
  2009. ACM.

\bibitem{cpu-gpu}
Thomas~B. Jablin, Prakash Prabhu, James~A. Jablin, Nick~P. Johnson, Stephen~R.
  Beard, and David~I. August.
\newblock {Automatic CPU-GPU Communication Management and Optimization}.
\newblock In {\em Proceedings of the 32nd ACM SIGPLAN Conference on Programming
  Language Design and Implementation}, PLDI '11, pages 142--151, New York, NY,
  USA, 2011. ACM.

\bibitem{rrip}
Aamer Jaleel, Kevin~B. Theobald, Simon~C. Steely, Jr., and Joel Emer.
\newblock {High Performance Cache Replacement Using Re-reference Interval
  Prediction (RRIP)}.
\newblock In {\em Proceedings of the 37th Annual International Symposium on
  Computer Architecture}, ISCA '10, pages 60--71, New York, NY, USA, 2010. ACM.

\bibitem{hmc}
J.~Jeddeloh and B.~Keeth.
\newblock {Hybrid Memory Cube: New DRAM architecture increases density and
  performance}.
\newblock In {\em VLSIT}, pages 87--88, June 2012.

\bibitem{ddr3-1600}
JEDEC.
\newblock {DDR3 SDRAM Standard, JESD79-3D}.
\newblock \url{http://www.jedec.org/sites/default/files/docs/JESD79-3D.pdf},
  2009.

\bibitem{spd}
{JEDEC}.
\newblock {Standard No. 21-C. Annex K: Serial Presence Detect (SPD) for DDR3
  SDRAM Modules}, 2011.

\bibitem{ddr3}
JEDEC.
\newblock {DDR3 SDRAM, JESD79-3F}, 2012.

\bibitem{ddr4}
JEDEC.
\newblock {DDR4 SDRAM Standard}.
\newblock \url{http://www.jedec.org/standards-documents/docs/jesd79-4a}, 2013.

\bibitem{pcm-compression}
Lei Jiang, Youtao Zhang, and Jun Yang.
\newblock {Mitigating Write Disturbance in Super-Dense Phase Change Memories}.
\newblock In {\em DSN}, pages 216--227, Washington, DC, USA, 2014. IEEE
  Computer Society.

\bibitem{bulk-copy-initialize}
Xiaowei Jiang, Yan Solihin, Li~Zhao, and Ravishankar Iyer.
\newblock {Architecture Support for Improving Bulk Memory Copying and
  Initialization Performance}.
\newblock In {\em PACT}, pages 169--180, Washington, DC, USA, 2009. IEEE
  Computer Society.

\bibitem{clams}
Adwait Jog, Onur Kayiran, Ashutosh Pattnaik, Mahmut~T. Kandemir, Onur Mutlu,
  Ravi Iyer, and Chita~R. Das.
\newblock {Exploiting Core Criticality for Enhanced Performance in GPUs}.
\newblock In {\em SIGMETRICS}, 2016.

\bibitem{skipcache}
Raghavendra K, Tripti~S. Warrier, and Madhu Mutyam.
\newblock {SkipCache: Miss-rate Aware Cache Management}.
\newblock In {\em Proceedings of the 21st International Conference on Parallel
  Architectures and Compilation Techniques}, PACT '12, pages 481--482, New
  York, NY, USA, 2012. ACM.

\bibitem{hstore}
R.~Kallman, H.~Kimura, J.~Natkins, A.~Pavlo, A.~Rasin, S.~Zdonik, E.~P.~C.
  Jones, S.~Madden, M.~Stonebraker, Y.~Zhang, J.~Hugg, and D.~J. Abadi.
\newblock {H-Store: a High-Performance, Distributed Main Memory Transaction
  Processing System}.
\newblock {\em VLDB}, 2008.

\bibitem{flexram}
Yi~Kang, Wei Huang, Seung-Moon Yoo, D.~Keen, Zhenzhou Ge, V.~Lam, P.~Pattnaik,
  and J.~Torrellas.
\newblock {FlexRAM: Toward an Advanced Intelligent Memory System}.
\newblock In {\em Proceedings of the 1999 IEEE International Conference on
  Computer Design}, ICCD '99, pages 192--, Washington, DC, USA, 1999. IEEE
  Computer Society.

\bibitem{dram-cd}
Brent Keeth, R.~Jacob Baker, Brian Johnson, and Feng Lin.
\newblock {\em DRAM Circuit Design: Fundamental and High-Speed Topics}.
\newblock Wiley-IEEE Press, 2nd edition, 2007.

\bibitem{efficacy-error-techniques}
Samira Khan, Donghyuk Lee, Yoongu Kim, Alaa~R. Alameldeen, Chris Wilkerson, and
  Onur Mutlu.
\newblock {The Efficacy of Error Mitigation Techniques for DRAM Retention
  Failures: A Comparative Experimental Study}.
\newblock In {\em The 2014 ACM International Conference on Measurement and
  Modeling of Computer Systems}, SIGMETRICS '14, pages 519--532, New York, NY,
  USA, 2014. ACM.

\bibitem{parbor}
Samira~M. Khan, Donghyuk Lee, and Onur Mutlu.
\newblock {PARBOR: An Efficient System-Level Technique to Detect Data-Dependent
  Failures in DRAM}.
\newblock In {\em DSN}, 2016.

\bibitem{stt2}
A~V Khvalkovskiy, D~Apalkov, S~Watts, R~Chepulskii, R~S Beach, A~Ong, X~Tang,
  A~Driskill-Smith, W~H Butler, P~B Visscher, D~Lottis, E~Chen, V~Nikitin, and
  M~Krounbi.
\newblock {Basic principles of STT-MRAM cell operation in memory arrays}.
\newblock {\em Journal of Physics D: Applied Physics}, 46(7):074001, 2013.

\bibitem{dynamic-memory-design}
Joohee Kim and Marios~C. Papaefthymiou.
\newblock {Block-based Multiperiod Dynamic Memory Design for Low Data-retention
  Power}.
\newblock {\em IEEE Trans. Very Large Scale Integr. Syst.}, 11(6):1006--1018,
  December 2003.

\bibitem{atlas}
Yoongu Kim, Dongsu Han, O.~Mutlu, and M.~Harchol-Balter.
\newblock {ATLAS: A scalable and high-performance scheduling algorithm for
  multiple memory controllers}.
\newblock In {\em IEEE 16th International Symposium on High Performance
  Computer Architecture}, pages 1--12, Jan 2010.

\bibitem{tcm}
Yoongu Kim, Michael Papamichael, Onur Mutlu, and Mor Harchol-Balter.
\newblock {Thread Cluster Memory Scheduling: Exploiting Differences in Memory
  Access Behavior}.
\newblock In {\em Proceedings of the 2010 43rd Annual IEEE/ACM International
  Symposium on Microarchitecture}, MICRO '43, pages 65--76, Washington, DC,
  USA, 2010. IEEE Computer Society.

\bibitem{salp}
Yoongu Kim, Vivek Seshadri, Donghyuk Lee, Jamie Liu, and Onur Mutlu.
\newblock {A Case for Exploiting Subarray-level Parallelism (SALP) in DRAM}.
\newblock In {\em Proceedings of the 39th Annual International Symposium on
  Computer Architecture}, ISCA '12, pages 368--379, Washington, DC, USA, 2012.
  IEEE Computer Society.

\bibitem{btt-knuth}
D.~E. Knuth.
\newblock {The Art of Computer Programming. Fascicle 1: Bitwise Tricks \&
  Techniques; Binary Decision Diagrams}, 2009.

\bibitem{execube}
Peter~M. Kogge.
\newblock {EXECUBE: A New Architecture for Scaleable MPPs}.
\newblock In {\em ICPP}, pages 77--84, Washington, DC, USA, 1994. IEEE Computer
  Society.

\bibitem{stt4}
E.~Kultursay, M.~Kandemir, A.~Sivasubramaniam, and O.~Mutlu.
\newblock {Evaluating STT-RAM as an energy-efficient main memory alternative}.
\newblock In {\em Performance Analysis of Systems and Software (ISPASS), 2013
  IEEE International Symposium on}, pages 256--267, April 2013.

\bibitem{amoeba}
Snehasish Kumar, Hongzhou Zhao, Arrvindh Shriraman, Eric Matthews, Sandhya
  Dwarkadas, and Lesley Shannon.
\newblock {Amoeba-Cache: Adaptive Blocks for Eliminating Waste in the Memory
  Hierarchy}.
\newblock In {\em Proceedings of the 2012 45th Annual IEEE/ACM International
  Symposium on Microarchitecture}, MICRO-45, pages 376--388, Washington, DC,
  USA, 2012. IEEE Computer Society.

\bibitem{snowflock}
Horacio~Andr{\'e}s Lagar-Cavilla, Joseph~Andrew Whitney, Adin~Matthew Scannell,
  Philip Patchin, Stephen~M. Rumble, Eyal de~Lara, Michael Brudno, and Mahadev
  Satyanarayanan.
\newblock {SnowFlock: Rapid Virtual Machine Cloning for Cloud Computing}.
\newblock In {\em Proceedings of the 4th ACM European Conference on Computer
  Systems}, EuroSys '09, pages 1--12, New York, NY, USA, 2009. ACM.

\bibitem{power6-prefetcher}
H.~Q. Le, W.~J. Starke, J.~S. Fields, F.~P. O'Connell, D.~Q. Nguyen, B.~J.
  Ronchetti, W.~M. Sauer, E.~M. Schwarz, and M.~T. Vaden.
\newblock {IBM POWER6 Microarchitecture}.
\newblock {\em IBM J. Res. Dev.}, 51(6):639--662, November 2007.

\bibitem{pcm1}
Benjamin~C. Lee, Engin Ipek, Onur Mutlu, and Doug Burger.
\newblock {Architecting Phase Change Memory As a Scalable DRAM Alternative}.
\newblock In {\em Proceedings of the 36th Annual International Symposium on
  Computer Architecture}, ISCA '09, pages 2--13, New York, NY, USA, 2009. ACM.

\bibitem{dram-aware-wb}
C.~J. Lee, V.~Narasiman, E.~Ebrahimi, O.~Mutlu, and Y.~N. Patt.
\newblock {DRAM-aware last-level cache writeback: Reducing write-caused
  interference in memory systems}.
\newblock Technical Report TR-HPS-2010-2, University of Texas at Austin, 2010.

\bibitem{prefetch-dram}
Chang~Joo Lee, Onur Mutlu, Veynu Narasiman, and Yale~N. Patt.
\newblock {Prefetch-Aware DRAM Controllers}.
\newblock In {\em Proceedings of the 41st Annual IEEE/ACM International
  Symposium on Microarchitecture}, MICRO 41, pages 200--209, Washington, DC,
  USA, 2008. IEEE Computer Society.

\bibitem{smla}
Donghyuk Lee, Saugata Ghose, Gennady Pekhimenko, Samira Khan, and Onur Mutlu.
\newblock {Simultaneous Multi-Layer Access: Improving 3D-Stacked Memory
  Bandwidth at Low Cost}.
\newblock {\em ACM Trans. Archit. Code Optim.}, 12(4):63:1--63:29, January
  2016.

\bibitem{al-dram}
Donghyuk Lee, Yoongu Kim, Gennady Pekhimenko, Samira~Manabi Khan, Vivek
  Seshadri, Kevin Kai-Wei Chang, and Onur Mutlu.
\newblock {{Adaptive-latency DRAM: Optimizing DRAM timing for the
  common-case}}.
\newblock In {\em {HPCA}}, pages 489--501. {IEEE}, 2015.

\bibitem{tl-dram}
Donghyuk Lee, Yoongu Kim, Vivek Seshadri, Jamie Liu, Lavanya Subramanian, and
  Onur Mutlu.
\newblock {Tiered-latency DRAM: A Low Latency and Low Cost DRAM Architecture}.
\newblock In {\em Proceedings of the 2013 IEEE 19th International Symposium on
  High Performance Computer Architecture (HPCA)}, HPCA '13, pages 615--626,
  Washington, DC, USA, 2013. IEEE Computer Society.

\bibitem{ddma}
Donghyuk Lee, Lavanya Subramanian, Rachata Ausavarungnirun, Jongmoo Choi, and
  Onur Mutlu.
\newblock {Decoupled Direct Memory Access: Isolating CPU and IO Traffic by
  Leveraging a Dual-Data-Port DRAM}.
\newblock In {\em Proceedings of the 2015 International Conference on Parallel
  Architecture and Compilation (PACT)}, PACT '15, pages 174--187, Washington,
  DC, USA, 2015. IEEE Computer Society.

\bibitem{dna-algo1}
Heng Li and Richard Durbin.
\newblock {Fast and accurate long-read alignment with Burrows--Wheeler
  transform}.
\newblock {\em Bioinformatics}, 26(5):589--595, 2010.

\bibitem{mcpat}
Sheng Li, Jung~Ho Ahn, Richard~D. Strong, Jay~B. Brockman, Dean~M. Tullsen, and
  Norman~P. Jouppi.
\newblock {McPAT: An Integrated Power, Area, and Timing Modeling Framework for
  Multicore and Manycore Architectures}.
\newblock In {\em Proceedings of the 42nd Annual IEEE/ACM International
  Symposium on Microarchitecture}, MICRO 42, pages 469--480, New York, NY, USA,
  2009. ACM.

\bibitem{value-locality}
Mikko~H. Lipasti, Christopher~B. Wilkerson, and John~Paul Shen.
\newblock {Value Locality and Load Value Prediction}.
\newblock In {\em Proceedings of the Seventh International Conference on
  Architectural Support for Programming Languages and Operating Systems},
  ASPLOS VII, pages 138--147, New York, NY, USA, 1996. ACM.

\bibitem{sectored-cache}
J.~S. Liptay.
\newblock {Structural Aspects of the System/360 Model 85: II the Cache}.
\newblock {\em IBM Syst. J.}, 7(1):15--21, March 1968.

\bibitem{data-retention}
Jamie Liu, Ben Jaiyen, Yoongu Kim, Chris Wilkerson, and Onur Mutlu.
\newblock {An Experimental Study of Data Retention Behavior in Modern DRAM
  Devices: Implications for Retention Time Profiling Mechanisms}.
\newblock In {\em Proceedings of the 40th Annual International Symposium on
  Computer Architecture}, ISCA '13, pages 60--71, New York, NY, USA, 2013. ACM.

\bibitem{raidr}
Jamie Liu, Ben Jaiyen, Richard Veras, and Onur Mutlu.
\newblock {RAIDR: Retention-Aware Intelligent DRAM Refresh}.
\newblock In {\em Proceedings of the 39th Annual International Symposium on
  Computer Architecture}, ISCA '12, pages 1--12, Washington, DC, USA, 2012.
  IEEE Computer Society.

\bibitem{3d-stacking}
Gabriel~H. Loh.
\newblock {3D-Stacked Memory Architectures for Multi-core Processors}.
\newblock In {\em Proceedings of the 35th Annual International Symposium on
  Computer Architecture}, ISCA '08, pages 453--464, Washington, DC, USA, 2008.
  IEEE Computer Society.

\bibitem{da-subarray}
Shih-Lien Lu, Ying-Chen Lin, and Chia-Lin Yang.
\newblock {Improving DRAM Latency with Dynamic Asymmetric Subarray}.
\newblock In {\em Proceedings of the 48th International Symposium on
  Microarchitecture}, MICRO-48, pages 255--266, New York, NY, USA, 2015. ACM.

\bibitem{hmean}
Kun Luo, J.~Gummaraju, and M.~Franklin.
\newblock {Balancing thoughput and fairness in SMT processors}.
\newblock In {\em Performance Analysis of Systems and Software, 2001. ISPASS.
  2001 IEEE International Symposium on}, pages 164--171, 2001.

\bibitem{enc1}
S.~A. Manavski.
\newblock {CUDA Compatible GPU as an Efficient Hardware Accelerator for AES
  Cryptography}.
\newblock In {\em IEEE International Conference on Signal Processing and
  Communications, \ 2007. ICSPC 2007}, pages 65--68, Nov 2007.

\bibitem{jsn}
Gokhan Memik, Glenn Reinman, and William~H. Mangione-Smith.
\newblock {Just Say No: Benefits of Early Cache Miss Determination}.
\newblock In {\em Proceedings of the 9th International Symposium on
  High-Performance Computer Architecture}, HPCA '03, pages 307--, Washington,
  DC, USA, 2003. IEEE Computer Society.

\bibitem{vlf1}
MemSQL.
\newblock {Datatypes}.
\newblock \url{http://docs.memsql.com/4.0/ref/datatypes/}, 2015.

\bibitem{micron-power}
Micron.
\newblock {DDR3 SDRAM system-power calculator}, 2011.

\bibitem{load-approx}
Joshua~San Miguel, Mario Badr, and Natalie~Enright Jerger.
\newblock {Load Value Approximation}.
\newblock In {\em Proceedings of the 47th Annual IEEE/ACM International
  Symposium on Microarchitecture}, MICRO-47, pages 127--139, Washington, DC,
  USA, 2014. IEEE Computer Society.

\bibitem{gp-simd}
Amir Morad, Leonid Yavits, and Ran Ginosar.
\newblock {GP-SIMD Processing-in-Memory}.
\newblock {\em ACM Trans. Archit. Code Optim.}, 11(4):53:1--53:26, January
  2015.

\bibitem{mpa}
Thomas Moscibroda and Onur Mutlu.
\newblock Memory performance attacks: Denial of memory service in multi-core
  systems.
\newblock In {\em Proceedings of 16th USENIX Security Symposium on USENIX
  Security Symposium}, SS'07, pages 18:1--18:18, Berkeley, CA, USA, 2007.
  USENIX Association.

\bibitem{effra}
Onur Mutlu.
\newblock {\em {Efficient Runahead Execution Processors}}.
\newblock PhD thesis, Austin, TX, USA, 2006.
\newblock AAI3263366.

\bibitem{stfm}
Onur Mutlu and Thomas Moscibroda.
\newblock {Stall-Time Fair Memory Access Scheduling for Chip Multiprocessors}.
\newblock In {\em Proceedings of the 40th Annual IEEE/ACM International
  Symposium on Microarchitecture}, MICRO 40, pages 146--160, Washington, DC,
  USA, 2007. IEEE Computer Society.

\bibitem{parbs}
Onur Mutlu and Thomas Moscibroda.
\newblock {Parallelism-Aware Batch Scheduling: Enhancing Both Performance and
  Fairness of Shared DRAM Systems}.
\newblock In {\em Proceedings of the 35th Annual International Symposium on
  Computer Architecture}, ISCA '08, pages 63--74, Washington, DC, USA, 2008.
  IEEE Computer Society.

\bibitem{runahead}
Onur Mutlu, Jared Stark, Chris Wilkerson, and Yale~N. Patt.
\newblock {Runahead Execution: An Alternative to Very Large Instruction Windows
  for Out-of-Order Processors}.
\newblock In {\em Proceedings of the 9th International Symposium on
  High-Performance Computer Architecture}, HPCA '03, pages 129--, Washington,
  DC, USA, 2003. IEEE Computer Society.

\bibitem{shadow-memory}
Vijay Nagarajan and Rajiv Gupta.
\newblock {Architectural Support for Shadow Memory in Multiprocessors}.
\newblock In {\em VEE}, pages 1--10, New York, NY, USA, 2009. ACM.

\bibitem{ghb}
K.~J. Nesbit and J.~E. Smith.
\newblock {Data Cache Prefetching Using a Global History Buffer}.
\newblock In {\em HPCA}, 2004.

\bibitem{shadow-mem-check}
Nicholas Nethercote and Julian Seward.
\newblock {How to Shadow Every Byte of Memory Used by a Program}.
\newblock In {\em Proceedings of the 3rd International Conference on Virtual
  Execution Environments}, VEE '07, pages 65--74, New York, NY, USA, 2007. ACM.

\bibitem{os-speculation-3}
Edmund~B. Nightingale, Peter~M. Chen, and Jason Flinn.
\newblock {Speculative Execution in a Distributed File System}.
\newblock In {\em Proceedings of the Twentieth ACM Symposium on Operating
  Systems Principles}, SOSP '05, pages 191--205, New York, NY, USA, 2005. ACM.

\bibitem{opt-dram-refresh}
Taku Ohsawa, Koji Kai, and Kazuaki Murakami.
\newblock {Optimizing the DRAM Refresh Count for Merged DRAM/Logic LSIs}.
\newblock In {\em Proceedings of the 1998 International Symposium on Low Power
  Electronics and Design}, ISLPED '98, pages 82--87, New York, NY, USA, 1998.
  ACM.

\bibitem{bmidc}
Elizabeth O'Neil, Patrick O'Neil, and Kesheng Wu.
\newblock {Bitmap Index Design Choices and Their Performance Implications}.
\newblock In {\em Proceedings of the 11th International Database Engineering
  and Applications Symposium}, IDEAS '07, pages 72--84, Washington, DC, USA,
  2007. IEEE Computer Society.

\bibitem{oracle}
{Oracle}.
\newblock {Using Bitmap Indexes in Data Warehouses}.
\newblock \url{https://docs.oracle.com/cd/B28359_01/server.111/b28313/indexe\
  s.htm}.

\bibitem{active-pages}
Mark Oskin, Frederic~T. Chong, and Timothy Sherwood.
\newblock {Active Pages: A Computation Model for Intelligent Memory}.
\newblock In {\em Proceedings of the 25th Annual International Symposium on
  Computer Architecture}, ISCA '98, pages 192--203, Washington, DC, USA, 1998.
  IEEE Computer Society.

\bibitem{stride-stream-buffer}
S.~Palacharla and R.~E. Kessler.
\newblock {Evaluating Stream Buffers As a Secondary Cache Replacement}.
\newblock In {\em ISCA}, 1994.

\bibitem{mesi}
Mark~S. Papamarcos and Janak~H. Patel.
\newblock {A Low-overhead Coherence Solution for Multiprocessors with Private
  Cache Memories}.
\newblock In {\em Proceedings of the 11th Annual International Symposium on
  Computer Architecture}, ISCA '84, pages 348--354, New York, NY, USA, 1984.
  ACM.

\bibitem{iram}
David Patterson, Thomas Anderson, Neal Cardwell, Richard Fromm, Kimberly
  Keeton, Christoforos Kozyrakis, Randi Thomas, and Katherine Yelick.
\newblock {A Case for Intelligent RAM}.
\newblock {\em IEEE Micro}, 17(2):34--44, March 1997.

\bibitem{toggle-compression}
Gennady Pekhimenko, Evgeny Bolotin, Nandita Vijaykumar, Onur Mutlu, Todd~C.
  Mowry, and Stephen~W. Keckler.
\newblock {A Case for Toggle-Aware Compression for GPU Systems}.
\newblock In {\em HPCA}, 2016.

\bibitem{camp}
Gennady Pekhimenko, Tyler Huberty, Rui Cai, Onur Mutlu, Phillip~B. Gibbons,
  Michael~A. Kozuch, and Todd~C. Mowry.
\newblock {Exploiting Compressed Block Size as an Indicator of Future Reuse}.
\newblock In {\em HPCA}, 2015.

\bibitem{lcp}
Gennady Pekhimenko, Vivek Seshadri, Yoongu Kim, Hongyi Xin, Onur Mutlu,
  Phillip~B. Gibbons, Michael~A. Kozuch, and Todd~C. Mowry.
\newblock {Linearly Compressed Pages: A Low-complexity, Low-latency Main Memory
  Compression Framework}.
\newblock In {\em MICRO}, pages 172--184, New York, NY, USA, 2013. ACM.

\bibitem{bdi}
Gennady Pekhimenko, Vivek Seshadri, Onur Mutlu, Phillip~B. Gibbons, Michael~A.
  Kozuch, and Todd~C. Mowry.
\newblock {Base-delta-immediate Compression: Practical Data Compression for
  On-chip Caches}.
\newblock In {\em Proceedings of the 21st International Conference on Parallel
  Architectures and Compilation Techniques}, PACT '12, pages 377--388, New
  York, NY, USA, 2012. ACM.

\bibitem{revive}
Milos Prvulovic, Zheng Zhang, and Josep Torrellas.
\newblock {ReVive: Cost-effective Architectural Support for Rollback Recovery
  in Shared-memory Multiprocessors}.
\newblock In {\em Proceedings of the 29th Annual International Symposium on
  Computer Architecture}, ISCA '02, pages 111--122, Washington, DC, USA, 2002.
  IEEE Computer Society.

\bibitem{avatar}
M.~K. Qureshi, D.~H. Kim, S.~Khan, P.~J. Nair, and O.~Mutlu.
\newblock {AVATAR: A Variable-Retention-Time (VRT) Aware Refresh for DRAM
  Systems}.
\newblock In {\em 2015 45th Annual IEEE/IFIP International Conference on
  Dependable Systems and Networks}, pages 427--437, June 2015.

\bibitem{dip}
Moinuddin~K. Qureshi, Aamer Jaleel, Yale~N. Patt, Simon~C. Steely, and Joel
  Emer.
\newblock {Adaptive Insertion Policies for High Performance Caching}.
\newblock In {\em Proceedings of the 34th Annual International Symposium on
  Computer Architecture}, ISCA '07, pages 381--391, New York, NY, USA, 2007.
  ACM.

\bibitem{pcm2}
Moinuddin~K. Qureshi, Vijayalakshmi Srinivasan, and Jude~A. Rivers.
\newblock {Scalable High Performance Main Memory System Using Phase-change
  Memory Technology}.
\newblock In {\em Proceedings of the 36th Annual International Symposium on
  Computer Architecture}, ISCA '09, pages 24--33, New York, NY, USA, 2009. ACM.

\bibitem{rambus-power}
Rambus.
\newblock {DRAM power model}, 2010.

\bibitem{dna-our-algo}
Kim~R Rasmussen, Jens Stoye, and Eugene~W Myers.
\newblock {Efficient q-gram filters for finding all $\varepsilon$-matches over
  a given length}.
\newblock {\em Journal of Computational Biology}, 13(2):296--308, 2006.

\bibitem{pr-interleaving}
B.~Ramakrishna Rau.
\newblock {Pseudo-randomly Interleaved Memory}.
\newblock In {\em Proceedings of the 18th Annual International Symposium on
  Computer Architecture}, ISCA '91, pages 74--83, New York, NY, USA, 1991. ACM.

\bibitem{mmio}
E.~D. Reilly.
\newblock {Memory-mapped I/O}.
\newblock In {\em Encyclopedia of Computer Science}, page 1152. John Wiley and
  Sons Ltd., Chichester, UK.

\bibitem{frfcfs}
Scott Rixner, William~J. Dally, Ujval~J. Kapasi, Peter Mattson, and John~D.
  Owens.
\newblock Memory access scheduling.
\newblock In {\em Proceedings of the 27th Annual International Symposium on
  Computer Architecture}, ISCA '00, pages 128--138, New York, NY, USA, 2000.
  ACM.

\bibitem{unitd}
B.~Romanescu, A.~R. Lebeck, D.~J. Sorin, and A.~Bracy.
\newblock {UNified Instruction/Translation/Data (UNITD) Coherence: One Protocol
  to Rule Them All}.
\newblock In {\em HPCA}, 2010.

\bibitem{dna-algo2}
Stephen~M Rumble, Phil Lacroute, Adrian~V Dalca, Marc Fiume, Arend Sidow, and
  Michael Brudno.
\newblock {SHRiMP: Accurate mapping of short color-space reads}.
\newblock 2009.

\bibitem{windows-security}
M.~E. Russinovich, D.~A. Solomon, and A.~Ionescu.
\newblock {\em {Windows Internals}}, page 701.
\newblock Microsoft Press, 2009.

\bibitem{fork-exp}
R.~F. Sauers, C.~P. Ruemmler, and P.~S. Weygant.
\newblock {\em {HP-UX 11i Tuning and Performance}}, chapter 8. Memory
  Bottlenecks.
\newblock Prentice Hall, 2004.

\bibitem{eraser}
Stefan Savage, Michael Burrows, Greg Nelson, Patrick Sobalvarro, and Thomas
  Anderson.
\newblock {Eraser: A Dynamic Data Race Detector for Multithreaded Programs}.
\newblock {\em ACM Trans. Comput. Syst.}, 15(4):391--411, November 1997.

\bibitem{value-prediction}
Yiannakis Sazeides and James~E. Smith.
\newblock {The Predictability of Data Values}.
\newblock In {\em Proceedings of the 30th Annual ACM/IEEE International
  Symposium on Microarchitecture}, MICRO 30, pages 248--258, Washington, DC,
  USA, 1997. IEEE Computer Society.

\bibitem{dna-overview}
Sophie Schbath, V{\'e}ronique Martin, Matthias Zytnicki, Julien Fayolle,
  Valentin Loux, and Jean-Fran{\c{c}}ois Gibrat.
\newblock {Mapping reads on a genomic sequence: An algorithmic overview and a
  practical comparative analysis}.
\newblock {\em Journal of Computational Biology}, 19(6):796--813, 2012.

\bibitem{dbi}
Vivek Seshadri, Abhishek Bhowmick, Onur Mutlu, Phillip~B. Gibbons, Michael~A.
  Kozuch, and Todd~C. Mowry.
\newblock {The Dirty-block Index}.
\newblock In {\em Proceeding of the 41st Annual International Symposium on
  Computer Architecuture}, ISCA '14, pages 157--168, Piscataway, NJ, USA, 2014.
  IEEE Press.

\bibitem{buddy-cal}
Vivek Seshadri, Kevin Hsieh, Amirali Boroumand, Donghyuk Lee, Michael~A.
  Kozuch, Onur Mutlu, Phillip~B. Gibbons, and Todd~C. Mowry.
\newblock {Fast Bulk Bitwise AND and OR in DRAM}.
\newblock {\em IEEE Comput. Archit. Lett.}, 14(2):127--131, July 2015.

\bibitem{rowclone}
Vivek Seshadri, Yoongu Kim, Chris Fallin, Donghyuk Lee, Rachata
  Ausavarungnirun, Gennady Pekhimenko, Yixin Luo, Onur Mutlu, Phillip~B.
  Gibbons, Michael~A. Kozuch, and Todd~C. Mowry.
\newblock {RowClone: Fast and Energy-efficient in-DRAM Bulk Data Copy and
  Initialization}.
\newblock In {\em Proceedings of the 46th Annual IEEE/ACM International
  Symposium on Microarchitecture}, MICRO-46, pages 185--197, New York, NY, USA,
  2013. ACM.

\bibitem{gsdram}
Vivek Seshadri, Thomas Mullins, Amirali Boroumand, Onur Mutlu, Phillip~B.
  Gibbons, Michael~A. Kozuch, and Todd~C. Mowry.
\newblock {Gather-scatter DRAM: In-DRAM Address Translation to Improve the
  Spatial Locality of Non-unit Strided Accesses}.
\newblock In {\em Proceedings of the 48th International Symposium on
  Microarchitecture}, MICRO-48, pages 267--280, New York, NY, USA, 2015. ACM.

\bibitem{eaf}
Vivek Seshadri, Onur Mutlu, Michael~A. Kozuch, and Todd~C. Mowry.
\newblock {The Evicted-address Filter: A Unified Mechanism to Address Both
  Cache Pollution and Thrashing}.
\newblock In {\em Proceedings of the 21st International Conference on Parallel
  Architectures and Compilation Techniques}, PACT '12, pages 355--366, New
  York, NY, USA, 2012. ACM.

\bibitem{page-overlays}
Vivek Seshadri, Gennady Pekhimenko, Olatunji Ruwase, Onur Mutlu, Phillip~B.
  Gibbons, Michael~A. Kozuch, Todd~C. Mowry, and Trishul Chilimbi.
\newblock {Page Overlays: An Enhanced Virtual Memory Framework to Enable
  Fine-grained Memory Management}.
\newblock In {\em Proceedings of the 42nd Annual International Symposium on
  Computer Architecture}, ISCA '15, pages 79--91, New York, NY, USA, 2015. ACM.

\bibitem{icp}
Vivek Seshadri, Samihan Yedkar, Hongyi Xin, Onur Mutlu, Phillip~B. Gibbons,
  Michael~A. Kozuch, and Todd~C. Mowry.
\newblock {Mitigating Prefetcher-Caused Pollution Using Informed Caching
  Policies for Prefetched Blocks}.
\newblock {\em ACM Trans. Archit. Code Optim.}, 11(4):51:1--51:22, January
  2015.

\bibitem{dscache}
A.~Seznec.
\newblock {Decoupled Sectored Caches: Conciliating Low Tag Implementation
  Cost}.
\newblock In {\em ISCA}, 1994.

\bibitem{memsql}
N.~Shamgunov.
\newblock {The MemSQL In-Memory Database System}.
\newblock In {\em VLDB}, 2014.

\bibitem{simpoints}
Timothy Sherwood, Erez Perelman, Greg Hamerly, and Brad Calder.
\newblock {Automatically Characterizing Large Scale Program Behavior}.
\newblock In {\em Proceedings of the 10th International Conference on
  Architectural Support for Programming Languages and Operating Systems},
  ASPLOS X, pages 45--57, New York, NY, USA, 2002. ACM.

\bibitem{indra}
Weidong Shi, Hsien-Hsin~S. Lee, Laura `Falk, and Mrinmoy Ghosh.
\newblock {An Integrated Framework for Dependable and Revivable Architectures
  Using Multicore Processors}.
\newblock In {\em ISCA}, pages 102--113, Washington, DC, USA, 2006. IEEE
  Computer Society.

\bibitem{fs-free-space}
A.~Silberschatz, P.~B. Galvin, and G.~Gagne.
\newblock {\em {Operating System Concepts}}, chapter 11. File-System
  Implementation.
\newblock Wiley, 2012.

\bibitem{macos-security}
A.~Singh.
\newblock {\em {Mac OS X Internals: A Systems Approach}}.
\newblock Addison-Wesley Professional, 2006.

\bibitem{multiscalar}
Gurindar~S. Sohi, Scott~E. Breach, and T.~N. Vijaykumar.
\newblock {Multiscalar Processors}.
\newblock In {\em Proceedings of the 22nd Annual International Symposium on
  Computer Architecture}, ISCA '95, pages 414--425, New York, NY, USA, 1995.
  ACM.

\bibitem{charm}
Young~Hoon Son, O.~Seongil, Yuhwan Ro, Jae~W. Lee, and Jung~Ho Ahn.
\newblock {Reducing Memory Access Latency with Asymmetric DRAM Bank
  Organizations}.
\newblock In {\em Proceedings of the 40th Annual International Symposium on
  Computer Architecture}, ISCA '13, pages 380--391, New York, NY, USA, 2013.
  ACM.

\bibitem{fdp}
Santhosh Srinath, Onur Mutlu, Hyesoon Kim, and Yale~N. Patt.
\newblock {Feedback Directed Prefetching: Improving the Performance and
  Bandwidth-Efficiency of Hardware Prefetchers}.
\newblock In {\em Proceedings of the 2007 IEEE 13th International Symposium on
  High Performance Computer Architecture}, HPCA '07, pages 63--74, Washington,
  DC, USA, 2007. IEEE Computer Society.

\bibitem{flashback}
Sudarshan~M. Srinivasan, Srikanth Kandula, Christopher~R. Andrews, and Yuanyuan
  Zhou.
\newblock {Flashback: A Lightweight Extension for Rollback and Deterministic
  Replay for Software Debugging}.
\newblock In {\em Proceedings of the Annual Conference on USENIX Annual
  Technical Conference}, ATEC '04, pages 3--3, Berkeley, CA, USA, 2004. USENIX
  Association.

\bibitem{sheaved-memory}
M.~E. Staknis.
\newblock {Sheaved Memory: Architectural Support for State Saving and
  Restoration in Pages Systems}.
\newblock In {\em Proceedings of the Third International Conference on
  Architectural Support for Programming Languages and Operating Systems},
  ASPLOS III, pages 96--102, New York, NY, USA, 1989. ACM.

\bibitem{tls}
J.~Greggory Steffan, Christopher~B. Colohan, Antonia Zhai, and Todd~C. Mowry.
\newblock {A Scalable Approach to Thread-level Speculation}.
\newblock In {\em Proceedings of the 27th Annual International Symposium on
  Computer Architecture}, ISCA '00, pages 1--12, New York, NY, USA, 2000. ACM.

\bibitem{lim-computer}
Harold~S. Stone.
\newblock {A Logic-in-Memory Computer}.
\newblock {\em IEEE Trans. Comput.}, 19(1):73--78, January 1970.

\bibitem{c-store}
Mike Stonebraker, Daniel~J. Abadi, Adam Batkin, Xuedong Chen, Mitch Cherniack,
  Miguel Ferreira, Edmond Lau, Amerson Lin, Sam Madden, Elizabeth O'Neil, Pat
  O'Neil, Alex Rasin, Nga Tran, and Stan Zdonik.
\newblock {C-store: A Column-oriented DBMS}.
\newblock In {\em Proceedings of the 31st International Conference on Very
  Large Data Bases}, VLDB '05, pages 553--564. VLDB Endowment, 2005.

\bibitem{elastic-refresh}
Jeffrey Stuecheli, Dimitris Kaseridis, Hillery C.Hunter, and Lizy~K. John.
\newblock {Elastic Refresh: Techniques to Mitigate Refresh Penalties in High
  Density Memory}.
\newblock In {\em Proceedings of the 2010 43rd Annual IEEE/ACM International
  Symposium on Microarchitecture}, MICRO '43, pages 375--384, Washington, DC,
  USA, 2010. IEEE Computer Society.

\bibitem{vwq}
Jeffrey Stuecheli, Dimitris Kaseridis, David Daly, Hillery~C. Hunter, and
  Lizy~K. John.
\newblock {The Virtual Write Queue: Coordinating DRAM and Last-level Cache
  Policies}.
\newblock In {\em Proceedings of the 37th Annual International Symposium on
  Computer Architecture}, ISCA '10, pages 72--82, New York, NY, USA, 2010. ACM.

\bibitem{bliss}
L.~Subramanian, D.~Lee, V.~Seshadri, H.~Rastogi, and O.~Mutlu.
\newblock {The Blacklisting Memory Scheduler: Achieving high performance and
  fairness at low cost}.
\newblock In {\em ICCD}, 2014.

\bibitem{bliss-tpds}
L.~Subramanian, D.~Lee, V.~Seshadri, H.~Rastogi, and O.~Mutlu.
\newblock {BLISS: Balancing Performance, Fairness and Complexity in Memory
  Access Schedyuling}.
\newblock {\em IEEE Transactions on Parallel and Distributed Systems}, 2016.

\bibitem{mise}
L.~Subramanian, V.~Seshadri, Yoongu Kim, B.~Jaiyen, and O.~Mutlu.
\newblock {MISE: Providing performance predictability and improving fairness in
  shared main memory systems}.
\newblock In {\em IEEE 19th International Symposium on High Performance
  Computer Architecture}, pages 639--650, Feb 2013.

\bibitem{asm}
Lavanya Subramanian, Vivek Seshadri, Arnab Ghosh, Samira Khan, and Onur Mutlu.
\newblock {The Application Slowdown Model: Quantifying and Controlling the
  Impact of Inter-application Interference at Shared Caches and Main Memory}.
\newblock In {\em Proceedings of the 48th International Symposium on
  Microarchitecture}, MICRO-48, pages 62--75, New York, NY, USA, 2015. ACM.

\bibitem{micropages}
Kshitij Sudan, Niladrish Chatterjee, David Nellans, Manu Awasthi, Rajeev
  Balasubramonian, and Al~Davis.
\newblock {Micro-pages: Increasing DRAM Efficiency with Locality-aware Data
  Placement}.
\newblock In {\em Proceedings of the Fifteenth Edition of ASPLOS on
  Architectural Support for Programming Languages and Operating Systems},
  ASPLOS XV, pages 219--230, New York, NY, USA, 2010. ACM.

\bibitem{data-access-opt-pim}
Zehra Sura, Arpith Jacob, Tong Chen, Bryan Rosenburg, Olivier Sallenave, Carlo
  Bertolli, Samuel Antao, Jose Brunheroto, Yoonho Park, Kevin O'Brien, and Ravi
  Nair.
\newblock Data access optimization in a processing-in-memory system.
\newblock In {\em Proceedings of the 12th ACM International Conference on
  Computing Frontiers}, CF '15, pages 6:1--6:8, New York, NY, USA, 2015. ACM.

\bibitem{moesi}
P.~Sweazey and A.~J. Smith.
\newblock {A Class of Compatible Cache Consistency Protocols and Their Support
  by the IEEE Futurebus}.
\newblock In {\em Proceedings of the 13th Annual International Symposium on
  Computer Architecture}, ISCA '86, pages 414--423, Los Alamitos, CA, USA,
  1986. IEEE Computer Society Press.

\bibitem{tlb-consistency-1}
Patricia~J. Teller.
\newblock {Translation-Lookaside Buffer Consistency}.
\newblock {\em Computer}, 23(6):26--36, June 1990.

\bibitem{rollback-vp}
Bradley Thwaites, Gennady Pekhimenko, Hadi Esmaeilzadeh, Amir Yazdanbakhsh,
  Onur Mutlu, Jongse Park, Girish Mururu, and Todd Mowry.
\newblock {Rollback-free Value Prediction with Approximate Loads}.
\newblock In {\em Proceedings of the 23rd International Conference on Parallel
  Architectures and Compilation}, PACT '14, pages 493--494, New York, NY, USA,
  2014. ACM.

\bibitem{mxt}
R.~B. Tremaine, P.~A. Franaszek, J.~T. Robinson, C.~O. Schulz, T.~B. Smith,
  M.~E. Wazlowski, and P.~M. Bland.
\newblock {IBM Memory Expansion Technology (MXT)}.
\newblock {\em IBM J. Res. Dev.}, 45(2):271--285, March 2001.

\bibitem{xor1}
P.~Tuyls, H.~D.~L. Hollmann, J.~H.~Van Lint, and L.~Tolhuizen.
\newblock {XOR-based Visual Cryptography Schemes}.
\newblock {\em Designs, Codes and Cryptography}, 37(1):169--186.

\bibitem{rethinking-dram}
Aniruddha~N. Udipi, Naveen Muralimanohar, Niladrish Chatterjee, Rajeev
  Balasubramonian, Al~Davis, and Norman~P. Jouppi.
\newblock {Rethinking DRAM Design and Organization for Energy-constrained
  Multi-cores}.
\newblock In {\em Proceedings of the 37th Annual International Symposium on
  Computer Architecture}, ISCA '10, pages 175--186, New York, NY, USA, 2010.
  ACM.

\bibitem{dash}
Hiroyuki Usui, Lavanya Subramanian, Kevin Kai-Wei Chang, and Onur Mutlu.
\newblock {DASH: Deadline-Aware High-Performance Memory Scheduler for
  Heterogeneous Systems with Hardware Accelerators}.
\newblock {\em ACM Trans. Archit. Code Optim.}, 12(4):65:1--65:28, January
  2016.

\bibitem{power-of-2}
Mateo Valero, Tom\'{a}s Lang, and Eduard Ayguad{\'e}.
\newblock Conflict-free access of vectors with power-of-two strides.
\newblock In {\em Proceedings of the 6th International Conference on
  Supercomputing}, ICS '92, pages 149--156, New York, NY, USA, 1992. ACM.

\bibitem{flexitaint}
G.~Venkataramani, I.~Doudalis, D.~Solihin, and M.~Prvulovic.
\newblock {FlexiTaint: A Programmable Accelerator for Dynamic Taint
  Propagation}.
\newblock In {\em HPCA}, 2008.

\bibitem{didi}
Carlos Villavieja, Vasileios Karakostas, Lluis Vilanova, Yoav Etsion, Alex
  Ramirez, Avi Mendelson, Nacho Navarro, Adrian Cristal, and Osman~S. Unsal.
\newblock {DiDi: Mitigating the Performance Impact of TLB Shootdowns Using a
  Shared TLB Directory}.
\newblock In {\em Proceedings of the 2011 International Conference on Parallel
  Architectures and Compilation Techniques}, PACT '11, pages 340--349,
  Washington, DC, USA, 2011. IEEE Computer Society.

\bibitem{esx-server}
Carl~A. Waldspurger.
\newblock {Memory Resource Management in VMware ESX Server}.
\newblock {\em SIGOPS Oper. Syst. Rev.}, 36(SI):181--194, December 2002.

\bibitem{checkpointing}
Yi-Min Wang, Yennun Huang, Kiem-Phong Vo, Pe-Yu Chung, and C.~Kintala.
\newblock {Checkpointing and Its Applications}.
\newblock In {\em Proceedings of the Twenty-Fifth International Symposium on
  Fault-Tolerant Computing}, FTCS '95, pages 22--, Washington, DC, USA, 1995.
  IEEE Computer Society.

\bibitem{threaded-module}
F.A. Ware and C.~Hampel.
\newblock {Improving Power and Data Efficiency with Threaded Memory Modules}.
\newblock In {\em ICCD}, 2006.

\bibitem{hacker-delight}
Henry~S. Warren.
\newblock {\em {Hacker's Delight}}.
\newblock Addison-Wesley Professional, 2nd edition, 2012.

\bibitem{dna-algo3}
David Weese, Anne-Katrin Emde, Tobias Rausch, Andreas D{\"o}ring, and Knut
  Reinert.
\newblock {Razers - fast read mapping with sensitivity control}.
\newblock {\em Genome research}, 19(9):1646--1654, 2009.

\bibitem{os-speculation-1}
Benjamin Wester, Peter~M. Chen, and Jason Flinn.
\newblock {Operating System Support for Application-specific Speculation}.
\newblock In {\em Proceedings of the Sixth Conference on Computer Systems},
  EuroSys '11, pages 229--242, New York, NY, USA, 2011. ACM.

\bibitem{legba}
A.~Wiggins, S.~Winwood, H.~Tuch, and G.~Heiser.
\newblock {Legba: Fast Hardware Support for Fine-Grained Protection}.
\newblock In Amos Omondi and Stanislav Sedukhin, editors, {\em Advances in
  Computer Systems Architecture}, volume 2823 of {\em Lecture Notes in Computer
  Science}, 2003.

\bibitem{mmp}
Emmett Witchel, Josh Cates, and Krste Asanovi\'{c}.
\newblock {Mondrian Memory Protection}.
\newblock In {\em ASPLOS}, pages 304--316, New York, NY, USA, 2002. ACM.

\bibitem{pcm3}
H.~S.~P. Wong, S.~Raoux, S.~Kim, J.~Liang, J.~P. Reifenberg, B.~Rajendran,
  M.~Asheghi, and K.~E. Goodson.
\newblock {Phase Change Memory}.
\newblock {\em Proceedings of the IEEE}, 98(12):2201--2227, Dec 2010.

\bibitem{bicompression}
Kesheng Wu, Ekow~J. Otoo, and Arie Shoshani.
\newblock {Compressing Bitmap Indexes for Faster Search Operations}.
\newblock In {\em Proceedings of the 14th International Conference on
  Scientific and Statistical Database Management}, SSDBM '02, pages 99--108,
  Washington, DC, USA, 2002. IEEE Computer Society.

\bibitem{dna-algo4}
Hongyi Xin, Donghyuk Lee, Farhad Hormozdiari, Samihan Yedkar, Onur Mutlu, and
  Can Alkan.
\newblock {Accelerating read mapping with FastHASH}.
\newblock {\em BMC genomics}, 14(Suppl 1):S13, 2013.

\bibitem{why-nothing-matters}
Xi~Yang, Stephen~M. Blackburn, Daniel Frampton, Jennifer~B. Sartor, and
  Kathryn~S. McKinley.
\newblock {Why Nothing Matters: The Impact of Zeroing}.
\newblock In {\em OOPSLA}, pages 307--324, New York, NY, USA, 2011. ACM.

\bibitem{ecc-fifo}
Doe~Hyun Yoon and Mattan Erez.
\newblock {Flexible Cache Error Protection Using an ECC FIFO}.
\newblock In {\em Proceedings of the Conference on High Performance Computing
  Networking, Storage and Analysis}, SC '09, pages 49:1--49:12, New York, NY,
  USA, 2009. ACM.

\bibitem{agms}
Doe~Hyun Yoon, Min~Kyu Jeong, and Mattan Erez.
\newblock {Adaptive Granularity Memory Systems: A Tradeoff Between Storage
  Efficiency and Throughput}.
\newblock In {\em ISCA}, 2011.

\bibitem{dgms}
Doe~Hyun Yoon, Min~Kyu Jeong, Michael Sullivan, and Mattan Erez.
\newblock {The Dynamic Granularity Memory System}.
\newblock In {\em ISCA}, 2012.

\bibitem{top-pim}
Dongping Zhang, Nuwan Jayasena, Alexander Lyashevsky, Joseph~L. Greathouse,
  Lifan Xu, and Michael Ignatowski.
\newblock {TOP-PIM: Throughput-oriented Programmable Processing in Memory}.
\newblock In {\em Proceedings of the 23rd International Symposium on
  High-performance Parallel and Distributed Computing}, HPDC '14, pages 85--98,
  New York, NY, USA, 2014. ACM.

\bibitem{impulse-journal}
L.~Zhang, Z.~Fang, M.~Parker, B.~K. Mathew, L.~Schaelicke, J.~B. Carter, W.~C.
  Hsieh, and S.~A. McKee.
\newblock {The Impulse Memory Controller}.
\newblock {\em IEEE TC}, November 2001.

\bibitem{half-dram}
Tao Zhang, Ke~Chen, Cong Xu, Guangyu Sun, Tao Wang, and Yuan Xie.
\newblock {Half-DRAM: A High-bandwidth and Low-power DRAM Architecture from the
  Rethinking of Fine-grained Activation}.
\newblock In {\em Proceeding of the 41st Annual International Symposium on
  Computer Architecuture}, ISCA '14, pages 349--360, Piscataway, NJ, USA, 2014.
  IEEE Press.

\bibitem{firm}
Jishen Zhao, Onur Mutlu, and Yuan Xie.
\newblock {FIRM: Fair and High-Performance Memory Control for Persistent Memory
  Systems}.
\newblock In {\em Proceedings of the 47th Annual IEEE/ACM International
  Symposium on Microarchitecture}, MICRO-47, pages 153--165, Washington, DC,
  USA, 2014. IEEE Computer Society.

\bibitem{copy-engine}
Li~Zhao, Laxmi~N. Bhuyan, Ravi Iyer, Srihari Makineni, and Donald Newell.
\newblock {Hardware Support for Accelerating Data Movement in Server Platform}.
\newblock {\em IEEE Trans. Comput.}, 56(6):740--753, June 2007.

\bibitem{ems}
Qin Zhao, Derek Bruening, and Saman Amarasinghe.
\newblock {Efficient Memory Shadowing for 64-bit Architectures}.
\newblock In {\em ISMM}, pages 93--102, New York, NY, USA, 2010. ACM.

\bibitem{umbra}
Qin Zhao, Derek Bruening, and Saman Amarasinghe.
\newblock {Umbra: Efficient and Scalable Memory Shadowing}.
\newblock In {\em Proceedings of the 8th Annual IEEE/ACM International
  Symposium on Code Generation and Optimization}, CGO '10, pages 22--31, New
  York, NY, USA, 2010. ACM.

\bibitem{stt3}
W.S. Zhao, Y.~Zhang, T.~Devolder, J.O. Klein, D.~Ravelosona, C.~Chappert, and
  P.~Mazoyer.
\newblock {Failure and reliability analysis of STT-MRAM}.
\newblock {\em Microelectronics Reliability}, 52:1848 -- 1852, 2012.

\bibitem{mini-rank}
Hongzhong Zheng, Jiang Lin, Zhao Zhang, Eugene Gorbatov, Howard David, and
  Zhichun Zhu.
\newblock {Mini-rank: Adaptive DRAM Architecture for Improving Memory Power
  Efficiency}.
\newblock In {\em MICRO}, 2008.

\bibitem{iwatcher}
Pin Zhou, Feng Qin, Wei Liu, Yuanyuan Zhou, and Josep Torrellas.
\newblock {iWatcher: Efficient Architectural Support for Software Debugging}.
\newblock In {\em Proceedings of the 31st Annual International Symposium on
  Computer Architecture}, ISCA '04, pages 224--, Washington, DC, USA, 2004.
  IEEE Computer Society.

\bibitem{pcm4}
Ping Zhou, Bo~Zhao, Jun Yang, and Youtao Zhang.
\newblock {A Durable and Energy Efficient Main Memory Using Phase Change Memory
  Technology}.
\newblock In {\em Proceedings of the 36th Annual International Symposium on
  Computer Architecture}, ISCA '09, pages 14--23, New York, NY, USA, 2009. ACM.

\bibitem{spmm-mul-lim}
Q.~Zhu, T.~Graf, H.~E. Sumbul, L.~Pileggi, and F.~Franchetti.
\newblock {Accelerating sparse matrix-matrix multiplication with 3D-stacked
  logic-in-memory hardware}.
\newblock In {\em High Performance Extreme Computing Conference (HPEC), 2013
  IEEE}, pages 1--6, Sept 2013.

\bibitem{frfcfs-patent}
W.~K. Zuravleff and T.~Robinson.
\newblock {Controller for a synchronous {DRAM} that maximizes throughput by
  allowing memory requests and commands to be issued out of order}.
\newblock Patent 5630096, 1997.

\end{thebibliography}

\end{document}